\documentstyle[preprint,aps,eqsecnum]{revtex}
\begin{document}
\tightenlines
\draft

\newcommand{\beq}{\begin{equation}}
\newcommand{\eeq}{\end{equation}}
\newcommand{\bea}{\begin{eqnarray}}
\newcommand{\eea}{\end{eqnarray}}
\newcommand{\cir}{{\buildrel \circ \over =}}

\title{The Rest-Frame Instant Form of Metric Gravity.}

\author{Luca Lusanna}

\address{
Sezione INFN di Firenze\\
L.go E.Fermi 2 (Arcetri)\\
50125 Firenze, Italy\\
E-mail LUSANNA@FI.INFN.IT}

\maketitle
\begin{abstract}

In a special class of globally hyperbolic, topologically trivial, 
asymptotically flat at spatial infinity spacetimes selected by the requirement 
of absence of supertranslations (compatible with Christodoulou-Klainermann
spacetimes) it is possible to define the {\it rest-frame instant form} of
ADM canonical gravity by using Dirac's strategy of adding ten extra variables 
at spatial infinity and ten extra first class constraints implying the gauge 
nature of these variables. The final canonical Hamiltonian is the weak ADM 
energy and a discussion of the Hamiltonian gauge transformations generated by 
the eight first class ADM constraints is given. When there is matter and the 
Newton constant is switched off, one recovers the description of the matter
on the Wigner hyperplanes of the rest-frame instant form of dynamics in 
Minkowski spacetime.

\vskip 1truecm
\noindent \today
\vskip 1truecm

\end{abstract}

\newpage

\vfill\eject

\section
{Introduction}

Our description of the four  (gravitational, electromagnetic, weak, strong,
with or without supersymmetry) interactions is based on action principles 
which, due to manifest Lorentz invariance, to local gauge invariance (minimal 
coupling) and/or diffeomorphism invariances make use of singular Lagrangians. 
This implies the use of Dirac-Bergmann theory of constraints
\cite{dirac,ber,lu1,haux}
for their Hamiltonian formulation. An open problem is the identification of 
the physical degrees of freedom hidden behind manifest gauge invariance 
and/or general covariance (Dirac observables). When this can be done in a 
global way,  classical physics can be reformulated only in terms of 
a canonical basis of Dirac observables and then quantized. This alternative 
to the traditional sequence {\it first quantize then reduce} \footnote{Based
on BRST observables invariant under infinitesimal gauge transformations modulo 
problems like the Gribov ambiguity.} is still unexplored. In particular with 
it one would avoid to quantize any timelike degree of freedom. See 
Refs.\cite{india,re} for the status of these topics.

As a consequence of these researches\cite{india}, in special relativity it is 
now possible to describe isolated systems (particles, strings, field 
configurations) in a way which implements the separation of the relativistic 
canonical center of mass from the relative degrees of freedom and allows to
make the canonical reduction to Dirac observables in a Wigner-covariant way.
Since it is known that the reduction to a completely fixed gauge breaks 
manifest Lorentz covariance, it turns out that in this approach the breaking 
can be concetrated in the non-covariance of the canonical center of mass (viz. 
that of the Newton-Wigner 3-position operator) independently from the 
system under consideration\footnote{As shown in Refs.\cite{india,re} the 
region of non-covariance of the canonical center of mass identifies an 
intrinsic classical unit of length, the M$\o$ller radius\cite{mol} (ratio of 
the spin to the mass of the isolated system), which is a natural candidate 
for a ultraviolet cutoff in quantization.}. 
A new form of dynamics\cite{dira2}, the
{\it one-time Wigner-covariant rest-frame instant form}\cite{lus1}, 
emerges from these investigations and it is reviewed in Appendix A.
In it each configuration of an isolated system with timelike conserved 
4-momentum is described on the spacelike Wigner hyperplanes orthogonal to the
4-momentum ({\it equal time} Cauchy surfaces), leaves of the foliation 
associated with a 3+1 splitting of Minkowski spacetime identified by the 
configuration itself. While the decoupled center of mass can be identified with
a point particle clock for the mathematical time describing the evolution
(the Hamiltonian is the invariant mass of the system configuration), all the 
dynamics is in the relative degrees of freedom (weak form of Mach principle).

Then the principal tool for the canonical reduction of every theory to a 
canonical basis of Dirac observables is the Shanmugadhasan canonical 
transformation\cite{sha} together with the associated multitemporal equations
(see the paper e) in Ref.\cite{lu1}). These canonical transformations allow
to find (in general only local) Darboux bases in which a subset of the new 
momenta carry the same information of the first class constraints, namely 
their vanishing identifies the same presymplectic submanifold
\cite{lich,go} of phase space 
as the original constraints\footnote{If second class constraints are present,
they are replaced by pairs of conjugate variables.}. The variables conjugate
to this subset of momenta (the Abelianized first class constraints) describe
the gauge degrees of freedom of the system, namely they give a parametrization
of the Hamiltonian gauge orbits. To find this set of Abelianized gauge 
variables one has to solve the multitemporal equations, namely the equations
describing the Hamiltonian gauge transformations viewed as functional 
equations in the gauge parameters. Their solution shows how the original 
canonical variables describing the system depends on the Abelianized gauge 
variables. The remaining pairs of conjugate variables in these Darboux bases
form a canonical basis of Dirac observables associated to the given 
Abelianization of the first class constraints: they are in strong involution
(namely they have zero Poisson brackets) with both the Abelianized 
constraints and gauge variables and in weak involution with respect to the 
original constraints. In flat spacetime one has also to take into account the 
stratification of the constraint presymplectic manifold induced by the 
Hamiltonian action of the Poincar\'e group: there are as many strata as 
allowed Poincar\'e orbits for the conserved total 4-momentum of the isolated 
system. Each stratum will have different Darboux-Shanmugadhasan bases adapted 
to the little group of the associated Poincar\'e orbit. In general this 
procedure works only locally, but when the configuration space is non-compact 
there can exist global Shanmugadhasan canonical transformations. For finite 
dimensional systems there are general theorems\cite{kulk} connected with the 
Lie theory of functions groups\cite{kulk1}
 which ensure the existence of local 
Shanmugadhasan canonical transformations. However till now there is no
extension of these theorems to gauge field theories, mainly because now some
of the first class constraints can be interpreted as elliptic equations, so
that, for most (but not all) of the choices of the function space for the 
fields, they can have zero modes (the Gribov ambiguity of Yang-Mills 
theories). Notwithstanding that, the heuristic search of Shanmugadhasan 
canonical trnasformations for gauge theories is the only existing method for
the individuation of possible canonical bases of Dirac observables. See Ref.
\cite{india} for a full account of what is known on these topics.  

After these developments the main question is whether the ADM Hamiltonian 
formulation\cite{adm} of gravity plus matter can be put in a form which 
reproduces this instant form of dynamics for the matter when the Newton 
constant is switched off\footnote{This is the
deparametrization problem of general relativity, only partially solved
in Ref.\cite{isha} by using coordinate gauge conditions.}. 
To try to implement this program, the allowed 
pseudo-Riemannian spacetimes must be restricted to be orientable, globally 
hyperbolic, topologically trivial (with the leaves of each 3+1 splitting 
diffeomorphic to $R^3$, so that they admit global coordinate charts) and 
asymptotically flat at spatial infinity. 

This last requirement implies the existence of the asymptotic ADM Poincar\'e 
charges\cite{reg,reg1}, which should reduce to the ten Poincar\'e generators 
of the isolated system (whose existence is fundamental for particle physics) 
when the Newton constant is switched off. However, it is known that at 
spatial infinity the group of asymptotic symmetries is the infinite 
dimensional SPI group\cite{p3,mcca}. Besides 
an invariant 4-dimensional subgroup of translations it contains an infinite 
number of Abelian supertranslations. This forbids the identification of a 
unique Lorentz subgroup\footnote{Only an abstract Lorentz group appears from 
the quotient of the SPI group with respect to the invariant subgroup of all 
translations and supertranslations.}. The presence of supertranslations is an 
obstruction to the definition of angular momentum in general relativity 
\cite{p10,wald} and there is no idea how to measure this infinite number of 
constants of motion if they are allowed to exist. Therefore, suitable 
boundary conditions at spatial infinity have to be assumed to kill the 
supertranslations. In this way the SPI group is reduced to a well defined 
asymptotic Poincar\'e group. As it will be shown in this paper, a convenient 
set of boundary conditions is obtained by 
assuming that the coordinate atlas of spacetime is restricted in such a way 
that the 4-metric always tends to the Minkowski metric in Cartesian 
coordinates at spatial infinity with the 3-metric on each spacelike 
hypersurface associated with the allowed 3+1 splittings becoming Euclidean
at spatial infinity in a direction independent way\footnote{Since the 
Hamiltonian formulation of gauge theories is still at a heuristic stage of
development, in this paper only coordinate-dependent statements will be used.
Moreover, working in the framework of variational principles, no statement 
can be made regarding null and timelike infinities. Hopefully, at some stage 
also the Hamiltonian theory will be reformulated in a convenient geometric 
coordinate-independent way like in the configuration space approaches using 
only Einstein's equations}. Then, this last property is assumed also for all 
the other Hamiltonian variables like the lapse and shift functions. These 
latter variables are assumed to be the sum of their asymptotic part (growing 
linearly in the 3-coordinates on the leave\cite{reg,reg1}) plus a bulk part 
with the quoted property. The final result of all these requirements is a set 
of boundary conditions compatible with Christodoulou-Klainermann spacetimes
\cite{ckl}.

As a consequence, the allowed 3+1 splittings of spacetime have all the 
spacelike leaves approaching Minkowski spacelike hyperplanes at spatial 
infinity in a direction-independnt way. It will be shown that these 
asymptotic hyperplanes are orthogonal to the weak (viz. the volume form of
the) ADM 4-momentum, for those spacetimes for which it is timelike. Therefore 
these hyperplanes reduce to the Wigner hyperplanes in Minkowski spacetime 
when the Newton constant is switched off. To arrive at these results Dirac's 
strategy\cite{p22,dirac} of adding ten extra degrees of freedom at spatial 
infinity and then to add ten first class constraints so that the new degrees 
of freedom are gauge variables, will be followed.

In this way the {\it rest-frame instant form of metric gravity} may be defined.
The weak ADM energy turns out to play the role of the canonical Hamiltonian 
for the evolution in the scalar mathematical time labelling the leaves of the 
3+1 splitting (consistently with Ref.\cite{fermi}). 
There will be a point near spatial infinity playing the role of
the decoupled canonical center of mass of the universe and which can be 
interpreted as a point-particle clock for the mathematical time. There will 
be three first class constraints implying the vanishing of the weak ADM 
3-momentum: they define the rest frame of the universe. Therefore, at spatial
infinity we have inertial observers whose unit 4-velocity is determined by
the timelike ADM 4-momentum. Modulo 3-rotations, these observers carry an
asymptotic tetrad adapted to the asymptotic spacelike hyperplanes: after a 
conventional choice of the 3-rotation this tetrad defines the {\it dynamical 
fixed stars} (standard of non-rotation). By using Frauendiener's reformulation
\cite{p28} of Sen-Witten equations\cite{sen,p27} for the triads and the 
adapted tetrads on a spacelike hypersurface of this kind, one can determine 
{\it preferred dynamical adapted tetrads} \footnote{They seem to be the 
natural realization of the {\it non-flat preferred observers} of Bergmann
\cite{be}.} in each point of the hypersurface (they are dynamical 
because the solution of Einstein's equations is needed to find them). 
Therefore, these special spacelike hypersurfaces can be named 
Wigner-Sen-Witten (WSW) hypersurfaces: as already said they reduce to the
Wigner hyperplanes of Minkowski spacetime when the Newton constant is
switched off. Since the weak ADM Poincar\'e charges vanish when the spacetime 
is restricted to the Minkowski spacetime with Cartesian coordinates, the
rest-frame instant form of ADM canonical gravity in presence of matter reduces
to the Minkowski rest-frame instant form desciption of the same matter when the
Newton constant is switched off.

Moreover, a M$\o$ller radius may be associated with each timelike universe of 
this kind: it opens the possibility of defining an intrinsic ultraviolet
cutoff in canonical quantization.

An open problem is to find the connection of this construction with the 
Friedrich's description\cite{rich,p15,frauen,bicak} of spacelike infinity for 
spacetimes which admit a conformal completion. The ten Dirac variables at 
spatial infinity should be connected with the arbitrary choice of coordinates 
and tetrads needed to define evolution (1+3 splitting or threading point of 
view) in this approach, in the special case in which the 1+3 splitting is 
also a 3+1 splitting (slicing point of view).

At this point one should study the possibility of implementing a 
Shanmugadhasan canonical transformation\footnote{These canonical 
transformations, belonging to the most general class among those defined in 
Ref.\cite{bk}, are related to the York map of the conformal Lichnerowicz-York
approach\cite{conf,york,yoyo,ciuf}, 
whose existence has been proved in Ref.\cite{yorkmap}, but which 
has never been constructed explicitly.}. A discussion and an interpretation
of the Hamiltonian gauge transformations generated by the eight first class
constraint of ADM canonical gravity will be given. Also the comparison between
the equivalence classes of spacetimes modulo the Hamiltonian gauge 
transformations versus the 4-geometries (equivalence classes of 4-metrics 
modulo spacetime diffeomorphisms) of the configuration space approach will be 
done: they agree only on the solution of Einstein's equations. However, in
this paper we will not study the canonical reduction of metric gravity. In a 
future paper, based on the partial results of Refs.\cite{russo1,russo2,russo3},
the canonical reduction of tetrad gravity to a completely fixed gauge will be 
studied: this will allow to have a formulation i) containing timelike 
observers as fundamental configurational variables; ii) adapted to the 
coupling to fermions; iii) allowing to induce the canonical reduction of 
metric gravity.

Finally there will be some comment on interpretational problems like how to 
identify spacetime points, notwithstanding general covariance destroyes their 
individuality\cite{einst,stachel,rov}, 
a posteriori with the Komar-Bergmann individuating fields
\cite{komar,be} and which is their relation to the non-generally covariant 
Dirac observables in a completely fixed gauge.

In Section II, after notational remarks, there is a review of the ADM canonical
formulation of metric gravity.

In Section III there is a review of the following topics: asymptotic flatness,
asymptotic symmetries, supertranslations and Hamiltonian gauge transformations.

In Section IV Dirac's approach to asymptotically flat (at spatial infinity)
metric gravity is discussed.

In Section V there is a study of the asymptotic ADM Poincar\'e charges of 
metric gravity.

In Section VI it is shown that the requirement of absence of supertranslations
identifies a class of spacetimes of the Christodoulou-Klainermann type.

In Section VII two possible scenarios for metric gravity are presented.

In Section VIII there is the definition of the rest-frame instant form of 
metric gravity.

In Section IX there the interpetation of the Hamiltonian gauge transformations
and their relation with the spacetime diffeomorphisms.

In Section X some interpretational problems regarding the observables of
metric gravity are discussed.

In Section XI there are some comments on the problem of time in metric gravity.

In Section XII there is the determination of the embedding in spacetime and 
of some of the properties of the Wigner-Sen-Witten spacelike hypersurfaces.

Some final remarks and some comments on the quantization of metric gravity in 
a completely fixed gauge are made in the Conclusions.

Appendix A contains a review of the rest-frame instant form of dynamics for 
isolated systems in Minkowski spacetime.

\vfill\eject

\section{ADM Hamiltonian Theory.}

In this Section after some mathematical preliminaries needed to fix the 
notations there will be a review of the ADM Hamiltonian formulation of metric 
gravity.

Let $M^4$ be a torsion-free, orientable, topologically trivial, globally 
hyperbolic, asymptotically flat at spatial infinity pseudo-Riemannian (or 
Lorentzian) 4-manifold with a $C^{\infty}$ atlas of coordinate charts $\{
x^{\mu} \}$. In each chart local coordinate bases for vector fields [$TM^4$] 
and one-forms [$T^{*}M^4$] are $e_{\mu}=\partial_{\mu}$ and
$dx^{\mu}$ respectively.The nondegenerate 4-metric tensor 
${}^4g_{\mu\nu}(x)$ has Lorentzian signature
$\epsilon (+,-,-,-)$ \footnote{Here $\epsilon =\pm 1$ according to particle 
physics and general relativity conventions respectively.
We shall follow the conventions of Refs.\cite{mtw,ciuf,wald} for $\epsilon
=-1$ and those of Ref.\cite{wei} for $\epsilon =+1$.}.
The covariant derivative is denoted ${}^4\nabla_{\mu}$ [or with a semicolon 
``;"].

The world indices will be denoted by Greek letters $\mu ,\nu ,..$ 
($\mu =0,1,2,3$)  while Greek letters inside round brackets
$(\alpha ), (\beta ),..,$ will denote flat Minkowski indices [the flat 
4-metric tensor in Cartesian coordinates is ${}^4\eta_{(\alpha )(\beta )}=
\epsilon (+,-,-,-)$]; analogously, $a, b,..,$ and
$(a), (b),..,$ [a=1,2,3], will denote world and flat 3-space indices
respectively. The summation convention over repeated indices of the same 
kind is used.

Let $M^4$ be foliated  (3+1 splitting or slicing) with spacelike Cauchy 
hypersurfaces $\Sigma_{\tau}$ through the embeddings $i_{\tau}:\Sigma
\rightarrow \Sigma_{\tau} \subset M^4$, $\vec \sigma \mapsto
x^{\mu}=z^{\mu}(\tau ,\vec
\sigma )$, of a 3-manifold $\Sigma$, assumed diffeomorphic to $R^3$, into 
$M^4$ \footnote{$\tau :M^4\rightarrow R$ is a
global, timelike, future-oriented function labelling the leaves of the
foliation; $x^{\mu}$ are local coordinates in a chart of $M^4$; $\vec
\sigma =\{ \sigma^r \}$, r=1,2,3, are coordinates in a global chart of
$\Sigma$, which is diffeomorphic to $R^3$; the notations $\sigma^A=
(\sigma^{\tau}=\tau ;\vec \sigma )$, 
$A=\tau ,r$,  for the coordinates of $M^4$ adapted to the 3+1 splitting
and $z^{\mu}(\sigma )=z^{\mu}(\tau ,\vec \sigma )$ will be used.}.

Let $n^{\mu}(\sigma )$ and $l^{\mu}(\sigma )= N(\sigma )
n^{\mu}(\sigma )$ be the controvariant timelike normal and unit normal
[${}^4g_{\mu\nu}(z(\sigma ))\, l^{\mu}(\sigma )
l^{\nu}(\sigma )=\epsilon$] to $\Sigma_{\tau}$ at the point $z(\sigma
)\in \Sigma_{\tau}$. The positive function $N(\sigma ) > 0$ is the lapse
function: $N(\sigma )d\tau$ measures the proper time interval at
$z(\sigma )\in \Sigma_{\tau}$ between $\Sigma_{\tau}$ and
$\Sigma_{\tau +d\tau}$. The shift functions $N^r(\sigma )$ are defined so 
that $N^r(\sigma )d\tau$ describes the horizontal shift on $\Sigma_{\tau}$ 
such that, if $z^{\mu}(\tau +d\tau ,\vec \sigma +d\vec \sigma )\in 
\Sigma_{\tau +d\tau}$, then $z^{\mu}(\tau
+d\tau ,\vec \sigma +d\vec \sigma )\approx z^{\mu}(\tau ,\vec \sigma )+N(\tau ,
\vec \sigma )d\tau l^{\mu}(\tau ,\vec \sigma )+[d\sigma^r+N^r(\tau ,\vec
\sigma )d\tau ]{{\partial z^{\mu}(\tau ,\vec \sigma )}\over {\partial \sigma^r}
}$; therefore, the so called evolution vector is
${{\partial z^{\mu}(\sigma )}\over {\partial \tau}}=
N(\sigma ) l^{\mu}(\sigma )+N^r(\sigma ) {{\partial z^{\mu}(\tau ,\vec
\sigma )}\over {\partial \sigma^r}}$. The covariant unit normal to 
$\Sigma_{\tau}$ is $l_{\mu}(\sigma )={}^4g
_{\mu\nu}(z(\sigma )) l^{\nu}(\sigma )=N(\sigma ) \partial_{\mu}\tau{|}
_{x=z(\sigma )}$, with $\tau =\tau (\sigma )$ a global timelike 
future-oriented function.

Instead of local coordinates $x^{\mu}$ for $M^4$,
use will be done of coordinates $\sigma^A$ on
$R\times \Sigma \approx M^4$ [$x^{\mu}=z^{\mu}(\sigma )$ with inverse 
$\sigma^A= \sigma^A(x)$], and of the associated $\Sigma_{\tau}$-adapted 
holonomic coordinate basis  $\partial_A={{\partial}\over {\partial 
\sigma^A}}\in T(R\times \Sigma ) \mapsto b^{\mu}_A(\sigma ) \partial_{\mu} 
={{\partial z^{\mu}(\sigma )}\over {\partial \sigma^A}} \partial_{\mu} \in 
TM^4$ for vector fields, and  $dx^{\mu}\in T^{*}M^4 \mapsto d\sigma^A=b^A
_{\mu}(\sigma )dx^{\mu}={{\partial \sigma^A(z)}\over {\partial z^{\mu}}} dx
^{\mu} \in T^{*}(R\times \Sigma )$ for differential one-forms. In flat 
Minkowski spacetime the transformation coefficients $b^A_{\mu}(\sigma )$ and 
$b^{\mu}_A(\sigma )$ become the flat orthonormal tetrads 
$\delta^{(\mu )}_{\mu} z^A_{(\mu )}(\sigma )=
{{\partial \sigma^A(x)}\over {\partial x^{\mu}}}{|}_{x=z(\sigma )}$ and
cotetrads $\delta^{\mu}_{(\mu )} z^{(\mu )}_A(\sigma )={{\partial 
z^{\mu}(\sigma )}\over {\partial \sigma^A}}$ of Ref.\cite{lus1} (see
Appendix A). 

The induced 4-metric and inverse 4-metric
become in the new basis 
\footnote{For the sake of simplicity the
notation ${}^4g_{AB}(\sigma )$ for ${}^4g^{'}_{AB}(\sigma )
={{\partial z^{\mu}(\sigma )}\over {\partial \sigma^A}}
{{\partial z^{\nu}(\sigma )}\over {\partial \sigma^B}}\,
{}^4g_{\mu\nu}(x=z(\sigma ))=b^{\mu}_A(\sigma )b^{\nu}_B(\sigma )\,
{}^4g_{\mu\nu}(z(\sigma ))$ will be used.}

\begin{eqnarray}
{}^4g(x) &=&  {}^4g_{\mu\nu}(x) dx^{\mu} \otimes dx^{\nu} =
{}^4g_{AB}(\sigma )d\sigma^A \otimes d\sigma^B,\nonumber \\
&&{}\nonumber \\ {}^4g_{\mu\nu}&=&b^A_{\mu}\, {}^4g_{AB} b^B_{\nu}
=\nonumber \\ &=&\epsilon \, (N^2-{}^3g_{rs}N^rN^s)\partial_{\mu}\tau
\partial_{\nu}\tau -
\epsilon \, {}^3g_{rs}N^s(\partial_{\mu}\tau \partial_{\nu}\sigma^r+
\partial_{\nu}\tau \partial_{\mu}\sigma^r)-\epsilon \, {}^3g_{rs}
\partial_{\mu}\sigma^r \partial_{\nu}\sigma^s=\nonumber \\
&=& \epsilon \, l_{\mu} l_{\nu} -\epsilon \, {}^3g_{rs} (\partial_{\mu}
\sigma^r +N^r\, \partial_{\mu}\tau ) (\partial_{\nu}\sigma^s+N^s\,
\partial_{\nu}\tau ),\nonumber \\
&&{}\nonumber \\
&\Rightarrow&{}^4g_{AB}=\lbrace {}^4g_{\tau\tau}=
\epsilon (N^2-{}^3g_{rs}N^rN^s); {}^4g_{\tau r}=-
\epsilon \, {}^3g_{rs}N^s; {}^4g_{rs}=-\epsilon \, {}^3g_{rs}\rbrace =
\nonumber \\
&=&\epsilon [ l_Al_B-{}^3g_{rs}(\delta^r_A+N^r\delta^{\tau}_A)(\delta^s_B+
N^s\delta^{\tau}_B)], \nonumber \\
&&{}\nonumber \\
{}^4g^{\mu\nu}&=& b^{\mu}_A {}^4g^{AB} b^{\nu}_B=\nonumber \\
&=&{{\epsilon}\over {N^2}} \partial_{\tau}z^{\mu}\partial_{\tau}z^{\nu}-
{{\epsilon \, N^r}\over {N^2}} (\partial_{\tau}z^{\mu}\partial_rz^{\nu}+
\partial_{\tau}z^{\nu}\partial_rz^{\mu}) -\epsilon ({}^3g^{rs}-{{N^rN^s}\over
{N^2}})\partial_rz^{\mu}\partial_sz^{\nu}=\nonumber \\
&=& \epsilon [\, l^{\mu} l^{\nu} - \, {}^3g^{rs} \partial_rz^{\mu}
\partial_sz^{\nu}],\nonumber \\
&&{}\nonumber \\
&\Rightarrow&{}^4g^{AB}=\lbrace {}^4g^{\tau\tau}=
{{\epsilon}\over {N^2}}; {}^4g^{\tau r}=-{{\epsilon \, N^r}
\over {N^2}}; {}^4g^{rs}=-\epsilon ({}^3g^{rs} - {{N^rN^s}\over {N^2}})
\rbrace =\nonumber \\
&=&\epsilon [l^Al^B -{}^3g^{rs}\delta^A_r\delta^B_s],\nonumber \\
&&{}\nonumber \\
l^A&=&l^{\mu} b^A_{\mu} = N \, {}^4g^{A\tau}={{\epsilon}\over N} (1; -N^r),
\nonumber \\
l_A&=&l_{\mu} b_A^{\mu} = N \partial_A \tau =N \delta^{\tau}_A = (N; \vec 0).
\label{II1}
\end{eqnarray}

Here, the 3-metric $\, {}^3g_{rs}=-\epsilon \, {}^4g_{rs}$, with signature 
(+++),  of $\Sigma_{\tau}$ was introduced.
If ${}^4\gamma^{rs}$ is the inverse
of the spatial part of the 4-metric [${}^4\gamma^{ru}\, 
{}^4g_{us}=\delta^r_s$],
the inverse of the 3-metric is ${}^3g^{rs}=-\epsilon \, {}^4\gamma^{rs}$
[${}^3g^{ru}\, {}^3g_{us}=\delta^r_s$]. ${}^3g_{rs}(\tau ,\vec \sigma )$
are the components of
the ``first fundamental form" of the Riemann 3-manifold $(\Sigma_{\tau},{}^3g)$
and the line element of $M^4$ is

\bea
 ds^2&=&{}^4g_{\mu\nu} dx^{\mu} dx^{\nu}=
\epsilon (N^2-{}^3g_{rs}N^rN^s) (d\tau )^2-2\epsilon
\, {}^3g_{rs}N^s d\tau d\sigma^r -\epsilon \, {}^3g_{rs} 
d\sigma^rd\sigma^s=\nonumber \\
&=&\epsilon \Big[ N^2(d\tau )^2 -{}^3g_{rs}(d\sigma^r+N^rd\tau
)(d\sigma^s+ N^sd\tau )\Big].
\label{II2}
\eea

It must be $\epsilon \, {}^4g_{oo} >0$,
$\epsilon \, {}^4g_{ij} < 0$, $\left| \begin{array}{cc} {}^4g_{ii}& {}^4g_{ij}
\\ {}^4g_{ji}& {}^4g_{jj} \end{array} \right| > 0$, $\epsilon \,
det\, {}^4g_{ij} > 0$.

Defining $g={}^4g=|\, det\, ({}^4g_{\mu\nu})\, |$ and $\gamma ={}^3g =|\,
det\, ({}^3g_{rs})\, |$, the lapse and shift functions assume the following 
form

\begin{eqnarray}
&&N=\sqrt{ {{}^4g\over {{}^3g}} }={1\over { \sqrt{{}^4g^{\tau\tau}} }} =
\sqrt{{g\over {\gamma}}}=
\sqrt{{}^4g_{\tau\tau}-\epsilon \, {}^3g^{rs}\, {}^4g_{\tau r}
{}^4g_{\tau s} },\nonumber \\
&&N^r=-\epsilon \, {}^3g^{rs}\, {}^4g_{\tau s} =
-{{{}^4g^{\tau r}}\over {{}^4g^{\tau\tau}}}
,\quad N_r={}^3g_{rs}N^s=-\epsilon \,\, {}^4g_{rs}N^s=-\epsilon {}^4g_{\tau r}.
\label{II3}
\end{eqnarray}

See Refs.\cite{in,mtw,ish} for the 3+1 decomposition of 4-tensors on $M^4$. The
horizontal projector ${}^3h^{\nu}_{\mu}=\delta^{\nu}_{\mu}-\epsilon \, l_{\mu}
l^{\nu}$ on $\Sigma_{\tau}$ defines the 3-tensor fields on $\Sigma_{\tau}$
starting from the 4-tensor fields on $M^4$.

In the standard (non-Hamiltonian) description of the 3+1 decomposition 
a $\Sigma_{\tau}$-adapted nonholonomic noncoordinate basis [$\bar A=(l;r)$]
is used

\begin{eqnarray}
{\hat b}^{\mu}_{\bar A}(\sigma ) &=&\lbrace {\hat b}^{\mu}_l(\sigma )=
\epsilon l^{\mu}
(\sigma ) =N^{-1}(\sigma ) [b^{\mu}_{\tau}(\sigma )- N^r(\sigma )b^{\mu}_r
(\sigma )];\nonumber \\
&&{\hat b}^{\mu}_r(\sigma ) = b^{\mu}_r(\sigma ) \rbrace ,\nonumber \\
{\hat b}^{\bar A}_{\mu}(\sigma ) &=& \lbrace {\hat b}^l_{\mu}(\sigma ) =
l_{\mu}(\sigma )= N(\sigma )b^{\tau}_{\mu}(\sigma )=N(\sigma )\partial_{\mu}
\tau (z(\sigma ));\nonumber \\
&&{\hat b}^r_{\mu}(\sigma ) = b^r_{\mu}(\sigma )+ N^r(\sigma )
b^{\tau}_{\mu}(\sigma ) \rbrace ,\nonumber \\
&&{}\nonumber \\
&&{\hat b}_{\mu}^{\bar A}(\sigma ) {\hat b}^{\nu}_{\bar A}(\sigma )=
\delta^{\nu}_{\mu},\quad {\hat b}^{\bar A}_{\mu}(\sigma ) {\hat b}^{\mu}
_{\bar B}(\sigma )=\delta^{\bar A}_{\bar B}, \nonumber \\
{}^4{\bar g}_{\bar A\bar B}(z(\sigma ))&=&{\hat b}^{\mu}_{\bar A}(\sigma )
{}^4g_{\mu\nu}(z(\sigma )) {\hat b}^{\nu}_{\bar B}(\sigma )=\nonumber \\
&&=\lbrace {}^4{\bar g}_{ll}(\sigma )=\epsilon ; {}^4{\bar g}_{lr}(\sigma )=0;
{}^4{\bar g}_{rs}(\sigma )=
{}^4g_{rs}(\sigma )=-\epsilon \, {}^3g_{rs}\rbrace ,\nonumber \\
{}^4{\bar g}^{\bar A\bar B}&=&\lbrace {}^4{\bar g}^{ll}=\epsilon ; {}^4{\bar g}
^{lr}=0; {}^4{\bar g}^{rs}={}^4\gamma^{rs}=-\epsilon {}^3g^{rs}\rbrace ,
\nonumber \\
&&{}\nonumber \\
&&X_{\bar A}={\hat b}^{\mu}_{\bar A}\partial_{\mu}=\{ X_l={1\over
N}(\partial_{\tau}- N^r\partial_r);\partial_r\},\nonumber \\
&&\theta^{\bar A}={\hat b}^{\bar A}_{\mu}dx^{\mu}=\{ \theta^l=Nd\tau ;
\theta^r=d\sigma^r+N^rd\tau \} ,\nonumber \\
&&{}\nonumber \\
&&\Rightarrow l_{\mu}(\sigma )b^{\mu}_r(\sigma )=0,\quad l^{\mu}(\sigma )b^r
_{\mu}(\sigma )=-N^r(\sigma )/N(\sigma ),\nonumber \\
&&{}\nonumber \\
l^{\bar A}&=& l^{\mu} {\hat b}_{\mu}^{\bar A} = (\epsilon ; l^r+N^rl^{\tau})=
(\epsilon ; \vec 0),\nonumber \\
l_{\bar A}&=& l_{\mu} {\hat b}^{\mu}_{\bar A} = (1; l_r) = (1; \vec 0).
\label{II4}
\end{eqnarray}

\noindent One has ${}^3h_{\mu\nu}={}^4g_{\mu\nu}-\epsilon l_{\mu}l_{\nu}=-
\epsilon \, {}^3g_{rs}(b^r_{\mu}+N^rb^{\tau}_{\mu})(b^s_{\mu}+N^sb^{\tau}
_{\mu})=-\epsilon \, {}^3g_{rs}{\hat b}^r_{\mu}{\hat b}^s_{\nu}$.
For a 4-vector ${}^4V^{\mu}={}^4V^{\bar A}{\hat b}^{\mu}_{\bar A}=
{}^4V^l l^{\mu}+{}^4V^r{\hat b}^{\mu}_r$ one gets ${}^3V^{\mu}={}^3V^r{\hat b}
^{\mu}_r={}^3h^{\mu}_{\nu}\, {}^4V^{\nu}$, ${}^3V^r={}^4V^r={\hat b}^r_{\mu}
\, {}^3V^{\mu}$.

The nonholonomic basis in $\Sigma_{\tau}$-adapted coordinates is

\bea
&&{\hat b}_A^{\bar A}={\hat b}^{\bar A}_{\mu}b^{\mu}_A= \{ {\hat b}^l_A=l_A;\,
{\hat b}^r_A=\delta^r_A+N^r \delta^{\tau}_A \},\nonumber \\
&&{\hat b}^A_{\bar A}={\hat b}^{\mu}_{\bar A}b^A_{\mu}= \{ {\hat b}^A_l
=\epsilon l^A;\, {\hat b}^A_r=\delta^A_r \}.
\label{II5}
\eea

The 3-dimensional covariant derivative [denoted ${}^3\nabla$ or with the
subscript ``$|$"] of a 3-dimensional tensor ${}^3T^{\mu_1..
\mu_p}{}_{\nu_1..\nu_q}$ of rank (p,q) is the  3-dimensional tensor
of rank (p,q+1)
${}^3\nabla_{\rho}\, {}^3T^{\mu_1..\mu_p}{}_{\nu_1..\nu_q}={}^3T^{\mu_1..\mu_p}
{}_{\nu_1..\nu_q | \rho}=
{}^3h^{\mu_1}_{\alpha_1}\cdots {}^3h^{\mu_p}_{\alpha_p}\, {}^3h^{\beta_1}
_{\nu_1}\cdots {}^3h^{\beta_q}_{\nu_q}\, {}^3h^{\sigma}_{\rho}\, {}^4\nabla
_{\sigma}\, {}^3T^{\alpha_1..\alpha_p}{}_{\beta_1..\beta_q}$.

The components of the ``second fundamental form" of
$(\Sigma_{\tau},{}^3g)$ describe its extrinsic curvature

\bea
{}^3K_{\mu\nu}&=&{}^3K_{\nu\mu}=-{1\over 2}{\cal L}_l\, {}^3g_{\mu\nu}=
{\hat b}^r_{\mu} {\hat b}^s_{\nu}\, {}^3K_{rs},\nonumber \\
&&{}\nonumber \\
 {}^3K_{rs}&=&{}^3K_{sr}={1\over {2N}}(N_{r|s}+N_{s|r}-{{\partial \,
{}^3g_{rs}}\over {\partial \tau}});
\label{II6}
\eea

\noindent
one has ${}^4\nabla_{\rho} \, l^{\mu}=\epsilon \, {}^3a^{\mu} l_{\rho}-
{}^3K_{\rho}{}^{\mu}$, with the acceleration ${}^3a^{\mu}={}^3a^r {\hat b}
^{\mu}_r$ of the observers travelling along the congruence of timelike curves
with tangent vector $l^{\mu}$ given by ${}^3a_r=\partial_r\, ln\, N$.

The information contained in the 20 independent components
${}^4R^{\alpha} {}_{\mu\beta\nu}= {}^4\Gamma^{\alpha}_{\beta\rho}\,
{}^4\Gamma^{\rho}_{\nu\mu} - {}^4\Gamma^{\alpha}_{\nu\rho}\,
{}^4\Gamma^{\rho}_{\beta\mu} +\partial_{\beta}\, {}^4\Gamma^{\alpha}_{\mu\nu}
-\partial_{\nu}\, {}^4\Gamma^{\alpha}_{\beta\mu}$ [with the associated Ricci 
tensor ${}^4R_{\mu\nu}= {}^4R^{\beta}{}_{\mu\beta\nu}$] of the curvature 
Riemann tensor of $M^4$ is replaced by its  three projections
given by Gauss, Codazzi-Mainardi and Ricci equations \cite{mtw}. In the
nonholonomic basis the Einstein tensor becomes
${}^4G_{\mu\nu}= {}^4R_{\mu\nu}-{1\over 2}\, {}^4g_{\mu\nu}\, {}^4R
=\epsilon \, {}^4{\bar G}_{ll}l_{\mu}l_{\nu}+\epsilon
\, {}^4{\bar G}_{lr}(l_{\mu}{\hat b}^r_{\nu}+l_{\nu}{\hat b}^r_{\mu})+
{}^4{\bar G}_{rs}{\hat b}^r_{\mu}{\hat b}^s_{\nu}$. The Bianchi
identities ${}^4G^{\mu\nu}{}_{;\nu}\equiv 0$ imply the following four 
contracted Bianchi identities 

\bea
&&{1\over N}\partial_{\tau}\, {}^4{\bar G}_{ll}-{{N^r}\over N}\partial_r\,
{}^4{\bar G}_{ll}-{}^3K\, {}^4{\bar G}_{ll}+\partial_r\, {}^4{\bar G}_l{}^r+
(2\, {}^3a_r+{}^3\Gamma^s_{sr}){}^4{\bar G}_l{}^r-{}^3K_{rs}\, 
{}^4{\bar G}^{rs}\equiv 0,\nonumber \\
 &&{}\nonumber \\
 &&{1\over N}\partial_{\tau}\, {}^4{\bar G}_l{}^r-{{N^s}\over N}\partial_s\,
{}^4{\bar G}_l{}^r+{}^3a^r\, {}^4{\bar G}_{ll}-(2\, {}^3K^r{}_s+\delta^r_s\,
{}^3K+{{\partial_sN^r}\over N}){}^4{\bar G}_l{}^s+\partial_s\, {}^4{\bar G}
^{rs}+\nonumber \\
 &+& ({}^3a_s+{}^3\Gamma^u_{us}){}^4{\bar G}^{rs}\equiv 0.
\label{II7}
\eea

The vanishing of ${}^4{\bar G}_{ll}$,
${}^4{\bar G}_{lr}$, corresponds to the four secondary constraints
(restrictions of Cauchy data) of the ADM Hamiltonian formalism (see
Chapter V and Appendix G). The four contracted Bianchi identities imply 
\cite{wald} that, if the
restrictions of Cauchy data are satisfied initially and the spatial
equations ${}^4G_{ij}\, {\buildrel \circ \over =}\, 0$ are satisfied 
everywhere, then the secondary constraints are satisfied also at later times 
[see Ref.\cite{cho,wald,rendal,hfr}
 for the initial value problem]. The four contracted Bianchi
identities plus the four secondary constraints imply that only two combinations
of the Einstein equations contain the accelerations (second time derivatives)
of the two (non tensorial)
independent degrees of freedom of the gravitational field and that these
equations can be put in normal form [this was one of the motivations behind the
discovery of the Shanmugadhasan canonical transformations \cite{sha}].

The ``intrinsic geometry" of $\Sigma_{\tau}$ is defined by the Riemannian 
3-metric ${}^3g_{rs}$ [it allows to evaluate the length of space curves], 
the Levi-Civita
affine connection, i.e. the Christoffel symbols ${}^3\Gamma^u_{rs}$, [for
the parallel transport of 3-dimensional tensors on $\Sigma_{\tau}$] and the
curvature Riemann tensor ${}^3R^r{}_{stu}$ [for the evaluation of the holonomy
and for the geodesic deviation equation].
The ``extrinsic geometry" of $\Sigma_{\tau}$ is defined by the lapse N and 
shift $N^r$ functions [which describe the ``evolution" of $\Sigma_{\tau}$ in 
$M^4$] and by the ``extrinsic curvature" ${}^3K_{rs}$ [it is needed to 
evaluate how much a 3-dimensional vector goes outside $\Sigma_{\tau}$ under 
spacetime parallel transport and to rebuild the spacetime curvature from the 
3-dimensional one].

Given an arbitrary 3+1 splitting of $M^4$, the ADM action\cite{adm} expressed 
in terms of the independent $\Sigma_{\tau}$-adapted variables
N, $N_r={}^3g_{rs}N^s$, ${}^3g_{rs}$ is 

\bea
S_{ADM}&=&\int d\tau \,
L_{ADM}(\tau )=
\int d\tau d^3\sigma {\cal L}_{ADM}(\tau ,\vec \sigma )=\nonumber \\
 &=&- \epsilon k\int_{\triangle \tau}d\tau  \, \int d^3\sigma \, \lbrace
\sqrt{\gamma} N\, [{}^3R+{}^3K_{rs}\, {}^3K^{rs}-({}^3K)^2]\rbrace (\tau ,\vec
\sigma ),
\label{II8}
\eea 

\noindent where $k={{c^3}\over {16\pi G}}$, with $G$ the Newton constant.

The Euler-Lagrange equations are\footnote{The symbol ${\buildrel \circ
\over =}$ means evaluated on the extremals of the variational principle,
namely on the solutions of the equation of motion.}

\begin{eqnarray}
L_N&=&{{\partial {\cal L}_{ADM}}\over {\partial N}}-\partial_{\tau}
{{\partial {\cal L}_{ADM}}\over {\partial \partial_{\tau}N}}-\partial_r
{{\partial {\cal L}_{ADM}}\over {\partial \partial_rN}}=\nonumber \\
&=&-\epsilon k
\sqrt{\gamma} [{}^3R-{}^3K_{rs}\, {}^3K^{rs}+({}^3K)^2]=-2\epsilon k\,
{}^4{\bar G}_{ll}\, {\buildrel \circ
\over =}\, 0,\nonumber \\
L^r_{\vec N}&=&{{\partial {\cal L}_{ADM}}\over {\partial N_r}}-\partial_{\tau}
{{\partial {\cal L}_{ADM}}\over {\partial \partial_{\tau}N_r}}-\partial_s
{{\partial {\cal L}_{ADM}}\over {\partial \partial_sN_r}}=\nonumber \\
&=&2\epsilon k
[\sqrt{\gamma}({}^3K^{rs}-{}^3g^{rs}\, {}^3K)]_{\, |s}=2k\, {}^4{\bar G}_l{}^r
\, {\buildrel \circ \over =}\, 0,\nonumber \\
L_g^{rs}&=& -\epsilon k \Big[
{{\partial}\over {\partial \tau}}[\sqrt{\gamma}({}^3K^{rs}-{}^3g
^{rs}\, {}^3K)]\, - N\sqrt{\gamma}({}^3R^{rs}-
{1\over 2} {}^3g^{rs}\, {}^3R)+\nonumber \\
&&+2N\, \sqrt{\gamma}({}^3K^{ru}\, {}^3K_u{}^s-{}^3K\, {}^3K^{rs})+{1\over 2}N
\sqrt{\gamma}[({}^3K)^2-{}^3K_{uv}\, {}^3K^{uv}){}^3g^{rs}+\nonumber \\
&&+\sqrt{\gamma} ({}^3g^{rs} N^{|u}{}_{|u}-N^{|r |s})\Big] =-\epsilon kN
\sqrt{\gamma}\, {}^4{\bar G}^{rs}\, {\buildrel \circ \over =}\,0,
\label{II9}
\end{eqnarray}

\noindent and correspond to the Einstein equations in the form ${}^4{\bar G}
_{ll}\, {\buildrel \circ \over =}\, 0$, ${}^4{\bar G}_{lr}\, {\buildrel \circ
\over =}\, 0$, ${}^4{\bar G}_{rs}\, {\buildrel \circ \over =}\, 0$,
respectively. The four contracted Bianchi identities imply
that only two of the six equations $L^{rs}_g\, {\buildrel \circ
\over =}\, 0$ are independent.

The canonical momenta (densities of weight -1) are

\begin{eqnarray}
&&{\tilde \pi}^N(\tau ,\vec \sigma )={{\delta S_{ADM}}\over {\delta
\partial_{\tau}N(\tau ,\vec \sigma )}} =0,\nonumber \\
&&{\tilde \pi}^r_{\vec N}(\tau ,\vec \sigma )={{\delta S_{ADM}}\over
{\delta \partial_{\tau} N_r(\tau ,\vec \sigma )}} =0,\nonumber \\
&&{}^3{\tilde \Pi}^{rs}(\tau ,\vec \sigma )={{\delta S_{ADM}}\over
{\delta \partial_{\tau} {}^3g_{rs}(\tau ,\vec \sigma )}}=\epsilon k\,
[
\sqrt{\gamma}({}^3K^{rs}-{}^3g^{rs}\, {}^3K)](\tau ,\vec \sigma ),\nonumber \\
&&{}\nonumber \\
&&\Downarrow \nonumber \\
&&{}\nonumber \\
&&{}^3K_{rs}={{\epsilon}\over {k\sqrt{\gamma}}} [{}^3{\tilde \Pi}_{rs}-{1\over
2}{}^3g_{rs}\, {}^3{\tilde \Pi}],\quad\quad {}^3\tilde \Pi ={}^3g_{rs}\,
{}^3{\tilde \Pi}^{rs}=-2\epsilon k\sqrt{\gamma}\, {}^3K,
\label{II10}
\end{eqnarray}

\noindent and satisfy the Poisson brackets

\begin{eqnarray}
&&\lbrace N(\tau ,\vec \sigma ),{\tilde \Pi}^N(\tau ,{\vec \sigma}^{'} )
\rbrace =\delta^3(\vec \sigma ,{\vec \sigma}^{'}),\nonumber \\
&&\lbrace N_r(\tau ,\vec \sigma ),{\tilde \Pi}^s_{\vec N}(\tau ,{\vec \sigma}
^{'} )\rbrace =\delta^s_r \delta^3(\vec \sigma ,{\vec \sigma}^{'}),\nonumber \\
&&\lbrace {}^3g_{rs}(\tau ,\vec \sigma ),{}^3{\tilde \Pi}^{uv}(\tau ,{\vec
\sigma}^{'}\rbrace = {1\over 2} (\delta^u_r\delta^v_s+\delta^v_r\delta^u_s)
\delta^3(\vec \sigma ,{\vec \sigma}^{'}).
\label{II11}
\end{eqnarray}

The Wheeler- De Witt supermetric is

\begin{equation}
{}^3G_{rstw}(\tau ,\vec \sigma )=[{}^3g_{rt}\, {}^3g_{sw}+{}^3g_{rw}\, {}^3g
_{st}-{}^3g_{rs}\, {}^3g_{tw}](\tau ,\vec \sigma ).
\label{II12}
\end{equation}

\noindent Its inverse is defined by the equations

\begin{eqnarray}
{}&&{1\over 2} {}^3G_{rstw}\, {1\over 2} {}^3G^{twuv} ={1\over 2}(\delta^u_r
\delta^v_s+\delta^v_r\delta^u_s),\nonumber \\
{}^3G^{twuv}(\tau ,\vec \sigma )&=&[{}^3g^{tu}\, {}^3g^{wv}+{}^3g^{tv}\, {}^3g
^{wu}-2\, {}^3g^{tw}\, {}^3g^{uv}](\tau ,\vec \sigma ),
\label{II13}
\end{eqnarray}

\noindent so that one gets

\begin{eqnarray}
{}^3{\tilde \Pi}^{rs}(\tau ,\vec \sigma )&=&{1\over 2}\epsilon k 
\sqrt{\gamma}\,
{}^3G^{rsuv}(\tau ,\vec \sigma )\, {}^3K_{uv}(\tau ,\vec \sigma ),\nonumber \\
{}^3K_{rs}(\tau ,\vec \sigma )&=&{{\epsilon}\over {2k\sqrt{\gamma}}}\,
{}^3G_{rsuv}(\tau ,\vec \sigma )\, {}^3{\tilde \Pi}^{uv}(\tau ,\vec \sigma ),
\nonumber \\
&&[{}^3K^{rs}\, {}^3K_{rs}-({}^3K)^2](\tau ,\vec \sigma )=\nonumber \\
&&=k^{-2}[\gamma^{-1}({}^3{\tilde \Pi}^{rs}\, {}^3{\tilde \Pi}_{rs}-{1\over 2}
({}^3{\tilde \Pi})^2](\tau ,\vec \sigma )=(2k)^{-1}[\gamma^{-1}\, {}^3G_{rsuv}
\, {}^3{\tilde \Pi}^{rs}\, {}^3{\tilde \Pi}^{uv}](\tau ,\vec \sigma ),
\nonumber \\
\partial_{\tau}\, {}^3g_{rs}(\tau ,\vec \sigma )&=&[N_{r|s}+N_{s|r}-{{\epsilon
N}\over {k\sqrt{\gamma}}}\, {}^3G_{rsuv}\, {}^3{\tilde \Pi}^{uv}](\tau ,
\vec \sigma ).
\label{II14}
\end{eqnarray}

Since ${}^3{\tilde \Pi}^{rs}\partial_{\tau}\, {}^3g_{rs}=$${}^3{\tilde \Pi}
^{rs} [N_{r | s}+N_{s | r}-{{\epsilon N}\over {k\sqrt{\gamma}}} {}^3G_{rsuv}
{}^3{\tilde \Pi}^{uv}]=$$-2 N_r {}^3{\tilde \Pi}^{rs}{}_{| s}-
{{\epsilon N}\over
{k\sqrt{\gamma}}}\, {}^3G_{rsuv}\, {}^3{\tilde \Pi}^{rs} {}^3{\tilde \Pi}^{uv}+
(2N_r\, {}^3{\tilde \Pi}^{rs})_{| s}$, we obtain the canonical Hamiltonian
\footnote{Since $N_r\, {}^3{\tilde \Pi}^{rs}$ is a vector density of weight 
-1, it holds ${}^3\nabla_s(N_r\, {}^3{\tilde \Pi}^{rs})=\partial_s(N_r\, 
{}^3{\tilde \Pi}^{rs})$.}

\begin{eqnarray}
H_{(c)ADM}&=& \int_Sd^3\sigma \, [{\tilde
\pi}^N\partial_{\tau}N+{\tilde
\pi}^r_{\vec N}\partial_{\tau}N_r+{}^3{\tilde \Pi}^{rs}\partial_{\tau}
{}^3g_{rs}](\tau ,\vec \sigma ) -L_{ADM}=\nonumber \\
&=&\int_Sd^3\sigma \, [\epsilon N(k\sqrt{\gamma}\, {}^3R-
{1\over {2k\sqrt{\gamma}}} {}^3G_{rsuv}{}^3{\tilde \Pi}^{rs} {}^3{\tilde
\Pi}^{uv})-2N_r\, {}^3{\tilde \Pi}^{rs}{}_{| s}](\tau ,\vec \sigma )+
\nonumber \\
&+&2\int_{\partial S}d^2\Sigma_s [N_r\, {}^3{\tilde \Pi}^{rs}\,\,
](\tau ,\vec \sigma ),
\label{II15}
\end{eqnarray}

\noindent In the following discussion the surface term will be omitted.

The Dirac Hamiltonian is \footnote{The $\lambda (\tau ,\vec \sigma )$'s 
are arbitrary Dirac multipliers.}

\begin{equation}
H_{(D)ADM}=H_{(c)ADM}+\int d^3\sigma \, [\lambda_N\, {\tilde \pi}^N +
\lambda
^{\vec N}_r\, {\tilde \pi}^r_{\vec N}](\tau ,\vec \sigma ).
\label{II16}
\end{equation}

The $\tau$-constancy of the primary constraints [$\partial_{\tau} {\tilde
\pi}^N(\tau ,\vec \sigma )\, {\buildrel \circ \over =}\, 
\lbrace {\tilde \pi}^N(\tau ,\vec \sigma ),H_{(D)
ADM}\rbrace \approx 0$, $\partial_{\tau} {\tilde \pi}^r_{\vec N}(\tau
,\vec \sigma )\, {\buildrel \circ \over =}\, 
\lbrace {\tilde \pi}^r_{\vec N}(\tau ,\vec \sigma ),H_{(D)ADM}
\rbrace \approx 0$] generates four secondary constraints 
(they  are densities
of weight -1) which correspond to the Einstein equations ${}^4{\bar
G}_{ll} (\tau ,\vec \sigma )\, {\buildrel \circ \over =}\, 0$,
${}^4{\bar G}_{lr} (\tau ,\vec \sigma )\, {\buildrel \circ \over =}\,
0$

\begin{eqnarray}
{\tilde {\cal H}}(\tau ,\vec \sigma )&=&\epsilon
[k\sqrt{\gamma}\, {}^3R-{1\over {2k
\sqrt{\gamma}}} {}^3G_{rsuv}\, {}^3{\tilde \Pi}^{rs}\, {}^3{\tilde \Pi}^{uv}]
(\tau ,\vec \sigma )=\nonumber \\
&=&\epsilon [\sqrt{\gamma}\, {}^3R-{1\over {k\sqrt{\gamma}}}({}^3{\tilde \Pi}
^{rs}\, {}^3{\tilde \Pi}_{rs}-{1\over 2}({}^3\tilde \Pi )^2)](\tau ,\vec
\sigma )=\nonumber \\
&=&\epsilon k \{ \sqrt{\gamma} [{}^3R-({}^3K_{rs}\, {}^3K^{rs}-({}^3K)^2 )]\}
(\tau ,\vec \sigma )\approx 0,
\nonumber \\
{}^3{\tilde {\cal H}}^r(\tau ,\vec \sigma )&=&-2\, {}^3{\tilde 
\Pi}^{rs}{}_{| s}
(\tau ,\vec \sigma )=-2[\partial_s\, {}^3{\tilde \Pi}^{rs}+{}^3\Gamma^r_{su}
{}^3{\tilde \Pi}^{su}](\tau ,\vec \sigma )=\nonumber \\
&=&-2\epsilon k \{ \partial_s[\sqrt{\gamma}({}^3K^{rs}-{}^3g^{rs}\, {}^3K)]+
{}^3\Gamma^r_{su}\sqrt{\gamma}({}^3K^{su}-{}^3g^{su}\, {}^3K) \}
(\tau ,\vec \sigma )\approx 0,
\label{II17}
\end{eqnarray}

\noindent so that the Hamiltonian becomes

\begin{equation}
H_{(c)ADM}= \int d^3\sigma [N\, {\tilde {\cal H}}+N_r\, {}^3{\tilde
{\cal H}}^r](\tau ,\vec \sigma )\approx 0,
\label{II18}
\end{equation}

\noindent with ${\tilde {\cal H}}(\tau ,\vec \sigma )\approx 0$ called the
superhamiltonian constraint and ${}^3{\tilde {\cal H}}^r(\tau ,\vec \sigma )
\approx 0$  the supermomentum constraints.
In ${\tilde {\cal H}}(\tau ,\vec \sigma )\approx 0$ one can say that
the term $-\epsilon k \sqrt{\gamma}({}^3K_{rs}\, {}^3K^{rs}-{}^3K^2)$
is the kinetic energy and $\epsilon k\sqrt{\gamma}\, {}^3R$ the
potential energy.

All the constraints are first class, because the only non-identically zero
Poisson brackets correspond to the so called universal Dirac algebra
\cite{dirac}:

\begin{eqnarray}
\lbrace {}^3{\tilde {\cal H}}_r(\tau ,\vec \sigma ),{}^3{\tilde {\cal H}}_s
(\tau ,{\vec \sigma}^{'})\rbrace  &=&{}^3{\tilde {\cal H}}_r(\tau
,{\vec
\sigma}^{'} )\, {{\partial \delta^3(\vec \sigma ,{\vec \sigma}^{'})}\over
{\partial \sigma^s}} + {}^3{\tilde {\cal H}}_s(\tau ,\vec \sigma ) {{\partial
\delta^3(\vec \sigma ,{\vec \sigma}^{'})}\over {\partial \sigma^r}},
\nonumber \\
\lbrace {\tilde {\cal H}}(\tau ,\vec \sigma ),{}^3{\tilde {\cal H}}_r(\tau ,
{\vec \sigma}^{'})\rbrace &=& {\tilde {\cal H}}(\tau ,\vec \sigma )
{{\partial \delta^3(\vec \sigma ,{\vec \sigma}^{'})}\over {\partial \sigma^r}},
\nonumber \\
\lbrace {\tilde {\cal H}}(\tau ,\vec \sigma ),{\tilde {\cal H}}(\tau ,{\vec
\sigma}^{'})\rbrace &=&[{}^3g^{rs}(\tau ,\vec \sigma ) {}^3{\tilde {\cal H}}_s
(\tau ,\vec \sigma )+\nonumber \\
&+&{}^3g^{rs}(\tau ,{\vec \sigma}^{'}){}^3{\tilde
{\cal H}}_s(\tau ,{\vec \sigma}^{'})]{{\partial \delta^3(\vec \sigma ,{\vec
\sigma}^{'})}\over {\partial \sigma^r}},
\label{II19}
\end{eqnarray}

\noindent with ${}^3{\tilde {\cal H}}_r={}^3g_{rs}\, {}^3{\tilde {\cal H}}^r$
as the combination of the supermomentum constraints satisfying the
algebra of 3-diffeomorphisms. In Ref.\cite{tei} it is shown that
Eqs.(\ref{II19}) are sufficient conditions for the embeddability of
$\Sigma_{\tau}$ into $M^4$. In the second paper in Ref.\cite{kuchar}
it is shown that the last two lines of the Dirac algebra are the
equivalent in phase space of the Bianchi identities
${}^4G^{\mu\nu}{}_{;\nu}\equiv 0$.

The Hamilton-Dirac equations are [$\cal L$ is the notation for the Lie 
derivative]

\begin{eqnarray}
\partial_{\tau}N(\tau ,\vec \sigma )\, &{\buildrel \circ \over =}\,&
\lbrace N(\tau ,\vec \sigma ),H_{(D)ADM}
\rbrace =\lambda_N(\tau ,\vec \sigma ),\nonumber \\
\partial_{\tau}N_r(\tau ,\vec \sigma )\, &{\buildrel \circ \over =}\,&
\lbrace N_r(\tau ,\vec \sigma ),
H_{(D)ADM}\rbrace =\lambda^{\vec N}_r(\tau ,\vec \sigma ),\nonumber \\
\partial_{\tau}\, {}^3g_{rs}(\tau ,\vec \sigma )\, 
&{\buildrel \circ \over =}\,&
\lbrace {}^3g_{rs}(\tau ,
\vec \sigma ),H_{(D)ADM}\rbrace =[N_{r | s}+N_{s | r}-{{2\epsilon N}\over
{k\sqrt{\gamma}}}({}^3{\tilde \Pi}_{rs}-{1\over 2}{}^3g_{rs}\, {}^3{\tilde
\Pi})](\tau ,\vec \sigma )=\nonumber \\
&=&[N_{r|s}+N_{s|r}-2N\, {}^3K_{rs}](\tau ,\vec \sigma ),\nonumber \\
\partial_{\tau}\, {}^3{\tilde \Pi}^{rs}(\tau ,\vec \sigma )
\, &{\buildrel \circ \over =}\,& \lbrace {}^3
{\tilde \Pi}^{rs}(\tau ,\vec \sigma ),H_{(D)ADM}\rbrace =\epsilon [N\,
k\sqrt{\gamma} ({}^3R^{rs}-{1\over 2}{}^3g^{rs}\, {}^3R)](\tau ,\vec \sigma )-
\nonumber \\
&-&2\epsilon [{N\over {k\sqrt{\gamma}}}({1\over 2}{}^3{\tilde \Pi}\, 
{}^3{\tilde \Pi}^{rs}-{}^3{\tilde \Pi}^r{}_u\, {}^3{\tilde \Pi}^{us})(\tau 
,\vec \sigma )-\nonumber \\
&-&{{\epsilon N}\over 2}
{{{}^3g^{rs}}\over {k\sqrt{\gamma}}}({1\over 2}{}^3{\tilde \Pi}^2-{}^3{\tilde
\Pi}_{uv}\, {}^3{\tilde \Pi}^{uv})](\tau ,\vec \sigma )+\nonumber \\
&+&{\cal L}_{\vec N}\, {}^3{\tilde \Pi}^{rs}(\tau ,\vec \sigma )+\epsilon
[k\sqrt{\gamma} (N^{| r | s}-{}^3g^{rs}\, N^{| u}{}_{| u})](\tau ,\vec 
\sigma ),\nonumber \\
&&{}\nonumber \\
&&\Downarrow \nonumber \\
&&{}\nonumber \\
\partial_{\tau}\, {}^3K_{rs}(\tau ,\vec \sigma )\, 
&{\buildrel \circ \over =}\,&
\Big( N [{}^3R_{rs}+{}^3K\, {}^3K_{rs}-2\, {}^3K_{ru}\, {}^3K^u{}_s]-
\nonumber \\
&-&N_{|s|r}+N^u{}_{|s}\, {}^3K_{ur}+N^u{}_{|r}\, {}^3K_{us}+N^u\, {}^3K_{rs|u}
\Big) (\tau ,\vec \sigma ),\nonumber \\
 &&{}\nonumber \\
 &&with \nonumber \\
 {\cal L}_{\vec N}\, {}^3{\tilde \Pi}^{rs}&=&-\sqrt{\gamma}\,
{}^3\nabla_u({{N^u}\over {\sqrt{\gamma}}} {}^3{\tilde \Pi}^{rs})+
{}^3{\tilde \Pi}^{ur}\, {}^3\nabla_u N^s+{}^3{\tilde \Pi}^{us}\,
{}^3\nabla_u N^r.
\label{II20}
\end{eqnarray}

The above equation for $\partial_{\tau}\, {}^3g_{rs}(\tau ,\vec \sigma
)$ shows that the generator of space pseudo-diffeomorphisms 
\footnote{The Haniltonian transformations generated by these
constraints are the extension to the 3-metric of  passive or 
pseudo- diffeomorphisms, namely changes of 
coordinate charts, of $\Sigma_{\tau}$ [$Diff\, \Sigma_{\tau}$].}
$\int d^3\sigma
N_r(\tau ,\vec \sigma )$ $\, {}^3{\tilde {\cal H}}^r(\tau ,\vec \sigma
)$ produces a variation, tangent to $\Sigma_{\tau}$, $\delta_{tangent}
{}^3g_{rs}={\cal L}
_{\vec N}\, {}^3g_{rs}=N_{r|s}+N_{s|r}$ in accord with the infinitesimal
pseudo-diffeomorphisms in $Diff\, \Sigma_{\tau}$. Instead, the gauge
transformations induced by the
superhamiltonian generator $\int d^3\sigma N(\tau ,\vec \sigma )\,$
${\tilde {\cal H}}(\tau ,\vec
\sigma )$ do not reproduce the infinitesimal diffeomorphisms in $Diff\, M^4$
normal to $\Sigma_{\tau}$ (see Ref.\cite{wa}). For the clarification of
the connection between spacetime diffeomorphisms and Hamiltonian gauge 
transformations see  Ref.\cite{ppons} and Section IX.

Finally, the canonical transformation  ${\tilde
\pi}^N\, dN+{\tilde
\pi}^r_{\vec N}\, dN_r+{}^3{\tilde \Pi}^{rs}\, d{}^3g_{rs} =
{}^4{\tilde \Pi}^{AB}\, d{}^4g
_{AB}$ allows to define the following momenta conjugated to ${}^4g_{AB}$

\begin{eqnarray}
{}^4{\tilde \Pi}^{\tau\tau}&=&{{\epsilon}\over {2N}} {\tilde
\pi}^N,\nonumber \\
 {}^4{\tilde \Pi}^{\tau r}&=&{{\epsilon}\over 2}
({{N^r}\over N} {\tilde \pi}
^N-{\tilde \pi}^r_{\vec N}),\nonumber \\
{}^4{\tilde \Pi}^{rs}&=&\epsilon ({{N^rN^s}\over {2N}} {\tilde \pi}^N-
{}^3{\tilde \Pi}^{rs}),
\nonumber \\
&&{}\nonumber \\
&&\lbrace {}^4g_{AB}(\tau ,\vec \sigma ),{}^4{\tilde \Pi}^{CD}(\tau ,{\vec
\sigma}^{'} )\rbrace ={1\over 2}(\delta^C_A\delta^D_B+\delta^D_A\delta^C_B)
\delta^3(\vec \sigma ,{\vec \sigma}^{'}),\nonumber \\
&&{}\nonumber \\ {\tilde \pi}^N&=& {{2\epsilon}\over {\sqrt{\epsilon
{}^4g^{\tau\tau}}}} {}^4{\tilde \Pi}^{\tau\tau},\nonumber \\
 {\tilde \pi}^r_{\vec N}&=&2\epsilon {{{}^4g^{\tau r}}\over {{}^4g^{\tau\tau}}}
{}^4{\tilde \Pi}^{\tau\tau}-2\epsilon {}^4{\tilde \Pi}^{\tau
r},\nonumber \\ {}^3{\tilde \Pi}^{rs}&=&\epsilon {{{}^4g^{\tau r}
{}^4g^{\tau S}}\over {({}^4g^{\tau\tau})^2}}\, {}^4{\tilde
\Pi}^{\tau\tau} -\epsilon {}^4{\tilde
\Pi}^{rs},
\label{II21}
\end{eqnarray}

\noindent which would emerge if the ADM action would be considered function
of ${}^4g_{AB}$ instead of N, $N_r$ and ${}^3g_{rs}$.

Let us add a comment on the structure of gauge-fixings for metric gravity. As 
said in Refs.\cite{giap,lusa},
in a system with only primary and secondary first class constraints (like
electromagnetism, Yang-Mills theory and both metric and tetrad gravity) the
Dirac Hamiltonian $H_D$ contains only the arbitrary Dirac multipliers 
associated
with the primary first class constraints. The secondary first class constraints
are already contained in the canonical Hamiltonian with well defined
coefficients [the temporal components $A_{ao}$ of the gauge potential in
Yang-Mills theory; the lapse and shift functions in metric and tetrad gravity
as evident from Eq.(\ref{II18});
in both cases, through the first half of the Hamilton equations, the Dirac
multipliers turn out to be equal to the $\tau$-derivatives of these quantities,
which, therefore, inherit an induced arbitrariness]. See the second paper in
Ref.\cite{sha} for a discussion of this point and for a refusal of Dirac's
conjecture\cite{dirac} according to which also the secondary first class
constraints must have arbitrary Dirac multipliers 
\footnote{In such a case one does not
recover the original Lagrangian by inverse Legendre transformation and one
obtains a different "off-shell" theory.}. In these cases one must adopt the
following gauge-fixing strategy: i) add gauge-fixing constraints
$\chi_a\approx 0$ to the secondary constraints; ii) their time constancy,
$\partial_{\tau} \chi_a \, {\buildrel \circ \over =}\, \lbrace \chi_a,H_D
\rbrace =g_a \approx 0$, implies the appearance of gauge-fixing constraints
$g_a\approx 0$ for the primary constraints; iii) the time constancy of the
constraints $g_a\approx 0$, $\partial_{\tau} g_a\, {\buildrel \circ \over =}\,
\lbrace g_a,H_D\rbrace \approx 0$, determines the Dirac multipliers in front
of the primary constraints [the $\lambda$'s in Eq.(\ref{II16})].

As shown in the second paper of Ref.\cite{lusa} for the electromagnetic case,
this method works also with covariant gauge-fixings: the electromagnetic
Lorentz gauge $\partial^{\mu}A_{\mu}(x) \approx 0$ may be rewritten in phase
space as a gauge-fixing constraint depending upon the Dirac multiplier; its
time constancy gives a multiplier-dependent gauge-fixing for $A_o(x)$ and the
time constancy of this new constraint gives the elliptic equation for the
multiplier with the residual gauge freedom connected with the kernel of the
elliptic operator.

In metric gravity, the covariant gauge-fixings analogous to the Lorentz
gauge are those determining the harmonic coordinates
(harmonic or De Donder gauge): $\chi^B={1\over
{\sqrt{{}^4g}}} \partial_A(\sqrt{{}^4g}\, {}^4g^{AB}) \approx 0$ in the
$\Sigma_{\tau}$-adapted holonomic coordinate basis. More explicitly, they are:
\hfill\break
\hfill\break
i) for $B=\tau$: $N \partial_{\tau} \gamma -\gamma \partial_{\tau} N -N^2
\partial_r({{\gamma N^r}\over N}) \approx 0$; \hfill\break
ii) for $B=s$: $N N^s \partial
_{\tau}\gamma +\gamma (N \partial_{\tau}N^s-N^s \partial_{\tau}N)+N^2
\partial_r[N\gamma ({}^3g^{rs}-{{N^rN^s}\over {N^2}})] \approx 0$.\hfill\break
\hfill\break
{}From the Hamilton-Dirac equations  we get
\begin{eqnarray}
\partial_{\tau}N    &{\buildrel \circ \over=}  & \lambda_N, \nonumber \\
\partial_{\tau} N_r &{\buildrel \circ \over =} & \lambda^{\vec N}_r
\qquad \mbox{and} \nonumber \\
\partial_{\tau} \gamma &=& {1\over 2}\gamma \, {}^3g^{rs}
\partial_{\tau}\, {}^3g_{rs}\, {\buildrel \circ \over =}\,
{1\over 2}\gamma [{}^3g^{rs}(N_{r|s}+N_{s|r})
 -{{5\epsilon N}\over {k\sqrt{\gamma}}}\,{}^3{\tilde \Pi}].
\label{II22}
\end{eqnarray}

Therefore, in phase space the harmonic coordinate gauge-fixings 
associated with the secondary superhamiltonian and supermomentum constraints
take the form

\beq
\chi^B={\bar \chi}^B(N,N_r,N_{r|s}, {}^3g_{rs}, {}^3{\tilde \Pi}^{rs},
\lambda_N, \lambda^{\vec N}_r) \approx 0.
\label{II23}
\eeq

The conditions $\partial_{\tau}
{\bar \chi}^B\, {\buildrel \circ \over =}\, \lbrace {\bar \chi}^B,H_D\rbrace
=g^B\approx 0$ give the gauge-fixings for the primary constraints ${\tilde
\pi}^N\approx 0$, ${\tilde \pi}^r_{\vec N}\approx 0$. 

The conditions $\partial
_{\tau} g^B\, {\buildrel \circ \over =}\, \lbrace g^B,H_D\rbrace \approx 0$
are partial differential equations for the Dirac multipliers $\lambda_N$,
$\lambda_r^{\vec N}$, implying a residual gauge freedom like it happens for
the electromagnetic Lorentz gauge.

\vfill\eject

\section{Asymptotic Flatness and Hamiltonian Gauge Transformations.}

In this Section after some comments on gauge field theories there will be a 
short review of the notion of isolated system in general relativity with the 
associated concepts of asymptotic flatness and asymptotic symmetries. It will 
be shown that the possible existence of asymptotic supertranslations is an
obstacle to define an asymptotic Poincar\'e group and to make contact with the
theory of isolated systems in Minkowski spacetime delineated in Appendix A
when the Newton constant is switched off. Then a discussion of the boundary
conditions needed to have a well defined Hamiltonian formalism and well 
defined Hamiltonian gauge transformations is given.

\subsection{Gauge Field Theories.}

In ADM canonical gravity there are 8 first class constraints, which are 
generators of Hamiltonian gauge transformations. Some general properties of 
these transformations will now be analyzed. In Section IX, after having 
interpreted the action of the Hamiltonian gauge transformations of metric 
gravity, they will be compared with the transformations induced by the
spacetime diffeomorphisms of the spacetime ($Diff\, M^4$).

In the Hamiltonian formulation of every gauge field theory one has to
make a choice of the boundary conditions of the canonical variables
and of the parameters of the gauge transformations \footnote{The infinitesimal
ones are generated by the first class constraints of the theory.} in
such a way to give a meaning to integrations by parts, to the
functional derivatives (and therefore to Poisson brackets) and to the
{\it proper} gauge transformations connected with the identity \footnote{The
{\it improper} ones, including the {\it rigid or global or first kind} gauge
transformations related to the non-Abelian charges, have to be treated
separately; when there are topological numbers like winding number,
they label disjoint sectors of gauge transformations and one speaks of
{\it large} gauge transformations.}. In particular, the boundary
conditions must be such that the variation of the final Dirac
Hamiltonian $H_D$ must be linear in the variations of the canonical
variables \footnote{The coefficients are the Dirac-Hamilton equations of
motion.} and this may require a redefinition of $H_D$, namely $H_D$ has
to be replaced by ${\tilde H}_D=H_D+H_{\infty}$, where $H_{\infty}$ is
a suitable integral on the surface at spatial infinity. When this is
accomplished, one has a good definition of functional derivatives and
Poisson brackets. Then, one must consider the most general generator
of gauge transformations of the theory (it includes $H_D$ as a special
case), in which there are arbitrary functions (parametrizing
infinitesimal gauge transformations) in front of all the first class
constraints and not only in front of the primary ones\footnote{These are the 
generalized Hamiltonian gauge transformations of the Dirac conjecture. As
said at the end of the previous Section they are not generated by the Dirac
Hamiltonian. However, their pullback to configuration space generates
local Noether transformations under which the ADM action (\ref{II8}) is 
quasi-invariant, in accord with the general theory of singular Lagrangians
\cite{lu1}. }. Also the variations
 of this generator must be linear in the variations of the canonical variables:
this implies that all the surface terms coming from integration by
parts must vanish with the given boundary conditions on the canonical
variables or must be compensated by the variation of $H_{\infty}$. In
this way, one gets boundary conditions on the parameters of the
infinitesimal gauge transformations identifying the {\it proper} ones,
which transform the canonical variables among themselves without
altering their boundary conditions. Let us remark that in this way one
is defining Hamiltonian boundary conditions which are not manifestly
covariant; however, in Minkowski spacetime a Wigner covariant
formulation is obtained by reformulating the theory on spacelike
hypersurfaces \cite{lus3,re} and then restricting it to spacelike
hyperplanes.

In the Yang-Mills case\cite{lusa}, with the Hamiltonian gauge transformations
restricted to go to the identity in an angle-independent way at spatial 
infinity, so to have well defined covariant non-Abelian charges, the 
{\it proper} gauge
transformations are those which are connected to the identity and generated by
the Gauss law first class constraints at the infinitesimal level. The
{\it improper} ones are a priori of four types: \hfill\break
\hfill\break
i) {\it global or rigid or first kind} ones (the gauge parameter fields
tend to constant at spatial infinity) connected with the group G
(isomorphic to the structure group of the Yang-Mills principal bundle)
generated by the {\it non-Abelian charges};
\hfill\break
ii) the global or rigid ones in the {\it center of the gauge group G}
[triality when G=SU(3)]; \hfill\break
 iii) gauge transformations with
non-vanishing winding number $n\in Z$ ({\it large} gauge transformations
not connected with the identity; zeroth homotopy group of the gauge
group); \hfill\break
 iv) other {\it improper non rigid} gauge
transformations. Since this last type of gauge transformations does
not play any role in Yang-Mills dynamics, it was assumed \cite{lusa}
that the choice of the function space for the gauge parameter fields
$\alpha_a(\tau ,\vec \sigma )$ (describing the component of the gauge
group connected with the identity) be such that for $r=|\vec \sigma |\, 
\rightarrow \infty$ one has

\beq
\alpha_a(\tau ,\vec \sigma )\, \rightarrow
\alpha_a^{(rigid)}+\alpha_a^{(proper)}(\tau ,\vec \sigma ),
\label{III1}
\eeq
with constant
$\alpha_a^{(rigid)}$ and with $\alpha_a^{(proper)}(\tau ,\vec \sigma )$ tending
to zero in a direction-independent way.

However, in gauge theories, in the framework of local quantum field
theory, one does not consider the Abelian and non-Abelian charges
generators of gauge transformations of first kind, but speaks of
supersection sectors determined by the charges. This is valid both for
the electric charge, which is a physical observable, and for the color
charge in QCD, where the hypothesis of quark confinement requires the
existence only of color singlets, namely: i) physical observables must
commute with the non-Abelian charges; ii) the SU(3) color charges of
isolated systems have to vanish themselves.

We will follow the same scheme in the analysis of the Hamiltonian
gauge transformations of metric gravity.

\subsection{Isolated Systems and Asymptotic Flatness.}

The definition of an {\it isolated system} in general relativity is a 
difficult problem (see Ref.\cite{p14} for a review), since there is neither a
background flat metric ${}^4\eta$ nor a natural global inertial coordinate
system allowing to define a preferred radial coordinate $r$ and a limit
${}^4g_{\mu\nu}\, \rightarrow \, {}^4\eta_{\mu\nu} +O(1/r)$ for $r\,
\rightarrow \, \infty$ along either spatial or null directions. Usually, one
considers an asymptotic Minkowski metric ${}^4\eta_{\mu\nu}$ in rectangular
coordinates and tries to get asymptotic statements with various types of
definitions of $r$. However, it is difficult to correctly specify the limits
for $r\, \rightarrow \, \infty$ in a meaningful, coordinate independent way.

This led to the introduction of coordinate independent definitions of
{\it asymptotic flatness} of a spacetime:

i) Penrose \cite{p2} \footnote{See also Ref.\cite{p0}) for definitions of
{\it asymptotically simple and weakly asymptotically simple} spacetimes,
intended to ensure that the asymptotic structure be globally the same
as that of Minkowski spacetime.} introduced the notions of {\it asymptotic 
flatness at null infinity} (i.e. along null geodesics) and of {\it asymptotic
simplicity} with his conformal completion approach. A smooth (time- and
-space orientable) spacetime $(M^4,{}^4g)$ is asymptotically simple if
there exists another smooth Lorentz manifold $({\hat M}^4, {}^4{\hat g})$
such that: i) $M^4$ is an open submanifold of ${\hat M}^4$ with smooth
boundary $\partial M^4 ={\cal S}$ smooth conformal boundary); ii) there 
exists a smooth scalar field $\Omega \geq 0$ on ${\hat M}^4$, such that 
${}^4{\hat g}=\Omega^2\, {}^4g$ on $M^4$ and $\Omega =0$, $d\Omega \not= 0$ 
on ${\cal S}$; iii) every null geodesic in $M^4$ acquires a future and past 
endpoint on ${\cal S}$. An asymptotically simple spacetime is {\it 
asymptotically flat} if vacuum Einstein equations hold in a neighbourhood of
${\cal S}$\footnote{In this case the conformal boundary ${\cal S}$ is a 
shear-free smooth null hypersurface with two connected components 
${\cal I}^{\pm}$ (scri-plus and -minus), each with
topology $S^2\times R$ and the conformal Weyl tensor vanishes on it. In the
conformal completion of Minkowski spacetime ${\cal S}$ if formed by the 
future ${\cal I}^{+}$ and past ${\cal I}^{-}$ null infinity, which join in a 
point $i^o$ representing the compactified spacelike infinity; ${\cal I}^{+}$ 
terminates in the future at a point $i^{+}$ (future timelike infinity), while 
${\cal I}^{-}$ terminates in the past at a point $i^{-}$ (past timelike 
infinity).}.

ii)  Geroch \cite{p1} introduced a definition of {\it asymptotic flatness
at spatial infinity} in terms of the {\it large distance} behaviour of
initial data on a Cauchy surface.

iii) In the projective approach \cite{p4}  a timelike hyperboloid is
introduced as the spacelike boundary of spacetime.

iv) The two definitions of asymptotic flatness at null and spatial
infinity were unified in the SPI formalism of Ashtekar and
Hanson\cite{p3}. Essentially, in the SPI approach, the spatial
infinity of the spacetime $M^4$ is compactified to a point $i^o$ and
fields on $M^4$ have direction-dependent limits at $i^o$ (this implies a 
peculiar differential structure on $\Sigma_{\tau}$ and awkward 
differentiability conditions of the 4-metric).

v) In Ref.\cite{p5}, a new kind of completion, neither conformal nor
projective, is developed by Ashtekar and Romano: now the boundary of
$M^4$ is a unit timelike hyperboloid like in the projective approach,
which, however, has a well defined contravariant normal in the
completion \footnote{There are different conformal rescalings of the 4-metric 
${}^4g \mapsto {}^4\tilde g=\Omega^2\, {}^4g$ ($\Omega \geq 0$, $\Omega =0$ is 
the boundary 3-surface of the
unphysical spacetime ${\tilde M}^4$) and
of the normal $n^{\mu} \mapsto {\tilde n}^{\mu}=
\Omega^{-4}\, n^{\mu}$.}; now, there is
no need  of awkward differentiability conditions. While in the SPI
framework each hypersurface $\Sigma_{\tau}$ has the sphere at spatial
infinity compactified at the same point $i^o$, which is the vertex for
both future ${\cal I}^{+}$ (scri-plus) and past ${\cal I}^{-}$
(scri-minus) null infinity, these properties are lost in the new
approach: each $\Sigma_{\tau}$ has as boundary at spatial infinity the
sphere $S^2_{\tau ,\infty}$ cut by $\Sigma_{\tau}$ in the timelike
hyperboliod; there is no relation between the timelike hyperboloid at
spatial infinity and ${\cal I}^{\pm}$. This new approach simplifies
the analysis of Ref.\cite{p6} of uniqueness (modulo the logarithmic
translations of Bergmann\cite{p7}) of the completion at spacelike
infinity.

vi) See Ref.\cite{rich,p15} and the recent reviews\cite{frauen,bicak}
for the status of Friedrich's conformal field
equations, derived from Einstein's equations, which arise in the study
of the compatibility of Penrose's conformal completion approach with
Einstein's equations. 
In the final description spacelike infinity is a cylinder  since  each
spacelike hypersurface has its point $i^o$ blown up to a 2-sphere$I^{o}$ at 
spatial infinity \footnote{It is interpretable as the space of spacelike 
directions at $i^o$, namely the set of the endpoints of spacelike geodesics.
This allows to define a regular initial value problem at spacelike
infinity with Minkowski data on $I^{o}$.}. The
cylinder meets future null infinity ${\cal I}^{+}$ in a sphere $I^{+}$ and
past null infinity ${\cal I}^{-}$ in a sphere $I^{-}$.
It is an open question whether the concepts of asymptotic
simplicity and conformal completion are too strong requirements. Other
reviews of the {\it problem of consistency}, i.e. whether the geometric
assumptions inherent in the existing definitions of asymptotic
flatness are compatible with Einstein equations, are given in
Refs.\cite{p14,p15}, while in Ref.\cite{p16} a review is given about
spacetimes with gravitational radiation (nearly all the results on
radiative spacetimes are at null infinity, where, for instance, the
SPI requirement of vanishing of the pseudomagnetic part of the Weyl
tensor to avoid supertranslations is too strong and destroys
radiation).

There are also  coordinate-dependent formalisms:

i) The one of Beig and Schmidt\cite{p8} \footnote{It was developed to avoid the
awkward differentiability conditions of the SPI framework and using
polar coordinates like the standard hyperbolic ones for Minkowski
spacetime and agreeing with them at {\it first order in 1/r}.}, whose
relation to the new completion is roughly the same as that between
Penrose's coordinate-independent approach to null infinity\cite{p2}
and Bondi's approach\cite{p9} based on null coordinates. The class of
spacetimes studied in Ref.\cite{p8} (called {\it radially smooth of order
m} at spatial infinity) have 4-metrics of the type

\beq
ds^2=d\rho^2(1+{{{}^1\sigma}\over {\rho}}+{{{}^2\sigma}\over
{\rho^2}}+..)^2+
\rho^2({}^oh_{rs}+{1\over {\rho}}\, {}^1h_{rs}+..)d\phi^rd\phi^s,
\label{III2}
\eeq

\noindent where
${}^oh_{rs}$ is the 3-metric on the unit timelike hyperboloid, which
completes $M^4$ at spatial infinity, and
${}^n\sigma$, ${}^nh_{rs}$, are functions on it. There are coordinate
charts $x^{\sigma}$ in $(M^4,{}^4g)$ where the 4-metric becomes

\beq
{}^4g_{\mu\nu}={}^4\eta_{\mu\nu}+
\sum^m_{n=1} {1\over {\rho^n}}\, {}^nl_{\mu\nu}({{x^{\sigma}}\over {\rho}})+
O(\rho^{-(m+1)}).
\label{III3}
\eeq

ii) The Christodoulou and Klainerman \cite{ckl} result on the
nonlinear gravitational stability of Minkowski spacetime implies a
peeling behaviour of the conformal Weyl tensor near null infinity
which is weaker than the peeling behaviour implied by asymptotic
simplicity [see Ref.\cite{p9,p2}] and this could mean that asymptotic
simplicity can be established only, if at all, with conditions
stronger than those required by these authors. In Ref.\cite{ckl} one
studies the existence of global, smooth, nontrivial solutions to
Einstein's equations without matter, which look, in the large, like
the Minkowski spacetime \footnote{These spacetimes are without singularities: 
since the requirements needed for the existence of a conformal completion
are not satisfied, it is possible to evade the singularity theorems.},
are close to Minkowski spacetime in all directions in a precise manner
(for developments of the initial data sets uniformly close to the
trivial one) and admit gravitational radiation in the Bondi sense.
These author's reformulate Einstein's
equations with the ADM variables (there are four constraint equations
plus the equations for $\partial_{\tau}\, {}^3g_{rs}$ and
$\partial_{\tau}\, {}^3K_{rs}$), put the shift functions equal to zero 
\footnote{The lapse
function is assumed equal to 1 at spatial infinity, but not everywhere
because, otherwise, one should have a finite time breakdown.} and add
the maximal slicing condition ${}^3K=0$. Then, they assume the
existence of a coordinate system $\vec \sigma$ near spatial infinity
on the Cauchy surfaces $\Sigma_{\tau}$ and of smoothness properties
for ${}^3g_{rs}$, ${}^3K_{rs}$, such that for $r=\sqrt{ {\vec
\sigma}^2}\, \rightarrow \, \infty$ the initial data set
$(\Sigma_{\tau},{}^3g_{rs},{}^3K_{rs})$ is {\it strongly asymptotically
flat}, namely \footnote{$f(\vec \sigma )$ is $o_m(r^{-k})$ if $\partial
^l f =o(r^{-k-l})$ for $l=0,1,..,m$ and $r\, \rightarrow \, \infty$.}

\begin{eqnarray}
{}^3g_{rs}&=&(1+{M\over r}) \delta_{rs} +o_4(r^{-3/2}),\nonumber \\
{}^3K_{rs}&=& o_3(r^{-5/2}),
\label{III4}
\end{eqnarray}

\noindent where the leading term in ${}^3g_{rs}$ is called the Schwarzschild
part of the 3-metric, also in absence of matter; this asymptotic
behaviour ensures the existence of the ADM energy and angular momentum
and the vanishing of the ADM momentum ({\it center-of-mass frame}). The
addition of a technical global smallness assumption on the strongly
asymptotically flat initial data leads to a unique, globally
hyperbolic, smooth and geodesically complete solution of Einstein's
equations without matter, which is globally asymptotically flat in the
sense that its Riemann curvature tensor approaches zero on any causal
or spacelike geodesic. It is also shown that the 2-dimensional space
of the dynamical degrees of freedom of the gravitational field at a
point (the reduced configuration space) is the space
of trace-free symmetric 2-covariant tensors on a 2-plane. A serious
technical difficulty \footnote{It requires the definition of an `optical
function' and reflecting the presence of gravitational radiation in
any nontrivial perturbation of Minkowski spacetime.} derives from the
`mass term' in the asymptotic Schwarzschild part of the 3-metric: it
has the long range effect of changing the asymptotic position of the
null geodesic cone relative to the maximal (${}^3K=0$) foliation
\footnote{These cones are expected to diverge logarithmically from their
positions in flat spacetime and to have their asymptotic shear
drastically different from that in Minkowski spacetime.}.

\subsection{Asymptotic Symmetries and Supertranslations.}

Let us now consider the problem of {\it asymptotic symmetries}
\cite{wald} and of the associated conserved asymptotic charges
containing the ADM Poincar\'e charges.

Like null infinity admits an infinite-dimensional group (the {\it BMS group}
\cite{p9}) of {\it asymptotic symmetries}, the SPI formalism admits an even 
bigger group, the {\it SPI group} \cite{p3}, of such symmetries. 
Both BMS and SPI algebras have an invariant 4-dimensional subalgebra of 
translations, but they also have invariant infinite-dimensional Abelian 
subalgebras (including the translation subalgebra) of so called 
{\it supertranslations} or {\it angle (or direction)-dependent
translations}. Therefore, there is an infinite number of copies of Poincar\'e
subalgebras in both BMS and SPI algebras, whose Lorentz parts are conjugate
through supertranslations \footnote{The quotient of BMS and SPI groups with 
respect to supertranslations is isomorphic to a Lorentz group.}. 
All this implies that there  is no unique definition of Lorentz generators 
and that in general relativity one cannot define intrinsically angular 
momentum and the Poincar\'e spin Casimir, so important for the classification 
of particles in Minkowski spacetime.
In Ref.\cite{mcca} it is shown that the only known Casimirs of the BMS group
are $p^2$ and one its generalization involving supertranslations. While
Poincar\'e asymptotic symmetries correspond to the ten Killing fields of the
Minkowski spacetime \footnote{An asymptotically flat spacetime tends
asymptotically to Minkowski spacetime in some way which depends on the chosen 
definition of asymptotic flatness.}, supertranslations are 
{\it angle-dependent translations}, which come just as close to satisfying 
Killing's equations asymptotically as any Poincar\'e transformation 
\cite{wald}. The problem seems to be that all known function spaces, used
for the 4-metric and for Klein-Gordon and electromagnetic fields, do not put
any restriction on the asymptotic angular behaviour of the fields, but only
restrict their radial decrease. Due to the relevance of the Poincar\'e group
for particle physics in Minkowski spacetime, and also to have a good definition
of angular momentum in general relativity\cite{p10,wald,p3}, one usually 
restricts the class of spacetimes with boundary
conditions such that supertranslations are not allowed to exist. In the SPI
framework \cite{p3}, one asks that the pseudomagnetic part of the limit of the
conformally rescaled Weyl tensor vanishes at $i^o$.
In Ref.\cite{p11} a 3+1 decomposition is made of the SPI
framework; after having reexpressed the conserved quantities at $i^o$ in terms
of canonical initial data, it is shown that to remove ambiguities connected 
with the supertranslations one must use stronger boundary conditions again 
implying the vanishing of the pseudomagnetic part of the Weyl tensor.

A related approach to these problems is given by Anderson in
Ref.\cite{reg2}. He proved a slice theorem for the action of spacetime
diffeomorphisms asymptotic to Poincar\'e transformations on the set of
asymptotically flat solutions of Einstein's equations in the context
of spatial infinity, maximal slicing and asymptotic harmonic
coordinates (as gauge conditions). There is a heuristic extension of
the momentum map method of reduction of dynamical systems with
symmetries to the diffeomorphism group. 

For metric general relativity the spatially {\it compact'} case has been 
solved in Ref.\cite{yorkmap}, with the result that, in absence of Killing 
vector fields,  the reduced phase space turns out to be a stratified 
symplectic ILH manifold\footnote{In this case the space of solutions of 
Einstein's equations is a fibered space, which is smooth at 
$({}^3g, {}^3{\tilde \Pi})$ if and only if the initial data $({}^3g,{}^3\tilde
\Pi )$ corresponds to a solution ${}^3g$ with no Killing field.}.

In the spatially {\it asymptotically flat} case, one considers
the group of those diffeomorphisms which preserve the conditions for
asymptotic flatness and the nature of this group depends strongly on
the precise asymptotic conditions. Apart from the compactification
schemes of Geroch\cite{p1} and of Ashtekar-Hansen\cite{p3}, 3 main
types of asymptotic conditions have been studied: i) the {\it finite
energy} condition of O'Murchadha\cite{p12}; ii) the York {\it quasi
isotropic} (QI) gauge conditions\cite{yoyo};
iii) the conditions of the type introduced by
Regge-Teitelboim\cite{reg} with the {\it parity conditions} introduced
by Beig-O'Murchadha\cite{reg1} plus the
gauge conditions of maximal slices and {\it 3-harmonic} asymptotic
coordinates (their existence was shown in Ref.\cite{p12}). 

These 3 types of asymptotic conditions have quite different properties. 

i) In the case of the finite energy conditions, one finds that the group
which leaves the asymptotic conditions invariant is a semidirect
product $S\, |\times L$, where L is the Lorentz group and S consists
of diffeomorphisms $\eta$ such that roughly $D^2\eta \in L^2$, i.e. it
is square integrable; S contains space- and time- translations. Under these 
conditions, it does not appear to be possible to talk about Hamiltonian 
dynamics. For a general element of the Lie algebra of $S\, |\times L$, the
corresponding momentum integral does not converge, although for the special 
case of space- and time-translations the ADM 4-momentum is well defined.

ii) QI gauge conditions of Ref.\cite{yoyo} have the desirable feature that 
{\it no supertranslations} are allowed, but a more detailed analysis reveals 
that without extra conditions, the transformations corresponding to boosts 
are not well behaved; in any case, the QI asymptotic conditions do not give a 
well defined angular and boost momentum and therefore are suitable only for 
the study of diffeomorphisms asymptotic to space- and time- translations.

iii) To get a well defined momentum for rotations and boosts, Anderson
defines asymptotic conditions which contain the parity conditions of
Ref.\cite{reg1}, but he replaces the 3-harmonic coordinates used in
this paper with York's QI conditions. The space of diffeomorphisms
$Diff_P\, M^4$, which leaves invariant the space of solutions of the
Einstein equations satisfying the parity conditions, is a semidirect
product $Diff_P\, M^4=Diff_S\, M^4\, |\times P$, where P is the
Poincar\'e group and $Diff_S\, M^4$ denotes the space of
diffeomorphisms which are asymptotic to supertranslations, which in
this case are O(1) with odd leading term. When the QI conditions are
added, the $Diff_S\, M^4$ part is restricted to $Diff_I\, M^4$, the
space of diffeomorphisms which tend to the identity at spatial
infinity \footnote{This result cannot be obtained with the finite energy
conditions\cite{p12} or from boost theorems \cite{chris}.}. 

In this way one obtains a realization of Bergmann's ideas based on his
criticism \cite{be} of general covariance: the group of coordinate 
transformations is restricted to contain an invariant Poincar\'e subgroup 
plus asymptotically trivial diffeomorphisms, analogously to what happens
with the gauge transformations of electromagnetism. 

It can be shown that the use of the parity conditions implies that the lapse 
and shift functions corresponding to the group of
supertranslations S have {\it zero momentum}. Thus, assuming the QI
conditions, the ADM momentum appears as the momentum map with respect
to the Poincar\'e group. Note that the classical form of the ADM
momentum is correct only using the restrictive assumption of parity
conditions, which are nontrivial restrictions {\it not only} on the gauge
freedom {\it but also} on the asymptotic dynamical degrees of freedom of
the gravitational field (this happens also with Ashtekar-Hansen
asymptotic condition on the Weyl tensor).

By assuming the validity of the conjecture on global existence of
solutions of Einstein's equations and of maximal slicing
\cite{york,yoyo,ciuf}  and working with
Sobolev spaces with {\it radial smoothness}, Anderson demonstrates a
{\it slice} theorem\footnote{ See Appendix B of Anderson's paper for the
definition of slice.}, according to which, assumed the parity and QI
conditions (which exclude the logarithmic translations of
Bergmann\cite{p7}), for every solution ${}^3g_o$ of Einstein's
equations one has that: i) the gauge orbit of ${}^3g_o$ is a closed
$C^1$ embedded submanifold of the manifold of solutions; ii) there
exists a submanifold containing ${}^3g_o$ which is a {\it slice for the
action of $Diff_I\, M^4$}.
York's QI conditions should be viewed as a slice condition which fixes
part of the gauge freedom at spatial infinity: i)  the $O(1/r^2)$ part
of the trace of ${}^3{\tilde \Pi}^{rs}$ must vanish; ii) if
${}^3g={}^3f+{}^3h$ (${}^3f$ is a flat metric) and if
${}^3h={}^3h_{TT}={}^3h_T+L_f(W)$ is the York decomposition\cite{york,yoyo} 
of ${}^3h$ with respect to ${}^3f$ , then the O(1) part of
the longitudinal quantity W must vanish. In this way, one selects a QI
asymptotically flat metric ${}^3g_{QI}$ and a {\it preferred frame at
spatial infinity} like in Ref.\cite{be}, i.e. preferred spacelike
hypersurfaces  corresponding to the intersections of the unit timelike
hyperboloid at spatial infinity by spatial hyperplanes in $R^4$.

Since there is no agreement among the various viewpoints on the
coordinate-independent definition of asymptotic flatness at spatial
infinity, since we are interested in the coupling of general
relativity to the standard SU(3)xSU(2)XU(1) model and since we wish to
recover the theory in Minkowski spacetime if the Newton constant is switched 
off, in this paper we shall use a coordinate-dependent approach and we shall 
work in the framework of Refs. \cite{reg,reg1}.

The boundary conditions and gauge-fixings, which will be chosen in the
next Sections, will imply an angle (i.e. direction)-independent
asymptotic limit of the canonical variables, just as it is needed in
Yang-Mills theory to have well defined covariant non-Abelian
charges\cite{ym,lusa} \footnote{As shown in Ref.\cite{lusa}, one needs a set of
Hamiltonian boundary conditions both for the fields and the gauge
transformations in the Hamiltonian gauge group,
implying angle-independent limits at spatial infinity; it is also
suggested that the elimination of Gribov ambiguity requires the use of
the following weighted Sobolev spaces\cite{moncr}
: ${\vec A}_a, {\vec E}_a \in W^{p,s-1,\delta -1}$, ${\vec B}_a
\in W^{p,s-2,\delta +2}$, ${\bar {\cal G}} \in W^{p,s,\delta}$, with
$p > 3$, $s \geq 3$, $0 \leq \delta \leq 1-{3\over p}$.}.
This is an important point for a future unified
description of general relativity and of the standard  model.

In particular, following Ref.\cite{hh}, we will
assume that at spatial infinity there is a 3-surface $S_{\infty}$ 
\footnote{It is not
necessarily a timelike hyperboloid but with outer unit (spacelike)
normal $n^{\mu}(\tau ,\vec \sigma )$, asymptotically parallel to the
spacelike hypersurfaces $\Sigma_{\tau}$.}, which intersects orthogonally the 
Cauchy surfaces $\Sigma_{\tau}$. The 3-surface $S_{\infty}$ is foliated by a 
family of 2-surfaces $S^2_{\tau ,\infty}$ coming from its intersection with the
slices $\Sigma_{\tau}$. The normals $l^{\mu}(\tau ,\vec
\sigma )$ to $\Sigma_{\tau}$ at spatial infinity, $l^{\mu}_{(\infty )}$, are
tangent to $S_{\infty}$ and normal to the corresponding $S^2_{\tau
,\infty}$. The vector $b^{\mu}_{\tau}=z^{\mu}_{\tau}=   N l^{\mu}
+N^rb^{\mu}_r$ is not in general asymptotically tangent to
$S_{\infty}$. We assume that, given a subset $U\subset M^4$ of spacetime, 
$\partial U$ consists of two slices, $\Sigma_{\tau_i}$ (the initial one) and
$\Sigma_{\tau_f}$ (the final one) with outer normals
$-l^{\mu}(\tau_i,\vec \sigma )$ and $l^{\mu}(\tau_f,
\vec \sigma )$ respectively, and of the surface $S_{\infty}$ near space 
infinity. Since we will identify special families of hypersurfaces
$\Sigma_{\tau}$ asymptotic to Minkowski hyperplanes at spatial
infinity, these families can be mapped onto the space of cross
sections of the unit timelike hyperboloid by using a lemma of Ref.
\cite{reg2}.

Let us add some information on the existence of the ADM
Lorentz boost generators:

a) In Ref.\cite{chris} on the boost problem in general relativity,
Christodoulou and O'Murchadha show (using weighted Sobolev spaces)
that a very large class of asymptotically flat initial data for
Einstein's equations have a development which includes complete
spacelike surfaces boosted relative to the initial surface.
Furthermore, the asymptotic fall off \footnote{${}^3g-{}^3f\in
W^{2,s,\delta+1/2} (\Sigma )$, ${}^3K\in W^{2,s-1,\delta +1/2}(\Sigma
)$, $s\geq 4$, $\delta > -2$.} is preserved along these boosted
surfaces and there exist a global system of harmonic coordinates on
such a development. As noted in Ref.\cite{p5}, the results of
Ref.\cite{chris} suffice to establish the existence of a large class
off spacetimes which are asymptotically flat at $i^o$ (in the sense of
Ref.\cite{p3}) in all spacelike directions along a family of Cauchy
surfaces related to one another by ``finite" boosts (it is hoped that
new results will allow to put control also on ``infinite" boosts). The
situation is unsettled with regard the existence of spacetimes
admitting both $i^o$ (in the sense of Ref.\cite{p3}) as well as smooth
${\cal I}^{\pm}$.

b) In Ref.\cite{p13}, Chru\'sciel says that for asymptotically flat
metrics ${}^3g={}^3f+O(r^{-\alpha})$, ${1\over 2} < \alpha \leq 1$, it
is not proved the {\it asymptotic symmetry conjecture} that, given any
two coordinate systems of the previous type, all twice-differentiable
coordinate transformations preserving these boundary conditions are of
type $y^{\mu}=\Lambda^{\mu}{}_{\nu}x^{\nu}+\zeta^{\mu}$ (a Lorentz 
transformation + a supertranslation $\zeta =O(r^{1-
\alpha})$): this would be needed for having the ADM 4-momentum Lorentz 
covariant. By defining $P_{\mu}$ in terms of Cauchy data on a 3-end N (a 
spacelike 3-surface $\Sigma$ minus a ball), on which ${}^3g={}^3f+
O(r^{-\alpha})$, one can evaluate the invariant mass $m(N)=
\sqrt{\epsilon P^{\mu}P_{\mu}}$. Then, provided the hypersurfaces $N_1\,\,$:
$x^o=const.$ , $N_2\,\,$: $y^o=const.$, lie
within a {\it finite} boost of each other or if the metric is a no-radiation
metric, one can show the validity of the {\it invariant mass conjecture}
$m(N_1)=m(N_2)$ for metrics satisfying vacuum Einstein equations. The main 
limitation is the lack of knowledge of long-time behaviour of Einstein's 
equations. Ashtekar-Hansen and Beig-O'Murchadha requirements are much 
stronger and restrictive than what is compatible with Einstein's equations.

\subsection{Hamiltonian Gauge Transformations in Metric Gravity.}

The counterpart of the Yang-Mills non-Abelian charges and also of the
Abelian electric charge are the asymptotic Poincar\'e charges
\cite{adm,reg,reg1,reg2,reg3,reg4}: in a natural way they should be
connected with gauge transformations of first kind (there is no
counterpart of the center of the gauge group in metric gravity).

However in Ref.\cite{p18} two alternative options are presented for the 
asymptotic Poincar\'e charges of asymptotically flat metric gravity :
\hfill\break
\hfill\break
i) There is the usual interpretation\cite{p19}, admitting gauge
transformations of first kind, in which  some observer is assumed to sit at or
just outside the boundary at spatial infinity but he/she is not explicitly
included in the action functional; this observer merely supplies a
coordinate chart on the boundaries (perhaps through his
{\it parametrization clock}), which may be used to fix the gauge of the
system at the boundary (the asymptotic lapse function). If one wishes,
this external observer may construct his clock to yield zero
Poincar\'e charges \footnote{In this way one recovers a Machian 
interpretation\cite{p20}
also in noncompact universes with boundary; there is a strong
similarity with the results of Einstein-Wheeler cosmology \cite{ciuf},
based on a closed compact universe without boundaries, for which
Poincar\'e charges are not defined.} so that every connection
with particle physics is lost;\hfill\break
\hfill\break
ii) Instead Marolf's proposal \cite{p18} is to consider the system in
isolation without the utilization of any structure outside the
boundary at spatial infinity and to consider, at the quantum level,
superselection rules for the asymptotic Poincar\'e Casimirs, in
particular for the ADM invariant mass (see Refs. \cite{p21} for
similar conclusions from different motivations). In this viewpoint the
Poincar\'e charges are {\it not} considered generators of first kind gauge
transformations and the open problem of boosts looses part of its
importance.\hfill\break
\hfill\break
In Ref.\cite{giul}, also Giulini considers a matter of physical interpretation
whether all 3-diffeomorphisms of $\Sigma_{\tau}$ into itself must be considered
as gauge transformations. In the asymptotically flat open case, he studies
{\it large} diffeomorphisms, but not the gauge transformations generated by 
the superhamiltonian constraint. After a 1-point compactification 
${\bar \Sigma}_{\tau}$ of $\Sigma_{\tau}$,
there is a study of the quotient space $Riem\, {\bar \Sigma}_{\tau}/ Diff_F\,
{\bar \Sigma}_{\tau}$, where $Diff_F\, {\bar \Sigma}_{\tau}$ are those
3-diffeomorphisms whose pullback goes to the identity at spatial infinity (the
point of compactification) where a {\it privileged oriented frame} is chosen.
Then there is a study of the decomposition of ${\bar \Sigma}_{\tau}$ into its 
prime factors as a 3-manifold, of the induced decomposition of $Diff_F\, {\bar
\Sigma}_{\tau}$ and of the evaluation of the homotopy groups of $Diff_f\,
{\bar \Sigma}_{\tau}$. The conclusion is that  the Poincar\'e charges are 
{\it not} considered as generators of gauge transformations.

We shall take the point of view that the asymptotic Poincar\'e charges
are {\it not generators of first kind gauge transformations} like in
Yang-Mills theory (the ADM energy will be shown to be the physical 
Hamiltonian for the evolution in $\tau$), that there are superselection sectors
labelled by the asymptotic Poincar\'e Casimirs and that the parameters
of the gauge transformations of ADM metric gravity have a clean
separation between a rigid part (differently from Yang-Mills theory, 
Eq.(\ref{III1}), it has both a constant and a term linear in $\vec \sigma$) 
and a proper one, namely we assume the absence of {\it improper non-rigid} 
gauge transformations like in Yang-Mills theory.

Let us now define the {\it proper} gauge transformations of the ADM
metric gravity. In Refs. \cite{witt,dew,hh} it is noted that, in
asymptotically flat spacetimes, the surface integrals arising in the
transition from the Hilbert action to the ADM action and, then, from
this to the ADM phase space action are connected with the ADM
energy-momentum of the gravitational field of the linearized theory of
metric gravity \cite{adm}, if the lapse and shift functions have
certain asymptotic behaviours at spatial infinity. Extra complications
for the differentiability of the ADM canonical Hamiltonian come from
the presence of the second spatial derivatives of the 3-metric inside
the ${}^3R$ term of the superhamiltonian constraint.  In
Ref.\cite{hh} it is also pointed out that the Hilbert action for
non-compact spacetimes is in general divergent and must be regularized
with a reference metric (static solution of Einstein's equations): for
spacetimes asymptotically flat at spatial infinity one chooses a flat
reference Minkowski metric (see later on Eqs.(\ref{V2}) and
(\ref{V3}) for the associated regularization of ADM Lorentz boosts).

By using the original ADM results\cite{adm}, Regge and
Teitelboim\cite{reg} wrote the expression of the ten conserved
Poincar\'e charges, by allowing the functions $N(\tau ,\vec \sigma )$,
$N_{\check r}(\tau ,\vec \sigma )$, to have a linear behaviour in
$\vec \sigma$ for $r=|\vec \sigma |\, \rightarrow \infty$. These charges are 
surface integrals at spatial infinity, which have to added to the Dirac
Hamiltonian so that it becomes differentiable. In Ref.\cite{reg} there
is a set of boundary conditions for the ADM canonical variables
${}^3g_{rs}(\tau ,\vec \sigma )$, ${}^3{\tilde
\Pi}^{rs}(\tau ,\vec \sigma )$, so that it is possible to define 10
surface integrals associated with the conserved Poincar\'e charges of
the spacetime (the translation charges are the ADM energy-momentum)
and to show that the functional derivatives and Poisson brackets are
well defined in metric gravity. There is no statement about gauge
transformations and supertranslations in this paper, but it is pointed
out that the lapse and shift functions have the following asymptotic
behaviour at spatial infinity (the notation used will be clarified in the
next Section)

\bea
N(\tau ,\vec \sigma )\, &\rightarrow&\, N_{(as)}(\tau ,\vec \sigma
)=-{\tilde \lambda}_{(\mu )}(\tau )l^{(\mu )}
_{(\infty )}-l^{(\mu )}_{(\infty )}{\tilde \lambda}_{(\mu )(\nu )}(\tau )
b^{(\nu )}_{(\infty) \check s}(\tau ) \sigma^{\check s}=\nonumber \\
&=&-{\tilde \lambda}_{\tau}(\tau )-{1\over 2}{\tilde \lambda}_{\tau \check
s}(\tau )\sigma^{\check s},\nonumber \\
 N_{\check r}(\tau ,\vec \sigma )\, &\rightarrow&\,
N_{(as) \check r}(\tau ,\vec \sigma )=-b^{(\mu )}_{(\infty ) \check
r}(\tau ) {\tilde \lambda}_{(\mu )}(\tau )-b^{(\mu )}_{(\infty )
\check r}(\tau ){\tilde
\lambda}_{(\mu )(\nu )}(\tau ) b^{(\nu )}_{(\infty ) \check s}(\tau ) \sigma
^{\check s}=\nonumber \\
&=&-{\tilde \lambda}_{\check r}(\tau )-{1\over 2}{\tilde
\lambda}_{\check r\check s}(\tau ) \sigma^{\check s},\nonumber \\
 &&{}\nonumber \\
 &&{\tilde \lambda}_A(\tau )={\tilde \lambda}_{(\mu )}(\tau )b^{(\mu )}
_{(\infty )A}(\tau ),\quad\quad {\tilde \lambda}_{(\mu )}(\tau )=b^A
_{(\infty )(\mu )}(\tau ){\tilde \lambda}_A(\tau ),\nonumber \\
&&{\tilde \lambda}_{AB}(\tau )={\tilde \lambda}_{(\mu )(\nu )}(\tau )
[b^{(\mu )}_{(\infty )A}b^{(\nu )}_{(\infty )B}-b^{(\nu )}
_{(\infty )A}b^{(\mu )}_{(\infty )B}](\tau )=2[{\tilde \lambda}_{(\mu )(\nu )}
b^{(\mu )}_{(\infty )A}b^{(\nu )}_{(\infty )B}](\tau ),\nonumber \\
&&{\tilde \lambda}_{(\mu )(\nu )}(\tau )={1\over 4}[b^A_{(\infty)(\mu )}b^B
_{(\infty )(\nu )}-b^B_{(\infty )(\mu )}b^A_{(\infty )(\nu )}](\tau ){\tilde
\lambda}_{AB}(\tau )=\nonumber \\
&&={1\over 2} [b^A_{(\infty )(\mu )}b^B_{(\infty )(\nu )}{\tilde \lambda}
_{AB}](\tau ),
\label{III5}
\end{eqnarray}

Let us remark that with this asymptotic behaviour any 3+1 splitting of
the spacetime $M^4$ is in some sense {\it ill-defined} because the
associated foliation with leaves $\Sigma_{\tau}$  has diverging proper
time interval $N(\tau ,\vec \sigma ) d\tau$ and  shift functions at
spatial infinity for each fixed $\tau$. Only in those gauges where
${\tilde \lambda}_{AB}(\tau )=-{\tilde \lambda}_{BA}(\tau )=0$ these
problems disappear. These problems are connected with  the boost
problem quoted in the previous Subsection: also they suggest that the 
asymptotic Lorentz algebra (and therefore also the Poincar\'e and SPI 
algebras) have not to be interpreted as generators of improper gauge 
transformations. 

A more complete analysis, including also a discussion of supertranslations in
the ADM canonical formalism, has been given by Beig and O'Murchadha\cite{reg1}
(extended to Ashtekar's formalism in Ref.\cite{reg3}). They consider
3-manifolds $\Sigma_{\tau}$ diffeomorphic to $R^3$ as in this paper, so that 
there exist {\it global coordinate systems}. If $\{ \sigma^{\check r}\}$ is 
one of these global coordinate systems on $\Sigma_{\tau}$, the 3-metric 
${}^3g_{\check r\check s}(\tau ,\sigma^{\check t})$, evaluated in this 
coordinate system, is assumed asymptotically Euclidean in the following 
sense: if $r=\sqrt{\delta_{\check r\check s}\sigma^{\check r}\sigma
^{\check s}}$ \footnote{One could put $r=
\sqrt{{}^3g_{\check r\check s}\sigma^{\check r}\sigma^{\check s}}$ and get the
same kind of decomposition.}, then one assumes

\begin{eqnarray}
{}^3g_{\check r\check s}(\tau ,\vec \sigma )&=&\delta_{\check r\check s}+
{1\over r}\,\, {}^3s_{\check r\check s}(\tau ,{{\sigma^{\check n}}\over r})+
{}^3h_{\check r\check s}(\tau ,\vec \sigma ),\quad r\, \rightarrow \infty ,
\nonumber \\
&&{}\nonumber \\
&&{}^3s_{\check r\check s}(\tau ,{{\sigma^{\check n}}\over r})={}^3s_{\check
r\check s}(\tau ,-{{\sigma^{\check n}}\over r}),\quad EVEN\, PARITY,
\nonumber \\
&&{}^3h_{\check r\check s}(\tau ,\vec \sigma )= O((r^{-(1+\epsilon )}),\quad
\epsilon > 0,\, for\, r\, \rightarrow \infty ,\nonumber \\
&&\partial_{\check u}\, {}^3h_{\check r\check s}(\tau ,\vec \sigma )=O(r^{-
(2+\epsilon )}).
\label{III6}
\end{eqnarray}

\noindent The functions ${}^3s_{\check r\check s}(\tau ,{{\sigma^{\check n}}
\over r})$ are $C^{\infty}$ on the sphere $S^2_{\tau ,\infty}$ at spatial
infinity of $\Sigma_{\tau}$; if they would be of odd parity, the ADM
energy would vanish. The difference ${}^3g_{\check r\check s}(\tau
,\vec
\sigma )-\delta_{\check r\check s}$ cannot fall off faster that 1/r,
because otherwise the ADM energy would be zero and the positivity
energy theorem \cite{p17} would imply that the only solution of the
constraints is flat spacetime. 

For the ADM momentum one assumes the following boundary conditions

\begin{eqnarray}
{}^3{\tilde \Pi}^{\check r\check s}(\tau ,\vec \sigma )&=&{1\over {r^2}}\,\,
{}^3t^{\check r\check s}(\tau ,{{\sigma^{\check n}}\over r})+{}^3k^{\check
r\check s}(\tau ,\vec \sigma ),\quad r\, \rightarrow \infty ,\nonumber \\
&&{}\nonumber \\
&&{}^3t^{\check r\check s}(\tau ,{{\sigma^{\check n}}\over r})=-{}^3t^{\check
r\check s}(\tau ,-{{\sigma^{\check n}}\over r}),\quad ODD\, PARITY,\nonumber \\
&&{}^3k^{\check r\check s}(\tau ,\vec \sigma )=O(r^{-(2+\epsilon )}),\quad
\epsilon > 0,\quad r\, \rightarrow \infty .
\label{III7}
\end{eqnarray}

\noindent If ${}^3{\tilde \Pi}^{\check r\check s}(\tau ,\vec \sigma )$ were to
fall off faster than $1/r^2$, the ADM linear momentum would vanish and we could
not consider Lorentz transformations. In this way, the integral $\int_{\Sigma
_{\tau}}d^3\sigma [{}^3{\tilde \Pi}^{\check r\check s} \delta \, {}^3g_{\check
r\check s}](\tau ,\vec \sigma )$ is well defined and finite: since the 
integrand is of order $O(r^{-3})$, a possible logarithmic divergence is 
avoided due to the odd parity of ${}^3t^{\check r\check s}$.

These boundary conditions imply that functional derivatives and Poisson
brackets  are well defined \cite{reg1}. In a more rigorous treatment one should
use appropriate weighted Sobolev spaces.

The supermomentum and superhamiltonian constraints, see
Eqs.(\ref{II17}), ${}^3{\tilde {\cal H}}^{\check r}(\tau ,\vec
\sigma )\approx 0$ and ${\tilde {\cal H}}(\tau ,
\vec \sigma )\approx 0$, are even functions of $\vec \sigma$ of order $O(r
^{-3})$. Their smeared version with the lapse and shift functions, appearing
in the canonical Hamiltonian (\ref{II18}), will give a finite and 
differentiable $H_{(c)ADM}$ if we assume\cite{reg1}

\begin{eqnarray}
N(\tau ,\vec \sigma )&=&m(\tau ,\vec \sigma )=s(\tau ,\vec \sigma
)+n(\tau ,\vec \sigma )\, {\rightarrow}_{r\, \rightarrow
\infty} \nonumber \\
&{\rightarrow}_{r\, \rightarrow \infty}\,& k(\tau ,{{\sigma^{\check
n}}\over r})+O(r^{-\epsilon}),\quad
\epsilon >0,\nonumber \\
N_{\check r}(\tau ,\vec \sigma )&=&m_{\check r}(\tau ,\vec \sigma
)=s_{\check r}(\tau ,\vec \sigma )+n_{\check r}(\tau ,\vec \sigma ) \,
{\rightarrow}_{r\, \rightarrow \infty} \nonumber \\
&{\rightarrow}_{r\, \rightarrow \infty} \,& k_{\check r}(\tau ,
{{\sigma^{\check n}}\over r})+O(r^{-\epsilon}),\nonumber \\
&&{}\nonumber \\
 &&s(\tau ,\vec
\sigma )=k(\tau ,{{\sigma^{\check n}}\over r})=-k(\tau ,-{{\sigma
^{\check n}}\over r}),\quad ODD\, PARITY,\nonumber \\
&&s_{\check r}(\tau ,\vec \sigma )=k_{\check r}(\tau ,{{\sigma^{\check n}}
\over r})=-k_{\check r}(\tau ,-{{\sigma^{\check n}}\over r}),\quad\quad
ODD\, PARITY,
\label{III8}
\end{eqnarray}

\noindent with $n(\tau ,\vec \sigma )$, $n_{\check r}(\tau ,\vec \sigma)$ 
going to zero for $r\, \rightarrow \infty$ like $O(r^{-\epsilon})$ in an
angle-independent way.

With these boundary conditions one gets differentiability, i.e.
$\delta H_{(c)ADM}$ is linear in $\delta \, {}^3g_{\check r\check s}$
and $\delta \, {}^3{\tilde \Pi}^{\check r\check s}$, with the
coefficients being the Dirac-Hamilton equations of metric gravity.
Therefore, since N and $N_{\check r}$ are a special case of the
parameter fields for the most general infinitesimal gauge
transformations generated by the first class constraints ${\tilde
{\cal H}}$, ${}^3{\tilde {\cal H}}^{\check r}$, with generator $G=\int
d^3\sigma [ \alpha {\tilde {\cal H}}+\alpha_{\check r}\, {}^3{\tilde
{\cal H}}^{\check r}](\tau ,\vec \sigma )$, the {\it proper} gauge
transformations preserving Eqs.(\ref{IV6}) and (\ref{IV7}) have the
multiplier fields $\alpha (\tau ,\vec \sigma )$ and $\alpha_{\check
r}(\tau ,\vec \sigma )$ with the same boundary conditions (\ref{IV8})
of $m(\tau ,\vec \sigma)$ and $m_{\check r}(\tau ,\vec \sigma )$.
Then, the Hamilton equations imply that also the Dirac multipliers
$\lambda_N(\tau ,\vec \sigma )$ and $\lambda^{\vec N}
_{\check r}(\tau ,\vec \sigma )$ have these boundary conditions  [$\lambda_N\,
{\buildrel \circ \over =}\, \delta N$, $\lambda^{\vec N}_{\check r}\, 
{\buildrel \circ \over =}\, \delta N_{\check r}$]. Instead, the momenta 
${\tilde \pi}^N(\tau ,\vec \sigma )$ and ${\tilde \pi}^{\check r}
_{\vec N}(\tau ,\vec \sigma )$, conjugate to N and $N_{\check r}$, must be of
$O(r^{-(3+\epsilon )})$ to have $H_{(D)ADM}$ finite.

The angle-dependent functions $s(\tau ,\vec \sigma )=k(\tau 
,{{\sigma^{\check n}}\over r})$ and $s_{\check r}(\tau ,\vec \sigma )=
k_{\check r}(\tau ,{{\sigma^{\check n}}\over r})$ on $S^2_{\tau ,\infty}$  
are called {\it odd time and space supertranslations}. The
piece $\int d^3\sigma [s\, {\tilde {\cal H}}+s_{\check r}\,
{}^3{\tilde {\cal H}}^{\check r} ](\tau ,\vec \sigma ) \approx 0$ of
the Dirac Hamiltonian is the Hamiltonian generator of
supertranslations (the {\it zero momentum} of supertranslations of
Ref.\cite{reg2}). Their contribution to gauge transformations is to
alter the angle-dependent asymptotic terms ${}^3s_{\check r\check s}$
and ${}^3t^{\check r\check s}$ in ${}^3g_{\check r\check s}$ and
${}^3{\tilde \Pi}^{\check r\check s}$. While Sachs\cite{p9} gave an
explicit form of the generators (including supertranslations) of the
algebra of the BMS group of asymptotic symmetries, no such form is
explicitly known for the generators of the SPI group.

With $N=m$, $N_{\check r}=m_{\check r}$ one can verify the validity of
the smeared form of the Dirac algebra (\ref{II19}) of the superhamiltonian and
supermomentum constraints:

\begin{eqnarray}
&&\lbrace H_{(c)ADM}[m_1, m_1^{\check r}], H_{(c)ADM}[m_2,m_2^{\check
r}]\rbrace = \nonumber \\
 &&{}\nonumber \\
 && =H_{(c)ADM} [m_2^{\check r}\,
{}^3\nabla_{\check r} m_1 -m_1^{\check r}\, {}^3\nabla
_{\check r} m_2,\, {\cal L}_{{\vec m}_2}\, m_1^{\check r}+m_2\,
{}^3\nabla^{\check r} m_1 -m_1\, {}^3\nabla^{\check r} m_2],\nonumber \\
 &&
\label{III9}
\end{eqnarray}

\noindent with $m^{\check r}={}^3g^{\check r\check s} m_{\check s}$ and with
$H_{(c)ADM}[m,m^{\check r}]=\int d^3\sigma [m {\tilde {\cal H}}+m^{\check r}\,
{}^3{\tilde {\cal H}}_{\check r}](\tau ,\vec \sigma )=
\int d^3\sigma [m {\tilde {\cal H}}+m_{\check r}\,
{}^3{\tilde {\cal H}}^{\check r}](\tau ,\vec \sigma )$.

When the functions $N(\tau ,\vec \sigma )$ and $N_{\check r}(\tau
,\vec \sigma )$ [and also $\alpha (\tau ,\vec \sigma )$,
$\alpha_{\check r}(\tau ,\vec \sigma )$] do not have the asymptotic
behaviour of $m(\tau ,\vec \sigma )$ and $m_{\check r}(\tau ,\vec
\sigma )$ respectively, one speaks of {\it improper} gauge
transformations, because $H_{(D)ADM}$ is not differentiable even at
the constraint hypersurface.

At this point one has identified:

a) Certain global coordinate systems $\{ \sigma^{\check r} \}$ on the 
spacelike 3-surface $\Sigma_{\tau}$, which hopefully define a minimal atlas
${\cal C}_{\tau}$ for the spacelike hypersurfaces $\Sigma_{\tau}$ foliating 
the asymptotically flat spacetime $M^4$. With the ${\cal C}_{\tau}$'s and the 
parameter $\tau$ as $\Sigma_{\tau}$-adapted coordinates of $M^4$ one should 
build an atlas ${\cal C}$ of allowed coordinate systems of $M^4$.

b) A set of boundary conditions on the fields on $\Sigma_{\tau}$ (i.e. a
function space for them) ensuring that the 3-metric on $\Sigma_{\tau}$ is
asymptotically Euclidean in this minimal atlas (modulo 3-diffeomorphisms, see
the next point).

c) A set of {\it proper} gauge transformations generated infinitesimally by the
first class constraints, which leave the fields on $\Sigma_{\tau}$ in the
chosen function space. Since the gauge transformations generated by the
supermomentum constraints ${}^3{\tilde {\cal H}}^{\check r}(\tau ,\vec \sigma
)\approx 0$ are the lift to the space of the tensor fields on $\Sigma_{\tau}$
(which contains the phase space of metric gravity) of the 3-diffeomorphisms
$Diff\, \Sigma_{\tau}$ of $\Sigma_{\tau}$ into itself, the restriction of
$N(\tau ,\vec \sigma )$, $N_{\check r}(\tau ,\vec \sigma )$ to $m(\tau ,\vec
\sigma )$, $m_{\check r}(\tau ,\vec \sigma )$, ensures that these
3-diffeomorphisms are restricted to be compatible with the chosen
minimal atlas for $\Sigma_{\tau}$ [this is the problem of the
coordinate transformations preserving Eq.(\ref{III6})]. The discussion of the 
meaning of the gauge transformations generated by the superhamiltonian 
constraint is delayed to Sections VI and IX.
Also the parameter fields $\alpha (\tau ,\vec \sigma )$,
$\alpha_{\check r}(\tau ,\vec
\sigma )$, of arbitrary (also improper) gauge transformations should
acquire this behaviour. 

Since the ADM Poincar\'e charges are not considered as {\it extra
improper gauge transformations} ({\it Poincar\'e transformations at
infinity}) but as numbers individuating superselection sectors, they cannot 
alter the assumed asymptotic behaviour of the fields.

Let us remark at this point that the addition of gauge-fixing
constraints to the superhamiltonian and supermomentum constraints must
happen in the chosen function space for the fields on $\Sigma_{\tau}$.
Therefore, the time constancy of these gauge-fixings will generate
secondary gauge-fixing constraints for the restricted lapse and shift
functions $m(\tau ,\vec \sigma )$, $m_{\check r}(\tau ,\vec \sigma )$.

These results, in particular Eqs. (\ref{III5}) and (\ref{III8}), suggest
to assume the following form for the lapse and shift functions 

\bea
N(\tau ,\vec \sigma ) &=& N_{(as)}(\tau ,\vec \sigma ) +\tilde N(\tau
,\vec \sigma ) + m(\tau ,\vec \sigma ),\nonumber \\
 N_{\check r}(\tau ,\vec \sigma ) &=& N_{(as)\check r}(\tau ,\vec \sigma ) +
 {\tilde N}_{\check r}(\tau ,\vec \sigma ) +m_{\check r}(\tau ,\vec \sigma ),
 \label{III10}
 \eea

 \noindent with $\tilde N$, ${\tilde N}_{\check r}$ describing improper
 gauge transformations not of first kind. Since, like in Yang-Mills theory,
 they do not play any role in the dynamics of metric gravity, we shall
 assume that they must be absent, so that [see Eq.(\ref{III1})]  we can
 parametrize the lapse and shift functions in the following form

\begin{eqnarray}
N(\tau ,\vec \sigma )\, &=& N_{(as)}(\tau
,\vec \sigma )+m(\tau ,\vec \sigma ),\nonumber \\
N_{\check r}(\tau ,\vec \sigma )\, &=&
N_{(as) \check r}(\tau ,\vec \sigma )+m_{\check r}(\tau ,\vec \sigma ),
\label{III11}
\end{eqnarray}

\noindent The improper parts $N_{(as)}$, $N_{(as) r}$, given in 
Eqs.(\ref{III5}),
behave as the lapse and shift functions associated with spacelike
hyperplanes in Minkowski spacetime in parametrized Minkowski theories.

\vfill\eject

\section{Dirac's Approach to Asymptotically Flat Metric Gravity.}

In Ref.\cite{p22} and in the book in Ref.\cite{dirac} (see also 
Ref.\cite{reg}), Dirac introduced asymptotic Minkowski Cartesian coordinates

\begin{equation}
z^{(\mu )}
_{(\infty )}(\tau ,\vec \sigma )=x^{(\mu )}_{(\infty )}(\tau )+b^{(\mu )}
_{(\infty )\, \check r}(\tau ) \sigma^{\check r}
\label{IV1}
\end{equation}

\noindent in $M^4$ at spatial infinity $S_{\infty}=\cup_{\tau} S^2_{\tau
,\infty}$ \footnote{Here $\{ \sigma^{\check r} \}$ are the previous global 
coordinate charts of the atlas ${\cal C}_{\tau}$ of $\Sigma_{\tau}$, not 
matching the spatial coordinates $z^{(i)}_{(\infty )}(\tau ,\vec \sigma )$.}.
For each value of $\tau$, the coordinates $x^{(\mu )}
_{(\infty )}(\tau )$ labels an arbitrary point, near spatial infinity
chosen as origin. On it there is a flat tetrad $b^{(\mu )}_{(\infty )\, A}
(\tau )= (\, l^{(\mu )}_{(\infty )}=b^{(\mu )}_{(\infty )\, \tau}=\epsilon
^{(\mu )}{}_{(\alpha )(\beta )(\gamma )} b^{(\alpha )}_{(\infty )\, \check 1}
(\tau )b^{(\beta )}_{(\infty )\, \check 2}(\tau )b^{(\gamma )}_{(\infty )\,
\check 3}(\tau );\, b^{(\mu )}_{(\infty )\, \check r}(\tau )\, )$, with
$l^{(\mu )}_{(\infty )}$ $\tau$-independent, satisfying $b^{(\mu
)}_{(\infty )\, A}\, {}^4\eta_{(\mu )(\nu )}\, b^{(\nu )}_{(\infty )\,
B}={}^4\eta_{AB}$ for every $\tau$. There will be transformation
coefficients $b^{\mu}_A(\tau ,\vec
\sigma )$ from the  adapted coordinates $\sigma^A=(\tau ,\sigma
^{\check r})$ to coordinates $x^{\mu}=z^{\mu}(\sigma^A)$ in an atlas of $M^4$,
such that in a chart at spatial infinity one has $z^{\mu}(\tau ,\vec \sigma )
\rightarrow \delta^{\mu}_{(\mu )} z^{(\mu )}(\tau ,\vec \sigma )$ and $b^{\mu}
_A(\tau ,\vec \sigma ) \rightarrow
\delta^{\mu}_{(\mu )} b^{(\mu )}_{(\infty )A}(\tau )$
\footnote{For $r\, \rightarrow \, \infty$ one has ${}^4g_{\mu\nu}\, 
\rightarrow \, \delta^{(\mu )}_{\mu}\delta^{(\nu )}_{\nu}{}^4\eta
_{(\mu )(\nu )}$ and ${}^4g_{AB}=b^{\mu}_A\, {}^4g_{\mu\nu} b^{\nu}_B\, 
\rightarrow \, b^{(\mu )}_{(\infty )A}\, {}^4\eta_{(\mu )(\nu )} 
b^{(\nu )}_{(\infty )B}= {}^4\eta_{AB}$.}. The atlas ${\cal C}$ of the 
allowed coordinate systems of $M^4$ is assumed to have this property.

Dirac\cite{p22} and, then, Regge and Teitelboim\cite{reg} proposed
that the asymptotic Minkowski Cartesian coordinates $z^{(\mu
)}_{(\infty )}(\tau ,\vec \sigma )=x^{(\mu )}_{(\infty )} (\tau
)+b^{(\mu )}_{(\infty ) \check r}(\tau )\sigma^{\check r}$ should
define 10 new independent degrees of freedom at the spatial boundary
$S_{\infty}$ (with ten associated conjugate momenta), as it
happens for Minkowski parametrized theories\cite{lus1,india,crater,iten,mate} 
(see Appendix A) when the extra
configurational variables $z^{(\mu )}(\tau ,\vec \sigma )$ are reduced
to 10 degrees of freedom by the restriction to spacelike hyperplanes,
defined by $z^{(\mu )}(\tau ,\vec \sigma )\approx x^{(\mu )}_s(\tau
)+b^{(\mu )}_{\check r}(\tau )\sigma^{\check r}$.

In Dirac's approach to metric gravity  the 20 extra variables of the
Dirac proposal can be chosen as the set: $x^{(\mu )}_{(\infty )}(\tau
)$, $p^{(\mu )}_{(\infty )}$, $b^{(\mu )}_{(\infty ) A}(\tau )$ \footnote{With
$b^{(\mu )}_{(\infty ) \tau }=l^{(\mu )}_{(\infty )}$
$\tau$-independent and coinciding with the asymptotic normal to
$\Sigma_{\tau}$, tangent to $S_{\infty}$.}, $S^{(\mu )(\nu )}_{(\infty
)}$, with the  Dirac brackets (\ref{a11}) implying the orthonormality
constraints $b^{(\mu )}_{(\infty ) A}\, {}^4\eta_{(\mu )(\nu )}
b^{(\nu )}_{(\infty ) B}={}^4\eta_{AB}$. Moreover, $p^{(\mu
)}_{(\infty )}$ and $J^{(\mu )(\nu )}_{(\infty )}=x^{(\mu )}_{(\infty
)}p^{(\nu )}_{(\infty )}- x^{(\nu )}_{(\infty )}p^{(\mu )}_{(\infty
)}+S^{(\mu )(\nu )}_{(\infty )}$ satisfy a Poincar\'e algebra. In
analogy with Minkowski parametrized theories restricted to spacelike
hyperplanes, one expects to have 10 extra first class constraints of
the type

\beq
p^{(\mu )}_{(\infty )}-P^{(\mu )}_{ADM}\approx 0,\quad\quad S^{(\mu
)(\nu )}_{(\infty )}-S^{(\mu )(\nu )}_{ADM}\approx 0,
\label{IV2}
\eeq

\noindent
with $P^{(\mu )}_{ADM}$, $S^{(\mu )(\nu )}_{ADM}$ related to the ADM
Poincar\'e charges $P^A_{ADM}$, $J^{AB}_{ADM}$ (which will be defined
in the next Section) in place of $P^{(\mu )}_{sys}$, $S^{(\mu )(\nu
)}_{sys}$ and 10 extra Dirac multipliers ${\tilde \lambda}
_{(\mu )}(\tau )$, ${\tilde \lambda}_{(\mu )(\nu )}(\tau )$, in front of them
in the Dirac Hamiltonian. The origin $x^{(\mu )}_{(\infty )}(\tau )$
is going to play the role of an {\it external} decoupled observer with
his parametrized clock. The main problem with respect to Minkowski
parametrized theory on spacelike hyperplanes is that it is not known
which could be the ADM spin part $S_{ADM}^{(\mu )(\nu )}$ of the ADM
Lorentz charge $J^{(\mu )(\nu )}_{ADM}$.

The way out from these problems is based on the following observation.
If we replace $p^{(\mu )}_{(\infty )}$ and $S^{(\mu )(\nu )}_{(\infty
)}$, whose Poisson algebra is the direct sum of an Abelian algebra of
translations and of a Lorentz algebra, with the new variables (with
indices adapted to $\Sigma_{\tau}$)

\beq
p^A_{(\infty )}=b^A_{(\infty )(\mu )}p^{(\mu )}
_{(\infty )}, \quad\quad
J^{AB}_{(\infty )}\, {\buildrel {def} \over =}\, b^A_{(\infty )(\mu
)}b^B_{(\infty )(\nu )} S^{(\mu )(\nu )}_{(\infty )} [\not=
b^A_{(\infty )(\mu )}b^B_{(\infty )(\nu )} J^{(\mu )(\nu )}_{(\infty
)}],
\label{IV3}
\eeq

\noindent
the Poisson brackets for $p^{(\mu )}
_{(\infty )}$, $b^{(\mu )}_{(\infty ) A}$, $S^{(\mu )(\nu )}_{(\infty )}$
\footnote{One has $\lbrace b^A_{(\infty )(\gamma )},S^{(\nu )(\rho )}
_{(\infty )}\rbrace =\eta^{(\nu )}_{(\gamma )}b_{(\infty )}^{A (\rho )}-
\eta^{(\rho )}_{(\gamma )}b_{(\infty )}^{A(\nu )}$.}, imply

\begin{eqnarray}
&&\lbrace p^A_{(\infty )},p^B_{(\infty )}\rbrace =0,\nonumber \\
&&\lbrace p^A_{(\infty )},J^{BC}_{(\infty )}\rbrace ={}^4g^{AC}_{(\infty )}p^B
_{(\infty )}-{}^4g^{AB}_{(\infty )} p^C_{(\infty )}, \nonumber \\
&&\lbrace J^{AB}
_{(\infty )},J^{CD}_{(\infty )}\rbrace =-(\delta^B_E\delta^C_F\, {}^4g^{AD}
_{(\infty )}+\delta^A_E\delta^D_F\, {}^4g^{BC}_{(\infty )}-\delta^B_E\delta^D
_F\, {}^4g^{AC}_{(\infty )}-\delta^A_E\delta^C_F\, {}^4g^{BD}_{(\infty )})J
^{EF}_{(\infty )}=\nonumber \\
&&=-C^{ABCD}_{EF} J^{EF}_{(\infty )},
\label{IV4}
\end{eqnarray}

\noindent where ${}^4g^{AB}_{(\infty )}=b^A_{(\infty )(\mu )}\,
{}^4\eta^{(\mu )(\nu )} b^B_{(\infty )(\nu )}={}^4\eta^{AB}$ since the
$b^{(\mu )}_{(\infty )A}$ are flat tetrad in both kinds of indices.
Therefore, we get the algebra of a realization of the Poincar\'e group (this
explains the notation $J^{AB}_{(\infty )}$) with all the structure constants
inverted in the sign (transition from a left to a right action).

This  implies that, after the transition to the asymptotic Dirac Cartesian 
coordinates the Poincar\'e generators ${P}^A_{ADM}$, ${J}^{AB}_{ADM}$
in $\Sigma_{\tau}$-adapted coordinates should  become a momentum
${P}^{(\mu )}_{ADM}=b^{(\mu )}_A  P^A_{ADM}$ and only an  ADM spin tensor 
${S}^{(\mu )(\nu )}_{ADM}$ \footnote{To define an angular
momentum tensor $J^{(\mu )(\nu )}_{ADM}$ one should find an ``external center
of mass of the gravitational field" $X^{(\mu )}_{ADM} [{}^3g,
{}^3{\tilde \Pi}]$ (see Ref.\cite{lon,mate}
for the Klein-Gordon case) conjugate to
$P^{(\mu )}_{ADM}$, so that $J^{(\mu )(\nu )}_{ADM}=X^{(\mu )}_{ADM}P^{(\nu )}
_{ADM}-X^{(\nu )}_{ADM}P^{(\mu )}_{ADM}+S^{(\mu )(\nu )}_{ADM}$.}.

As a consequence of the previous results we shall assume the existence
of a global coordinate system $\{ \sigma^{\check r} \}$ on
$\Sigma_{\tau}$ , in which we have

\begin{eqnarray}
N(\tau ,\vec \sigma )\, &=& N_{(as)}(\tau
,\vec \sigma )+m(\tau ,\vec \sigma ),\nonumber \\
N_{\check r}(\tau ,\vec \sigma )\, &=&
N_{(as) \check r}(\tau ,\vec \sigma )+m_{\check r}(\tau ,\vec \sigma ),
\nonumber \\
&&{}\nonumber \\
N_{(as)}(\tau ,\vec \sigma )&=&-{\tilde \lambda}_{(\mu )}(\tau )l^{(\mu )}
_{(\infty )}-l^{(\mu )}_{(\infty )}{\tilde \lambda}_{(\mu )(\nu )}(\tau )
b^{(\nu )}_{(\infty) \check s}(\tau ) \sigma^{\check s}=\nonumber \\
&=&-{\tilde \lambda}_{\tau}(\tau )-{1\over 2}{\tilde \lambda}_{\tau \check
s}(\tau )\sigma^{\check s},\nonumber \\
N_{(as) \check r}(\tau ,\vec \sigma )&=&-b^{(\mu )}_{(\infty ) \check r}(\tau )
{\tilde \lambda}_{(\mu )}(\tau )-b^{(\mu )}_{(\infty ) \check r}(\tau ){\tilde
\lambda}_{(\mu )(\nu )}(\tau ) b^{(\nu )}_{(\infty ) \check s}(\tau ) \sigma
^{\check s}=\nonumber \\
&=&-{\tilde \lambda}_{\check r}(\tau )-{1\over 2}{\tilde
\lambda}_{\check r\check s}(\tau ) \sigma^{\check s},
\label{IV5}
\end{eqnarray}

\noindent with $m(\tau ,\vec \sigma )$, $m_{\check r}(\tau ,\vec \sigma )$,
given by Eqs.(\ref{IV8}): they still contain odd supertranslations.

This very strong assumption implies that one is selecting
asymptotically at spatial infinity only coordinate systems in which the lapse
and shift functions have behaviours similar to those of
Minkowski spacelike hyperplanes, so that the allowed foliations of the 3+1
splittings of the spacetime $M^4$ are restricted to have the leaves $\Sigma
_{\tau}$ approaching these Minkowski hyperplanes at spatial infinity in a way
independent from the direction. But this is coherent with Dirac's
choice of asymptotic Cartesian coordinates (modulo 3-diffeomorphisms
not changing the nature of the coordinates, namely tending to the identity
at spatial infinity like in Ref.\cite{reg2}) and with the assumptions
used to define the asymptotic Poincar\'e charges. It is  also needed
to eliminate coordinate transformations not becoming the identity at
spatial infinity, which are not associated with the gravitational
fields of isolated systems \cite{ll}.

By replacing the ADM configuration variables $N(\tau ,\vec \sigma )$ and 
$N_{\check r}(\tau ,\vec \sigma )$ with the new ones
${\tilde \lambda}_A(\tau )=\{ {\tilde \lambda}_{\tau}(\tau );
{\tilde \lambda}_{\check r}(\tau ) \}$, ${\tilde \lambda}_{AB}(\tau )=
-{\tilde \lambda}_{BA}(\tau )$, $n(\tau ,\vec \sigma )$, $n_{\check r}(\tau
,\vec \sigma )$ inside the ADM Lagrangian, one only gets the replacement of the
primary first class constraints of ADM metric gravity

\beq
{\tilde \pi}^N(\tau ,\vec \sigma )\approx 0,\quad\quad {\tilde \pi}
_{\vec N}^{\check r}(\tau ,\vec \sigma )\approx 0,
\label{IV6}
\eeq

\noindent
with the new first class constraints

\beq
{\tilde \pi}^n(\tau ,\vec \sigma )\approx 0,\quad {\tilde \pi}
_{\vec n}^{\check r}(\tau ,\vec \sigma )\approx 0,\quad {\tilde \pi}^A(\tau )
\approx 0,\quad {\tilde \pi}^{AB}(\tau )=-{\tilde \pi}^{BA}(\tau )\approx 0,
\label{IV7}
\eeq

\noindent
corresponding to the vanishing of the canonical momenta ${\tilde
\pi}^A$, ${\tilde \pi}^{AB}$ conjugate to the new configuration variables
\footnote{We assume the Poisson
brackets $\{ {\tilde \lambda}_A(\tau ),{\tilde \pi}^B(\tau )\}
=\delta^B_A$, $\{ {\tilde \lambda}_{AB}(\tau ), {\tilde \pi}^{CD}(\tau
) \} =\delta^C_A \delta^D_B-\delta^D_A \delta^C_B$.}. The only change in the
Dirac Hamiltonian of metric gravity $H_{(D)ADM}=H_{(c)ADM}+\int
d^3\sigma [\lambda_N{\tilde \pi}^N+\lambda^{\vec N}_{\check r}{\tilde
\pi}^{\check R}_{\vec N}](\tau ,\vec \sigma )$, $H_{(c)ADM}=\int d^3\sigma 
[N{\tilde {\cal H}}+N_{\check r}{\tilde {\cal H}}^{\check r}](\tau ,\vec \sigma
)$ of Eq.(\ref{II16})  is

\beq
\int d^3\sigma [\lambda_N {\tilde \pi}^N+\lambda
^{\vec N}_{\check r} {\tilde \pi}_{\vec N}^{\check r}](\tau ,\vec \sigma )\,
\mapsto \,
\zeta_A(\tau ) {\tilde \pi}^A(\tau )+\zeta_{AB}(\tau ) {\tilde \pi}^{AB}(\tau )
+\int d^3\sigma [\lambda_n {\tilde \pi}^n+\lambda^{\vec n}_{\check r} {\tilde
\pi}_{\vec n}^{\check r}](\tau ,\vec \sigma ),
\label{IV8}
\eeq

\noindent
with $\zeta_A(\tau )$, $\zeta_{AB}(\tau )$ Dirac's multipliers.

\vfill\eject

\section{Surface Terms and Asymptotic Poincar\'e Charges in Metric Gravity.}

The  presence of the terms $N_{(as)}$, $N_{(as)\check r}$ in
Eq.(\ref{IV5}) makes $H_D$ not differentiable.

In Refs,\cite{reg,reg1}, following Refs.\cite{witt,dew}, it is shown
that the differentiability of the ADM canonical Hamiltonian
[$H_{(c)ADM}\, \rightarrow {\hat H}_{(c)ADM}+H_{\infty}$] requires the
introduction of the following surface term

\begin{eqnarray}
H_{\infty}=-
\int_{S^2_{\tau ,\infty}} d^2\Sigma_u &&\{ \epsilon
k\sqrt{\gamma}\, {}^3g^{uv}\, {}^3g^{rs}
[N (\partial_r\, {}^3g_{vs}-\partial_v\, {}^3g_{rs})+ \nonumber \\
&+&\partial_uN ({}^3g_{rs}-\delta_{rs})
-\partial_rN ({}^3g_{sv}-\delta_{sv})] -2 N_r\, {}^3{\tilde \Pi}^{ru} \}
(\tau ,\vec \sigma )=\nonumber \\
 &=&-\int_{S^2_{\tau ,\infty}} d^2\Sigma_u \{ \epsilon
k\sqrt{\gamma}\, {}^3g^{uv}\, {}^3g^{rs} [N_{(as)} (\partial_r\,
{}^3g_{vs}-\partial_v\, {}^3g_{rs})+ \nonumber \\
&+&\partial_uN_{(as)} ({}^3g_{rs}-\delta_{rs})
-\partial_rN_{(as)} ({}^3g_{sv}-\delta_{sv})] -2 N_{(as)r}\, 
{}^3{\tilde \Pi}^{ru} \}
(\tau ,\vec \sigma )=\nonumber \\
 &=& {\tilde \lambda}_A(\tau ) P^A_{ADM} + {1\over 2} 
{\tilde \lambda}_{AB}(\tau )
 J^{AB}_{ADM}.
\label{V1}
\end{eqnarray}

 Indeed, by putting $N=N_{(as)}$, $N_{\check r}=N_{(as)\check r}$ in
the surface integrals,  the added term $H_{\infty}$ becomes the given
linear combination of the {\it strong ADM Poincar\'e charges} $P^A_{ADM}$,
$J^{AB}_{ADM}$ \cite{reg,reg1} first identified in the linearized
theory \cite{adm}:

\begin{eqnarray}
P^{\tau}_{ADM}&=&\epsilon k
\int_{S^2_{\tau ,\infty}}d^2\Sigma_{\check u}
[\sqrt{\gamma}\,\, {}^3g^{\check u\check v}\, {}^3g^{\check r\check s}
(\partial_{\check r}\, {}^3g_{\check v\check s}-\partial_{\check v}\, {}^3g
_{\check r\check s})](\tau ,\vec \sigma ),\nonumber \\
P^{\check r}_{ADM}&=&-2 \int_{S^2_{\tau ,\infty}}d^2\Sigma
_{\check u} \, {}^3{\tilde \Pi}^{\check r\check u}(\tau ,\vec \sigma ),
\nonumber \\
J_{ADM}^{\tau \check r}&=&\epsilon k
\int_{S^2_{\tau ,\infty}}d^2\Sigma_{\check u} \sqrt{\gamma}\,\,
{}^3g^{\check u\check v}\, {}^3g^{\check n\check s}\cdot \nonumber \\
&\cdot& [\sigma^{\check r} (\partial_{\check n}\, {}^3g_{\check v\check s}-
\partial_{\check v}\, {}^3g_{\check n\check s})+\delta^{\check r}_{\check v}
({}^3g_{\check n\check s}-\delta_{\check n\check s})-\delta^{\check r}
_{\check n}({}^3g_{\check s\check v}-\delta_{\check s\check v})]
(\tau ,\vec \sigma ),\nonumber \\
J_{ADM}^{\check r\check s}&=&\int_{S^2_{\tau ,\infty}}d^2\Sigma_{\check u}
[\sigma^{\check r}\, {}^3{\tilde \Pi}^{\check s\check u}-
\sigma^{\check s}\, {}^3{\tilde \Pi}^{\check r\check u}]
(\tau ,\vec \sigma ),\nonumber \\
 &&{}\nonumber \\
 &&{}\nonumber \\
 P^{(\mu )}_{ADM}&=&
l^{(\mu )} P^{\tau}_{ADM}+b^{(\mu )}_{(\infty ) \check r}(\tau ) P^{\check
r}_{ADM}=b^{(\mu )}_{(\infty )A}(\tau ) P^A_{ADM},\nonumber \\
S^{(\mu )(\nu )}_{ADM}&=&
[l^{(\mu )}_{(\infty )}b^{(\nu )}_{(\infty
) \check r}(\tau )-l^{(\nu )}_{(\infty )}b^{(\mu )}_{(\infty ) \check r}(\tau
)] J^{\tau \check r}_{ADM}+\nonumber \\
&+&[b^{(\mu )}_{(\infty ) \check r}(\tau )b^{(\nu )}_{(\infty ) \check s}(\tau
)-b^{(\nu )}_{(\infty ) \check r}(\tau )b^{(\mu )}_{(\infty ) \check s}(\tau )]
J^{\check r\check s}_{ADM}=\nonumber \\
&=&[b^{(\mu )}_{(\infty )A}(\tau )b^{(\nu )}_{(\infty )B}(\tau )-b^{(\nu )}
_{(\infty )A}(\tau )b^{(\mu )}_{(\infty )B}(\tau )] J^{AB}_{ADM}.
\label{V2}
\end{eqnarray}

\noindent Here $J^{\tau \check r}_{ADM}=-J^{\check r\tau}_{ADM}$ by definition
and the inverse asymptotic tetrads are defined by $b^A_{(\infty )(\mu )}
b^{(\nu )}_{(\infty )B}=\delta^A_B$, $b^A_{(\infty )(\mu )}b^{(\nu )}
_{(\infty )A}=\delta^{(\nu )}_{(\mu )}$.

As shown in Ref.\cite{reg,reg1}, the parity conditions of
Eqs.(\ref{III6}) and (\ref{III7})are necessary to have a well defined
and finite 3-angular-momentum $J_{ADM}
^{\check r\check s}$: in Appendix B of Ref.\cite{reg1} there is an explicit
example of initial data satisfying the constraints but not the parity 
conditions, for which the 3-angular-momentum is infinite 
\footnote{Moreover, it is shown that the
conditions of the SPI formalism to kill supertranslations and pick out a
unique asymptotic Poincar\'e group (the vanishing of the first-order asymptotic
part of the pseudomagnetic Weyl tensor) may give infinite 3-angular-momentum
if the parity conditions are not added.}.

The definition of the boosts $J_{ADM}^{\tau \check r}$ given in Ref.
\cite{reg1} is not only differentiable like the one in Ref.\cite{reg},
but also finite. As seen in Section III, the problem of boosts is still
open. However, for any isolated system the boost part of the conserved
Poincar\'e group cannot be an independent variable \footnote{Only the
Poincar\'e Casimirs (giving the invariant mass and spin of the system)
are relevant and not the Casimirs of the Lorentz subgroup.}: as shown in
Appendix A for the parametrized Minkowski theories on the Wigner hyperplane
of the rest-frame instant form, the {\it internal} boosts are gauge 
variables. At the end of this Section this point will be clarified by giving 
the explicit realization of the {\it external} Poincar\'e generators in the 
rest-frame instant form (they are independent from the ADM boosts).

The surface term $H_{\infty}$ arises from a suitable splitting of the
superhamiltonian and supermomentum constraints (\ref{II17}). By using
${}^3\Gamma^s_{sr}={1\over {\sqrt{\gamma}}} \partial_r \sqrt{\gamma},$
${}^3g^{uv}\, {}^3\Gamma^r_{uv}=-{1\over {\sqrt{\gamma}}} \partial_s
(\sqrt{\gamma}\, {}^3g^{rs}),$\hfill\break ${}^3R={}^3g^{uv}
({}^3\Gamma^r_{us}\, {}^3\Gamma^s_{vr}- {}^3\Gamma^r_{uv}\,
{}^3\Gamma^s_{sr})+ {1\over {\sqrt{\gamma}}} \partial_r [\sqrt{\gamma}
({}^3g^{uv}\, {}^3\Gamma^r_{uv}-{}^3g^{ur}\,
{}^3\Gamma^v_{vu})],$\hfill\break and ${}^3g^{rs}\,
{}^3\Gamma^u_{rs}-{}^3g^{uv}\, {}^3\Gamma^s_{sv}={}^3g^{rs}\,
{}^3g^{uv}(\partial_r\, {}^3g_{vs}-\partial_v\, {}^3g_{rs})$, we get
[it is valid also with $N$ and $N_r$ replaced by $N_{(as)}$ and
$N_{(as)r}$]

\begin{eqnarray}
&&\int d^3\sigma [ N {\tilde {\cal H}}+N_r\, {}^3{\tilde {\cal H}}^r]
(\tau ,\vec \sigma )=\nonumber \\
&&=\int d^3\sigma \{ \epsilon k N \partial_u[\sqrt{\gamma}({}^3g^{rs}\,
{}^3\Gamma^u_{rs}-{}^3g^{uv}\, {}^3\Gamma^s_{sv})]-2N_r\partial_u\,
{}^3{\tilde \Pi}^{ru}\} (\tau ,\vec \sigma )+\nonumber \\
&&+\int d^3\sigma \{ \epsilon kN[\sqrt{\gamma}\, {}^3g^{rs}({}^3\Gamma^u_{rv}\,
{}^3\Gamma^v_{su}-{}^3\Gamma^u_{rs}\, {}^3\Gamma^v_{vu})-\nonumber \\
&&-{1\over {2k\sqrt{\gamma}}}{}^3G_{rsuv}\, {}^3{\tilde \Pi}^{rs}\, {}^3{\tilde
\Pi}^{uv}]-2 N_r\, {}^3\Gamma^r_{su}\, {}^3{\tilde \Pi}^{su} \} (\tau ,\vec
\sigma )=\nonumber \\
&&=\int_{S^2_{\tau ,\infty}}d^2\Sigma_u
\{ \epsilon k N \sqrt{\gamma}\, {}^3g^{rs}\,
{}^3g^{uv} (\partial_r\, {}^3g_{vs}-\partial_v\, {}^3g_{rs})-2 N_r\,
{}^3{\tilde \Pi}^{ru} \} (\tau ,\vec \sigma )+\nonumber \\
&&+\int d^3\sigma \{ \epsilon
kN [\sqrt{\gamma}\, {}^3g^{rs}({}^3\Gamma^u_{rv}\,
{}^3\Gamma^v_{su}-{}^3\Gamma^u_{rs}\, {}^3\Gamma^v_{vu})-\nonumber \\
&&-{1\over {2k\sqrt{\gamma}}}{}^3G_{rsuv}\, {}^3{\tilde \Pi}^{rs}\, {}^3{\tilde
\Pi}^{uv} ]
-\epsilon k \partial_uN \sqrt{\gamma}\, {}^3g^{rs}\, {}^3g^{uv}(\partial_r\,
{}^3g_{vs}-\partial_v\, {}^3g_{rs})-\nonumber \\
&&-2 N_r\, {}^3\Gamma^r_{su}\, {}^3{\tilde \Pi}^{su}+2 \partial_uN_r\,
{}^3{\tilde \Pi}^{ru} \} (\tau ,\vec \sigma ).
\label{V3}
\end{eqnarray}

In Ref.\cite{reg1} it is noted that, with the boundary conditions of
Refs. \cite{reg,reg1}, the term in $\partial_uN$ in the volume
integral diverges. The following (non-tensorial) regularization is
proposed: $\partial_r\, {}^3g_{vs}-\partial_v\, {}^3g_{rs}=\partial_r({}^3g
_{vs}-\delta_{vs})-\partial_v({}^3g_{rs}-\delta_{rs})$ \footnote{It is the
subtraction of the static background metric of Ref. \cite{hh}; in this spirit 
one could think to use static background metrics ${}^3f_{rs}$ different from 
$\delta_{rs}$: $\partial_r\, {}^3g_{vs}-\partial_v\, {}^3g_{rs} \mapsto
\partial_r({}^3g_{vs}-{}^3f_{vs})- \partial_v({}^3g_{rs}-{}^3f_{rs})\not= 
\partial_r\, {}^3g_{vs}-\partial_v\, {}^3g_{rs}$.}. If we make a further 
integration by parts of the volume term containing $\partial_uN$, we get the 
identity \footnote{With $N$ and $N_r$ replaced by $N_{(as)}$ and $N_{(as)r}$ 
we have $H_{\infty}$ appearing in the first two lines; only $J^{\tau
\check r}_{ADM}$, but not $P^{\tau}_{ADM}$, depends on the
regularization in Eqs.(\ref{V2}).}

\begin{eqnarray}
&&-\int_{S^2_{\tau ,\infty}} d^2\Sigma_u \{ \epsilon
k\sqrt{\gamma}\, {}^3g^{uv}\, {}^3g^{rs}
[N (\partial_r\, {}^3g_{vs}-\partial_v\, {}^3g_{rs})+ \partial_uN ({}^3g_{rs}-
\delta_{rs})-\nonumber \\
&&-\partial_rN ({}^3g_{sv}-\delta_{sv})] -2 N_r\, {}^3{\tilde
\Pi}^{ru} \} (\tau ,\vec \sigma )+\int d^3\sigma [ N
{\tilde {\cal H}}+N_r\, {}^3{\tilde {\cal H}}^r] (\tau ,\vec \sigma
)=\nonumber \\ &&{}\nonumber \\ &&=+\int d^3\sigma \{ \epsilon N [k
\sqrt{\gamma}\, {}^3g^{rs} ({}^3\Gamma^u
_{rv}\, {}^3\Gamma^v_{su}-{}^3\Gamma^u_{rs}\, {}^3\Gamma^v_{vu})-{1\over
{2k\sqrt{\gamma}}}{}^3G_{rsuv}\, {}^3{\tilde \Pi}^{rs}\, 
{}^3{\tilde \Pi}^{uv} ]
+\nonumber \\
&&+\epsilon k ({}^3g_{vs}-\delta_{vs}) \partial_r[\sqrt{\gamma} \partial_uN
({}^3g^{rs}\, {}^3g^{uv}-{}^3g^{ru}\, {}^3g^{sv})]-\nonumber \\
&&-2 N_r\, {}^3\Gamma^r_{su}\, {}^3{\tilde \Pi}^{su} +2 \partial_uN_r\,
{}^3{\tilde \Pi}^{ru} \} (\tau ,\vec \sigma ).
\label{V4}
\end{eqnarray}

By using Eqs.(\ref{IV5}) the modified canonical Hamiltonian becomes

\begin{eqnarray}
{\hat H}_{(c)ADM}&=&\int d^3\sigma [N {\tilde {\cal H}}+N_{\check r}\,
{}^3{\tilde {\cal H}}^{\check r}](\tau ,\vec \sigma ) =\nonumber \\
&=&\int d^3\sigma [(N_{(as)}+m) {\tilde {\cal H}}+(N_{(as)\check r}+
m_{\check r})\, {}^3{\tilde {\cal H}}^{\check r}](\tau ,\vec \sigma )
\mapsto\nonumber \\
\mapsto {\hat H}^{'}_{(c)ADM}&&={\hat H}^{'}_{(c)ADM}[N,N^{\check r}]=
{\hat H}_{(c)ADM}+H_{\infty}=\nonumber \\
 &=&\int d^3\sigma
[(N_{(as)}+m) {\tilde {\cal H}}+(N_{(as)\check r}+ m_{\check r})\,
{}^3{\tilde {\cal H}}^{\check r}](\tau ,\vec \sigma )+
\nonumber \\
&+&{\tilde \lambda}_{(\mu )}(\tau ) P^{(\mu )}_{ADM}+
{\tilde \lambda}_{(\mu )(\nu )}(\tau ) S^{(\mu )(\nu )}_{ADM}=\nonumber \\
&=&\int d^3\sigma [(N_{(as)}+m) {\tilde {\cal H}}+(N_{(as)\check r}+
m_{\check r})\, {}^3{\tilde {\cal H}}^{\check r}](\tau ,\vec \sigma )+
\nonumber \\
&+&{\tilde \lambda}_A(\tau )P^A_{ADM}+{1\over 2}{\tilde \lambda}_{AB}(\tau )
J^{AB}_{ADM} \approx \nonumber \\
&\approx& {\tilde \lambda}_A(\tau )
P^A_{ADM}+{1\over 2}{\tilde \lambda}_{AB}(\tau ) J^{AB}_{ADM}.
\label{V5}
\end{eqnarray}

The analysis of Ref. \cite{hh} shows that $H_{\infty}$, with the
quoted regularization, is equivalent to the sum of the surface term
distinguishing the ADM action from the regularized Hilbert action and
of the one appearing in making the Legendre transformation from the
ADM action to the ADM canonical Hamiltonian, see Eqs.(\ref{II15}).

The terminology {\it strong} derives from Ref.\cite{lusa}, where
there is the definition of the {\it weak} and {\it strong} improper conserved
non-Abelian charges in Yang-Mills theory and their derivation from the
Noether identities implied by the second Noether theorem. In this case
one gets (see Ref.\cite{lu1} b) for the general theory):\hfill\break
 i) {\it strong conserved improper currents} (their conservation is an identity
independent from the Euler-Lagrange equations), whose {\it strong
conserved improper charges} are just surface integrals at spatial
infinity;\hfill\break
 ii) {\it weak conserved improper
currents} (their conservation implies the Euler-Lagrange equations; it
is a form of first Noether theorem hidden in the second one), whose
{\it weak conserved improper charges} are volume integrals;\hfill\break
 iii) the two kinds of charges differ by the volume integral of the
 Gauss law first class constraints and, therefore, coincide when use is done 
of the acceleration-independent Euler-Lagrange equations, i.e. the secondary
first class Gauss law constraints.

In ADM metric gravity it is difficult to check explicitly these
statements due to the presence  of the lapse and shift
functions. In this paper we shall adopt the terminology
{\it strong} and {\it weak} Poincar\'e charges to refer to {\it surface} and
{\it volume} integrals respectively, even if the strong charges are not
strongly conserved improper charges but only weakly conserved ones
like the weak charges.

Then Eqs.(\ref{IV5}), (\ref{V4}) and (\ref{II17}) imply

\begin{eqnarray}
{\hat H}^{'}_{(c)ADM}&=&\int d^3\sigma [ m {\tilde {\cal H}}+m_{\check r}\,
{}^3{\tilde {\cal H}}^{\check r}](\tau ,\vec \sigma )+\nonumber \\
&+&{\tilde \lambda}_{\tau}(\tau ) [-\int d^3\sigma {\tilde {\cal H}}(\tau
,\vec \sigma )+P^{\tau}_{ADM}]+{\tilde \lambda}_{\check r}(\tau )[-\int
d^3\sigma \,
{}^3{\tilde {\cal H}}^{\check r}(\tau ,\vec \sigma )+P^{\check r}_{ADM}]
+\nonumber \\
&+&{\tilde \lambda}_{\tau \check r}(\tau )[-{1\over 2}\int d^3\sigma
\sigma^{\check r}\, {\tilde {\cal H}}(\tau ,\vec \sigma )+J^{\tau \check r}
_{ADM}]+\nonumber \\
&+&{\tilde \lambda}_{\check r\check s}(\tau )[-{1\over 2}\int d^3\sigma
\sigma^{\check s}\, {}^3{\tilde {\cal H}}^{\check r}(\tau ,\vec \sigma )+
J^{\check r\check s}_{ADM}]=\nonumber \\
&=&\int d^3\sigma [ m {\tilde {\cal H}}+m_{\check r}\,
{}^3{\tilde {\cal H}}^{\check r}](\tau ,\vec \sigma )+\nonumber \\
&+&\int d^3\sigma \{ \epsilon N_{(as)} [k\sqrt{\gamma}\, {}^3g^{\check r\check
s} ({}^3\Gamma^{\check u}_{\check r\check v}\, {}^3\Gamma^{\check v}_{\check
s\check u}-{}^3\Gamma^{\check u}_{\check r\check s}\, {}^3\Gamma^{\check v}
_{\check v\check u})-\nonumber \\
&-&{1\over {2k\sqrt{\gamma} }} {}^3G_{\check r\check s\check u\check v}\,
{}^3{\tilde \Pi}^{\check r\check s}\,
{}^3{\tilde \Pi}^{\check u\check v}]+\nonumber \\
&+&\epsilon k ({}^3g_{\check v\check s}-\delta_{\check v\check s}) \partial_
{\check r}[\sqrt{\gamma} \partial_{\check u}N_{(as)} ({}^3g^{\check r\check s}
\, {}^3g^{\check u\check v}-{}^3g^{\check u\check r}\, {}^3g^{\check v\check
s})]-\nonumber \\
&-&2 N_{(as)\check r}\, {}^3\Gamma^{\check r}_{\check s\check u}\, {}^3{\tilde
\Pi}^{\check s\check u}+2 \partial_{\check u}N_{(as) \check r}\, {}^3{\tilde
\Pi}^{\check r\check u} \} (\tau ,\vec \sigma )=\nonumber \\
&=& \int d^3\sigma [ m {\tilde {\cal H}}+m_{\check
r} {}^3{\tilde {\cal H}}^{\check r}](\tau ,\vec \sigma )+
{\tilde \lambda}_{(\mu )}(\tau ) {\hat P}^{(\mu )}_{ADM}+{\tilde \lambda}
_{(\mu )(\nu )}(\tau ) {\hat S}^{(\mu )(\nu )}_{ADM} =\nonumber \\
&=&\int d^3\sigma [ m {\tilde {\cal H}}+m_{\check r}\,
{}^3{\tilde {\cal H}}^{\check r}](\tau ,\vec \sigma )+
{\tilde \lambda}_A(\tau ) {\hat P}^A_{ADM}+{1\over 2}{\tilde \lambda}
_{AB}(\tau ){\hat J}^{AB}_{ADM}\approx \nonumber \\
&\approx& {\tilde \lambda}_A(\tau ) {\hat P}^A_{ADM}+{1\over 2}{\tilde
\lambda}_{AB}(\tau ) {\hat J}^{AB}_{ADM},\nonumber \\
{\hat H}^{'}_{(D)ADM}&=&{\hat H}^{'}_{(c)ADM}[m,m^{\check r}]+\nonumber \\
&+&\int d^3\sigma [\lambda_n {\tilde
\pi}^n+\lambda^{\vec n}_r {\tilde \pi}^r_{\vec n}](\tau ,\vec \sigma )+
\zeta_A(\tau ) {\tilde \pi}^A(\tau )+\zeta_{AB}(\tau ) {\tilde \pi}
^{AB}(\tau ),
\label{V6}
\end{eqnarray}

\noindent with the following {\it weak conserved improper Poincar\'e charges}
${\hat P}^A_{ADM}$, ${\hat J}_{ADM}^{AB}$ \footnote{These volume expressions
(the analogue of the weak Yang-Mills non Abelian charges) for the ADM
4-momentum are used in Ref.\cite{positive}
in the study of the positiviteness of the energy;
the weak charges are Noether charges.}

\begin{eqnarray}
{\hat P}^{\tau}_{ADM}&=&
\int d^3\sigma \epsilon
[k\sqrt{\gamma}\,\, {}^3g^{\check r\check s}({}^3\Gamma^{\check u}
_{\check r\check v}\, {}^3\Gamma^{\check v}_{\check
s\check u}-{}^3\Gamma^{\check u}_{\check r\check s}\, {}^3\Gamma^{\check v}
_{\check v\check u})-\nonumber \\
&-&{1\over {2k\sqrt{\gamma} }} {}^3G_{\check r\check s\check u\check v}\,
{}^3{\tilde \Pi}^{\check r\check s}\,
{}^3{\tilde \Pi}^{\check u\check v}](\tau ,\vec \sigma ),\nonumber \\
{\hat P}^{\check r}_{ADM}&=&-
2\int d^3\sigma \, {}^3\Gamma
^{\check r}_{\check s\check u}(\tau ,\vec \sigma )\, {}^3{\tilde \Pi}^{\check
s\check u}(\tau ,\vec \sigma ),\nonumber \\
{\hat J}^{\tau \check r}_{ADM}&=&-{\hat J}^{\check r\tau}_{ADM}=
\int d^3\sigma  \epsilon \{ \sigma^{\check r}\nonumber \\
&&[k\sqrt{\gamma}\,\,  {}^3g^{\check n\check s}({}^3\Gamma^{\check u}_{\check
n\check v}\, {}^3\Gamma^{\check v}_{\check s\check u}-{}^3\Gamma^{\check u}
_{\check n\check s}\, {}^3\Gamma^{\check v}_{\check v\check u})-{1\over
{2k\sqrt{\gamma}}} {}^3G_{\check n\check s\check u\check v}\, {}^3{\tilde
\Pi}^{\check n\check s}\, {}^3{\tilde \Pi}^{\check u\check v}]+\nonumber \\
&+& k \delta^{\check r}_{\check u}({}^3g_{\check v\check s}-\delta
_{\check v\check s}) \partial_{\check n}[\sqrt{\gamma}({}^3g^{\check n\check s}
\, {}^3g^{\check u\check v}-{}^3g^{\check n\check u}\, {}^3g^{\check s\check
v})] \} (\tau ,\vec \sigma ),\nonumber \\
{\hat J}^{\check r\check s}_{ADM}&=&
\int d^3\sigma  [(\sigma^{\check r}\, {}^3\Gamma^{\check s}
_{\check u\check v}-\sigma^{\check s}\, {}^3\Gamma^{\check r}_{\check u\check
v})\, {}^3{\tilde \Pi}^{\check u\check v}](\tau ,\vec \sigma ),\nonumber \\
&&{}\nonumber \\
{\hat P}^{(\mu )}_{ADM}&=&
l^{(\mu )}_{(\infty )} {\hat P}^{\tau}_{ADM} + b^{(\mu )}_{(\infty ) \check
r}(\tau ) {\hat P}^{\check r}_{ADM}=b^{(\mu )}_{(\infty )A}(\tau ) {\hat P}^A
_{ADM},\nonumber \\
{\hat S}^{(\mu )(\nu )}_{ADM}&=&
[l^{(\mu )}_{(\infty )}b^{(\nu )}_{(\infty ) \check r}(\tau )-l^{(\nu )}
_{(\infty )}b^{(\mu )}_{(\infty ) \check r}(\tau )] {\hat J}^{\tau \check r}
_{ADM} +\nonumber \\
&+&[b^{(\mu )}_{(\infty ) \check r}(\tau )b^{(\nu )}_{(\infty ) \check
s}(\tau )-b^{(\nu )}_{(\infty ) \check r}(\tau )b^{(\mu )}_{(\infty ) \check
s}(\tau )] {\hat J}_{ADM}^{\check r\check s}=\nonumber \\
&=&[b^{(\mu )}_{(\infty )A}b^{(\nu )}_{(\infty )B}-b^{(\nu )}_{(\infty )A}b
^{(\mu )}_{(\infty )B}](\tau ) {\hat J}^{AB}_{ADM}.
\label{V7}
\end{eqnarray}

In both Refs.\cite{reg,reg1} it is shown that the canonical
Hamiltonian ${\hat H}^{'}_{(c)ADM}[N,N^{\check r}]$ of Eq.(\ref{V6})
with arbitrary $N$, $N^{\check r}={}^3g^{\check r\check s} N_{\check
s}$ (therefore including the ones of Eqs.(\ref{IV5})), has the same
Poisson brackets as in Eq.(\ref{III9}) for $N=m$, $N^{\check
r}=m^{\check r}$ ({\it proper} gauge transformations)

\begin{eqnarray}
&&\{ {\hat H}^{'}_{(c)ADM}[N_1,N_1^{\check r}],
{\hat H}^{'}_{(c)ADM}[N_2,N_2^{\check r}] \}=
{\hat H}^{'}_{(c)ADM}[N_3,N_3^{\check r}],\nonumber \\
&&{}\nonumber \\
 if&&
N_i(\tau ,\vec \sigma )=m_i(\tau ,\vec \sigma )-{\tilde \lambda}
_{i\tau}(\tau )-{1\over 2}
{\tilde \lambda}_{i\tau \check u}(\tau ) \sigma^{\check u},\quad
i=1,2,\nonumber \\
and&& N_{i\check r}(\tau ,\vec \sigma )=m_{i\check r}(\tau ,\vec \sigma )-
{\tilde \lambda}_{i\check r}(\tau )-{1\over 2}
{\tilde \lambda}_{i\check r\check u}(\tau )
 \sigma^{\check u},\quad i=1,2,\nonumber \\
&&\Downarrow \nonumber \\
 N_3&=&N_2^{\check r}\partial_{\check r} N_1-N_1^{\check r}\partial_{\check
r} N_2=
 m_3-{\tilde \lambda}_{3\tau}-{1\over 2}{\tilde \lambda}_{3\tau \check u}
\sigma^{\check u},\nonumber \\
N_3^{\check r}&=&
{\cal L}_{{\vec N}_2} N_1^{\check r}+N_2
\partial^{\check r} N_1-N_1
\partial^{\check r} N_2=\nonumber \\
 &=&-N_1^{\check
s}\partial_{\check s} N_2^{\check r}+N_2^{\check s}\partial
_{\check s} N_1^{\check r}+N_2 \partial^{\check r}
N_1-N_1 \partial^{\check r} N_2=\nonumber \\
 &=&-\epsilon {}^3g^{\check r\check s}[m_{3\check s}- {\tilde
\lambda}_{3\check s}-{1\over 2} {\tilde \lambda}_{3\check s\check
u}\sigma^{\check u}],\nonumber \\ &&{}\nonumber \\ &&with\nonumber \\
&&{}\nonumber \\ {\tilde \lambda}_{3\tau}&=&-{{\epsilon}\over 2}
\delta^{\check r\check s}[{\tilde \lambda}
_{1\check r}{\tilde \lambda}_{2\tau \check s}-{\tilde \lambda}_{2\check r}
{\tilde \lambda}_{1\tau \check s}],\nonumber \\
{\tilde \lambda}_{3\tau \check u}&=&-{{\epsilon}\over 2}
\delta^{\check r\check s}[{\tilde \lambda}
_{1\check r\check u}{\tilde \lambda}_{2\tau \check s}-{\tilde \lambda}_{2\check
r\check u}{\tilde \lambda}_{1\tau \check s}],\nonumber \\
m_3&=&-\epsilon
{}^3g^{\check r\check s}\Big(m_{2\check s}[\partial_{\check r}m_1-
{\tilde \lambda}_{2\tau \check r}]-m_{1\check s}[\partial_{\check r}m_2-
{1\over 2}{\tilde \lambda}_{2\tau \check r}]+\nonumber \\
&&+\partial_{\check r}m_2[{\tilde \lambda}_{1\check s}+
{1\over 2}{\tilde \lambda}_{1\check
s\check u}\sigma^{\check u}]-\partial_{\check r}m_1
[{\tilde \lambda}_{2\check s}
+{1\over 2}{\tilde \lambda}_{2\check s\check u}\sigma^{\check u}]\Big) -
\nonumber \\
&&-{{\epsilon}\over 2}
({}^3g^{\check r\check s}-\delta^{\check r\check s})\Big( {\tilde \lambda}
_{1\tau \check r}[{\tilde \lambda}_{2\check s}+
{1\over 2}{\tilde \lambda}_{2\check s\check
u}\sigma^{\check u}]-{\tilde \lambda}_{2\tau \check r}
[{\tilde \lambda}_{1\check
s}+{1\over 2}{\tilde \lambda}_{1\check s\check u}\sigma^{\check u}]\Big) ,
\nonumber \\
{\tilde \lambda}_{3\check r}&=&{1\over 2}\Big(
{\tilde \lambda}_{1\tau}{\tilde \lambda}_{2\tau
\check r}-{\tilde \lambda}_{2\tau}{\tilde \lambda}_{1\tau \check r}-\epsilon
\delta^{\check m\check n}[{\tilde \lambda}_{1\check r\check m}{\tilde \lambda}
_{2\check n}-{\tilde \lambda}_{2\check r\check m}{\tilde \lambda}_{1\check n}]
\Big) ,\nonumber \\
{\tilde \lambda}_{3\check r\check u}&=&{1\over 2}\Big(
{\tilde \lambda}_{1\tau \check u}{\tilde
\lambda}_{2\tau \check r}-{\tilde \lambda}_{2\tau \check u}{\tilde \lambda}
_{1\tau \check r}-\epsilon
\delta^{\check m\check n}[{\tilde \lambda}_{1\check r\check m}
{\tilde \lambda}_{2\check n\check u}-{\tilde \lambda}_{2\check r\check m}
{\tilde \lambda}_{1\check n\check u}]\Big) ,\nonumber \\
m_{3\check r}&=&m_2[\partial_{\check r}m_1-
{1\over 2}{\tilde \lambda}_{1\tau \check r}]-
m_1[\partial_{\check r}m_2-
{1\over 2}{\tilde \lambda}_{2\tau \check r}]+\nonumber \\
&&+\partial_{\check r}m_2[{\tilde \lambda}_{1\tau}+
{1\over 2}{\tilde \lambda}_{1\tau
\check u}\sigma^{\check u}]-\partial_{\check r}m_1[{\tilde \lambda}_{2\tau}+
{1\over 2}{\tilde \lambda}_{2\tau \check u}\sigma^{\check u}]-\nonumber \\
&&-{{\epsilon}\over 2}
({}^3g^{\check m\check n}-\delta^{\check m\check n})\Big([{\tilde \lambda}_{1
\check m}+
{1\over 2}{\tilde \lambda}_{1\check m\check u}\sigma^{\check u}]{\tilde 
\lambda}
_{2\check r\check n}-[{\tilde \lambda}_{2\check m}+
{1\over 2}{\tilde \lambda}_{2\check m
\check u}\sigma^{\check u}]{\tilde \lambda}_{1\check r\check n}\Big) -
\nonumber \\
&&-\epsilon
{}^3g^{\check m\check n}\Big( m_{1\check m}[\partial_{\check n}m_{2\check r}
-{1\over 2}
{\tilde \lambda}_{2\check r\check n}]-m_{2\check m}[\partial_{\check n}
m_{1\check r}-{1\over 2}{\tilde \lambda}_{1\check r\check n}]+\nonumber \\
&&+{1\over 2}
[\partial_{\check m}m_{1\check r}{\tilde \lambda}_{2\check n\check u}-
\partial_{\check m}m_{2\check r}{\tilde \lambda}_{1\check n\check u}]\sigma
^{\check u} \Big) -\nonumber \\
&&-\epsilon
{}^3g_{\check r\check s}\, {}^3g^{\check t\check n}\, \partial_{\check t}\,
{}^3g^{\check s\check m} \Big( [m_{1\check n}-{\tilde \lambda}_{1\check n}-
{1\over 2}{\tilde \lambda}_{1\check n\check u}\sigma^{\check u}] 
[m_{2\check m}-
{\tilde \lambda}_{2\check m}-{1\over 2}{\tilde \lambda}_{2\check m\check u}
\sigma^{\check u}]-\nonumber \\
&&-[m_{2\check n}-{\tilde \lambda}_{2\check n}-
{1\over 2}{\tilde \lambda}_{2\check n\check u}\sigma^{\check u}] 
[m_{1\check m}-
{\tilde \lambda}_{1\check m}-{1\over 2}{\tilde \lambda}_{1\check m\check u}
\sigma^{\check u}]\Big) ,\nonumber \\
&&{}\nonumber \\
\int d^3\sigma&& [m_3{\tilde {\cal H}}+m_3^{\check r}\, {}^3{\tilde {\cal H}}
_{\check r}](\tau ,\vec \sigma )+{\tilde \lambda}_{3A}(\tau ){\hat P}^A_{ADM}+
{1\over 2}{\tilde \lambda}_{3AB}(\tau ){\hat J}^{AB}_{ADM}=\nonumber \\
&=&\int d^3\sigma_1 d^3\sigma_2 \Big[ m_1(\tau ,{\vec \sigma}_1)m_2(\tau ,{\vec
\sigma}_2) \{ {\tilde {\cal H}}(\tau ,{\vec \sigma}_1),{\tilde {\cal H}}(\tau
,{\vec \sigma}_2) \} +\nonumber \\
&+&[m_1(\tau ,{\vec \sigma}_1)m_2^{\check r}(\tau ,{\vec \sigma}_2)-m_2(\tau
,{\vec \sigma}_1)m_1^{\check r}(\tau ,{\vec \sigma}_2)] \{ {\tilde {\cal H}}
(\tau ,{\vec \sigma}_1),{}^3{\tilde {\cal H}}_{\check r}(\tau ,{\vec \sigma}_2)
\} +\nonumber \\
&+&m_1^{\check r}(\tau ,{\vec \sigma}_1)m_2^{\check s}(\tau ,{\vec \sigma}_2)
\{ {}^3{\tilde {\cal H}}_{\check r}(\tau ,{\vec \sigma}_1),{}^3{\tilde 
{\cal H}}
_{\check s}(\tau ,{\vec \sigma}_2) \} \Big] +\nonumber \\
&+&\int d^3\sigma \Big[ \Big( {\tilde \lambda}_{1A}(\tau )m_2(\tau ,\vec \sigma
)-{\tilde \lambda}_{2A}(\tau )m_1(\tau ,\vec \sigma )\Big) \{ {\hat P}^A_{ADM},
{\tilde {\cal H}}(\tau ,\vec \sigma ) \} +\nonumber \\
&+&\Big( {\tilde \lambda}_{1A}(\tau )m_2^{\check r}(\tau ,\vec \sigma )-{\tilde
\lambda}_{2A}(\tau )m_1^{\check r}(\tau ,\vec \sigma )\Big) \{ 
{\hat P}^A_{ADM},
{}^3{\tilde {\cal H}}_{\check r}(\tau ,\vec \sigma ) \} +\nonumber \\
&+&{1\over 2}
\Big( {\tilde \lambda}_{1AB}(\tau )m_2(\tau ,\vec \sigma )-{\tilde \lambda}
_{2AB}(\tau )m_1(\tau ,\vec \sigma )\Big) \{ {\hat J}^{AB}_{ADM}, {\tilde {\cal
H}}(\tau ,\vec \sigma ) \} +\nonumber \\
&+&{1\over 2}
\Big( {\tilde \lambda}_{1AB}(\tau )m_2^{\check r}(\tau ,\vec \sigma )-
{\tilde \lambda}_{2AB}(\tau )m_1^{\check r}(\tau ,\vec \sigma )\Big) \{ {\hat
J}^{AB}_{ADM}, {}^3{\tilde {\cal H}}_{\check r}(\tau ,\vec \sigma ) \} \Big] +
\nonumber \\
&+&{\tilde \lambda}_{1A}(\tau ){\tilde \lambda}_{2B}(\tau ) \{ 
{\hat P}^A_{ADM},
{\hat P}^B_{ADM} \} +
{1\over 4}{\tilde \lambda}_{1AB}(\tau ){\tilde \lambda}_{2CD}(\tau )
\{ {\hat J}^{AB}_{ADM},{\hat J}^{CD}_{ADM} \} +\nonumber \\
&+&{1\over 2}
\Big( {\tilde \lambda}_{1A}(\tau ){\tilde \lambda}_{2CD}(\tau )-
{\tilde \lambda}_{2A}(\tau ){\tilde \lambda}_{1CD}(\tau )\Big) \{ {\hat P}^A
_{ADM},{\hat J}^{CD}_{ADM} \} .
\label{V8}
\end{eqnarray}

\noindent
This implies:
\hfill\break
i) the Poisson brackets of two {\it proper} gauge transformations [${\tilde 
\lambda}_{iA}={\tilde \lambda}_{iAB}=0$, i=1,2] is a {\it proper} gauge 
transformation [${\tilde \lambda}_{3A}={\tilde \lambda}_{3AB}=0$], see
Eq.(\ref{III9});\hfill\break 
 ii) if $N_2=m_2$, $N_{2\check
r}=m_{2\check r}$ [${\tilde \lambda}_{2A}= {\tilde \lambda}_{2AB}=0$]
correspond to a {\it proper} gauge transformation and $N_1,N_{1\check r}$
[$m_1=m_{1\check r}=0$] to an {\it improper} one, then we get a {\it proper} 
gauge transformation\hfill\break
\hfill\break
${\tilde \lambda}_{3A}={\tilde \lambda}_{3AB}=0$,\hfill\break
$m_3=-\epsilon {}^3g^{\check r\check s}(-{1\over 2}
m_{2\check s}{\tilde \lambda}_{1\tau \check r}+
\partial_{\check r}m_2[{\tilde \lambda}_{1\check s}+
{1\over 2}{\tilde \lambda}_{1\check s
\check u}\sigma^{\check u}])$,\hfill\break
$m_{3\check r}=-{1\over 2}
m_2{\tilde \lambda}_{1\tau \check r}+\partial_{\check r}m_2
[{\tilde \lambda}_{1\tau}+
{1\over 2}{\tilde \lambda}_{1\check s\check u}\sigma^{\check u}]
-{{\epsilon}\over 2}
{}^3g^{\check m\check n}(m_{2\check m}{\tilde \lambda}_{1\check r\check n}-
\partial_{\check m}m_{2\check r}{\tilde \lambda}_{1\check n\check u}\sigma
^{\check u})-$\hfill\break
$-\epsilon
{}^3g_{\check r\check s}\, {}^3g^{\check t\check n}\partial_{\check t}\,
{}^3g^{\check s\check m}(m_{2\check n}[{\tilde \lambda}_{1\check m}+
{1\over 2}{\tilde
\lambda}_{1\check m\check u}\sigma^{\check u}]-m_{2\check m}[{\tilde \lambda}
_{1\check n}+
{1\over 2}{\tilde \lambda}_{1\check n\check u}\sigma^{\check u}])$,
\hfill\break
\hfill\break
and Eqs.(\ref{V8}) may be interpreted as saying that the 10
Poincar\'e charges are {\it gauge invariant} and {\it Noether constants of
motion}

\begin{eqnarray}
&&\lbrace {\hat P}^{\tau}_{ADM},{\tilde {\cal H}}(\tau ,\vec \sigma )\rbrace
=-\partial_{\check r}\, {}^3{\tilde {\cal H}}^{\check r}(\tau ,\vec \sigma )
\approx 0,\nonumber \\
&&\lbrace {\hat P}^{\tau}_{ADM},{}^3{\tilde {\cal H}}_{\check r}(\tau ,\vec
\sigma )\rbrace =0,\nonumber \\
&&\lbrace {\hat P}^{\check r}_{ADM},{\tilde {\cal H}}(\tau ,\vec \sigma )
\rbrace =\epsilon
\partial_{\check s}[{}^3g^{\check r\check s}\, {\tilde {\cal H}}
(\tau ,\vec \sigma )]\approx 0,\nonumber \\
&&\lbrace {\hat P}^{\check r}_{ADM},{}^3{\tilde {\cal H}}_{\check s}(\tau ,\vec
\sigma )\rbrace =-\epsilon
\partial_{\check s}\, {}^3g^{\check r\check t}(\tau ,\vec
\sigma )\, {}^3{\tilde {\cal H}}_{\check t}(\tau ,\vec \sigma )+\nonumber \\
&&+\epsilon
{}^3g^{\check r\check t}(\tau ,\vec \sigma )\, {}^3g_{\check s\check w}(\tau
,\vec \sigma )\partial_{\check t}\, {}^3g^{\check w\check u}(\tau ,\vec 
\sigma )
\, {}^3{\tilde {\cal H}}_{\check u}(\tau ,\vec \sigma )\approx 0,\nonumber \\
&&\lbrace {\hat J}^{\tau \check r}_{ADM},{\tilde {\cal H}}(\tau ,\vec \sigma )
\rbrace =2\, {}^3{\tilde {\cal H}}^{\check r}(\tau ,\vec \sigma )-
2\partial_{\check
s}[\sigma^{\check r}\, {}^3{\tilde {\cal H}}^{\check s}(\tau ,\vec \sigma )]
\approx 0,\nonumber \\
&&\lbrace {\hat J}^{\tau \check r}_{ADM},{}^3{\tilde {\cal H}}_{\check s}(\tau
,\vec \sigma )\rbrace =- \delta^{\check r}_{\check s}\, {\tilde {\cal
H}}(\tau ,\vec \sigma )\approx 0,\nonumber \\
&&\lbrace {\hat J}^{\check r\check s}_{ADM},{\tilde {\cal H}}(\tau ,\vec
\sigma )\rbrace =\epsilon \partial_{\check u}\Big([{}^3g^{\check r\check u}
\sigma^{\check s}-{}^3g^{\check s\check u}\sigma^{\check r}]{\tilde {\cal H}}
\Big) (\tau ,\vec \sigma )\approx 0,\nonumber \\
&&\lbrace {\hat J}^{\check r\check s}_{ADM},{}^3{\tilde {\cal H}}_{\check
w}(\tau ,\vec \sigma )\rbrace =\Big( (\delta^{\check r}_{\check u}
\delta^{\check s}_{\check w}-\delta^{\check r}_{\check w}\delta^{\check s}
_{\check u}){}^3{\tilde {\cal H}}^{\check u}(\tau ,\vec \sigma )+\nonumber \\
&&+\sigma^{\check s}\Big[
-\epsilon \partial_{\check w}\, {}^3g^{\check r\check t}\,
{}^3{\tilde {\cal H}}_{\check t}+{}^3g_{\check w\check v}\, {}^3g^{\check
r\check m}\partial_{\check m}\, {}^3{\tilde {\cal H}}^{\check v}\Big]
(\tau ,\vec \sigma )-\nonumber \\
&&-\sigma^{\check r}\Big[
-\epsilon \partial_{\check w}\, {}^3g^{\check s\check t}\,
{}^3{\tilde {\cal H}}_{\check t}+{}^3g_{\check w\check v}\, {}^3g^{\check
s\check m}\partial_{\check m}\, {}^3{\tilde {\cal H}}^{\check v}\Big]
(\tau ,\vec \sigma ) \Big) \approx 0,\nonumber \\
&&\Downarrow \nonumber \\
&&\partial_{\tau}\, {\hat P}^A_{ADM}\, {\buildrel \circ \over =}\,
\lbrace {\hat P}^A_{ADM}, {\hat H}^{'}_{(D)ADM} \rbrace =
\lbrace {\hat P}^A_{ADM}, {\hat H}^{'}_{(c)ADM} \rbrace \approx 0,\nonumber \\
&&\partial_{\tau}\, {\hat J}^{AB}_{ADM}\, {\buildrel \circ \over
=}\, \lbrace {\hat J}^{AB}_{ADM}, {\hat H}^{'}_{(D)ADM} \rbrace =
\lbrace {\hat J}^{AB}_{ADM}, {\hat H}^{'}_{(c)ADM} \rbrace \approx 0.
\label{V9}
\end{eqnarray}

From Eqs.(\ref{V9}) we see that also the strong Poincar\'e charges
are constants of motion \footnote{It is not clear how it could be  shown that 
they are conserved independently from the first class constraints.}

\begin{eqnarray}
P^{\tau}_{ADM}&=&{\hat P}^{\tau}_{ADM}+\int d^3\sigma {\tilde {\cal H}}(\tau
,\vec \sigma ),\nonumber \\
P^{\check r}_{ADM}&=&{\hat P}^{\check r}_{ADM}+\int d^3\sigma \, {}^3{\tilde
{\cal H}}^{\check r}(\tau ,\vec \sigma ),\nonumber \\
J^{\tau \check r}_{ADM}&=&{\hat J}^{\tau \check r}_{ADM}+{1\over 2} \int
d^3\sigma \sigma^{\check r}\, {\tilde {\cal H}}(\tau ,\vec \sigma ),
\nonumber \\
J^{\check r\check s}_{ADM}&=&{\hat J}^{\check r\check s}_{ADM}+\int d^3\sigma
[\sigma^{\check s}\, {}^3{\tilde {\cal H}}^{\check r}(\tau ,\vec \sigma )-
\sigma^{\check r}\, {}^3{\tilde {\cal H}}^{\check s}(\tau ,\vec \sigma )],
\nonumber \\
&&\Downarrow \nonumber \\
&&\partial_{\tau}\, P^A_{ADM} \approx 0,\nonumber \\
&&\partial_{\tau}\, J^{AB}_{ADM} \approx 0;
\label{V10}
\end{eqnarray}

iii) the Poisson bracket of two {\it improper} gauge transformations
[$m_i=m_{i\check r}=0$, i=1,2] is an {\it improper} gauge transformation with 
the previous ${\tilde \lambda}_{3A}$, ${\tilde \lambda}_{3AB}$ and with\hfill
\break
\hfill\break
$m_3=-{{\epsilon}\over 2}
({}^3g^{\check r\check s}-\delta^{\check r\check s})({\tilde \lambda}
_{1\tau \check r}[{\tilde \lambda}_{2\check s}+
{1\over 2}{\tilde \lambda}_{2\check s\check
u}\sigma^{\check u}]-{\tilde \lambda}_{2\tau \check r}[{\tilde \lambda}
_{1\check s}+
{1\over 2}{\tilde \lambda}_{1\check s\check u}\sigma^{\check u}])$,
\hfill\break
$m_{3\check r}=-{{\epsilon}\over 2}
({}^3g^{\check m\check n}-\delta^{\check m\check n})([{\tilde
\lambda}_{1\check m}+
{1\over 2}{\tilde \lambda}_{1\check m\check u}\sigma^{\check u}]
{\tilde \lambda}_{2\check r\check n}-[{\tilde \lambda}_{2\check m}+{1\over 2}
{\tilde \lambda}_{2\check m\check u}\sigma^{\check u}]{\tilde \lambda}
_{1\check r\check n})$.\hfill\break
\hfill\break
This implies that the 10 strong Poincar\'e charges (and, therefore, also the
weak ones) satisfy the Poincar\'e algebra
modulo the first class constraints, namely modulo the Hamiltonian group 
of gauge transformations

\begin{eqnarray}
\{ {\hat P}^{\tau}_{ADM},{\hat J}^{\tau \check r}_{ADM} \} &=& -\epsilon
{\hat P}^{\check r}_{ADM},\nonumber\\
\{ {\hat P}^{\tau}_{ADM},{\hat J}^{\check r\check s}_{ADM} \} &=&0,\nonumber \\
\{ {\hat P}^{\check u}_{ADM}, {\hat J}^{\tau \check r}_{ADM} \} &=& -\epsilon
\delta^{\check u\check r}{\hat P}^{\tau}_{ADM}+\epsilon
\int d^3\sigma [({}^3g^{\check u\check
r}-\delta^{\check u\check r}) {\tilde {\cal H}}](\tau ,\vec \sigma ),
\nonumber \\
\{ {\hat P}^{\check u}_{ADM}, {\hat J}^{\check r\check s}_{ADM} \} &=& 
-\epsilon
\Big[ \delta^{\check u\check s}{\hat P}^{\check r}_{ADM}-\delta
^{\check u\check r}{\hat P}^{\check s}_{ADM}+\nonumber \\
&+&\int d^3\sigma [({}^3g^{\check u\check s}-\delta^{\check u\check
s}) {}^3{\tilde {\cal H}}^{\check r}-({}^3g^{\check u\check
r}-\delta^{\check u\check r}) {}^3{\tilde {\cal H}}^{\check s}](\tau
,\vec \sigma )\Big] ,\nonumber \\
\{ {\hat J}^{\tau \check r}_{ADM},{\hat J}^{\tau \check s}_{ADM} \} &=&\epsilon
{\hat J}^{\check r\check s}_{ADM},\nonumber \\
\{ {\hat J}^{\tau \check r}_{ADM}, {\hat J}^{\check u\check v}_{ADM} \} &=&
\epsilon \Big[ \delta^{\check r\check u}{\hat J}^{\tau \check v}_{ADM}
-\delta^{\check r\check v}{\hat J}^{\tau \check u}_{ADM}-\nonumber \\
&-&\int d^3\sigma [\Big( \sigma^{\check v}({}^3g^{\check r\check u}-\delta
^{\check r\check u})-\sigma^{\check u}({}^3g^{\check r\check v}-\delta^{\check
r\check v})\Big) {\tilde {\cal H}}](\tau ,\vec \sigma )\Big],\nonumber \\
\{ {\hat J}^{\check r\check s}_{ADM},{\hat J}^{\check u\check v}_{ADM} \} &=&
-\epsilon
[\delta^{\check r\check u}{\hat J}^{\check s\check v}_{ADM}+\delta^{\check
s\check v}{\hat J}_{ADM}^{\check r\check u}-\delta^{\check r\check v}{\hat J}
_{ADM}^{\check s\check u}-\delta^{\check s\check u}{\hat J}_{ADM}^{\check
r\check v}]+\nonumber \\
&+&\epsilon
\int d^3\sigma \Big[ \Big( \sigma^{\check s}({}^3g^{\check r\check v}-\delta
^{\check r\check v})-\sigma^{\check r}({}^3g^{\check s\check v}-\delta^{\check
s\check v})\Big) {}^3{\tilde {\cal H}}^{\check u}+\nonumber \\
&+&\Big( \sigma^{\check u}({}^3g^{\check v\check s}-\delta
^{\check v\check s})-\sigma^{\check v}({}^3g^{\check u\check s}-\delta^{\check
u\check s})\Big) {}^3{\tilde {\cal H}}^{\check r}-\nonumber \\
&-&\Big( \sigma^{\check s}({}^3g^{\check r\check u}-\delta
^{\check r\check u})-\sigma^{\check r}({}^3g^{\check s\check u}-\delta^{\check
s\check u})\Big) {}^3{\tilde {\cal H}}^{\check v}-\nonumber \\
&-&\Big( \sigma^{\check u}({}^3g^{\check r\check v}-\delta
^{\check r\check v})-\sigma^{\check v}({}^3g^{\check u\check r}-\delta^{\check
u\check r})\Big) {}^3{\tilde {\cal H}}^{\check s}\Big] (\tau ,\vec \sigma ),
\nonumber \\
&&\Downarrow\nonumber \\
\lbrace {\hat P}^A_{ADM},{\hat P}^B_{ADM} \rbrace &=& 0,
\nonumber \\
\lbrace {\hat P}^A_{ADM},{\hat J}^{BC}_{ADM} \rbrace &\approx& {}^4\eta^{AC}
{\hat P}^B_{ADM}-{}^4\eta^{AB} {\hat P}^C_{ADM},
\nonumber \\
\lbrace {\hat J}^{AB}_{ADM},{\hat J}^{CD}_{ADM}
\rbrace &\approx& - C^{ABCD}_{EF} {\hat J}^{EF}_{ADM},
\label{V11}
\end{eqnarray}

\begin{eqnarray}
&&\Downarrow \nonumber \\
&&{}\nonumber \\
&&\lbrace P^A_{ADM},P^B_{ADM} \rbrace \approx 0,\nonumber \\
&&\lbrace P^A_{ADM}, J^{BC}_{ADM} \rbrace \approx {}^4\eta^{AC} P^B_{ADM}-
{}^4\eta^{AB} P^C_{ADM},\nonumber \\
&&\lbrace J^{AB}_{ADM}, J^{CD}_{ADM} \rbrace \approx - C^{ABCD}_{EF} J^{EF}
_{ADM},
\label{V12}
\end{eqnarray}

\noindent in accord with Eqs. (\ref{IV4}).

In Ref.\cite{reg1} it is noted that the terms depending on the
constraints in Eq.(\ref{V11}) contain the Hamiltonian version of the
supertranslation ambiguity. Indeed, these terms depend on
${}^3g^{\check r\check s}(\tau ,\vec
\sigma )-\delta^{\check r\check s}$ and, by using Eq.(\ref{III6}), this
quantity may be rewritten as $-{1\over r}\, {}^3{\tilde s}^{\check r\check s}
(\tau ,{{\sigma^{\check n}}\over r})+{}^3{\tilde g}^{\check r\check s}(\tau
,\vec \sigma )$ with ${}^3{\tilde g}^{\check r\check s}(\tau ,\vec \sigma )$
going to zero at spatial infinity faster than $1/r$. Now the objects $\int
d^3\sigma \, {1\over r}\, {}^3{\tilde s}^{\check r\check s}
(\tau ,{{\sigma^{\check n}}\over r}) {\tilde {\cal H}}(\tau ,\vec \sigma )
\approx 0$,.... are generators of supertranslation gauge transformations with 
zero momentum generalizing those appearing in the Dirac Hamiltonian,
i.e. $\int d^3\sigma [s {\tilde {\cal H}}+s_{\check r}\,
{}^3{\tilde {\cal H}}](\tau ,\vec \sigma )$.

To remove this gauge ambiguity in the Poincar\'e algebra and
simultaneously to kill the supertranslations, which forbid the
existence of a unique Poincar\'e group, the strategy of
Ref.\cite{reg1} is to add four gauge fixings to the secondary first
class constraints ${\tilde {\cal H}}(\tau ,\vec \sigma )\approx 0$,
${}^3{\tilde {\cal H}}^{\check r}(\tau ,\vec \sigma )\approx 0$ to fix
a coordinate system and therefore to build a realization of the
reduced phase space. In Ref.\cite{reg1}
one uses the maximal slice condition and harmonic 3-coordinates.

Anderson's paper \cite{reg2} shows that to have ``zero momentum" for
the supertranslations \footnote{Namely vanishing supertranslation charges
arising from the parts $s(\tau ,\vec \sigma )$, $s_{\check r}(\tau
,\vec \sigma )$ of $n(\tau ,\vec \sigma)$, $n_{\check r}(\tau ,\vec
\sigma )$.} and also to have well defined Lorentz charges, one needs
the parity conditions in suitable function spaces, which do not imply
a strong Poincar\'e algebra, and a class ${\cal C}$ of coordinate
systems of $M^4$ including the gauges corresponding to York QI gauge
conditions. In that paper it is also shown that to preserve the
boundary conditions containing the parity conditions, one has to
restrict $Diff\, M^4$ to the allowed transformations $Diff_I\, M^4
\times P$ \footnote{Namely to pseudo-diffeomorphisms tending to the identity in
a direction-independent way at spatial infinity plus the Poincar\'e
group.}.

However, the presence of
supertranslations is intertwined with the problem of the choice of the
3+1 splittings of $M^4$ with associated foliations well defined
geometrically at spatial infinity (one needs a restriction to gauges
where ${\tilde \lambda}_{AB}(\tau )=0$, as said also in Appendix A) and with 
the rejection of the interpretation
of the asymptotic Poincar\'e charges (and of the supertranslation
generators in the SPI algebra) as generators of improper gauge
transformations.

\vfill\eject

\section{Absence of Supertranslations: Christodoulou-Klainermann 
Spacetimes.}

Due to all these problems, instead of adding gauge fixings, it will be
assumed: i) the existence of a restricted class ${\cal C}$ of coordinate
systems for $M^4$ associated with Eqs.(\ref{IV4}), i.e. with $m=n$,
$m_r=n_r$; ii) that the gauge transformations are so restricted that
we cannot leave this class ${\cal C}$. The four gauge fixings then
allow to choose a particular coordinate system in the class ${\cal C}$
and to get a strong Poincar\'e algebra in Eq.(\ref{V11}). With this
restriction the (unknown) supertranslation generators in the SPI
algebra should vanish.

Since supertranslations must be absent to have a unique Poincar\'e
algebra, it must be

\bea
 &&{}^3t^{\check r\check s}(\tau , {{\sigma^{\check n}}\over r})=0,\quad\quad
 {}^3s_{\check r\check s}(\tau , {{\sigma^{\check n}}\over r}) = 
M \delta_{\check r\check s},
\nonumber \\
 &&and \nonumber \\
 &&s(\tau ,\vec \sigma )=s_{\check r}(\tau ,\vec \sigma )=0,\nonumber \\
 &&\Downarrow \nonumber \\
 &&m(\tau ,\vec \sigma )=n(\tau ,\vec \sigma ),\quad \quad
m_{\check r}(\tau ,\vec \sigma )=n_{\check r}(\tau ,\vec \sigma ),
 \label{VI1}
 \eea

\noindent in every allowed coordinate system (they are connected by 
{\it proper} gauge transformations which do not introduce asymptotic
angle-dependence).

This suggests that, in a suitable class ${\cal C}$ of coordinate
systems for $M^4$ \footnote{Then transformed to  coordinates adapted to the 3+1
splitting of $M^4$ with a foliation with spacelike leaves
$\Sigma_{\tau}$, whose allowed coordinates systems are in the
previously defined atlas ${\cal C}_{\tau}$.} asymptotic to Minkowski
coordinates and with the general coordinate transformations suitably
restricted at spatial infinity so that it is not possible to go out
this class, one should have the following direction-independent
boundary conditions for the ADM variables for $r\, \rightarrow \,
\infty$ [$\epsilon > 0$]

\begin{eqnarray}
{}^3g_{\check r\check s}(\tau ,\vec \sigma )&=&(1+{M\over r})
\delta_{\check r\check s}+{}^3h_{\check r\check s}(\tau ,\vec \sigma ),
\quad\quad {}^3h_{\check r\check s}(\tau ,\vec \sigma )=O(r^{-(1+\epsilon )}),
\nonumber \\
{}^3{\tilde \Pi}^{\check r\check s}(\tau ,\vec \sigma )&=&{}^3k^{\check
r\check s}(\tau ,\vec \sigma )=O(r^{-(2+\epsilon )}),\nonumber \\
&&{}\nonumber \\
N(\tau ,\vec \sigma )&=& N_{(as)}(\tau ,\vec \sigma )
+n(\tau ,\vec \sigma ),\quad\quad n(\tau ,\vec
\sigma )\, = O(r^{-(2+\epsilon )}),\nonumber \\
N_{\check r}(\tau ,\vec \sigma )&=&N_{(as)\check r}(\tau ,\vec \sigma )+
n_{\check r}(\tau ,\vec \sigma ),\quad\quad
n_{\check r}(\tau ,\vec \sigma )\, = O(r^{-\epsilon}),\nonumber \\
&&{}\nonumber \\
N_{(as)}(\tau ,\vec \sigma )&=&
-{\tilde \lambda}_{\tau}(\tau )-{1\over 2}{\tilde \lambda}_{\tau \check
s}(\tau )\sigma^{\check s},\nonumber \\
N_{(as)\check r}(\tau ,\vec \sigma )&=&
-{\tilde \lambda}_{\check r}(\tau )-{1\over 2}{\tilde \lambda}_{\check r\check
s}(\tau ) \sigma^{\check s},\nonumber \\
\Rightarrow&& N_{(as)A}(\tau ,\vec \sigma )\, {\buildrel {def} \over =}\,
(N_{(as)}\, ;\, N_{(as) \check r}\, )(\tau ,\vec \sigma )
=-{\tilde \lambda}_A(\tau )-{1\over 2}{\tilde \lambda}_{A\check s}(\tau )
\sigma^{\check s},
\label{VI2}
\end{eqnarray}

\noindent in accord with Regge-Teitelboim\cite{reg} and Beig-O'Murchadha
\cite{reg1}.

We have assumed the angle-independent behaviour ${}^3s_{\check r\check
s}(\tau ,{{\sigma^{\check n}}\over r})=M\delta_{\check r\check s}$,
${}^3t^{\check r\check s}(\tau ,{{\sigma^{\check n}}\over r})= 0$.
Since this implies the vanishing of the ADM momentum, $P^{\check
r}_{ADM}=0$, we see that the elimination of supertranslations is
connected with a definition of {\it rest frame} in the asymptotic Dirac
coordinates $z^{(\mu )}_{(\infty )}(\tau ,\vec \sigma )$. Therefore,
the previous boundary conditions on ${}^3g$, ${}^3{\tilde \Pi}$, are
compatible and can be replaced with the Christodoulou-Klainermann ones
of Eq.(\ref{III4}), but in general with {\it non vanishing shift functions}.
To have a non-vanishing ADM momentum one should have ${}^3t^{\check
r\check s}(\tau ,\vec \sigma )= const.\, \delta^{\check r\check s}$ in 
Eqs.(\ref{III7}) violating the parity conditions and creating
problems with supertranslations.

The vanishing of the strong ADM 3-momentum $P^{\check r}_{ADM}=0$ and 
Eq.(\ref{V10}) imply

\begin{eqnarray}
{\hat P}^{\check r}_{ADM}&\approx& 0,\nonumber \\
 &&{}\nonumber \\
P^{(\mu )}_{ADM}&=&b^{(\mu )}_{(\infty )\tau} P^{\tau}_{ADM}=l^{(\mu )}
_{(\infty )} P^{\tau}_{ADM},\nonumber \\
{\hat P}^{(\mu )}_{ADM}&\approx& l^{(\mu )}_{(\infty )} {\hat P}^{\tau}_{ADM}.
\label{VI3}
\end{eqnarray}

Therefore, the boundary conditions (\ref{VI2}) require three first class
constraints implying the vanishing of the weak ADM 3-momentum as a
{\it rest frame} condition.

Therefore, to have a formulation of metric gravity in which all the
fields and the gauge transformations have an angle-independent limit
at spatial infinity we have to add 6 gauge fixings on the $b^{(\mu
)}_{(\infty )A}(\tau )$ [see later on Eqs.(\ref{VIII1})] like we do in
parametrized Minkowski theory for going from arbitrary spacelike
hyperplanes to the Wigner ones (orthogonal to $p^{(\mu )}_s\approx
P^{(\mu )}_{sys}$) as it is explained in Appendix A: only on them we get the 
constraints ${\vec P}_{sys}\approx 0$ giving the rest-frame conditions. Let 
us call Wigner-Sen-Witten (WSW) the selected spacelike hypersurfaces
$\Sigma^{(WSW)}_{\tau}$ (the reason of the name will be clear in
Section XII). Since $b^{(\mu )}_{(\infty )A}={{\partial z^{(\mu
)}_{(\infty )}(\sigma )}\over {\partial \sigma^A}}$, this is a strong
restriction on the coordinate systems $x^{\mu}=z^{\mu}(\tau ,\vec
\sigma )\, \rightarrow \, \delta^{\mu}_{(\mu )} z^{(\mu )}
_{(\infty )}(\tau ,\vec \sigma )$ of $M^4$, which can be reached
from the $\Sigma^{WSW)}_{\tau}$-adapted coordinates $\sigma^A=(\tau ,\vec
\sigma )$ without introducing asymptotic angle dependence (namely 
supertranslations). The time constancy of these six gauge fixings will give
${\tilde \lambda}_{AB}(\tau )=0$, solving the problems with the
geometrical definition of the foliations at spatial infinity since
${\tilde \lambda}_{AB}(\tau )$ discussed in Appendix A and  implying
3+1 splittings of $M^4$ with leaves well defined  at spatial infinity.

With these assumptions one has from Eqs.(\ref{II2}) the following form
of the line element

\begin{eqnarray}
ds^2&=& \epsilon \Big( [N_{(as)}+n]^2 - [N_{(as)\check r}+n_{\check r}]
{}^3g^{\check r\check s}[N_{(as)\check s}+n_{\check s}] \Big) (d\tau )^2-
\nonumber \\
&-&2\epsilon [N_{(as)\check r}+n_{\check r}] d\tau d\sigma^{\check r} -
\epsilon \, {}^3g_{\check r\check s} d\sigma^{\check r} 
d\sigma^{\check s}\Big) =
\nonumber \\
&=&\epsilon \Big( [N_{(as)}+n]^2 (d\tau )^2- \nonumber \\
&-&{}^3g^{\check r\check s} [{}^3g_{\check r\check u}d\sigma^{\check
u}+(N_{(as)\check r}+n_{\check r}) d\tau ] [{}^3g_{\check s\check
v}d\sigma^{\check v}+(N_{(as)\check s}+n_{\check s}) d\tau ]\Big) .
\label{VI4}
\end{eqnarray}

Let us remark that asymptotically at spatial infinity the line element 
$ds^2$ of Eq.(\ref{VI4}) becomes

\begin{eqnarray}
ds^2_{(as)}&=&\epsilon \Big( [N^2_{(as)}-{\vec N}^2_{(as)}] (d\tau )^2 -2{\vec
N}_{(as)}\cdot d\tau d\vec \sigma -d{\vec \sigma}^2 \Big) +O(r^{-1})=
\nonumber \\
&=&\epsilon \Big( \Big[ {\tilde \lambda}^2_{\tau}-{\vec {\tilde
\lambda}}^2+({\tilde
\lambda}_{\tau}{\tilde \lambda}_{\tau s}-{\tilde \lambda}_r{\tilde \lambda}
_{rs})\sigma^s+{1\over 4}({\tilde \lambda}_{\tau u}{\tilde \lambda}_{\tau v}-
{\tilde \lambda}_{ru}{\tilde \lambda}_{rv})\sigma^u\sigma^v\Big] (d\tau )^2+
\nonumber \\
&+&2({\tilde \lambda}_r+{1\over 2}{\tilde \lambda}_{rs}\sigma^s) 
d\tau d\sigma^r
-d{\vec \sigma}^2 \Big) +O(r^{-1})=\nonumber \\
&=&\epsilon \Big( [{\tilde \lambda}^2_{\tau}-{\vec {\tilde \lambda}}+2({\tilde
\lambda}_{\tau}{{a_s}\over {c^2}}+\epsilon_{sru}{\tilde \lambda}_r{{\omega^u}
\over c})\sigma^s+{1\over {c^2}}({{a^ua^v}\over {c^2}}+\omega^u\omega^v-\delta
^{uv}{\vec \omega}^2)\sigma^u\sigma^v ] (d\tau )^2+\nonumber \\
&+&2[{\tilde \lambda}_r-\epsilon_{rsu}\sigma^s{{\omega^u}\over c}]
d\tau d\sigma^r - d{\vec \sigma}^2 \Big) +O(r^{-1})=\nonumber \\
 &=& \epsilon \Big( [{\tilde \lambda}^2_{\tau}(\tau )-
 {\vec {\tilde \lambda}}^2(\tau )] (d\tau )^2 +2{\tilde \lambda}_r(\tau )
 d\tau d\sigma^r-d{\vec \sigma}^2 \Big) +\nonumber \\
 &+&{1\over {c^2}} [2\vec a\cdot
\vec \sigma +({{a^ua^v}\over {c^2}}+\omega^u\omega^v-
\delta^{uv}{\vec \omega}^2)\sigma^u\sigma^v] (d\tau )^2-2\epsilon^{rsu}\sigma^s
{{\omega^u}\over c} d\tau d\sigma^u +O(r^{-1}),\nonumber \\
&&{}\nonumber \\
 &&{\tilde \lambda}_{\tau r}(\tau )=2{{a_r(\tau
)}\over {c^2}},\quad acceleration ,\nonumber \\ &&{\tilde
\lambda}_{rs}(\tau )=-2\epsilon_{rsu}{{\omega^u(\tau )}\over c},\quad
angular\, velocity\, of\, rotation;\nonumber \\
 &&{}\nonumber \\
  &&for\, {\tilde \lambda}_{AB}(\tau )=0,\quad (absence\, of\, 
supertranslations)  \nonumber \\
 &&{}\nonumber \\
 ds^2_{(as)}&=& \epsilon \Big( [{\tilde \lambda}^2_{\tau}(\tau )-
 {\vec {\tilde \lambda}}^2(\tau )] (d\tau )^2 +2{\tilde \lambda}_r(\tau )
 d\tau d\sigma^r-d{\vec \sigma}^2 \Big) +O(r^{-1}),\nonumber \\
 &&{}\nonumber \\
 && for {\tilde \lambda}_{\tau}(\tau )=\epsilon ,\quad {\tilde 
\lambda}_r(\tau )=0 \nonumber \\
 &&{}\nonumber \\
 ds^2_{(as)}&=& {}^4\eta_{AB} d\sigma^A d\sigma^B + O(r^{-1}).
\label{VI5}
\end{eqnarray}

Since we have ${\dot x}^{(\mu )}_{(\infty )}(\tau )\, {\buildrel \circ
\over =}\, b^{(\mu )}_{(\infty )A}(\tau ) {\tilde \lambda}^A(\tau )$, it 
follows that for ${\tilde \lambda}_{\tau}(\tau )=\epsilon$, ${\tilde 
\lambda}_r(\tau )=0$, the point ${\tilde x}^{(\mu )}_{(\infty )}(\tau )$
moves with 4-velocity $(\epsilon ;\vec 0)$ and has attached an
accelerated rotating coordinate system\cite{stephani}, which becomes
inertial when ${\tilde \lambda}_{AB}(\tau )=0$,namely when the
foliations become geometrically well defined at spatial infinity.

The conclusion of this discussion is a qualitative indication on which
type of  atlas ${\cal C}$ of coordinate systems is allowed on the
4-manifolds of spacetimes $M^4$ without supertranslations and on which
type of function space ${\cal W}$ (an appropriate weighted Sobolev
space as for Yang-Mills theory\cite{lusa}) is needed for the field
variables ${}^3g_{\check r\check s}(\tau ,\vec \sigma )$, ${}^3{\tilde
\Pi}^{\check r\check s}(\tau ,\vec
\sigma )$, $n(\tau ,\vec \sigma )$, $n_{\check r}(\tau ,\vec \sigma )$ and for
the parameters $\alpha (\tau ,\vec \sigma )$, $\alpha_{\check r}(\tau ,\vec
\sigma )$ [of which $n(\tau ,\vec \sigma )$, $n_{\check r}(\tau ,\vec \sigma
)$ are special cases] of allowed proper gauge transformations connected to
the identity \footnote{The rigid improper ones have been eliminated and 
replaced by the new canonical variables ${\tilde \lambda}_A(\tau )$, ${\tilde
\lambda}_{AB}(\tau )$ but with ${\tilde \lambda}_{AB}(\tau )=0$ to avoid
supertranslations and geometrical problems with the foliations at
spatial infinity.}, generated by the secondary first class constraints.

We must have:\hfill\break

i) The allowed 3+1 splittings of these spacetimes $M^4$ must have the
leaves, i.e. the Cauchy spacelike hypersurfaces $\Sigma_{\tau}$,
approaching Minkowski hyperplanes at spatial infinity in a
direction-independent way and being asymptotically orthogonal to the
weak ADM 4-momentum. The leaves $\Sigma_{\tau}\approx R^3$ have an
atlas ${\cal C}_{\tau}$ containing the global coordinate systems $\{
\sigma^{\check r} \}$ in which Eq.(\ref{VI2}) holds. Starting from the adapted
coordinates $(\tau ,\vec \sigma )$ in $R_{\tau} \times {\cal
C}_{\tau}$ we build coordinates for $M^4$ which asymptotically tend to
$\delta^{\mu}_{(\mu )} z^{(\mu )}_{(\infty )}(\tau ,\vec
\sigma )$ of Eq.(\ref{IV1}).\hfill\break

ii) The atlas ${\cal C}$ for these spacetimes $M^4$ contains only
coordinate systems approaching the Dirac asymptotic Minkowski
rectangular coordinates of Eq.(\ref{IV1}) at spatial infinity in a
direction-independent way and restricted by the 6 gauge fixings [see
later on Eq.(\ref{VIII1})] needed to get ${\tilde \lambda}_{AB}(\tau
)=0$.\hfill\break

When anyone of these conditions i) and ii) is not respected, some
quantity becomes asymptotically angle-dependent and one looses control
on supertranslations (the angle-dependence is propagated by the gauge
transformations).\hfill\break
\hfill\break

iii)As a consequence of what has been said and of Eqs.(\ref{VI2}),
the space ${\cal W}$ should be defined by angle (or
direction)-independent boundary conditions for the field variables for
$r\, \rightarrow
\infty$ of the following form:

\begin{eqnarray}
&&{}^3g_{\check r\check s}(\tau ,\vec \sigma )\, {\rightarrow}_{r\,
\rightarrow \infty}\, (1+{M\over r})
\delta_{\check r\check s}+{}^3h_{\check r\check s}(\tau
,\vec \sigma )=(1+{M\over r})
\delta_{\check r\check s}+O(r^{-3/2}),\nonumber \\
&&{}^3{\tilde \Pi}^{\check r\check s}(\tau ,\vec \sigma )\, {\rightarrow}
_{r\, \rightarrow \infty}\, {}^3k^{\check r\check s}(\tau ,\vec \sigma )=
O(r^{-5/2}),\nonumber \\ &&n(\tau ,\vec \sigma )\, {\rightarrow}_{r\,
\rightarrow \infty}\, O(r^{-(2+\epsilon )}),\quad \epsilon > 0,\nonumber \\
&&n_{\check r}(\tau ,\vec \sigma )\,
{\rightarrow}_{r\, \rightarrow \infty}\, O(r^{-\epsilon}),\quad
\epsilon > 0,\nonumber \\
&&{\tilde \pi}_n(\tau ,\vec \sigma )\, {\rightarrow}_{r\,
\rightarrow \infty}\, O(r^{-3}),\nonumber \\
&&{\tilde \pi}^{\check r}_{\vec n}(\tau
,\vec \sigma )\, {\rightarrow}_{r\, \rightarrow \infty}\, O(r^{-3}),
\nonumber \\
&&\lambda_n(\tau ,\vec \sigma )\, {\rightarrow}_{r\, \rightarrow \infty}\,
O(r^{-(3+\epsilon )}),\nonumber \\
&&\lambda^{\vec n}_{\check r}(\tau ,\vec \sigma )\, {\rightarrow}_{r\,
\rightarrow \infty}\, O(r^{-\epsilon}),\nonumber \\
&&\alpha (\tau ,\vec \sigma )\, {\rightarrow}_{r\, \rightarrow \infty}\,
O(r^{-(3+\epsilon )}),\nonumber \\
&&\alpha_{\check r}(\tau ,\vec \sigma )\, {\rightarrow}_{r\, \rightarrow
\infty}\, O(r^{-\epsilon}),\nonumber \\
&&\Downarrow \nonumber \\
&&{\tilde {\cal H}}(\tau ,\vec \sigma )\, {\rightarrow}_{r\, \rightarrow
\infty}\, O(r^{-3}),\nonumber \\
&&{}^3{\tilde {\cal H}}^{\check r}(\tau ,\vec \sigma )\, {\rightarrow}_{r\,
\rightarrow \infty}\, O(r^{-3}).
\label{VI6}
\end{eqnarray}

With these boundary conditions we have $\partial_{\check u}\, {}^3g_{\check
r\check s}=O(r^{-2})$ and not $O(r^{-(1+\epsilon )})$ 
\footnote{Note that with this last
condition and $\epsilon < 1/2$ it is shown in Ref.\cite{fadde} that the ADM
action (but in the first order formulation)
becomes meaningless since the spatial integral diverges (in this
reference it is also noted that with these boundary conditions adapted to
asymptotic flatness at spatial infinity the Hilbert action may not produce a
consistent and finite variational principle).}; this is compatible with
the definition of gravitational radiation given by Christodoulou and
Klainermann, but not with the one of Ref.\cite{trautm}.

In this function space ${\cal W}$ supertranslations are not allowed by
definition and proper gauge transformations generated by the secondary
constraints map ${\cal W}$ into itself. A coordinate-independent
characterization of ${\cal W}$ (see Ref.\cite{reg4} for an attempt)
should be given through an intrinsic definition of a minimal atlas of
coordinate charts ${\cal C}_{\tau}$ of $\Sigma_{\tau}$ such that the
lifts to 3-tensors on $\Sigma_{\tau}$ in ${\cal W}$ of the
3-diffeomorphisms in $Diff\, \Sigma_{\tau}$  maps them into them. 

Therefore, a unique asymptotic Poincar\'e group, modulo gauge 
transformations, is selected. Moreover, in accord with Anderson\cite{reg2} 
also $Diff\, M^4$ is restricted to $Diff_I\, M^4
\times P$, so to map the class ${\cal C}$ of coordinate systems into
itself. Now in $Diff_I\, M^4\times P$ the allowed  proper
pseudo-diffeomorphisms $Diff_I\, M^4$ are a normal subgroup (they go
to the identity in an angle-independent way at spatial infinity),
while the Poincar\'e group $P$ describes the rigid improper gauge
transformations (the non-rigid improper ones are assumed to be absent)
as in Bergmann''s proposal\cite{be}. Finally, following Marolf\cite{p18}, the
Poincar\'e group is not interpreted as a group of improper gauge
transformations but only as a source of superselection rules, which
however seem to be consistent only in the rest frame $P^{\check
r}_{ADM}=0$, if we insist on the absence of supertranslations so to
have the possibility to define the ADM spin Casimir.

Since in Section IX it will be shown that the gauge transformations generated 
by the superhamiltonian constraint produce a change in the extrinsic
curvature of the spacelike hypersurface $\Sigma_{\tau}$ trnasformaing it in a 
different spacelike hypersurface, one has the indication
that, in absence of supertranslations, the functions $N$, $\alpha$,
$\lambda_N$, should go like $O(r^{-(2+\epsilon )})$ and not like
$O(r^{-\epsilon })$ (in the case of proper gauge transformations).

iv) All the previous discussion points toward assuming the following Dirac
Hamiltonian

\begin{eqnarray}
{\hat H}^{"}_{(c)ADM} &=&\int d^3\sigma \Big[ (N_{(as)}+ n) {\tilde
{\cal H}}+(N_{(as)\check r}+ n_{\check r})\, {}^3{\tilde {\cal
H}}^{\check r}\Big] (\tau ,\vec \sigma )+
\nonumber \\
&+&{\tilde \lambda}_A(\tau ) P^A_{ADM}+{1\over 2}{\tilde
\lambda}_{AB}(\tau ) J^{AB}_{ADM}= \nonumber \\
 &=&\int d^3\sigma \Big[ n
{\tilde {\cal H}}+n_{\check r}\, {}^3{\tilde {\cal H}}^{\check r}\Big]
(\tau ,\vec \sigma )+ {\tilde \lambda}_A(\tau ) {\hat
P}^A_{ADM}+{1\over 2}{\tilde \lambda}
_{AB}(\tau ){\hat J}^{AB}_{ADM}, \nonumber \\
{\hat H}^{"}_{(D)ADM}&=&{\hat H}^{"}_{(c)ADM}+\int d^3\sigma [\lambda_n {\tilde
\pi}^n+\lambda^{\vec n}_r {\tilde \pi}^r_{\vec n}](\tau ,\vec \sigma )+
\zeta_A(\tau ) {\tilde \pi}^A(\tau )+\zeta_{AB}(\tau ) {\tilde \pi}
^{AB}(\tau ),
\label{VI7}
\end{eqnarray}

\noindent but with suggestion that it becomes well defined and without 
supertranslations only when ${\tilde \lambda}_{AB}(\tau )=0$.

\vfill\eject

\section{Two Scenarios for Hamiltonian Metric Gravity.}

At this point, after the two modifications of Hamiltonian
metric gravity connected  with i) the addition of the surface integrals 
and ii) the change (\ref{IV7}) of the primary constraints resulting
from the assumed splitting (\ref{IV5}) of the lapse and shift functions, two 
possible scenarios can be imagined (for the second one  the Lagrangian is 
unknown):

A) Consider as configurational variables

\beq
n_A(\tau ,\vec \sigma )=(n\, ;\, n_{\check r}\, )(\tau ,\vec \sigma
),\quad {\tilde \lambda}_A(\tau ),\quad {\tilde
\lambda}_{AB}(\tau ),\quad {}^3g_{\check r\check s}(\tau ,\vec \sigma ),
\label{VII1}
\eeq

\noindent with conjugate momenta

\beq
{\tilde \pi}^A_n(\tau ,\vec \sigma )=({\tilde \pi}^n\, ;\, {\tilde
\pi}^{\check r}_{\vec n}\, )(\tau ,\vec \sigma )\approx 0,\quad {\tilde
\pi}^A(\tau )\approx 0,\quad {\tilde \pi}^{AB} (\tau )\approx 0,\quad
{}^3{\tilde \Pi}^{\check r\check s} (\tau ,\vec \sigma ).
\label{VII2}
\eeq

The vanishing momenta are assumed to be the primary
constraints and one considers the following finite and
differentiable  Dirac Hamiltonian   as the defining Hamiltonian:

\begin{eqnarray}
{\hat H}^{(1)}_{(D)ADM}&=&\int d^3\sigma [n_A\, {\tilde {\cal H}}^A+\lambda
_{n\, A} {\tilde \pi}^A_n](\tau ,\vec \sigma )+{\tilde \lambda}_A(\tau )
{\hat P}^A_{ADM}+
{1\over 2}{\tilde \lambda}_{AB}(\tau ) {\hat J}^{AB}_{ADM}+\nonumber \\
&+&\zeta_A(\tau ) {\tilde \pi}^A(\tau )+
\zeta_{AB}(\tau ) {\tilde \pi}^{AB}(\tau ),
\label{VII3}
\end{eqnarray}

\noindent where $n_A=(n; n_{\check r})$,
${\tilde {\cal H}}^A=({\tilde {\cal H}}\, ;\, {}^3{\tilde
{\cal H}}^{\check r}\, )$ and where $\lambda_{n\, A}(\tau ,\vec \sigma )=
(\lambda_n\, ;\, \lambda^{\vec n}_{\check r}\, )(\tau ,\vec \sigma )$,
$\zeta_A(\tau )$, $\zeta_{AB}(\tau )$, are
Dirac multipliers associated with the primary constraints.\hfill\break

The time constancy of the primary constraints implies the following
secondary ones

\bea
&&{\tilde {\cal H}}^A(\tau ,\vec \sigma )\approx 0\,  \nonumber \\
 &&{\hat P}^A_{ADM}\approx 0,\quad\quad {\hat J}^{AB}_{ADM}\approx 0,
\label{VII4}
\eea

While the ${\tilde {\cal H}}^A(\tau ,\vec \sigma )$ are generators
of proper gauge transformations, the other ten either are generators of 
{\it improper} gauge transformations
(in this case 10 conjugate degrees of freedom in the 3-metric
are extra gauge variables) or, following Marolf's proposal \cite{p18},
define a superselection sector (like it happens for the vanishing of the
color charges for the confinement of quarks). All the constraints
(\ref{VII4}) are constants of the motion. 

All the constraints are first class, so that:
\hfill\break
\hfill\break
i) ${\tilde \lambda}_A(\tau )$, ${\tilde \lambda}_{AB}(\tau )$ are arbitrary
gauge variables conjugate to ${\tilde \pi}^A(\tau )\approx 0$, ${\tilde
\pi}^{AB}(\tau )\approx 0$ \footnote{Six gauge fixings to the constraints 
${\hat J}^{AB}_{ADM} \approx 0$ are needed to get the induced result 
${\tilde \lambda}_{AB}(\tau )=0$ which ensures foliations well defined at 
spatial infinity.};
\hfill\break ii) the physical reduced phase space of canonical metric
gravity is restricted to have {\it zero asymptotic Poincar\'e charges} so
that there is no natural Hamiltonian for the evolution in $\tau$, since 
Eq.(\ref{VII4}) implies ${\hat H}^{(1)}_{(D)ADM} \approx 0$.
\hfill\break
\hfill\break
This is the natural interpretation of ADM metric gravity which leads
to the Wheeler-De Witt equation after quantization (see Section  XI
for the problem of time in this scenario) and, in a sense, it is a
Machian formulation of an asymptotically flat noncompact (with
boundary $S_{\infty}$) spacetime $M^4$ in the same spirit of Barbour's
approach\cite{p20} and of the closed (without boundary)
Einstein-Wheeler universes. However, in this case there is no solution
to the problem of deparametrization of metric gravity and no
connection with parametrized Minkowski theories restricted to
spacelike hyperplanes.\hfill\break
\hfill\break
Note that the scenario A)
corresponds to the exceptional orbit ${\hat P}^A_{ADM}=0$
of the asymptotic Poincar\'e group.\hfill\break
\hfill\break

B) According to the suggestion of Dirac, modify ADM metric gravity by
adding the 10 new canonical pairs $x^{(\mu )}_{(\infty )}(\tau )$,
$p^{(\mu )}_{(\infty )}$, $b^{(\mu )}_{(\infty ) A}(\tau )$, $S^{(\mu
)(\nu )}_{\infty}$ \footnote{With the Dirac brackets (\ref{a11}) implying the
orthonormality constraints for the b's.} to the metric gravity phase
space with canonical basis $n_A(\tau ,\vec \sigma )
=(n\, ;\, n_{\check r}\, )(\tau ,\vec \sigma )$, ${\tilde \pi}^A_n(\tau ,\vec
\sigma )=({\tilde \pi}^n; {\tilde \pi}^{\check r}_{\vec n})
\approx 0$ (the primary constraints), ${}^3g_{\check r\check s}(\tau
,\vec \sigma )$, ${}^3{\tilde \Pi}^{\check r\check s}(\tau ,\vec
\sigma )$, and then: \hfill\break i) add the 10 new primary
constraints

\bea
\chi^A &=& p^A_{(\infty )}-{\hat P}^A_{ADM}=
b^A_{(\infty )(\mu )}(\tau ) [p^{(\mu )}_{(\infty )}-b^{(\mu )}
_{(\infty )B}(\tau ) {\hat P}^B_{ADM}] \approx 0,\nonumber \\
 \chi^{AB} &=& J^{AB}_{(\infty )}-{\hat J}^{AB}_{ADM}=b^A_{(\infty )
(\mu )}(\tau ) b^B_{(\infty )(\nu )}(\tau ) [S^{(\mu )(\nu )}_{(\infty )}-
b^{(\mu )}_{(\infty )C}(\tau ) b^{(\nu )}_{(\infty )D}(\tau ) {\hat J}^{CD}
_{ADM}] \approx 0,\nonumber \\
 &&{}\nonumber \\
&&\{ \chi^A, \chi^{BC} \} \approx {}^4\eta^{AC} \chi^B - {}^4\eta^{AB}
\chi^C \approx 0,\quad\quad \{ \chi^A,\chi^B \} \approx 0,\nonumber \\
&&\{ \chi^{AB}, \chi^{CD} \} \approx - C^{ABCD}_{EF} \chi^{EF}\approx
0,\nonumber \\
 &&\{ \chi^A(\tau ), {\tilde \pi}^D_n(\tau ,\vec \sigma )\}=\{ 
\chi^{AB}(\tau ),
{\tilde \pi}^D_n(\tau ,\vec \sigma )\}=0,\nonumber \\
 &&\{ \chi^A(\tau ),
{\tilde {\cal H}}^D(\tau ,\vec \sigma )\} \approx 0,\quad
\quad \{ \chi^{AB}(\tau ), {\tilde {\cal H}}^D(\tau ,\vec \sigma )\} \approx
0,
\label{VII5}
\eea

\noindent where
$p^A_{(\infty )}=b^A_{(\infty )(\mu )}p^{(\mu )}_{(\infty )}$, $J^{AB}_{(\infty
)}=b^A_{(\infty )(\mu )}b^B_{(\infty )(\nu )}S^{(\mu )(\nu )}_{(\infty )}$
\footnote{Remember that $p^A_{(\infty )}$ and $J^{AB}_{(\infty )}$ satisfy a 
Poincar\'e algebra.};\hfill\break
ii) consider ${\tilde \lambda}_A(\tau )$, ${\tilde \lambda}_{AB}(\tau )$, as
Dirac multipliers [like the $\lambda_{n A}(\tau ,\vec \sigma )$'s] for these 10
new primary constraints, and not as configurational
(arbitrary gauge) variables coming from the
lapse and shift functions \footnote{Therefore there are no conjugate momenta
${\tilde \pi}^A(\tau )$, ${\tilde \pi}^{AB}(\tau )$ and no associated Dirac
multipliers $\zeta_A(\tau )$, $\zeta_{AB}(\tau )$.}, in the
assumed finite and differentiable Dirac Hamiltonian

\begin{eqnarray}
H_{(D)ADM}&=& \int d^3\sigma [ n_A {\tilde {\cal H}}^A+\lambda_{n A} {\tilde
\pi}^A_n](\tau ,\vec \sigma )-\nonumber \\
&-&{\tilde \lambda}_A(\tau ) [p^A_{(\infty )}-{\hat P}^A
_{ADM}]-{1\over 2}{\tilde \lambda}_{AB}(\tau )[J^{AB}_{(\infty )}-
{\hat J}^{AB}_{ADM}]\approx 0,
\label{VII6}
\end{eqnarray}

\noindent further restricted to ${\tilde \lambda}_{AB}(\tau )=0$ to eliminate
supertranslations.

The reduced phase space is the ADM one and there is consistency with
Marolf's proposal\cite{p18} regarding superselection sectors: on the ADM
variables there are only the secondary first class constraints
${\tilde {\cal H}}^A(\tau ,\vec \sigma )
\approx 0$, generators of proper gauge transformations, because the other
first class constraints $p^A_{(\infty )}-{\hat P}^A_{ADM}\approx 0$, $J^{AB}
_{(\infty )}-{\hat J}^{AB}_{ADM}\approx 0$ do not generate improper gauge
transformations but eliminate 10 of the extra 20 variables.  One has an
asymptotically flat at spatial infinity
noncompact (with boundary $S_{\infty}$) spacetime $M^4$
with non-vanishing asymptotic Poincar\'e
charges  and the possibility to deparametrize metric gravity so to obtain the
connection with parametrized Minkowski theories restricted to spacelike
 Wigner hyperplanes, due to the rest-frame condition
$P^{\check r}_{ADM}=0$ forced by the elimination of supertranslations.

Scenario B) contains the rest-frame instant form of ADM metric
gravity.

\vfill\eject

\section{The Rest-Frame Instant Form of Metric Gravity.}

While the gauge fixings for the secondary constraints ${\tilde {\cal
H}}^A(\tau ,\vec \sigma )\approx 0$ and the resulting ones for the
primary ones ${\tilde \pi}^A_n(\tau ,\vec
\sigma )\approx 0$ , implying the determination of the
$\lambda_{n A}(\tau ,\vec \sigma )$, follow the scheme outlined at the
end of Section II, one has to clarify the meaning of the gauge fixings
for the extra 10 first class constraints.

The explicit absence of supertranslations requires the six gauge fixings to the
constraints $\chi^{AB}\approx 0$ so to have ${\tilde
\lambda}_{AB}(\tau )=0$. Let us remark that the Hamiltonian (\ref{VII6}) is
formally defined on more general spacelike hypersurfaces [those with 
 ${\tilde \lambda}_{AB}(\tau )\not= 0$], whose boundary conditions allow
a certain class of supertranslations. However, formally the constraints 
$\chi^{AB}(\tau )\approx 0$ generate gauge transformations, which make these
hypersurfaces equivalent to the WSW ones.

To go to the Wigner-Sen-Witten hypersurfaces \footnote{The analogue of the
Minkowski Wigner hyperplanes with the asymptotic normal $l^{(\mu
)}_{(\infty )}=l^{(\mu )}_{(\infty )\Sigma}$ parallel to ${\hat P}
^{(\mu )}_{ADM}$ (i.e. $l^{(\mu )}_{(\infty )}={\hat b}^{(\mu )}_{(\infty ) l}=
{\hat P}^{(\mu )}_{ADM}/\sqrt{\epsilon {\hat P}^2_{ADM}}$); see
Eqs.(\ref{VI3}).} one follows the procedure defined for Minkowski
spacetime: \hfill\break

i) one restricts oneself to spacetimes with $\epsilon p^2_{(\infty )}
={}^4\eta_{(\mu )(\nu )} p^{(\mu )}_{(\infty )}p^{(\nu )}
_{(\infty )} > 0$ \footnote{This is possible, because the positivity theorems 
for the ADM energy imply that one has only timelike or light-like
orbits of the asymptotic Poincar\'e group.}; \hfill\break

ii) one boosts at rest $b^{(\mu )}_{(\infty )A}(\tau )$ and $S^{(\mu
)(\nu )}_{(\infty )}$ with the Wigner boost $L^{(\mu )}{}_{(\nu )}
(p_{(\infty )}, {\buildrel \circ \over p}_{(\infty )})$; \hfill\break

iii) one adds the gauge fixings [with $u^{(\mu )}(p_{(\infty )})=p^{(\mu )}
_{(\infty )}/ \pm \sqrt{\epsilon p^2_{(\infty )} }$]

\bea
b^{(\mu )}_{(\infty )A}(\tau ) &\approx& L^{(\mu )}{}_{(\nu )=A}
(p_{(\infty )}, {\buildrel \circ \over p}_{(\infty )})=\epsilon ^{(\mu
)}_A (u(p_{(\infty )})), \nonumber \\
 &&{}\nonumber \\
  &&implying\quad\quad {\tilde \lambda}_{AB}(\tau ) =0,
\label{VIII1}
\eea

\noindent  and goes to Dirac brackets.\hfill\break
 In this way one gets

\bea
S^{(\mu )(\nu )}_{(\infty )} &\equiv& \epsilon^{(\mu )}_C(u(p_{(\infty
)}))\epsilon_D^{(\nu )}(u(p_{(\infty )})) {\hat J}^{CD}_{ADM}=S^{(\mu
)(\nu )}_{ADM},\nonumber \\
 z^{(\mu )}_{(\infty )}(\tau ,\vec \sigma )&=&x^{(\mu )}_{(\infty )}(\tau
)+\epsilon^{(\mu )}_r(u(p_{(\infty )})) \sigma^r,
\label{VIII2}
\eea

\noindent so that $z^{(\mu )}_{(\infty )}(\tau ,\vec \sigma )$ becomes
equal to the embedding identifying a Wigner hyperplane in Minkowski
spacetime [see Eq.(\ref{a13})].

The origin $x^{(\mu )}_{(\infty )}$ is now replaced by the not
covariant {\it external} center-of-mass-like canonical variable

\beq
{\tilde x}^{(\mu )}_{(\infty )}=x^{(\mu )}_{(\infty )}+{1\over 2}
\epsilon^A_{(\nu )}(u(p_{(\infty )})) \eta_{AB} {{\partial \epsilon^B_{(\rho )}(u(p
_{(\infty )}))}\over {\partial p_{(\infty )(\mu )}}} S^{(\nu )(\rho )}
_{(\infty )},
 \label{VIII3}
 \eeq

\noindent and one has

\beq
J^{(\mu )(\nu )}_{(\infty )}= {\tilde x}^{(\mu )}_{(\infty )}p^{(\nu
)}_{(\infty )}-{\tilde x}^{(\nu )}
_{(\infty )}p^{(\mu )}_{(\infty )}+{\tilde S}^{(\mu )(\nu )}_{(\infty )},
\label{VIII4}
\eeq

\noindent
with ${\tilde S}^{(\mu )(\nu )}_{(\infty )}=S^{(\mu )(\nu )}_{(\infty )}-
{1\over 2} \epsilon^A_{(\rho )}(u(p_{(\infty )})) \eta_{AB} ({{\partial
\epsilon^B_{(\sigma )}(u(p_{(\infty )}))}\over {\partial p_{(\infty )(\mu )}}}
p^{(\nu )}_{(\infty )}-{{\partial \epsilon^B_{(\sigma )}(u(p_{(\infty )}))}
\over {\partial p_{(\infty )(\nu )}}} p^{(\mu )}_{(\infty )} ) S^{(\rho
)(\sigma )}_{(\infty )}$. \hfill\break
\hfill\break
As in the Minkowski case one defines

\beq
{\bar S}^{AB}_{(\infty )}=\epsilon^A_{(\mu )} (u(p_{(\infty
)}))\epsilon^B_{(\nu )}(u(p_{(\infty )})) {\tilde S}^{(\mu ) (\nu
)}_{(\infty )},
 \label{VIII5}
 \eeq

\noindent and one obtains at the level of Dirac brackets

\begin{eqnarray}
{\bar S}^{\check r\check s}_{(\infty )}&\equiv& {\hat J}^{\check r\check s}
_{ADM},\nonumber \\
&&{}\nonumber \\
{\tilde \lambda}_{AB}(\tau )&=&0,\nonumber \\
&&{}\nonumber \\
-{\tilde \lambda}_A(\tau ) \chi^A &=&-{\tilde
\lambda}_A(\tau )\epsilon^A_{(\mu )}(u(p_{(\infty )})) [p^{(\mu )}_{(\infty )}
-\epsilon^{(\mu )}_B(u(p_{(\infty )})) {\hat P}^B_{AM}]=\nonumber \\
&=&-{\tilde \lambda}_A(\tau )\epsilon^A_{(\mu )}(u(p_{(\infty )})) [u^{(\mu )}
(p_{(\infty )}) (\epsilon_{(\infty )}-{\hat P}^{\tau}_{ADM})-\epsilon^{(\mu )}
_{\check r}(p_{(\infty )}){\hat P}^{\check r}_{ADM}]=\nonumber \\
&=&-{\tilde \lambda}_{\tau}(\tau ) [\epsilon_{(\infty )}-{\hat P}^{\tau}_{ADM}]
+{\tilde \lambda}_{\check r}(\tau ) {\hat P}^{\check r}_{ADM},\nonumber \\
&&{}\nonumber \\
\Rightarrow&& \epsilon_{(\infty )}-{\hat P}^{\tau}_{ADM} \approx 0,\quad\quad
{\hat P}^{\check r}_{ADM}\approx 0,\nonumber \\
 &&{}\nonumber \\
 H_{(D)ADM}&=& \int d^3\sigma \Big[ n_A {\cal H}^A +\lambda_{n A}
 {\tilde \pi}^A_n\Big] (\tau ,\vec \sigma ) -{\tilde \lambda}_{\tau}(\tau )
 [\epsilon_{(\infty )}-{\hat P}^{\tau}_{ADM}] +
 {\tilde \lambda}_{\check r}(\tau ) {\hat P}^{\check r}_{ADM},
\label{VIII6}
\end{eqnarray}

\noindent in accord with Eqs.(\ref{VI3}). Only after this reduction
supertranslations are absent, there are no ill-defined quantities and
there are only proper gauge transformations going to the identity
asymptotically at spatial infinity.

Therefore, on the Wigner-Sen-Witten hypersurfaces, the
remaining four extra constraints are:

\bea
&&{\hat P}^{\check r}_{ADM}\approx 0,\nonumber \\
 &&\epsilon_{(\infty )}=-\epsilon \sqrt{\epsilon p^2
_{(\infty )}}\approx {\hat P}^{\tau }_{ADM} \approx -\epsilon M_{ADM}=
-\epsilon \sqrt{\epsilon {\hat P}^2_{AM}}.
\label{VIII7}
\eea

 Now the spatial indices have become spin-1 Wigner indices \footnote{They
transform with Wigner rotations under asymptotic Lorentz
transformations. The Wigner indices will be denoted $(\tau ; r)$ instead of
$(\tau ; \check r)$.}. As said for parametrized theories in Minkowski
spacetime, in this special gauge 3 degrees of freedom of the
gravitational field become gauge variables, while ${\tilde x}^{(\mu
)}_{(\infty )}$ becomes a decoupled observer with his clock near
spatial infinity. These 3 degrees of freedom represent an {\it internal}
center-of-mass 3-variable ${\vec \sigma}_{ADM}[{}^3g,{}^3{\tilde
\Pi}]$ inside the Wigner-Sen-Witten hypersurface; $\sigma^{\check
r}=\sigma^{\check r}_{ADM}$ is a variable representing the {\it 3-center
of mass} of the 3-metric of the slice $\Sigma_{\tau}$ of the
asymptotically flat spacetime $M^4$ and is obtainable from the weak
Poincar\'e charges with the group-theoretical methods of
Ref.\cite{pauri} as it is done in Ref.\cite{mate} for the Klein-Gordon
field on the Wigner hyperplane. Due to ${\hat P}^r_{ADM}\approx 0$ we
have

\bea
\sigma^r_{ADM} &=& -{{ {\hat J}^{\tau r} }\over { \sqrt{({\hat P}^{\tau}
_{ADM})^2-({\hat {\vec P}}_{ADM})^2} }}+\nonumber \\
 &+& {{ ({\hat {\vec J}}_{ADM} \times {\hat
{\vec P}}_{ADM})^r }\over { \sqrt{({\hat P}^{\tau}_{ADM})^2- ({\hat
{\vec P}}_{ADM})^2} ({\hat P}^{\tau}_{ADM} +\sqrt{({\hat
P}^{\tau}_{ADM})^2- ({\hat {\vec P}}_{ADM})^2})}}+ \nonumber \\
 &+& {{({\hat J}^{\tau s}_{ADM} {\hat P}^s_{ADM}) {\hat P}^r_{ADM}}\over
 {{\hat P}^{\tau}_{ADM}\sqrt{({\hat P}^{\tau}_{ADM})^2-
({\hat {\vec P}}_{ADM})^2} ({\hat P}^{\tau}_{ADM} +\sqrt{({\hat
P}^{\tau}_{ADM})^2- ({\hat {\vec P}}_{ADM})^2})}}\approx \nonumber \\
&\approx& -{\hat J}^{\tau r}_{ADM}/ {\hat P}^{\tau}_{ADM},\nonumber \\
&&{}\nonumber \\
 &&\{ \sigma^r_{ADM},\sigma^s_{ADM} \} =0,\quad
\{ \sigma^r_{ADM},{\hat P}^s_{ADM} \} = \delta^{rs},
\label{VIII8}
\eea

\noindent
so that ${\vec \sigma}_{ADM}\approx 0$ is equivalent to the
requirement that the ADM boosts vanish: this is the way out from the
boost problem quoted in Section III in the framework of the
rest-frame instant form.

When $\epsilon {\hat P}^2_{ADM} > 0$, with the asymptotic Poincar\'e
Casimirs ${\hat P}^2_{ADM}$, ${\hat W}^2_{ADM}$ one can build the
 M\o ller radius $\rho_{AMD}=\sqrt{-\epsilon {\hat W}^2_{ADM}}/{\hat
P}^2_{ADM}c$, which is an intrinsic classical unit of length like in
parametrized Minkowski theories, to be used as an ultraviolet cutoff
in a future attempt of quantization.

By going from ${\tilde x}^{(\mu )}_{(\infty )}$ 
\footnote{The non-covariant variable replacing $x^{(\mu )}_{(\infty )}$
after going to Dirac brackets with respect to the previous six pairs
of second class constraints.} and $p^{(\mu )}_{(\infty )}$ to the
canonical basis \cite{lus1}

\bea
&&T_{(\infty )}=p_{(\infty )(\mu )}{\tilde x}^{(\mu )}_{(\infty
)}/\epsilon_{(\infty )}=p_{(\infty )(\mu )}x^{(\mu )}_{(\infty
)}/\epsilon_{(\infty )} {}{}{},\nonumber \\
 && \epsilon_{(\infty )},\nonumber \\
 &&z^{(i)}_{(\infty )}=\epsilon_{(\infty )} ({\tilde x}^{(i)}_{(\infty
)}-p^{(i)}_{(\infty )}{\tilde x}^{(o)}_{(\infty )}
/p^{(o)}_{(\infty )}), \nonumber \\
 &&k^{(i)}_{(\infty )}=p^{(i)}_{(\infty )}/\epsilon
_{(\infty )}=u^{(i)}(p^{(\rho )}_{(\infty )}),
\label{VIII9}
\eea

\noindent
one finds that the final reduction requires the gauge fixings

\beq
T_{(\infty )}-\tau \approx 0,\quad\quad \sigma^{\check r}_{ADM}\approx
0\quad (or\, {\hat J}^{\tau r}_{ADM}\approx 0).
\label{VIII10}
\eeq

Since $\{ T_{(\infty )},\epsilon_{(\infty )} \}=-\epsilon$, with the
gauge fixing $T_{(\infty )}-\tau \approx 0$ one gets ${\tilde
\lambda}_{\tau} (\tau )\approx \epsilon$, $\epsilon_{(\infty )}\equiv 
{\hat P}^{\tau}_{ADM}$
and $H_{(D)ADM}={\tilde \lambda}_{\check r}(\tau ) {\hat P}^{\check
r}_{ADM}$. This is the frozen picture of the reduced phase space, like
it happens in the standard Hamilton-Jacobi theory: there is no time
evolution. To reintroduce an evolution in $T_{(\infty )}\equiv \tau$
we must use   the energy $M_{ADM}=-\epsilon {\hat P}^{\tau}_{ADM}$
(the ADM mass of the universe) as the natural physical Hamiltonian
\footnote{See Ref.\cite{ppons} for another derivation of this result.}.
Therefore the final Dirac Hamiltonian is

\begin{equation}
H_D=M_{ADM}+{\tilde \lambda}_{\check r}(\tau ) {\hat P}^{\check r}_{ADM},\quad
\quad M_{ADM}= -\epsilon {\hat P}^{\tau}_{ADM}.
\label{VIII11}
\end{equation}

\noindent That $M_{ADM}$ is the correct Hamiltonian for getting a 
$\tau$-evolution
equivalent to Einstein's equations in spacetimes asymptotically flat
at spatial infinity is also shown in Ref.\cite{fermi}. In the
rest-frame the time is identified with the parameter $\tau$ labelling
the leaves $\Sigma_{\tau}$ of the foliation of $M^4$. See Section XI
for comments on the problem of time in general relativity.

The final gauge fixings $\sigma^{\check r}_{ADM}\approx 0$   [or
${\hat J}^{\tau r}_{ADM}\approx 0$] imply ${\tilde \lambda}_{\check
r}(\tau )\approx 0$, $H_D=M_{ADM}$ and a reduced theory with the
{\it external} center-of-mass variables $z^{(i)}_{(\infty )}$,
$k^{(i)}_{(\infty )}$ decoupled \footnote{Therefore the choice of the origin
$x^{(\mu )}_{(\infty )}$ becomes irrelevant.} and playing the role of a
{\it point particle clock} for the time $T_{(\infty )}
\equiv \tau$. There would be a weak form of
Mach's principle, because only relative degrees of freedom would be present.

The condition ${\tilde \lambda}_{AB}(\tau )=0$ with ${\tilde
\lambda}_{\tau} (\tau )=\epsilon$, ${\tilde \lambda}_r(\tau )=0$ means
that at spatial infinity there are no local (direction dependent)
accelerations and/or rotations [$\vec a=\vec \omega =0$]. The
asymptotic line element of Eqs.(\ref{VI5}) for ${\vec {\tilde
\lambda}}(\tau )=0$ reduces to the line element of an inertial system
near spatial infinity ({\it preferred asymptotic inertial observers}, for
instance the {\it fixed stars} \cite{soffel}).

While the asymptotic {\it internal} realization of the Poincar\'e
algebra has the weak Poincar\'e charges ${\hat P}^{\tau}_{ADM}\approx
-\epsilon M_{ADM}$, ${\hat P}^r_{ADM}\approx 0$, ${\hat J}^{rs}_{ADM}$, ${\hat
K}_{ADM}^r={\hat J}^{\tau  r}_{ADM}\approx 0$ as generators, the
rest-frame instant form asymptotic {\it external} realization of the
Poincar\'e generators becomes (no more reference to the boosts ${\hat
J}^{\tau r}_{ADM}$)

\begin{eqnarray}
&&\epsilon_{(\infty )}=M_{ADM},\nonumber \\
&&p^{(i)}_{(\infty )},\nonumber \\
&&J^{(i)(j)}_{(\infty )}={\tilde x}^{(i)}_{(\infty )}p^{(j)}_{(\infty )}-
{\tilde x}^{(j)}_{(\infty )} p^{(i)}_{(\infty )} +\delta^{(i)\check r}\delta
^{(j)\check s}{\hat J}^{\check r\check s}_{ADM},\nonumber \\
&&J^{(o)(i)}_{(\infty )}=p^{(i)}_{(\infty )} {\tilde x}^{(o)}_{(\infty )}-
\sqrt{M^2_{ADM}+{\vec p}^2_{(\infty )}} {\tilde x}^{(i)}_{(\infty )}-
{ {\delta^{(i)\check r}{\hat J}^{\check r\check s}_{ADM} \delta^{(\check s(j)}
p^{(j)}_{(\infty )} }\over
{M_{ADM}+\sqrt{M^2_{ADM}+{\vec p}^2_{(\infty )}} } } .
\label{VIII12}
\end{eqnarray}

The  line element is

\begin{eqnarray}
ds^2&=& \epsilon \Big( [N_{(as)}+n]^2 - [N_{(as) r}+n_r] {}^3g^{rs}
[N_{(as) s}+n_s] \Big) (d\tau )^2-\nonumber \\
&-&2\epsilon [N_{(as) r}+n_r] d\tau d\sigma^r -\epsilon \,
{}^3g_{rs} d\sigma^r d\sigma^s.
\label{VIII13}
\end{eqnarray}

\vfill\eject

\section{The Interpretation of Hamiltonian Gauge Transformations.}

In this Section the interpretation of the Hamiltonian gauge transformations 
generated by the first class constraints of ADM metric gravity will be given.
Then there will be a comparison of the Hamiltonian gauge transformations with 
spacetime diffeomorphisms, which are both local Noether symmetries of the
Hilbert action and dynamical symmetries of the Einstein's equations.

\subsection{The Superhamiltonian Constraint as a Generator of
         Gauge Transformations.}

While it is geometrically trivial to give an interpretation of the
gauge transformations generated by the primary and supermomentum constraints in
metric gravity\footnote{The supermomentum constraints generate 
3-pseudo-diffeomorphisms corresponding the changes of 3-coordinates on 
the spacelike hypersurfaces $\Sigma_{\tau}$. The vanishing momenta of the 
lapse and shift functions generate gauge transformations which respectively
modify: i) how densely the spacelike hypersurfaces are distributed in the 
spacetime; ii) the convention of synchronization of clocks and the associated
gravitomagnetic precessional effecs (dragging of inertial frames)
\cite{ciuf,bini}.}, it is 
not clear which is the meaning of the gauge transformations generated by the
superhamiltonian constraint (see for instance Refs.\cite{wa}).
In Ref.\cite{tei} the superhamiltonian and supermomentum constraints
of ADM metric  gravity are interpreted as the generators of the change
of the canonical data ${}^3g_{rs}$, ${}^3{\tilde
\Pi}^{rs}$, under the normal and tangent deformations of the spacelike
hypersurface $\Sigma_{\tau}$ which generate $\Sigma_{\tau +d\tau}$
\footnote{One thinks to $\Sigma_{\tau}$ as determined by a cloud of observers,
one per space point; the idea of bifurcation and reencounter of the
observers is expressed by saying that the data on $\Sigma_{\tau}$
(where the bifurcation took place) are propagated to some final
$\Sigma_{\tau + d\tau}$ (where the reencounter arises) along different
intermediate paths, each path being a monoparametric family of
surfaces that fills the sandwich in between the two surfaces;
embeddability of $\Sigma_{\tau}$ in $M^4$ becomes the synonymous with
path independence; see also Ref.\cite{gr13} for the connection with
the theorema egregium of Gauss.}. Therefore, the algebra (\ref{II19}) of the
supermomentum and superhamiltonian constraints reflects the
embeddability of $\Sigma_{\tau}$ into $M^4$.

However, the lacking pieces of information are: i) which is the variable 
determined by the superhamiltonian constraint? ii) which is the conjugate
{\it free gauge} variable?

A) Let us consider first {\it compact} spacetimes. As a consequence of the
previous geometrical property, in the case of compact spacetimes without 
boundary the superhamiltonian constraint is interpreted as a {\it 
time-dependent Hamiltonian} for general relativity in some  {\it internal} time
variable defined in terms of the canonical variables \footnote{See for instance
Ref.\cite{beig} and the so called {\it internal intrinsic many-fingered
time} \cite{kku}.}.

The two main proposals for an {\it internal time} are:

i) The {\it intrinsic internal time} : it is the conformal factor $q(\tau
,\vec \sigma )={1\over 6} ln\, det\, {}^3g_{rs}$ or $\phi (\tau ,\vec
\sigma )= e^{{1\over 2}q(\tau ,\vec \sigma )}  =({}^3g)^{1/12}
> 0$ of the 3-metric, ${}^3g_{rs}=e^{2q}\, {}^3\sigma_{rs}=\phi^4\,
{}^3\sigma_{rs}$, $det\, {}^3\sigma_{rs}=1$. It is not a scalar and is
proportional to Misner's time $\Omega=-{1\over 3}\, ln\,
\sqrt{\hat \gamma}$ \cite{misner} for asymptotically flat spacetimes
(see Appendix C of Ref.\cite{russo2} for more details):  
$q=-{1\over 2} \Omega$.

ii) York's {\it extrinsic internal time} ${\cal T}=-{4\over 3}\epsilon k
\, {}^3K={2\over {3\sqrt{\gamma}}}{}^3{\tilde \Pi}$. In
Ref.\cite{beig} there is a review of the known results with York's
extrinsic internal time, Ref.\cite{yoyo} contains the comparison of 
York cosmic time with proper time, while in Refs.\cite{ish,kuchar1} there are
 more general reviews about the problem of time in general relativity (see
also Section XI).

There are two interpretations of the superhamiltonian constraint in
this framework:

a) either as a generator of time evolution (being a time-dependent
Hamiltonian) like in the commonly accepted viewpoint based on the
Klein-Gordon interpretation of the quantized superhamiltonian
constraint, i.e. the Wheeler-DeWitt equation \footnote{See Kuchar in
Ref.\cite{ku1} and Wheeler's evolution of 3-geometries in superspace
in Ref.\cite{dc11,mtw} ; see Ref.\cite{paren} for the cosmological
implications.}.

b) or as a quantum Hamilton-Jacobi  equation without any time. In this
case one can introduce a concept of evolution, somehow connected with an 
{\it effective time}, only in a WKB sense \cite{kiefer}.

A related problem, equivalent to the
transition from a Cauchy problem to a Dirichlet one and requiring a
definition of which time parameter has to be used (see for instance
the review in Ref.\cite{kuchar1}), is the validity of the {\it full or
thick sandwich conjecture} \cite{dc11,mtw} \footnote{Given two nearby 3-metrics
on Cauchy surfaces $\Sigma_{\tau_1}$ and $\Sigma_{\tau_2}$, there is a
unique spacetime $M^4$, satisfying Einstein's equations, with these
3-metrics on those Cauchy surfaces.} and of the {\it thin sandwich
conjecture} \footnote{Given ${}^3g$ and $\partial_{\tau}\, {}^3g$ on
$\Sigma_{\tau}$, there is a unique spacetime $M^4$ with these initial
data satisfying Einstein's equations; doing so, the rangian version of the
constraints is interpreted as a set of equations for the lapse and shift 
functions\cite{york1} logic of the
Hamiltonian constraint theory, in which the constraints do not depend on
these functions.}: see Ref. \cite{bafo} (and also
Ref.\cite{giusa}) for the non validity of the {\it full} case and for the
restricted validity (and its connection with constraint theory) of the
{\it thin} case.

B) Let us now consider the problem of which variable is the unknown in the
superhamiltonian constraint.

Since the superhamiltonian  constraint is quadratic in the momenta,
one is naturally driven to make a comparison with the free scalar
relativistic particle described by the first class constraint
$p^2-\epsilon m^2\approx 0$. As shown in Refs.\cite{lus1,dc1}, the
constraint manifold in phase space  (the two disjointed branches of the 
mass-hyperboloid) has 1-dimensional gauge orbits; the $\tau$-evolution
generated by the Dirac Hamiltonian $H_D=\lambda (\tau ) (p^2-\epsilon
m^2)$ gives the parametrized solution $x^{\mu}(\tau )$. Instead, if one
goes to the reduced phase space by adding the non-covariant gauge fixing
$x^o-\tau \approx 0$ and eliminating the pair of canonical variables
$x^o\approx \tau$, $p^o \approx
\pm \sqrt{{\vec p}^2+m^2}$, one gets a frozen Jacobi data
description in terms of independent Cauchy data, in which the same Minkowski
trajectory of the particle can be recovered in the non-covariant form
$\vec x(x^o)$ by introducing as Hamiltonian the energy generator
$\pm \sqrt{{\vec p}^2+m^2}$ of the Poincar\'e group 
\footnote{With the variables of
Ref.\cite{longhi}, one adds the covariant gauge-fixing $p\cdot x/\sqrt{p^2}
-\tau \approx 0$ and eliminates the pair $T=p\cdot x/
\sqrt{p^2}$, $\epsilon =\eta \sqrt{p^2} \approx \pm m$; now, since the
invariant mass is constant, $\pm m$, the non-covariant Jacobi data $\vec z=
\epsilon (\vec x-\vec p x^o/p^o)$, $\vec k=\vec p/\epsilon$ cannot be made to
evolve.}.

This comparison would suggest to solve the superhamiltonian constraint
in one component of the ADM canonical momenta ${}^3{\tilde \Pi}^{rs}$,
namely in one component of the extrinsic curvature.
But, differently from the scalar particle, the solution of the
superhamiltonian constraint does not define the weak ADM energy,
which, instead, is connected with an integral over 3-space of that
part of the superhamiltonian constraint dictated by the associated
Gauss law, see Eqs.(\ref{V4}), (\ref{V2}). 

Indeed, the
superhamiltonian constraint, being a secondary first class constraint
of a field theory, has an associated {\it Gauss law} (see Eq.(\ref{V4})
with $N=\epsilon$ and $N_r=0$ ) like the supermomentum constraints. In
every Gauss law, the piece of the secondary first class constraint
corresponding to a divergence and giving the {\it strong} form of the
conserved charge (the strong ADM energy in this case) as the flux
through the surface at infinity of a corresponding density depends on
the variable ${\cal X}$ which has to be eliminated  by using the constraint
in the process of canonical reduction (as a consequence the  variable 
${\cal Y}$ conjugate to ${\cal X}$ is the {\it gauge} variable).
Once the constraint is solved in the variable ${\cal X}$, it can be put inside
the volume expression of the {\it weak} form of the conserved charge to
obtain its expression in terms of the remaining canonical variables and 
eventually of the
gauge variable ${\cal Y}$. Now the {\it strong} ADM
energy is the only known charge, associated with a constraint bilinear
in the momenta, depending only on the coordinates (${}^3g_{rs}$) and not on the
momenta (${}^3{\tilde \Pi}^{rs}$), so that this implies that the 
superhamiltonian constraint has to be solved in one of the components of 
the 3-metric.

As a consequence  the right approach to the superhamiltonian constraint
is the one of Lichnerowicz\cite{conf} leading to the conformal
approach to the reduction of ADM metric gravity
\cite{york,cho,yoyo,ciuf} \footnote{See Ref.\cite{ciuf} for its review.
In Appendix C of Ref.\cite{russo2} there is a recollection of
notions on mean extrinsic curvature slices and on the TT (transverse
traceless)-decomposition.}. In this approach the superhamiltonian
constraint supplemented with the gauge fixing ${}^3K(\tau ,\vec \sigma
)\approx 0$ \footnote{Or ${}^3K(\tau ,\vec \sigma )\approx const.$. It is a 
condition on the internal
extrinsic York time defining the constant mean extrinsic curvature (CMC)
hypersurfaces.}, named {\it maximal slicing condition}, is considered
as an elliptic equation (the {\it Lichnerowicz equation}) to be solved in
the conformal factor $\phi (\tau ,\vec \sigma )= e^{{1\over 2}q(\tau
,\vec \sigma )} > 0$ of the 3-metric \footnote{Namely in its determinant, 
which can be extracted from it in a 3-covariant way.}. Therefore, the
momentum  conjugate to the conformal factor of the 3-metric is the {\it free 
gauge} variable associated with the superhamiltonian constraint.

Lichnerowicz has shown that the superhamiltonian and supermomentum
constraints  plus the maximal slicing condition of ADM metric gravity
form a system of 5 elliptic differential equations which has one and only one 
solution; moreover, with this condition
Schoen and Yau \cite{p17} have shown that the ADM 4-momentum is
timelike (i.e. the ADM energy is positive or zero for Minkowski
spacetime). Moreover, Schoen-Yau have shown in their last proof of the
positivity of the ADM energy that one can relax the maximal slicing
condition. See the reviews\cite{cho,beig} with their rich
bibliography.

In the conformal approach one put ${}^3g_{rs}=\phi^4\,
{}^3\sigma_{rs}$ [$det\, {}^3\sigma_{rs} =1$] and ${}^3{\tilde
\Pi}^{rs}=\phi^{-10}\, {}^3{\tilde \Pi}^{rs}_A+{1\over 3} \, {}^3g^{rs}\,
{}^3\tilde \Pi$ [${}^3g_{rs}\, {}^3{\tilde \Pi}^{rs}_A=0$]. Then, one
makes the TT-decomposition ${}^3{\tilde \Pi}^{rs}_A={}^3{\tilde
\Pi}^{rs}_{TT} + {}^3{\tilde \Pi}^{rs}_L$ (the TT-part is the conformally 
rescaled {\it distortion tensor}) with ${}^3{\tilde \Pi}^{rs}_L=(L
W_{\pi})^{rs}=W^{r|s}_{\pi} + W^{s|r}_{\pi} -{2\over 3}\, {}^3g^{rs}\,
W^u_{\pi \, |u}$, where $W^r_{\pi}$ is {\it York gravitomagnetic vector
potential}. The superhamiltonian and supermomentum constraints 
are interpreted as coupled quasilinear elliptic equations for $\phi$ and
$W^r_{\pi}$ (the four conjugate variables are free gauge variables),
which decouple with the maximal slicing condition ${}^3K=0$; the two
physical degrees of freedom are hidden in ${}^3{\tilde
\Pi}^{rs}_{TT}$ (and in two conjugate variables).

In the conformal approach one uses York's TT-variables \cite{york},
because most of the work on the Cauchy problem for Einstein's
equations in metric gravity is done by using spacelike hypersurfaces
$\Sigma$ of constant mean extrinsic curvature (CMC surfaces) in the
compact case (see Refs.\cite{cho,i1,i2}) and in particular with the maximal 
slicing
condition ${\cal T}(\tau ,\vec \sigma )=0$. It may be extended to non
constant ${\cal T}$ in the asymptotically free case \footnote{See also
Ref.\cite{im} for recent work in the compact case with non constant
${\cal T}$ and Ref.\cite{brill1} for solutions of Einstein's equations
in presence of matter which do not admit constant mean extrinsic
curvature slices.}.

Let us remark that in Minkowski spacetime ${}^3K(\tau ,\vec \sigma
)=0$ are the hyperplanes, while ${}^3K(\tau ,\vec \sigma )=const.$ are
the mass hyperboloids (the hyperboloidal data), corresponding to the instant 
and point form of the dynamics according to Dirac\cite{dd} respectively (see
Refs.\cite{gaida} for other types of foliations).

In Ref.\cite{yorkmap} (see for instance  Eq.(C7) in Appendix C of 
Ref.\cite{russo2}) it is shown that given the non-canonical basis
${\cal T}=-{4\over 3}\epsilon k
\, {}^3K={2\over {3\sqrt{\gamma}}}{}^3{\tilde \Pi}$,
${\cal P}_{\cal T}=-det\, {}^3g_{rs}=- \phi^{12}$,
${}^3\sigma_{rs}={}^3g_{rs}/(det\, {}^3g)^{1/3}$, ${}^3{\tilde
\Pi}^{rs}_A$ , there exists a canonical basis
 hidden in the variables ${}^3\sigma_{rs}$,
${}^3{\tilde \Pi}_A^{rs}$ (but it has never been found explicitly) and
that one can define the reduced phase space (the conformal superspace)
${\tilde {\cal S}}$, in which one goes to the quotient with
respect to the space diffeomorphisms and to the conformal rescalings.
\footnote{The {\it conformal superspace} ${\tilde {\cal S}}$ may be defined 
as the space of conformal 3-geometries on {\it closed} manifolds and can be
identified in a natural way with the space of conformal 3-metrics (the
quotient of ordinary superspace by the group $Weyl\, \Sigma_{\tau}$ of
conformal Weyl rescalings) modulo space diffeomorphisms, or,
equivalently, with the space of Riemannian 3-metrics modulo space
diffeomorphisms and conformal transformations of the form
${}^3g_{rs}\mapsto \phi^4\, {}^3g_{rs}$, $\phi > 0$. Instead, the 
{\it ordinary superspace} ${\cal S}$ is the space of Lorentzian 4-metrics 
modulo spacetime diffeomorphisms. The {\it phase superspace} is the phase 
space over ${\cal S}$: it is the quotient of the ADM phase space with respect 
to the primary constraints, the space  pseudo-diffeomorphisms and
the gauge transformations generated by the superhamiltonian
constraint. In this way a bridge is built towards the phase
superspace, which is mathematically connected with the Moncrief
splitting theorem\cite{mo,cho} valid for closed $\Sigma_{\tau}$. See
Ref.\cite{cho} for what is known in the asymptotically flat
case by using weighted Sobolev spaces. See Refs.\cite{whe,fis,ing,arms} for
the mathematical structure of superspace.}
It is also shown that one can define a {\it York map} from this reduced
phase space to the subset of the standard phase superspace defined by the 
gauge fixing ${}^3K=const.$.

C) Let us now consider asymptotically free spacetimes. In them there exists a 
time evolution in the mathematical time parametrizing the leaves
$\Sigma_{\tau}$ of the 3+1 splitting of $M^4$ governed by the weak ADM
energy\cite{fermi}. The superhamiltonian constraint is
not connected with time evolution: the strong and weak ADM energies
are only integrals of parts of this constraint. Instead it is a
{\it generator of Hamiltonian gauge transformations}.

As a constraint it determines the conformal factor $\phi$
of the 3-metric as a functional of ${}^3\sigma_{rs}$ and ${}^3{\tilde
\Pi}^{rs}$. But this means that the associated gauge variable is the
canonical momentum $\pi_{\phi}$ conjugate to the conformal factor. This 
variable, and not York time, parametrizes the normal deformation of the
embeddable spacelike hypersurfaces $\Sigma_{\tau}$. Now, since
different $\Sigma_{\tau}$ corresponds to different 3+1 splittings of
$M^4$ \footnote{In the class of the allowed ones going in an angle-independent
way to Minkowski spacelike hyperplanes.} one gets that the gauge
transformations generated by the superhamiltonian constraint
correspond to the transition from an allowed 3+1 splitting to another
one: this is the gauge orbit in the phase space over superspace.
Therefore the theory is independent from the choice of the 3+1
splitting like in parametrized Minkowski theories.

This leads to the conclusion that neither York's internal extrinsic
time nor Misner's internal intrinsic time are to be used as time
parameters: Misner's time (the conformal factor) is determined by the
Lichnerowicz equation while York's time (the trace of the extrinsic
curvature) by the gauge-fixing.

As a matter of fact a gauge fixing for the superhamiltonian constraint is a 
choice of a particular 3+1 splitting and this is done by fixing the momentum
$\pi_{\phi}$ conjugate to the conformal factor \footnote{A non-local 
information on the extrinsic curvature of $\Sigma_{\tau}$, which becomes the 
York time, or the maximal slicing condition, only with the special canonical
basis identified by the York map.}.

Since the solution of the Lichnerowicz equation gives the conformal
factor $\phi = e^{q/2}=({}^3g)^{1/12}$ as a function of its conjugate
momentum and of the remaining canonical variables as in the compact
case, also in the asymptotically free case only the conformal
3-geometries contain the physical degrees of freedom, whose functional
form depends on the other gauge fixings, in particular on the choice
of the 3-coordinates.

Therefore it is important to study the Shanmugadhasan canonical bases
of metric  gravity which have the following structure ($\bar a =1,2$ are 
non-tensorial indices for the Dirac observables)

\bea
\begin{minipage}[t]{3cm}
\begin{tabular}{|l|l|l|} \hline
$n$ & $n_r$ & ${}^3g_{rs}$ \\ \hline 
${\tilde \pi}^n$ & ${\tilde \pi}_{\vec n}^r$ & 
${}^3{\tilde \Pi}^{rs}$ \\ \hline
\end{tabular}
\end{minipage} &&\ {\longrightarrow \hspace{.2cm}} \
\begin{minipage}[t]{4 cm}
\begin{tabular}{|ll|l|l|l|} \hline
$n$ & $n_r$ & $\xi^{r}$ & $\phi$ & $r_{\bar a}$\\ \hline
 ${\tilde \pi}^n$ & ${\tilde \pi}_{\vec n}^r$    
& ${\tilde \pi}^{{\vec {\cal H}}}_r$ &
 $\pi_{\phi}$ & $\pi_{\bar a}$ \\ \hline
\end{tabular}
\end{minipage} \nonumber \\
 &&{}\nonumber \\
&& {\longrightarrow \hspace{.2cm}} \
\begin{minipage}[t]{4 cm}
\begin{tabular}{|ll|l|l|l|} \hline
$n$ & $n_r$ & $\xi^{r}$ & $Q_{\cal H}$ & $r^{'}_{\bar a}$\\ \hline
 ${\tilde \pi}^n$ & ${\tilde \pi}^r_{\vec n}$   
& ${\tilde \pi}^{{\vec {\cal H}}}_r$ &
 $\Pi_{\cal H}$ & $\pi^{'}_{\bar a}$ \\ \hline
\end{tabular}
\end{minipage}.
\label{IX1}
\eea

The first canonical transformation has seven first class constraints replaced 
by Abelian momenta ($\xi^r$ are the gauge parameters of the 
3-pseudo-diffeomorphisms generated by the supermomentum constraints) and 
has the conformal factor $\phi$ of the 3-metric as a configuration variable.
Note that it is a point canonical transformation. This is a 
quasi-Shanmugadhasan canonical transformation because the superhamiltonian 
constraint has not been Abelianized. The second canonical transformation is 
a real Shanmugadhasan canonical transformation with $Q_{{\cal H}}(\tau ,\vec 
\sigma ) \approx 0$ the Abelianization of the superhamiltonian constraint
\footnote{ If $\tilde \phi [r_{\bar a}, \pi_{\bar a}, \xi^r, \pi_{\phi}]$
is the solution of the Lichnerowicz equation, then $Q_{{\cal H}}=\phi -
\tilde \phi \approx 0$. Other forms of this canonical transformation
should correspond to the extension of the York map to
asymptotically flat spacetimes: in it the momentum conjugate to the
conformal factor is just York time and one can add the maximal slicing
condition as a gauge fixing.}. The variables $n$, $n_r$, $\xi^r$, $\pi_{\phi}$ 
are the Abelianized Hamiltonian gauge variables and $r^{'}_{\bar a}$, 
$\pi^{'}_{\bar a}$ the Dirac observables. Since it is not known how to solve 
the Lichnerowicz equation, the best which can be achieved is to find the
quasi-Shanmugadhasan canonical transformation. It has the relevant
property that in the special gauge $\pi_{\phi}(\tau ,\vec \sigma ) \approx 0$
the variables $r_{\bar a}$, $\pi_{\bar a}$ form a canonical basis of Dirac
observables for the gravitational field even if the solution $\tilde \phi$ of 
the Lichnerowicz equation is not known.

How to do the canonical reduction of metric gravity to a completely fixed
gauge by building the quasi-Shanmugadhasan canonical transformation will be the
subject of a future paper, in which this program will be realized for tetrad 
gravity following the preliminary papers of Refs.\cite{russo1,russo2,russo3}
and then extended to metric gravity.

In particular there will be a study of the family of the 3-orthogonal gauges 
on the WSW hypersurfaces $\Sigma^{(WSW)}_{\tau}$, since they are the nearest 
ones to the standards of measurement used in the (generically accelerated) 
laboratories,
which corresponds to completely fixed gauges of metric gravity. The special
3-orthogonal gauge with $\pi_{\phi}(\tau ,\vec \sigma ) \approx 0$ will be 
the equivalent of the radiation gauge in classical electrodynamics (like the 
harmonic gauge is the equivalent of the Lorentz gauge).

\subsection{Einstein's Equations versus Constraint Theory.}

First of all, let us interpret metric  gravity according to
Dirac-Bergmann theory of constraints (the presymplectic approach).
Given a mathematical noncompact, topologically trivial, manifold $M^4$
with a maximal $C^{\infty}$-atlas A, its diffeomorphisms in $Diff\,
M^4$ are interpreted in passive sense (pseudo-diffeomorphisms): chosen
a reference atlas (contained in A) of $M^4$, each pseudo-diffeomorphism
identifies another possible atlas contained in A. The
pseudo-diffeomorphisms are assumed to tend to the identity at spatial
infinity in  the way discussed in Section VI. Then we add an
arbitrary $C^{\infty}$ metric structure on $M^4$, we assume that
$(M^4,{}^4g)$ is globally hyperbolic and asymptotically flat at
spatial infinity and we arrive at a family of Lorentzian spacetimes
$(M^4,{}^4g)$ over $M^4$.

On $(M^4,{}^4g)$ one usually defines \cite{mtw,mw} the standards of
length and time, by using some material bodies, with the help of
mathematical structures like the line element $ds^2$, timelike
geodesics (trajectories of test particles) and null geodesics
(trajectories of photons), without any reference to Einstein's
equations \footnote{See the conformal, projective, affine and metric structures
hidden in $(M^4, {}^4g)$ according to Ref.\cite{pirani}, which replace
at the mathematical level the {\it material reference frame} concept
\cite{rov,brown,wittt} with its {\it test} objects.}; only the equivalence
principle (statement about test particles in an external given
gravitational field) is used to emphasize the relevance of geodesics.
Let $\tilde Diff\, M^4$ be the extension of $Diff\, M^4$ to the space
of tensors over $M^4$. Since the Hilbert action of metric gravity is
invariant under the combined action of $Diff\, M^4$ and $\tilde Diff\,
M^4$, one says that the relevant object in gravity is the set of all
4-geometries over $M^4$ [$(M^4,{}^4g)$ modulo $Diff\, M^4$, i.e. the
{\it superspace}  ${\cal S}=Riem\,
M^4/Diff\, M^4$] and that the relevant quantities (generally covariant
observables) associated with it are the invariants under
diffeomorphisms like the curvature scalars. From the point of view of
dynamics, one has to select those special 4-geometries whose
representatives $(M^4,{}^4g)$ satisfy Einstein's equations, which are
invariant in form under diffeomorphisms (general covariance). The
variation of a solution ${}^4g_{\mu\nu}(x)$ of Einstein's equations
under infinitesimal spacetime diffeomorphisms, namely ${\cal
L}_{\xi^{\rho}\partial
_{\rho}}\, {}^4g_{\mu\nu}(x)$, satisfies the Jacobi equations associated with
Einstein's equations or linearized Einstein equations \footnote{See Refs.
\cite{fmm,monc,fermi}; with our assumptions we are in the noncompact
case (like Ref.\cite{fermi}) without Killing vectors [more exactly
without 3-Killing vectors on the Riemannian manifolds
$(\Sigma^{(WSW)}_{\tau}$, ${}^3g_{rs})$]: in this case it is known
that near Minkowski spacetime the Einstein empty space equations are
linearization stable.}: therefore these {\it Noether (gauge) symmetries} of
the Hilbert action are also {\it dynamical symmetries} of Einstein
equations.

One can say that a {\it kinematical gravitational field} is a 4-geometry
(an element of $Riem\, M^4/Diff\, M^4$), namely an
equivalence class of 4-metrics modulo $Diff\, M^4$, and that an
{\it Einstein or dynamical gravitational field} (or Einstein 4-geometry
or equivalence class of Einstein spacetimes) is a kinematical
gravitational field which satisfies Einstein's equations.

However, the fact that the ten Einstein equations are not a hyperbolic
system of differential equations and cannot be put in normal form
is only considered in connection with the
initial data problem. Instead, the ADM action \footnote{Needed as the starting
point to define the canonical formalism since it has a well posed
variational problem.} contains the extra input of a 3+1 splitting of
$M^4$: this allows the identification of the surface term containing
the second time derivatives of the 4-metric to be discarded from the
Hilbert action.

As a consequence the ADM action is quasi-invariant under the pullback
of the Hamiltonian group of gauge transformations generated by the
first class constraints (as every singular Lagrangian; see Appendix A
of Ref.\cite{russo3}) and this group is not $Diff\, M^4$ plus its extension 
${\tilde Diff}\, M^4$, but the Hamiltonian group of gauge transformations.

However, the ADM action is not invariant under diffeomorphisms in
$Diff\, M^4$ skew with respect to the foliation of $M^4$ associated to
the chosen 3+1 splitting, even if the ADM theory is independent from
the choice of the 3+1 splitting (see Appendix A of Ref.\cite{russo3}).
The results of Refs.\cite{ppons} show that the infinitesimal spacetime 
diffeomorphisms $\delta x^{\mu}=\epsilon^{\mu}$ are projectable to 
Hamiltonian gauge transformations if and only if $\epsilon^{\mu}= {{\xi^o}\over
{N}}l^{\mu}+\delta^{\mu}_i(\xi^i -{{N^i}\over N} \xi^o)$ with the $\xi^{\mu}$
independent from the lapse and shift functions. In the 
$\Sigma_{\tau}$-adapted coordinates [$\xi^A=(\xi^{\tau}; \xi^r)$, 
$N^A=(N;N^r)$, ${\tilde \pi}_A=({\tilde \pi}_N; {\tilde \pi}^{\vec N}_r)$,
${\tilde {\cal H}}_A-({\tilde {\cal H}}; {\tilde {\cal H}}_r)$] the
Hamiltonian generators of the projectable spacetime diffeomorphisms are
$G(\tau ) =\int d^3\sigma \Big[ \xi^A {\tilde {\cal H}}_A+ \xi^A N^B C^C_{AB} 
{\tilde \pi}_C+ \partial_{\tau}\xi^A {\tilde \pi}_A\Big] (\tau ,\vec \sigma )$
with the $C^C_{AB}$'s being the structure functions in Eqs.(\ref{II19}).

Since the ADM action generates the same equations of motion as the Hilbert
action, i.e. Einstein's equations, the space of the dynamical symmetries
of the equations of motion is the same in the two theories. For more details
see Appendix A of Ref.\cite{russo3}\footnote{See Appendix D of 
Ref.\cite{russo3} for a review of the second Noether theorem in
the case of the Hilbert action and for the consequences of its
4-diffeomorphism invariance like the Komar superpotential and the
energy-momentum pseudotensors.}. However, since the infinitesimal spacetime
pseudodiffeomorphisms of a 4-metric solution of Einstein's equations
(i.e. ${\cal L}_{\xi^{\rho}\partial_{\rho}}\, {}^4g_{\mu\nu}(x)$) are
solutions to the Jacobi equations in the Hilbert form, it turns out
that among the dynamical symmetries of Einstein's equations there are
both allowed strictly Hamiltonian gauge transformations, under which
the ADM action is quasi-invariant, and generalized transformations
under which the ADM action is not invariant (see Appendix A of 
Ref.\cite{russo3}). This derives from the fact that the Noether symmetries
of an action and the dynamical symmetries of its Euler-Lagrange
equations have an overlap but do not coincide. In conclusion, the
allowed gauge transformations are the subset of spacetime
diffeomorphisms under which the ADM action is quasi-invariant; the
other spacetime diffeomorphisms are dynamical symmetries of the
equations of motion but not Noether symmetries of the ADM action.

Regarding the 10 Einstein equations, the Bianchi identities imply that
four equations are linearly dependent on the other six ones and their
gradients. Moreover, the four combinations of Einstein's equations
projectable to phase space (where they become the secondary first
class superhamitonian and supermomentum constraints of canonical
metric and tetrad gravity) are independent from the accelerations and
are only restrictions on the Cauchy data. As a consequence, the
solutions of Einstein's equations have the ten components
${}^4g_{\mu\nu}$ of the 4-metric depending on only two dynamical  
non-tensorial degrees of freedom (defining the physical gravitational field) 
and  on eight undetermined degrees of freedom 
\footnote{More exactly the four components of
the 4-metric corresponding to the lapse and shift functions and on the
four functions depending on the gradients of the 4-metric (generalized
velocities) corresponding, through the first half of Hamilton
equations, to the four arbitrary Dirac multipliers in front of the
primary constraints (vanishing of the momenta conjugate to lapse and
shift functions) in the Dirac Hamiltonian \cite{lu1}.}.

\subsection{Gauge Variables and Gauge Fixings.}

This transition from the ten components ${}^4g
_{\mu\nu}$ of the tensor ${}^4g$ in some atlas of $M^4$ to the 2
(deterministic)+8 (undetermined) degrees of freedom breaks general
covariance, because the physical degrees of freedom  in general are neither 
tensors nor invariants under spacetime diffeomorphisms: their functional form
is atlas dependent in a way dictated by the 3+1 splittings of $M^4$
needed for defining the canonical formalism. This is manifest in the
canonical approach to metric gravity:\hfill\break
 i) choose an atlas for $M^4$, a WSW 3+1 splitting
$M^{3+1}$ of $M^4$ (with the WSW leaves $\Sigma^{(WSW)}_{\tau}$ of the
foliation assumed diffeomorphic to $R^3$), go to coordinates adapted
to the 3+1 splitting [atlas for $M^{3+1}$ with coordinate charts
$(\sigma^A)=(\tau ,\vec \sigma )$, connected to the $M^4$ atlas by the
transition functions $b^{\mu}_A(\tau ,\vec \sigma )$] and replace
$Diff\, M^4$ with $Diff\, M^{3+1}$ (the diffeomorphisms respecting the
3+1 splitting); \hfill\break
 ii) the ten components ${}^4g_{AB}$ of
the 4-metric in the  adapted coordinates are non covariantly replaced
with $n$, $n^r$, ${}^3g_{rs}$, whose conjugate momenta are ${\tilde
\pi}_n$, ${\tilde \pi}^{\vec n}_r$, ${}^3{\tilde \Pi}^{rs}$. We have assumed
$N=N_{(as)}+n$, $N_r=N_{(as)r}+n_r$ with the asymptotic parts equal to
the lapse and shift functions of Minkowski spacelike hyperplanes
further restricted to Wigner hyperplanes and  with their bulk parts
$n$, $n_r$ going to zero at spatial infinity in an angle-independent
way; therefore, ${\tilde \pi}_N$ and ${\tilde \pi}_r^{\vec N}$ have been 
replaced by ${\tilde \pi}_n$ and ${\tilde \pi}_r^{\vec n}$ respectively;
\hfill\break
iii) there are four primary [${\tilde \pi}_n\approx 0$, ${\tilde
\pi}^{\vec n}_r\approx 0$] and four secondary [${\tilde {\cal H}}\approx 0$, 
${\tilde {\cal
H}}^r\approx 0$] first class constraints;\hfill\break iv) therefore,
the twenty canonical variables have to be replaced (with a
Shanmugadhasan canonical transformation) with two pairs of genuine
physical degrees of freedom (Dirac's observables), with eight
gauge variables and with eight 
abelianized first class constraints;
\hfill\break
v) this separation is dictated by the Hamiltonian group ${\bar {\cal
G}}$ of gauge transformations which has eight 
generators and is not connected with ${\tilde Diff}\,
M^{3+1}$ (except for spatial diffeomorphisms $Diff\, \Sigma_{\tau}
\subset Diff\, M^{3+1}$), which has only four generators and whose
invariants are not Dirac observables \footnote{The so called
time-diffeomorphisms are replaced by the 5 gauge transformations
generated by ${\tilde \pi}_n$, ${\tilde
\pi}_r^{\vec n}$, and the superhamiltonian constraint.};\hfill\break
vi) as said at the end of Section II, the eight gauge variables
should be fixed by giving only four gauge fixings for the secondary
constraints (the same number of conditions needed to fix a
diffeomorphisms), because their time constancy determines the four
secondary gauge fixings for the primary constraints 
\footnote{Then their
time constancy determines the Dirac multipliers (four velocity
functions not determined by Einstein equations) in front of the
primary constraints in the Dirac Hamiltonian. This is in accord with the
results of Ref.\cite{ppons} that the projectable spacetime diffeomorphisms
depend only on four arbitrary functions and their time derivatives.}.

Since no one has solved the metric gravity secondary constraints till
now, it is not clear what is undetermined inside ${}^3g_{rs}$ (see
Appendix C of II for what is known from the conformal approach) and,
therefore, which is the physical meaning (with respect to the
arbitrary determination of the standards of length and time) of the
first four gauge-fixings. Instead, the secondary four gauge-fixings
(induced by the gauge fixings to the secondary constraints) determine
the lapse and shift functions, namely they determine how the leaves
$\Sigma_{\tau}^{(WSW)}$ are packed in the foliation 
\footnote{The gauge nature
of the shift functions, i.e. of ${}^4g_{oi}$, is connected with the
conventionality of simultaneity \cite{simul,soffel}.}. 

Let us remark
that the invariants under spacetime diffeomorphisms are in general not
Dirac observables, because they depend on the eight gauge variables
not determined by Einstein's equations. Therefore, all the curvature
scalars are gauge quantities at least at the kinematical level.

In this paper we have clarified the situation in the case of metric
gravity. The original 20 canonical variables $n$,
$n_r$, ${}^3g_{rs}$, ${\tilde \pi}_n$, ${\tilde
\pi}^{\vec n}_r$, ${}^3{\tilde \Pi}^{rs}$ have been replaced by the Dirac's
observables $r_{\bar a}$, $\pi_{\bar a}$ [the gravitational field],
by 8 first class constraints  and by 8
gauge variables: $n$, $n_r$, $\xi^r$, $\pi_{\phi}$ . Now we have to add 4 
primary gauge fixings:\hfill\break
\hfill\break
 i) 3 gauge fixings for the parameters $\xi^r$ of the
spatial pseudo-diffeomorphisms generated by the secondary constraints
${}^3{\tilde {\cal H}}_r\approx 0$: they correspond to the choice of an
atlas of coordinates on $\Sigma^{(WSW)}_{\tau}$ (chosen as
conventional origin of pseudo-diffeomorphisms) and, therefore, by
adding the parameter $\tau$, labelling the leaves of the foliation, of
an atlas on $M^{3+1}$. The gauge fixings on $\xi^r$, whose time
constancy produces the gauge fixings for the shift functions $n_r$ and,
therefore, a choice of simultaneity convention in $M^4$ (the choice of
how to synchronize clocks), can be interpreted as a fixation of 3
standards of length by means of the choice of a coordinate system on
$\Sigma_{\tau}^{(WSW)}$;
\hfill\break
 ii) a gauge fixing for $\pi_{\phi}$, which, being a momentum,
carries an information about the extrinsic curvature of
$\Sigma_{\tau}^{(WSW)}$ embedded in $M^4$ \footnote{It replaces the York
extrinsic time ${}^3K$ of the Lichnerowicz-York conformal approach and
is a parametrization of the normal deformations of the
$\Sigma^{(WSW)}_{\tau}$.}, for the superhamiltonian constraint. The
gauge-fixing on $\pi_{\phi}$ has nothing to do with a standard of time 
\footnote{The evolution is parametrized by the mathematical parameter $\tau$ 
of the induced coordinate system $(\tau ,\vec
\sigma )$ on $M^4$.}, but it is a fixation of the form of 
$\Sigma^{(WSW)}_{\tau}$
\footnote{It is a nonlocal statement about the extrinsic curvature of a
$\Sigma_{\tau}^{(WSW)}$ embedded in $M^4$.} and, therefore, it amounts
to the choice of one of the allowed 3+1 splittings of $M^4$. Let us
remember that the Poisson algebra (\ref{II19}) of the superhamiltonian and
supermomentum constraints reflects the embeddability properties of
$\Sigma_{\tau}^{(WSW)}$; the superhamiltonian constraint generates the
deformations normal to $\Sigma^{(WSW)}_{\tau}$, which partially
{\it replace} the $\tau$-diffeomorphisms.  
The dependence on $\pi_{\phi}$ is one
of the sources of the gauge dependence at the kinematical level of the
curvature scalars of $M^4$ (the other sources are the lapse and shift
functions and their gradients). The natural interpretation of the
gauge transformations generated by the superhamiltonian constraint is
to change the 3+1 splitting of $M^4$ by varying the gauge variable
$\pi_{\phi}(\tau ,\vec \sigma )$, so to make the theory {\it independent} from
the choice of the original 3+1 splitting of $M^4$, as it happens with
parametrized Minkowski theories. However, since the time constancy of
the gauge fixing on $\pi_{\phi}$ determines the gauge fixing for the lapse
function n (which says how the $\Sigma^{(WSW)}_{\tau}$ are packed in
$M^4$), there is a connection with the choice of the standard of local
proper time. Let us remark that only the gauge fixing $\pi_{\phi}(\tau
,\vec \sigma )\approx 0$  leaves the Dirac observables $r_{\bar a}$, 
$\pi_{\bar a}$, canonical; with other gauge fixings the canonical
degrees of freedom of the gravitational field have to be redefined.

Let us remark \cite{soffel} that the {\it reference standards} of time
and length correspond to units of {\it coordinate time and length} and
not to proper times and proper lengths: this is not in contradiction
with general covariance, because the laboratory in which one defines
the reference standards corresponds to a particular {\it completely fixed
gauge}.

Therefore, according to constraint theory, given an atlas on a 3+1
splitting $M^{3+1}$ of $M^4$, the phase space content of the 8
nondynamical Einstein equations is equivalent to the determination of
the Dirac observables (namely a kinematical gravitational field not
yet solution of the 2 dynamical Einstein equations, i.e. of the final
Hamilton equations with the ADM energy as Hamiltonian), whose
functional form in terms of the original variables depends on choice
of the atlas on $M^{3+1}$ and on a certain  information about the
extrinsic curvature of $\Sigma_{\tau}^{(WSW)}$.

\subsection{Hamiltonian Kinematical and Dynamical Gravitational Fields.}

Let us define  a {\it Hamiltonian kinematical gravitational field}  as the
quotient of the set of Lorentzian spacetimes $(M^{3+1},{}^4g)$ with a 3+1
splitting with respect to the Hamiltonian
gauge group ${\bar {\cal G}}$ with 8  generators
[$Riem\, M^{3+1}/{\bar {\cal G}}$]: while
space diffeomorphisms in $Diff\, M^{3+1}$ coincide with those in $Diff\, \Sigma
_{\tau}$, the {\it $\tau$-diffeomorphisms} in $Diff\, M^{3+1}$ are replaced by
the 5 gauge freedoms associated with $\pi_{\phi}$, $n$ and $n_r$.

A representative of a {\it Hamiltonian kinematical gravitational field} in a
given gauge equivalence class is
parametrized by $r_{\bar a}$, $\pi_{\bar a}$ and is an element of a
gauge orbit $\Gamma$ spanned by the gauge variables  
$\xi^r$, $\pi_{\phi}$, $n$, $n_r$. Let us consider the reduced gauge
orbit $\Gamma^{'}$ obtained from $\Gamma$ by going to the quotient
with respect to  $\xi^r$. The solution
$\phi = e^{q/2}$ of the reduced Lichnerowicz equation is
$\pi_{\phi}$-dependent, so that the gauge orbit $\Gamma^{'}$ contains one
conformal 3-geometry (conformal gauge orbit), or a family of conformal
3-metrics if the $\pi_{\phi}$-dependence of the solution $\phi$ does not
span all the Weyl rescalings. In addition $\Gamma^{'}$ contains the
lapse and shift functions. Now, each 3-metric in the conformal gauge
orbit has a different 3-Riemann tensor and different 3-curvature
scalars. Since 4-tensors and  4-curvature scalars depend : i) on the
lapse and shift functions (and their gradients); ii) on $\pi_{\phi}$ both
explicitly and implicitly through the solution of the Lichnerowicz
equation,  and this influences the 3-curvature scalars, most of
these objects are in general gauge variables from the Hamiltonian
point of view at least at the kinematical level. The simplest relevant
scalars of $Diff\, M^4$, where to visualize these effects, are
Komar-Bergmann's individuating fields (see later on) and/or the
bilinears ${}^4R_{\mu\nu\rho\sigma}\, {}^4R^{\mu\nu\rho\sigma}$,
${}^4R
_{\mu\nu\rho\sigma}\, \epsilon^{\mu\nu\alpha\beta}\, {}^4R_{\alpha\beta}{}
^{\rho\sigma}$.
Therefore, generically the elements of the gauge
orbit $\Gamma^{'}$ are, from the point of view of $M^4$ based on the Hilbert
action, associated with different 4-metrics belonging to different 4-geometries
(the standard {\it kinematical gravitational fields}).

Therefore, according to the gauge interpretation based on constraint
theory, a {\it Hamiltonian kinematical gravitational field} is an
equivalence class of 4-metrics modulo the pullback of the Hamiltonian
group of gauge transformations, which contains all the 4-geometries
connected by them and a well defined conformal 3-geometry. This is a
consequence of the different invariance properties of the ADM and
Hilbert actions, even if they generate the same equation of motion.

Let us define an {\it Hamiltonian Einstein or dynamical
gravitational field} as a Hamiltonian kinematical
gravitational field which satisfies the final Hamilton equations with the ADM
energy as Hamiltonian (equivalent to the two dynamical equations hidden in the
Einstein equations).

These Hamiltonian dynamical gravitational fields correspond to special
gauge equivalence classes, which contain only one 4-geometry whose
representative 4-metrics satisfy Einstein's equations, so that they
{\it coincide} with the standard dynamical gravitational fields. This
highly nontrivial statement is contained in the results of
Refs.\cite{fmm,fermi,monc} (in particular see Ref.\cite{fermi} for the
noncompact asymptotically free at spatial infinity case). On the space
of the solutions of the Hamilton-Dirac equations \footnote{They, together with
the first class constraints, are equivalent to Einstein's equations.}
the kinematical Hamiltonian gauge transformations are restricted to be
dynamical symmetries \footnote{Maps of solutions onto solutions; with them
there is not necessarily an associated constant of the motion like
with the Noether symmetries of an action.} of Einstein's equations in
the ADM presentation and this implies that the allowed Hamiltonian
gauge transformations must be equivalent to or contained in the
spacetime pseudodiffeomorphisms of $M^4$.

The allowed infinitesimal 
Hamiltonian gauge transformations on the space of solutions of the
Hamilton-Dirac equations must be solutions of the Jacobi equations
\footnote{The linearized constraints and the linearized evolution equations;
see Refs. \cite{monc} for their explicit expression.} and this excludes
most of the kinematically possible Hamiltonian gauge transformations
(all those generating a transition from a 4-geometry to another one).
The only allowed ones are restricted to coincide with the projectable spacetime
diffeomorphisms of Ref.\cite{ppons}, previously quoted.
In the allowed Hamiltonian gauge transformations the gauge parameters
$n$, $n_r$, $\xi^r$, $\pi_{\phi}$ are not independent but restricted by the
condition that the resulting gauge transformation must be a spacetime
pseudodiffeomorphisms.

This is the way in which on the space of solutions of Einstein's
equations spacetime diffeomorphisms are reconciled with the allowed
Hamiltonian gauge transformations adapted to the 3+1 splittings of the
ADM formalism. The kinematical freedom of the 8 independent types of
Hamiltonian gauge transformations of metric gravity is reduced to 4
dynamical types like for $Diff\, M^4$; partially, this was anticipated
at the kinematical level by the fact that in the original Dirac
Hamiltonian there are only 4 arbitrary Dirac multipliers, and that the
gauge-fixing procedure starts with the gauge fixings of the secondary
constraints, which generate those for the primary ones , which in turn
lead to the determination of the Dirac multipliers.

On the space of solutions of Einstein's equations in every completely
fixed Hamiltonian gauge we get a different canonical basis of
{\it dynamical Dirac observables} (with weakly vanishing Poisson
brackets with the original constraints, but with strongly vanishing
ones with the Abelianized constraints): being a dynamical symmetry a
spacetime diffeomorphism becomes a mapping of the dynamical Dirac
observables in one gauge onto the dynamical Dirac observables in
another gauge (selected by the the new coordinates defined by the
diffeomorphism).
These Dirac observables correspond to dynamical gravitational fields
(namely the invariants under the kinematical Hamiltonian gauge
transformations restricted to the solutions of Einstein's equations
and without any a priori tensorial character under $Diff\, M^4$). On
the other hand a spacetime quantity $Q$ scalar under spacetime
diffeomorphisms and restricted to the solutions of Einstein's
equations, becomes a well defined (gauge-dependent) function $Q_G$ of
the dynamical Dirac observables of a completely fixed Hamiltonian
gauge $G$. Since the dynamical Dirac observables change with a change
of the Hamiltonian gauge, also the functional form of the function
$Q_G$ will change with a change of the Hamiltonian gauge.

Regarding the understanding of possible tensorial properties of the
dynamical Dirac observables,  the first step would be to find the
connection of the Dirac observables $r_{\bar a}(\tau ,\vec
\sigma )$ in a completely fixed gauge with the symmetric traceless
2-tensors on 2-planes, which are the independent gravitational degrees
of freedom according to Christodoulou and Klainermann \cite{ckl}, and
with the, in some way connected, Newman-Penrose formalism.

\vfill\eject

\section{Interpretational Problems regarding the Observables in General 
Relativity.}

Our approach breaks the general covariance of general relativity completely by
going to the special 3-orthogonal gauge with $\pi_{\phi}(\tau ,\vec \sigma )
\approx 0$. But this is done in a way naturally associated with presymplectic 
theories (i.e. theories with first  class
constraints like all formulations of general relativity and the standard model
of elementary particles with or without supersymmetry): the global
Shanmugadhasan canonical transformations (when they exist; for instance they
do not exist when the configuration space is compact
like in closed spacetimes) correspond to privileged
Darboux charts for presymplectic manifolds. Therefore, the gauges identified by
these canonical transformations should have a special (till now unexplored)
role also in generally covariant theories, in which traditionally one
looks for observables invariant under
spacetime diffeomorphisms (but no complete basis is
known for them in general relativity) and not for (not generally covariant)
Dirac observables. While in electromagnetism and in Yang-Mills theories the
physical interpretation of Dirac observables is clear, in generally covariant
theories there is a lot of interpretational problems and ambiguities.

Therefore, let us make some considerations on interpretational
problems, whose relevance has been clearly pointed out in
Ref.\cite{stachel}.

\subsection{Interpretational Problems with Dirac's Observables.}

In generally covariant theories (without background fields)
the interpretational difference with respect
to the Dirac observables of Yang-Mills theories, is that one has to make a
complete gauge-fixing to give a meaning to ``space and time" (in the above
sense) before being able to identify the functional form of the Dirac
observables for the gravitational field 
\footnote{Regarding other approaches to the observables in general relativity
see also Refs.\cite{dc8}: the ``perennials" introduced
in this approach are essentially our Dirac observables.
See Ref.\cite{gr13} for the difficulties in observing perennials
experimentally at the classical and quantum levels and in their
quantization. See also Ref.\cite{torre} on the non existence of observables for
the vacuum gravitational field in a closed universe, built as spatial
integrals of local functions of Cauchy data and their first derivatives.}
and moreover we have to formulate the
problem only for the solutions of Einstein's equations (this is not necessary
for Yang-Mills theory).

This deep difference between the interpretations based on constraint
theory and on general covariance respectively is reflected in the two
viewpoints about what is observable in general relativity (and, as a
consequence, in all generally covariant theories) as one can clearly
see in Ref.\cite{rov} and in its  bibliography:\hfill\break
\hfill\break
 i) The
{\it non-local point of view} of Dirac \cite{dirac}, according to which
determinism implies that only gauge-invariant quantities, i.e.Dirac's
observables  can be
measured. The {\it hole argument} of Einstein\cite{einst} (see
Refs.\cite{rov,stachel} for its modern treatment) supports this
viewpoint: points of spacetime are not a priori distinguishable 
\footnote{Their
individuality is washed out by general covariance, i.e. by the
invariance under spacetime diffeomorphisms.}, so that, for instance,
${}^4R(\tau ,\vec \sigma )$ \footnote{A scalar under diffeomorphisms, but not a
Dirac observable at the kinematical level.} is not an observable
quantity. Even if ${}^4R(\tau ,\vec \sigma )\, {\buildrel \circ \over
=}\, 0$ in absence of matter, the other curvature scalars are non
vanishing after having used Einstein equations and, due to the lack of
known solutions without Killing vectors, it is not possible to say
which is their connection with Dirac observables. More in general, the
4-metric tensor ${}^4g_{\mu\nu}$ is a not observable gauge variable.
As said in Ref.\cite{stachel} an Einstein  spacetime manifold
corresponds to a dynamical gravitational field, but a dynamical
gravitational field corresponds to an equivalence class of spacetimes.
The metrical structure forms part of the set of dynamical variables,
which must be determined before the points of spacetime have any
physical properties. Therefore, one cannot assume in general
relativity what is valid in special relativity, namely that the
individuation of the points of Minkowski spacetime is established by a
framework of rigid rods and clocks.

Fixing the gauge freedoms in general relativity means to determine the
functional form of the 4-metric tensor ${}^4g_{\mu\nu}$: this is a
definition of the angle and distance properties of the material
bodies, which form the reference system (rods and clocks). At the
kinematical level the standard procedures of defining measures of
length and time \cite{ll,mtw,soffel} are gauge dependent as already
said, because the line element $ds^2$ is gauge dependent and
determined only after a complete gauge fixing and after the
restriction to the solutions of Einstein's equations \footnote{Note that in
textbooks these procedures are always defined without any reference to
Einstein's equations.}: only now the curvature scalars of $M^4$ become
measurable, like the electromagnetic vector potential in the radiation
gauge. Only now the procedure for measuring the Riemann tensor
described in Ref.\cite{ciuf} becomes completely meaningful. Moreover,
let us remember that the {\it standard} of unit of time is a
{\it coordinate time} \cite{soffel} and not a proper time and that in
astronomy and in the theory of satellites the unit of time is replaced
by a unit of coordinate  length ({\it ephemerid time}).

The measuring apparatuses should  also be described by the gauge
invariant Dirac observables associated with the given gauge (namely
identified by the Shanmugadhasan canonical transformation associated
with that gauge),
after the introduction of matter, since an experimental laboratory
corresponds by definition to a completely fixed gauge.

See also Ref.\cite{but} for the relevance of the {\it hole argument} in
the discussions on the nature of spacetime and for the attempts to
formulate quantum gravity. Even if the standard canonical (either
metric or tetrad) gravity approach presents serious problems in
quantization due to the intractable Lichnerowicz equation \footnote{So that
research turned towards either Ashtekar's approach or superstring
theory with its bigger general covariance group.}, still the problem of
what is observable at the classical level in generally covariant
theories is considered open.

ii) The {\it local point of view}, according to which the spacetime manifold 
$M^4$ is the manifold of physically determined `events' (like in special 
relativity), namely spacetime points are physically distinguishable, because 
any measurement is performed in the frame of a given reference system. The 
gauge freedom of generally covariant theories reflects the freedom of choosing
coordinate systems, i.e. reference systems. Therefore, the evolution is not
uniquely determined (since the reference systems are freely chosen)
and, for instance, ${}^4R(\tau ,\vec \sigma )$ is an
observable quantity, like the 4-metric tensor ${}^4g_{\mu\nu}$. See Ref.
\cite{feng} for a refusal of Dirac's observables in general relativity based on
the local point of view.

In Ref.\cite{rov} the non-local point of view is accepted and there is a
proposal for using some special kind of matter to define a {\it material
reference system} (not to be confused with a coordinate system) to localize
points in $M^4$, so to recover the local point of view in some approximate way
\footnote{The main approximations are: 1) to neglect, in Einstein equations, 
the energy-momentum tensor of the matter forming the material reference system
(it's similar to what happens for test particles); 2) to neglect, in the
system of dynamical equations, the entire set of equations determining the
motion of the matter of the reference system (this introduces some 
indeterminism in the evolution of the entire system).}, 
since in the analysis of
classical experiments both approaches tend to lead to the same conclusions.
See also Refs.\cite{brown,ish,kuchar1} for a complete review of {\it material 
clocks} and {\it reference fluids}. However, we think that one has to 
consider the use of {\it test objects} as an idealization for the attempt to 
approximate with realistic
dynamical objects the conformal, projective, affine and metric structures
\cite{pirani} of Lorentzian manifolds, which are used to define the {ideal
geodesic clocks}\cite{mtw} and the basis of the theory of measurement.

Let us remark that in applications, for instance in the search of gravitational
waves, one is always selecting a background reference metric and the associated
(Minkowski like) theory of measurement: the conceptual framework becomes the
same as in special relativity. The same happens for every string theory due to
necessity (till now) of a background metric in their formulation.

\subsection{Identification of the Physical Points of Spacetime.}

 Let us remark that our ADM tetrad formulation assumed the existence
of a mathematical abstract 4-manifold, the spacetime $M^4$, to which
we added 3+1 splittings with spacelike leaves
$\Sigma^{(WSW)}_{\tau}\approx R^3$. The mathematical points of $M^4$
have no physical meaning and are coordinatized with
$\Sigma_{\tau}^{(WSW)}$-adapted coordinates $(\tau ,\vec \sigma )$.
All fields (also matter fields when present) depend on these
mathematical coordinates for $M^4$, but till now there is no
justification for saying that the points (or events) of the spacetime
have any physical meaning (instead in special relativity they are
physical points by hypothesis).

Is it possible to label the points of $M^4$ in terms of Dirac's
observables a posteriori by introducing {\it physical points}? As already
said, once all gauge freedoms have been eliminated this can be done,
in analogy to what happens with the vector potential of
electromagnetism which becomes measurable in a completely fixed gauge
like the Coulomb one.

Regarding how to give a meaning to the mathematical points of the
abstract 4-manifold, we accept the proposal of Komar and
Bergmann\cite{komar,be} of identifying the physical points of a
spacetime $(M^4, {}^4g)$ without Killing vectors, solution of the
Einstein's equations, only {\it a posteriori} in a way invariant under
spacetime diffeomorphisms, by using four invariants bilinear and
trilinear in the Weyl tensors \footnote{As shown in Ref.\cite{gede} there are
14 algebraically independent curvature scalars for $M^4$, which are
reduced to four when Einstein equations without matter are used.},
called {\it individuating fields}, which do not depend on the lapse and
shift functions but only on the ADM canonical variables. These individuating 
fields depend on $r_{\bar a}$, $\pi_{\bar
a}$ and on the gauge parameters $\xi^r$ (choice of 3-coordinates on
$\Sigma_{\tau}$) and $\pi_{\phi}$ (replacing York's internal extrinsic time
${}^3K$) since the freedom in the choice of the mathematical
coordinates $\sigma^A$ is replaced by the gauge freedom in the choice
of $\xi^r$ and $\pi_{\phi}$ \footnote{Since these are three coordinates and 
one momentum one has the Lorents signature coming out from the 
quasi-Shanmugadhasan canonical transformation.}: note the difference from the 
proposal of Refs.\cite{ish,dc11} of using $\xi^r$ and $q=2 ln\, \phi$ for 
this aim. The 4-metric in this {\it physical 4-coordinate grid}, obtained from
${}^4g_{AB}$ by making a coordinate transformation from the adapted
coordinates $\sigma^A= (\tau ,\vec
\sigma )$, depends on the same variables and also on the lapse and
shift functions.

These individuating fields are not Dirac observables at the kinematical
level. On the solutions of Einstein's equations they become
gauge-dependent functions of the dynamical Dirac observables of a
completely fixed Hamiltonian gauge. In every complete gauge (choice of
the coordinate systems on $\Sigma^{(WSW)}_{\tau}$ and on $M^{3+1}$)
after the fixation of $\xi^r$ and $\pi_{\phi}$ they describe a special
gauge-dependent coordinate system for $M^4$, in which the dynamical
gravitational field degrees of freedom in that gauge can be used (at
least in some finite region) to characterize distinct points of $M^4$,
as also remarked by Stachel\cite{stachel} in connection with
Einstein's hole argument (but without taking into account constraint
theory). In this way we get a physical 4-coordinate grid on the
mathematical 4-manifold $M^4$ dynamically determined by tensors over
$M^4$ itself with a rule which is invariant under $Diff\, M^4$ but
with the functional form of the map {\it $\sigma^A=(\tau ,\vec
\sigma ) \mapsto \, physical\, 4-coordinates$} depending on the chosen
complete gauge: the {\it local point of view} is justified a posteriori
in every completely fixed gauge. In conclusion the physical content of
the gravitational field in absence of matter is just the
identification of the points of Einstein spacetimes by means of its
four independent phase space degrees of freedom.

Finally, let us remember that Bergmann\cite{be} made the following
critique of general covariance: it would be desirable to restrict the
group of coordinate transformations (spacetime diffeomorphisms) in
such a way that it could contain an invariant subgroup describing the
coordinate transformations that change the frame of reference of an
outside observer \footnote{These transformations could be called Lorentz
transformations; see also the comments in Ref.\cite{ll} on the
asymptotic behaviour of coordinate transformations.}; the remaining
coordinate transformations would be like the gauge transformations of
electromagnetism. This is what we have done with the redefinition of
lapse and shift functions. However, to avoid supertranslations we
refused the interpretation of the asymptotic Poincar\'e charges as
generators of improper gauge transformations, and we accepted Marolf's
proposal\cite{p18} about superselection sectors. In this way {\it preferred}
asymptotic coordinate systems will emerge, which, as
said by Bergmann, are {\it non-flat}: while the inertial coordinates are
determined experimentally by the observation of trajectories of
force-free bodies, these intrinsic coordinates can be determined only
by much more elaborate experiments (for instance precessional effects
on gyroscopes) with respect to fixed stars, since they depend, at
least, on the inhomogeneities of the ambient gravitational fields.

See also Ref.\cite{elma} for other critics to general covariance: very often
to get physical results one uses preferred coordinates not merely for
calculational convenience, but also for understanding. In Ref.\cite{zala} this
fact has been formalized as the {\it principle of restricted covariance}. In 
our case the choice of the gauge-fixings has been dictated by the 
Shanmugadhasan canonical transformations, which produce generalized radiation 
gauges, in which one can put in normal form the Hamilton equations for the 
canonical variables of the gravitational field, and, therefore,  also 
the two associated combinations of the Einstein equations which
depend on the accelerations.

\vfill\eject

\section{The Problem of Time.}

 Let us add some comments on time in general relativity in the case
of globally hyperbolic asymptotically flat at spatial infinity
spacetimes.

In general relativity, Isham\cite{ish} and Kuchar\cite{kuchar1} have
made a complete review of the problem of time (see Ref.\cite{but} for
a recent contribution to the problem), showing that till now there is
no consistent quantization procedure for it. See also:  i) Rovelli's
point of view in Refs.\cite{ro1} and Lawrie and Epp comments\cite{ro2}
on the limitations of his treatment of clocks; ii) Rovelli's proposal
of a thermodynamical origin of time \cite{thermo}.

A scheme in which time is identified before quantization has been used
in this paper.
The unphysical mathematical 1-time of the rest-frame instant form of
dynamics on WSW hypersurfaces discussed in the previous Sections is
the rest-frame {\it global time} $T_{(\infty )}
=p_{(\infty )}\cdot {\tilde x}_{(\infty )}/\sqrt{\epsilon
p^2_{(\infty )}}={\hat P}_{ADM}\cdot x_{(\infty )}/\sqrt{\epsilon
{\hat P}^2_{ADM}}=\tau$ \footnote{Let us note that this is
possible for globally hyperbolic, asymptotically flat at spatial infinity,
spacetimes; instead a global time does not exist, even with
a finite number of degrees of freedom, when the configuration space is compact;
see for instance Refs.\cite{dc8}.} and not an internal time.
It is the gauge-fixing $T_{(\infty )}-\tau\approx 0$ to the extra Dirac
constraint $\epsilon_{(\infty )}-\sqrt{\epsilon
{\hat P}^2_{ADM}}\approx 0$
which identifies the foliation parameter with the rest-frame time.
The evolution in $T_{(\infty )}=\tau$ of the two canonical pairs of
gravitational degrees of freedom is governed by the weak ADM energy
${\hat P}^{\tau}_{ADM}$.

The positions of the non-covariant {\it external}
center-of-mass variable ${\tilde x}^{(\mu )}
_{(\infty )}(\tau )$, replacing the arbitrary origin $x^{(\mu )}_{(\infty )}$
of the coordinates on the WSW hypersurfaces,
and of this origin are irrelevant, because,
as already said, at the end the 6 variables ${\vec z}_{(\infty )}$, ${\vec k}
_{(\infty )}$ of Eqs.(\ref{VIII9}) are decoupled: they describe the
{\it external} 3-center of mass of the isolated universe or equivalently
a decoupled external observer with his {\it point particle clock}
\footnote{Therefore one does not need {\it matter clocks and reference
fluids}\cite{kuchar1,brk}.}. They are not to be quantized because they
can be said to belong to the classical part of the Copenhagen
interpretation, but their non-covariance is fundamental in defining
the classical M\o ller radius $|{\hat {\vec S}}_{ADM}|/ {\hat
P}^{\tau}_{ADM}$ \footnote{Due to ${\hat {\vec P}}_{ADM}\approx 0$, one
has $|{\hat {\vec S}}_{ADM}|=
\sqrt{-\epsilon W^2_{ADM}}/
{\hat P}^{\tau}_{ADM}$ with $W^A_{ADM}$ the asymptotic Pauli-Lubanski
4-vector.} to be used as a ultraviolet cutoff also in metric gravity.

The {\it internal} center-of-mass 3-variable 
${\vec \sigma}_{ADM}[r_{\bar a},\pi_{\bar a}]$ \footnote{It is built in terms 
of the weak Poincar\'e charges as it is done
for the Klein-Gordon field on the Wigner hyperplane in Ref.\cite{mate};
due to ${\hat P}^r_{ADM}\approx 0$ we have $\sigma^r_{ADM}\approx -{\hat J}
^{\tau r}_{ADM}/ {\hat P}^{\tau}_{ADM}$.}
of the universe inside a WSW hypersurface identifies the 3
gauge-fixings ${\vec \sigma}_{ADM}\approx 0$ [i.e. ${\hat J}^{\tau
r}_{ADM}\approx 0$] to be added to ${\hat {\vec P}}_{ADM}[r_{\bar
a},\pi_{\bar a}]\approx 0$. With these gauge fixings this point
coincides with the arbitrary origin $x^{(\mu )}
_{(\infty )}(\tau )$. With ${\vec \sigma}_{ADM} \approx 0$ the origin
$x^{(\mu )}_{(\infty )}(\tau )$ becomes simultaneously
\cite{mate} the Fokker-Price
center of inertia, the Dixon center of mass and Pirani and Tulczjyew centroids
of the universe, while the non-covariant {\i external}
center-of-mass variable ${\tilde x}^{(\mu )}
_{(\infty )}(\tau )$ is the analog of the Newton-Wigner position operator.

Our final picture of the reduced phase space has similarities  with
the frozen Jacobi picture of Barbour\cite{dc12} and his proposal to
substitute time with  the astronomical ephemeris time\cite{dc120} \footnote{In
his timeless and frameless approach based on Ref.\cite{dc11} the local
ephemeris time coincides with the local proper time.} may be a starting
point for correlating local physical clocks with the mathematical time
parameter $\tau =T_{(\infty )}$ of the foliation (and not for defining
a timeless theory by using Jacobi's principle). We think that scenario
A) of Section VII, used for the description of void spacetimes without
matter, is a realization of the fully Machian approach of Barbour
which, however, seems possible only in absence of matter. Instead the
scenario B) with a decoupled free {\it external} center-of-mass variable
is Machian only in the fact that there are only dynamical relative
variables left both in asymptotically flat general relativity and in
parametrized Minkowski theories.

Let us remark that the interpretation of the superhamiltonian
constraint as a generator of gauge transformations given in  Section
X with natural gauge-fixing $\pi_{\phi}(\tau ,\vec \sigma )\approx 0$
(at least in 3-orthogonal coordinates) leads to the conclusion that
neither York's internal extrinsic time nor Misner's internal intrinsic
time are to be used as time parameters: Misner's time (the conformal
factor) is determined by the Lichnerowicz equation while York's time
(the trace of the extrinsic curvature) by the natural gauge-fixing
$\pi_{\phi} \approx 0$.  Instead, the gauge variable conjugate to the
conformal factor of the 3-metric [$\pi_{\phi}(\tau ,\vec
\sigma )$ in the 3-orthogonal gauges] describes the normal deformations
of the spacelike hypersurface $\Sigma_{\tau}$ which allow the
transition from one allowed 3+1 splitting of the spacetime $M^4$ to
another one.

Let us remember that in Ref.\cite{paur} the nonrelativistic limit of
the ADM action for metric gravity was considered: it allowed the
identification of a singular Lagrangian density with general Galileo
covariance depending on 27 fields (coming from the development in
series of powers of $1/c^2$ of $N$, $N_r$, ${}^3g_{rs}$) describing
Newton gravity in arbitrary coordinates. This theory has first class
constraints connected with inertial forces and second class
constraints, determining the static Newton potential in arbitrary
frames of reference when massive particles are present (see
Ref.\cite{paur1} for alternative nonrelativistic gravity theories).
This implies that it will be possible to consider the nonrelativistic
limit of our modified metric gravity and establish its connections
with the post-Newtonian approximations \cite{dc9}, in particular the
recent one of Ref.\cite{dam,dam1}.

Now, at the nonrelativistic level there is an absolute time t and the evolution
in this numerical parameter of every system is described by the Hamilton
equations associated with a Hamiltonian function H describing the energy of
the system \footnote{It is a generator of the kinematical (extended) Galileo 
group when the system is isolated.}. Alternatively, one can use a parametrized
reformulation of the system by enlarging phase space with a canonical pair
$t,\, E$ [$\{ t,E \} =\epsilon =\pm 1$, if $\epsilon$ is the signature of the
time axis], by adding the first class constraint $\chi =E-H\approx 0$, so that
the Dirac Hamiltonian is $H_D=\lambda (\tau ) \chi$, and by calling $\tau$
the scalar parameter associated with the canonical transformations generated by
$\chi$. The parameter $\tau$ labels the leaves of a foliation of Galilei
spacetime; the leaves are (rest-frame)
hyperplanes, which are the limit of Wigner
hyperplanes in parametrized Minkowski theories for $c \rightarrow \infty$.
One gets a parametric description of the same physics with t and the
solutions of the original Hamilton equations now expressed as functions of the
new time parameter $\tau$. If one adds the gauge-fixing $t-\tau \approx 0$
, one gets a frozen reduced phase space (equal to the
original one) like in the Jacobi theory, in which one reintroduce an evolution
by using the energy E=H for the evolution in $t=\tau$. However, with more
general gauge-fixings $t-f(\tau ,...)\approx 0$,
where dots mean other canonical
variables, the associated Hamiltonian is no more the energy
(see Ref.\cite{dc1}).

In the standard nonrelativistic quantization of the system one defines a
Hilbert space and writes a Schroedinger equation in which t is a parameter and
in which the t-evolution is governed by an operator obtained by quantizing the
Hamiltonian function corresponding to the energy 
\footnote{See Ref.\cite{dc2} for a
discussion of this point and of the associated ambiguities and problems.}.
Instead, in the parametrized theory, one should quantize also the pair $t, E$
\footnote{One introduces a unphysical Hilbert space in which the t-dependence 
of wave functions is restricted to be square integrable.} and
write a Schroedinger equation in $\tau$ with the quantum Dirac Hamiltonian 
\footnote{See Ref.\cite{dc3} on this point
and on the problem of the unphysical and physical scalar products.}
and then impose the constraint to identify the physical states. This procedure
is ambiguous, because in this way the energy operator has no lower bound for
its spectrum in the unphysical Hilbert space and it is delicate to recover the
physical Hilbert space from the quotient of the unphysical one with respect to
the quantum unitary gauge transformations generated by the quantum constraint.
In particular, physical states have infinite unphysical norm 
\footnote{Usually the zero eigenvalue belongs to the continuum spectrum of 
the constraint operators.} and the construction of the physical scalar 
product for physical states (without any
restriction on the t-dependence) depends on the form of the constraint
(see Ref.\cite{longhi} for a relativistic example).

Moreover, the absolute time t, which labels the Euclidean leaves of
the absolute foliation of Galileo spacetime, is unrelated to physical
clocks. As shown in Ref.\cite{dc4} (see also Ref.\cite{ish}), in the
physical Hilbert space there is no operator such that: i) it can be
used as a {\it perfect clock}, in the sense that, for some initial state,
its observed values increase monotonically with t; ii) is canonically
conjugate to the Hamiltonian operator (this would imply that this
operator is not definite positive). All this is also related with
Rovelli's proposal\cite{ro1} of replacing t (in the nonrelativistic
case) with an {\it evolving constant of the motion}, i.e. a t-dependent
function of operators commuting with the Hamiltonian (in a framework
with spatially compact spacetimes). This proposal can be done either
in the standard or in the parametrized version of the theory (see also
Ref.\cite{dc7}). Among others\cite{dc6}, Kuchar\cite{kuchar1} critics
it for the ambiguities coming from the operator ordering problem.
Lawrie and Epp \cite{ro2} notice that in the toy models with two
oscillators with fixed total energy, in which an oscillator is
supposed to be used as a clock for the other oscillator, there is no
physical degree of freedom left for the clock after having done the
canonical reduction to the reduced phase space (so that again one gets
an evolution of the surviving oscillator in a unobservable
mathematical time). In any case there are all the previously mentioned
problems and also the fact that the conjugate variables of these
evolving constants of motion generically have nothing to do with the
energy and can have spectra and symmetries of every type (see
Ref.\cite{dc2}).

All the proposals of replacing the parameter t with some physical time function
(or operator) show that this is the main unsolved problem: how to identify (at
least locally, possibly globally) the leaves of the foliation of Galileo
spacetime with {\it physical clocks}, i.e. with an apparatus described in the 
given
either phase or Hilbert space. See again in this connection Barbour\cite{dc12}
who uses as local time functions special space coordinates (the astronomical
ephemeris time\cite{dc120} or some its relativistic extension).

Now, in the approach based on parametrized special relativistic theories in
Minkowski spacetime, the final result is that every isolated system (or better
all its configurations with a timelike total 4-momentum) identifies a Wigner
foliation of Minkowski spacetime. Its leaves (the Wigner hyperplanes) are
labelled by a scalar parameter $T_s=\tau$ (the center-of-mass time in the rest
frame) in the {\it rest-frame Wigner-covariant 1-time instant form} with the
evolution in this parameter governed by the invariant mass of the system. There
is also a decoupled non-covariant center-of-mass point with free motion.
The quantization of this instant form produces a 1-time Schroedinger equation
as in the standard unparametrized nonrelativistic case with the Newtonian 
time t replaced by the Lorentz-scalar rest frame time $T_s$.

In our modified metric gravity the same picture appears in
the generalized rest-frame instant form with WSW foliations. Therefore, in this
unified approach to general relativity, special relativity and Newton-Galileo
theories one is {\it never going to quantize any time variable} and the 
problem of time is replaced by the problem of how to correlate  
physical clocks with the mathematical time parameter labelling the leaves of 
the 3+1 splitting of spacetime.

After the addition of matter to tetrad gravity (this will be done in
future papers starting with
perfect fluids\cite{nowak}) and the canonical reduction to the
rest-frame instant form with the evolution in $\tau \equiv T_{(\infty
)}$ governed by the ADM energy ${\hat P}^{\tau}_{ADM}$, one should
identify a matter subsystem with a physical clock, to take one Dirac
observable $A(\tau )$ of the clock as a physical definition of time
(the idea behind the ephemeris' time), to invert it to get $\tau
\equiv T_{(\infty )} = \tau (A)$ (in general this will be possible
only for a finite interval of $\tau$) and to transform the
$\tau$-dependence of all the Dirac observables into an $A$-dependence.
At this stage, but only for the resulting finite interval of $\tau$,
we can think to replace the ADM energy with an effective Hamiltonian
giving the $A$-evolution. However, it is completely unclear what would happen 
in any attempt to quatize the theory.

\vfill\eject

\section{The Embedding into Spacetime of the Wigner-Sen-Witten
         Hypersurfaces.}

It will be shown in this Section that the special Wigner-Sen-Witten
spacelike hypersurfaces $\Sigma_{\tau}^{(WSW)}$, needed for the
rest-frame instant form of tetrad gravity in the class of spacetimes,
asymptotically flat at spatial infinity and without supertranslations,
and corresponding to the Wigner hyperplanes orthogonal to the
4-momentum of an isolated system, can be defined by general embeddings
$z^{\mu}(\tau ,\vec \sigma )$ 
\footnote{Generalizing the embeddings $z^{(\mu
)}(\tau ,\vec \sigma )=x^{(\mu )}(\tau )+\epsilon^{(\mu )}_r(u(p_s))
\sigma^r$ for Minkowski Wigner spacelike hyperplanes.}.

It will be clear that the WSW hypersurfaces enjoy the same formal properties of
spacelike hyperplanes in Minkowski spacetime,  namely that, given an
origin on each one of them and an adapted tetrad at this origin, there
is a natural parallel transport so that one can uniquely define the
adapted tetrads in all points of the hyperplane starting from the
given adapted one at the origin. Namely due to the property of tending
asymptotically to Minkowski Wigner spacelike hyperplanes in a
direction-independent way at spatial infinity, the WSW spacelike
hypersurfaces allow the definition of asymptotic (angle-independent)
adapted tetrads with the timelike component parallel to the weak ADM
4-momentum. Then an adaptation to tensors of the Sen-Witten spinorial
equation based on the Sen connection allows to define preferred
adapted tetrads in each point of $\Sigma^{(WSW)}_{\tau}$ tending to
the given ones at spatial infinity: this can be reinterpreted as a
special form of {\it parallel transport} generalizing the trivial
Euclidean one on Minkowski spacelike hyperplanes.

These preferred tetrads  correspond to the {\it non-flat preferred
observers} of Bergmann\cite{be}: they are a set of {\it privileged
observers} (privileged tetrads adapted to $\Sigma^{(WSW)}_{\tau}$) of
{\it geometrical nature}, since they depend on the intrinsic and
extrinsic geometry of $\Sigma_{\tau}^{(WSW)}$, and not of
{\it static nature} like in the approaches of M\o ller\cite{p23}, Pirani
\cite{p24} and Goldberg\cite{gold}. On the solutions of
Einstein's equations they also acquire a {\it dynamical nature} depending
on the configuration of the gravitational field itself.
These privileged observers are
associated with the existence of the asymptotic Poincar\'e charges,
since their asymptotic 4-velocity is determined by the weak ADM
4-momentum. A posteriori, namely after having solved Einstein's
equations, one could try to use these {\it geometrical and dynamical}
privileged observers \footnote{Privileged non-holonomic coordinate systems
replacing the rectangular Minkowski coordinates of the flat case.} in
the same way as, in metric gravity, are used the {\it bimetric theories},
like the one of Rosen\cite{p25}, with a set of privileged static
non-flat background metrics. This congruence of timelike preferred
observers\footnote{With asymptotic inertial observers in the rest-frame instant
form with ${\tilde \lambda}_A(\tau )= (\epsilon ;\vec 0)$ and ${\tilde
\lambda}_{AB}(\tau )=0$.} is a
non-Machian element of these noncompact spacetimes. The asymptotic
worldlines of the congruence may replace the static concept of {\it fixed
stars} in the study of the precessional effects of gravitomagnetism on
gyroscopes (dragging of inertial frames) and seem to be naturally
connected with the definition of post-Newtonian coordinates
\cite{mtw,soffel}.

Ashtekar and Horowitz\cite{p26} pointed out the existence in metric
gravity of a preferred family of lapse and shift functions, which can
be extracted by the spinorial demonstration of Witten\cite{p27} of the
positivity of the ADM energy, and, therefore, of a
set of preferred spacelike hypersurfaces. Then, Frauendiener\cite{p28}
translated this fact in terms of privileged geometric adapted tetrads
on each $\Sigma_{\tau}$ of this set, enjoying the same properties of
tetrads adapted to Minkowski spacelike hyperplanes: he starts from the
Sen-Witten equation\cite{p27,sen,spinor,spinor1,rindler} and uses
ideas based on the Sparling 3-form \cite{p29,p30}.

Let us review these statements in more detail.

 i) In his demonstration of the positivity energy theorem Witten \cite{p27}
 introduced SU(2) spinor fields on $\Sigma_{\tau}$ [see also 
Refs.\cite{p35a,fadde}: one gets $P_{ADM,(\mu )} n^{(\mu )}\geq 0$ for all 
future pointing null vectors
($n^2=0$) and  this implies $P_{ADM,(\mu )} n^{(\mu )}\geq 0$ for all
future pointing asymptotic either timelike or null translations with
$n^{(\mu )}$ obtained from some SU(2) spinor field on $\Sigma_{\tau}$.
In the reformulation using the so called Nester-Witten 2-form $F$
\cite{p35a}, defined on the total space of the spin bundle over $M^4$,
one can show that $P_{ADM,(\mu )}n^{(\mu )}=lim_{r\, \rightarrow
\infty} 2k \int_{\Sigma_{\tau}} F(\xi ) =2k \int_{S^2_{\tau ,\infty}} 
dF(\xi )$.
As first noted by Sparling\cite{p29} (see also the last chapter of
Vol.2 of Ref.\cite{rindler}) there is a 3-form $\Gamma$ on the spin
bundle, the so called Sparling 3-form, such that $\Gamma = dF-{1\over
2} n^{\mu}\, {}^4G_{\mu\nu} X^{\nu}$ [$\, X
^{\mu}={1\over 6} \epsilon^{\mu}{}_{\alpha\beta\gamma} dx^{\alpha}\wedge
dx^{\beta}\wedge dx^{\gamma}$]; therefore, the vacuum Einstein equations can
be characterized by $d\Gamma =0$. In presence of matter Einstein equations
give $\Gamma \, {\buildrel \circ \over =}\, dF-{k\over 2} n^{\mu}\, {}^4T
_{\mu\nu} X^{\nu}$, so that
$P_{ADM,(\mu )}n^{(\mu )}\, {\buildrel \circ \over =}\,
2k \int_{S^2_{\tau ,\infty}} (\Gamma +{k\over 2} n^{\mu}\, {}^4T_{\mu\nu} X
^{\nu})$. Using the dominant energy condition\cite{he} for the positivity of
the second term, one can arrive at the result $P_{ADM,(\mu )} n^{(\mu
)} \geq 0$ if the SU(2) spinor determining $n^{(\mu )}$ satisfies the
elliptic Sen-Witten equation for the noncompact hypersurface
$\Sigma_{\tau}$

\begin{equation}
{}^3{\cal D}_{\tilde A\tilde B}\psi^{\tilde B}={}^3{\tilde \nabla}_{\tilde
A\tilde B}\psi^{\tilde B}+{1\over {2\sqrt{2}}}\, {}^3K \psi_{\tilde A} =0.
\label{XII1}
\end{equation}

In this equation ${}^3{\tilde
\nabla}_{\tilde A\tilde B}$ is the extension of ${}^3\nabla_{\mu}$ to
spatial SU(2) spinors on $\Sigma^{(WSW)}_{\tau}$ \footnote{It is the
torsion-free Levi-Civita connection of ${}^3g_{rs}$  and depends only
on the intrinsic geometry of $\Sigma^{(WSW)}_{\tau}$.}, while
${}^3{\cal D}_{\tilde A\tilde B}={}^3{\cal D}_{\tilde B\tilde A}$,
called the {\it Sen connection}, is the true {\it spatial} derivative acting
on spatial SU(2) spinors \footnote{It is an extension (depending on the
extrinsic geometry of $\Sigma^{(WSW)}_{\tau}$) of the pull-back to
$\Sigma^{(WSW)}_{\tau}$ of ${}^4\nabla_{\mu}$; it is torsion-free but
it is not the Levi-Civita connection of ${}^3g_{rs}$.}.

As stressed by Frauendiener and Mason\cite{p30}, the Sparling 3-form is a
Hamiltonian density for canonical general relativity (see also Ref.
\cite{p35b} on this point), while, when used quasi-locally, the 2-form F gives
rise to Penrose's formula\cite{p31}  for the angular momentum twistor of the
quasi-local mass construction.
These ideas are required for treatment of conserved quantities in
general relativity, since the Sparling 3-form can be extended to be
one of a collection of 3-forms {\it on the bundle of general linear
frames} which, when pulled back to spacetime, give rise to classical
formulas for the {\it pseudo-energy-momentum tensor} of the gravitational
field \cite{p32} \footnote{See Ref.\cite{emp} for the Einstein complex,
Ref.\cite{ll} for the Landau-Lifschitz one and Ref.\cite{empt} for a
review.}. See also Ref.\cite{p33}, where the Sparling 3-form is studied
in arbitrary dimension and where it is contrasted with Yang-Mills
theory. In Ref.\cite{p34} there is the relationship of the Sparling
3-form to the spin coefficient formalism. These papers show the
connection of the Poincar\'e charges with the standard theory of the
Komar superpotentials and of the energy-momentum pseudotensors, which
is reviewed in Appendix D of Ref.\cite{russo3}.

See Refs.\cite{soluz,p26} for the existence of solutions of the
Sen-Witten equation on noncompact spacelike hypersurfaces \footnote{For
non-spacelike ones see the last chapter of Vol.2 in
Ref.\cite{rindler}, its references and Ref. \cite{horotod}.} and
Refs.\cite{bergq} for the non-unicity of Witten's positivity proof as
first noted in Ref.\cite{horope}: other equations different from the
Sen-Witten one can be used in variants of the proof.

In particular, in the paper of Reula in Ref.\cite{soluz}, used in
Ref.\cite{p28}, the problem of the existence of solutions of the
Sen-Witten equation (\ref{XII1}) has been formalized in the following
way. An {\it initial data set} $(\Sigma_{\tau},{}^3g_{rs},{}^3K_{rs})$
for Einstein's equations consists of a 3-dimensional manifold
$\Sigma_{\tau}$ without boundary equipped with a positive definite
3-metric ${}^3g_{rs}$ and a second rank, symmetric tensor field
${}^3K_{rs}$. For simplicity it is assumed that $\Sigma_{\tau}$ is
diffeomorphic to $R^3$ and that ${}^3g
_{rs}$ and ${}^3K_{rs}$ are smooth tensor fields on
$\Sigma_{\tau}$. An initial data set is said to satisfy the {\it local
energy condition} if $\mu \geq {|\, J^{\mu}J_{\mu}\, |}^{1/2}$ with
$\mu ={1\over 2}[{}^3R+{}^3K_{rs}\, {}^3K
^{rs}-({}^3K)^2]\approx {}^3R$ and $J_{\mu}=\partial^{\nu}[{}^3K_{\mu\nu}-{}^3g
_{\mu\nu}\, {}^3K]$. An initial data set is {\it asymptotically flat} if one 
can introduce an asymptotically Euclidean coordinate system such that
${}^3g_{rs}-\delta_{rs}= O(r^{-1})$ and $\partial_u\,
{}^3g_{rs}=O(r^{-2})$ for $r\, \rightarrow \infty$ and, moreover,
${}^3K_{rs}=O(r^{-2})$ and ${}^3R_{rs}=O(r^{-3})$ for $r\,
\rightarrow \infty$ \footnote{They are compatible with 
Christodoulou-Klainermann Eqs.(\ref{III4}).}.
Then one has the following existence theorem(see also Ref.\cite{p26}):
If $(\Sigma_{\tau}, {}^3g_{rs},{}^3K_{rs})$ is an initial data set
that satisfies the local energy condition and is asymptotically flat,
then for any spinor field $\psi
^{\tilde A}_o$ that is {\it asymptotically constant} 
\footnote{I.e. $\partial_r \psi_o
^{\tilde A}=0$ outside a compact subset of $\Sigma_{\tau}$; see also Ref.
\cite{horotod}) there exists a spinor field $\psi^{\tilde A}$
satisfying the Sen-Witten equation (\ref{XII1}).} and such that
$\psi^{\tilde A}=\psi_o
^{\tilde A}+O(r^{-1})$ at spatial infinity.\hfill\break

ii)
In Ref.\cite{p26}, Ashtekar and Horowitz note that the Sen-Witten equation
enables one to {\it transport rigidly} constant spinors at infinity to the 
interior of the 3-manifold on which the initial data are defined. By 
{\it taking squares} of
the Sen-Witten spinors one can construct a {\it preferred} family of lapse and
shifts and interpret them as the projections of 4-dimensional null evolution
vector fields $z^{\mu}_{\tau}(\tau ,\vec \sigma )=[Nl^{\mu}+N^{\mu}](\tau
,\vec \sigma )$, $N^{\mu}(\tau ,\vec \sigma )=[z^{\mu}_rN^r](\tau ,\vec
\sigma )$, $[l^{\mu}N_{\mu}](\tau ,\vec \sigma )=0$, $z^2_{\tau}(\tau ,\vec
\sigma )=0$, obtained by transporting rigidly the spacetime asymptotic
translations at spatial infinity. The preferred family corresponds to
a {\it gauge fixing prescription} for lapse and shift functions. Next it
is shown that, on the phase space of general relativity, one can
compute Hamiltonians corresponding to these lapse and shifts. Although
these Hamiltonians have a complicated form  in terms of the usual
canonical variables (involving volume and surface integrals), they are
simply the volume integrals of squares of derivatives of the Witten
spinors. In particular, the Hamiltonians generating Witten-time
translations are manifestly positive and differentiable. These
expressions are essentially spinorial, i.e. they depend on the phases
of the individual spinors whereas the original lapse-shift vector did
not . It is essential for a coherent point of view, therefore, to
regard the spinors as fundamental, and the lapse-shift vector as
derived (this requires supergravity, which motivated Witten, but is
not justified in ordinary gravity). The Witten argument required that
the phases of the spinors making up the null lapse-shift vector be
assumed to be asymptotically constant along with the lapse-shift
vector: without this, the argument fails.

In terms of vectors, given a {\it tetrad} at infinity, it is noted in Ref.
\cite{p26}that the SL(2,C) Sen-Witten equation then provides us with a {\it 
tetrad field everywhere on $\Sigma_{\tau}$}. If we rotate the
tetrad at infinity, the entire field rotates rigidly by the same amount; the
freedom is that of global rather than local Lorentz transformations. It is in
this sense that we have a {\it gauge fixation procedure}. Note, however, that 
the preferred tetrad fields depend on the choice of the variables $({}^3g_{rs},
{}^3K_{rs})$ on $\Sigma_{\tau}$; if we change the metric
${}^3g_{rs}$ near $\Sigma_{\tau}$, the tetrad fields change.
It can also be shown\cite{p26} that if ${}^3T^{\mu}$ is a vector field tangent
to $\Sigma_{\tau}$ (not necessarily spacelike)
with asymptotic value ${}^3T_{(\infty )}^{\mu}$, then
${}^3T^{\mu}$ is timelike (respectively, null, spacelike) everywhere,
if ${}^3T_{(\infty )}^{\mu}$ is
timelike (respectively, null, spacelike) at infinity.

Then, in Ref.\cite{p26} it is noted that, if $(M^4,{}^4\eta_{(\mu )(\nu )})$ is
the Minkowski spacetime, then, since the constant spinor fields in it
automatically satisfy Sen-Witten equation,
for any choice of $\Sigma_{\tau}$, the transport of
translations at infinity yields the translational Killing fields everywhere on
$M^4$. In a generic spacetime, however, the transport is tied to the choice of
$\Sigma_{\tau}$. Thus, it is only when we are given a foliation of a generic
spacetime that we can obtain 4 vector fields everywhere on the spacetime, and
they {\it depend} on the choice of the foliation. The transport is well 
suited to the canonical framework, however, because in this framework one 
deals only with 3-surfaces.\hfill\break

In the approach of this paper the preferred lapse and shift functions of
Ref.\cite{p26} have to be replaced  with the asymptotic parts of the lapse 
and shift
functions of Eqs.(\ref{VI2}) when ${\tilde
\lambda}_{AB}(\tau )=0$ so that $N_{(as) A}(\tau ,
\vec \sigma )=N_{(as) A}(\tau )=- {\tilde \lambda}_A(\tau )$. Our 4 arbitrary
functions ${\tilde \lambda}_A(\tau )$ give the same multiplicity as in
the previous spinorial construction without relying on the special
null evolution vectors needed in it (the evolution vectors
$\partial_{\tau}z^{\mu}(\tau ,\vec \sigma )$ are now arbitrary).
Therefore, in this approach, the {\it gauge-fixing prescription} for
selecting the preferred family of lapse and shifts becomes the
requirement of absence of supertranslations according to
Eqs.(\ref{VI3}), i.e. ${\tilde
\lambda}_{AB}(\tau )=0$. But this implies ${\hat P}^{(\mu )}_{ADM}
\approx l^{(\mu )}_{(\infty )} {\hat P}^{\tau}_{ADM}$ and,
as a consequence, the allowed foliations and their leaves, i.e. the
spacelike hypersurfaces $\Sigma_{\tau}^{(WSW)}$, could be called
``Wigner-Sen-Witten" (WSW) foliations and spacelike hypersurfaces,
being the analogues of the Wigner foliations and spacelike hyperplanes
of the parametrized Minkowski theories.\hfill\break

iii) In Ref. \cite{p28} Frauendiener , exploiting the fact that there
is a unique 2-1 (up to a global sign) correspondence between a SU(2)
spinor and a triad on a spacelike hypersurface, derives the necessary
and sufficient conditions that have to be satisfied by a triad in
order to correspond to a spinor that satisfies the Sen-Witten
equation. In this way it is possible to eliminate completely any
reference to spinors and to speak only of triads
${}^3e^{(WSW)}{}^r_{(a)}$ on $\Sigma^{(WSW)}_{\tau}$ and
$\Sigma^{(WSW)}_{\tau}$-adapted tetrads on $M^4$.

These triads ${}^3e^{(WSW)}{}^r_{(a)}$ are built in terms of the SU(2)
spinors solutions of the Sen-Witten equation and, as a consequence of
this equation, they are shown \cite{p28} to satisfy the following
equations

\begin{eqnarray}
&&{}^3\nabla_r\, {}^3e^{(WSW) r}_{(1)}={}^3\nabla_r\, {}^3e^{(WSW)
r}_{(2)}=0,\nonumber \\
 &&{}^3\nabla_r\, {}^3e^{(WSW) r}_{(3)}=-\alpha
{}^3K,\nonumber \\
 &&{}^3e^{(WSW) r}_{(1)}\, {}^3e^{(WSW) s}_{(3)}\,
{}^3\nabla_r\, {}^3e^{(WSW)}_{(2)s} +{}^3e^{(WSW) r}_{(3)}\,
 {}^3e^{(WSW) s}_{(2)}\, {}^3\nabla_r\, {}^3e^{(WSW)}_{(1)s}+\nonumber \\
 &+&{}^3e^{(WSW) r}_{(2)}\, {}^3e^{(WSW) s}_{(1)}\, {}^3\nabla_r\,
{}^3e^{(WSW)}_{(3)s}=0.
\label{XII2}
\end{eqnarray}

Therefore, these triads are formed by 3 vector field with the
properties: i) two vector fields are divergence free; ii) the third
one has a non-vanishing divergence proportional to the trace of the
extrinsic curvature of $\Sigma^{(WSW)}_{\tau}$ \footnote{On  a maximal slicing
hypersurface (${}^3K=0$) all three vectors would be divergence free.};
iii) the vectors satisfy a cyclic condition.
The 4-dimensional freedom in the choice of a spinor at one point (at
spatial infinity) implies that a triad satisfying  Eqs.(\ref{XII2}) is
unique up to global frame {\it rotations} and {\it homotheties}.

In Ref.\cite{p28} it is shown: 1) these triads do not exist for
compact $\Sigma_{\tau}$; 2) with nontrivial topology for
$\Sigma_{\tau}$ there can be less than 4 real solutions and the triads
cannot be build; 3) the triads exist for asymptotically null surfaces
(hyperboloidal data), but the corresponding tetrad will be degenerate in the 
limit of null infinity.

Moreover, in Ref.\cite{p28}, using the results of Ref.\cite{p32}, it
is noted that the Einstein energy-momentum pseudo-tensor\cite{emp} is
a canonical object only in the frame bundle over $M^4$, where it
coincides with the Sparling 3-form. In order to bring this 3-form back
to a 3-form (and then to an energy-momentum tensor) over the spacetime
$M^4$, one needs a section (i.e. a tetrad) in the frame bundle. Only
with the 3+1 decomposition of $M^4$ with WSW foliations one gets that
(after imposition of Einstein's equations together with the local
energy condition) one has a preferred, geometrical and dynamical,
adapted tetrad on the initial surface $\Sigma^{(WSW)}_{\tau}$.

By assuming that these triads on $\Sigma^{(WSW)}_{\tau}$ have the asymptotic 
behaviour ${}^3e^{(WSW)}{}^r_{(a)}(\tau ,\vec \sigma )\, {\rightarrow}_{r\,
\rightarrow \infty}\, (1-{M\over {2r}})\delta^r_{(a)}+O(r^{-3/2})
\rightarrow \delta^r_{(a)}= {}^3e^{(WSW)}_{(\infty )}{}^r_{(a)}$,
one can select the solutions of Eqs.(\ref{XII2})  relevant for the rest-frame
instant form of metric gravity.

This initial data set  determines {\it uniquely}
a {\it triad} on $\Sigma^{(WSW)}_{\tau}$ and hence, taking into account
 the normal $l^{\mu}$ to $\Sigma^{(WSW)}_{\tau}$, an {\it adapted
tetrad} ${}^4_{(\Sigma )}{\check E}^{(W) (\mu )}_A$ in spacetime.

Therefore, we can define the $\Sigma^{(WSW)}_{\tau}$-adapted
{\it preferred tetrads} of the rest-frame instant form

\begin{eqnarray}
{}^4_{(\Sigma )}{\check {\tilde E}}^{(WSW)}{}^A_{(o)}(\tau ,\vec \sigma )
&=& {1\over {-\epsilon +n(\tau ,\vec \sigma )}} (1; -n^r(\tau
,\vec \sigma )),\nonumber \\
  {}^4_{(\Sigma )}{\check {\tilde E}}^{(WSW)}{}^A_{(a)}(\tau ,\vec \sigma ) 
&=& (0; {}^3e^{(WSW)}{}^r_{(a)})(\tau ,\vec \sigma ),\nonumber \\
  &&{}\nonumber \\
  {}^4_{(\Sigma )}{\check E}^{(WSW)}{}^{\mu}_{(o)}(\tau ,\vec \sigma ) 
&=& b^{\mu}_A(\tau ,\vec \sigma )\, {}^4_{(\Sigma )}{\check {\tilde E}}
{}^{(WSW) A}_{(o)}(\tau ,\vec \sigma )=
l^{\mu}(\tau ,\vec \sigma ),\nonumber \\
  {}^4_{(\Sigma )}{\check E}^{\mu}_{(a)}(\tau ,\vec \sigma ) 
&=&  b^{\mu}_A(\tau ,\vec \sigma )\, {}^4_{(\Sigma )}{\check {\tilde E}}
{}^{(WSW) A}_{(a)}(\tau ,\vec \sigma )=b^{\mu}_s(\tau ,\vec \sigma )\,\,\, 
{}^3e^{(WSW)}{}^s_{(a)}(\tau ,\vec \sigma ).
\label{XII3}
\end{eqnarray}

They replace static concepts like the {\it fixed stars} in the study of
the dragging of inertial frames. Since the WSW hypersurfaces and the
3-metric on them are dynamically determined \footnote{The solution of Einstein
equations is needed to find the physical 3-metric, the allowed WSW
hypersurfaces and the Sen connection.}, one has neither a static
background on system-independent hyperplanes like in parametrized
Newton theories nor a static one on the system-dependent Wigner
hyperplanes like in parametrized Minkowski theories. Now both the WSW
hyperplanes and the metric on it are system dependent.

iv) For completeness let us quote the formulation of general
relativity as a {\it teleparallel} theory done by Nester in
Refs.\cite{p35c} in order to prove the positivity of gravitational
energy with purely tensorial methods. It could be connected  with a
different notion of parallel transport on the WSW hypersurfaces.

Nester shows that by imposing certain gauge conditions on tetrads one
can obtain positivity of the ADM energy. His conditions  are closely
related to Eqs.(\ref{XII2}). Specifically, he also imposes the cyclic
condition but on global cotriads rather than on global triads.
Clearly, a global triad defines a connection on an initial surface, by
requiring that a parallel vector field has constant coefficients with
respect to the triad. This connection will be metric compatible and
integrable since it preserves the triad. Therefore, its curvature will
be zero, but the torsion will be nonzero. We see from the present
result, that on an initial data set satisfying the local energy
conditions (needed to prove the existence of Sen-Witten spinors) there
exists a {\it preferred absolute parallelism}.

While the orthonormal coframe ${}^3\theta^{(a)}={}^3e^{(a)}_rd\sigma^r$ 
determines the metric and the Riemannian geometry, a
given Riemannian geometry determines only an equivalence class of
orthonormal coframes: coframes are defined only modulo
position-dependent rotations and, under these gauge transformations,
the spin connection transforms as a SO(3) gauge potential. A
gauge-fixing for the rotation freedom usually means a choice of a
representative of the spin connection inside its gauge orbit (like the
Coulomb gauge for the electromagnetic vector gauge potential $\vec
A$): this would induce a choice of an associated coframe with respect
to some standard origin. However, since coframes ${}^3\theta^{(a)}$
are more elementary of the Levi-Civita spin connection
${}^3\omega^{(a)}{}_{(b)}$, which is built in terms of them, it is
possible to define gauge-fixings directly at the level of coframes
[see Ref.\cite{p35c},papers b)]. The idea of these papers is that the
choice of a preferred coframe ${}^3\theta^{(a)}_{(P)}$ on the
Riemannian parallelizable 3-manifold $(\Sigma_{\tau}^{(WSW)},{}^3g)$
\footnote{With its associated metric compatible Levi-Civita connection and
parallel transport and vanishing torsion.} may be associated with the
definition of a new kind of parallel transport on
$\Sigma_{\tau}^{(WSW)}$, i.e. of a {\it teleparallel} (or {\it Weitzenb\"ock
or distant parallelism}) geometry on $\Sigma^{(WSW)}_{\tau}$,
according to which a covariant vector is parallely transported along a
curve in $\Sigma_{\tau}^{(WSW)}$ if in each point q of the curve it
has the same components with respect to the local coframe
${}^3\theta^{(a)}_{(P)}{|}_q$. The special coframe ${}^3\theta^{(a)}
_{(P)}$ is said {\it orthoteleparallel} (OT) coframe.
With this structure $(\Sigma^{(WSW)}_{\tau},
\delta_{(a)(b)})$ is a 3-manifold with flat metric \footnote{The curvature 
vanish because this parallel transport is path-independent (absolute
parallelism) like in Euclidean geometry. The OT coframe
${}^3\theta^{(a)}_{(P)}$ is the special coframe in which by
construction also all the spin connection coefficients vanish.}, but
with a nonvanishing {\it torsion}, which completely characterizes this kind
of geometry. The Riemannian geometry $(\Sigma^{(WSW)}_{\tau},{}^3g)$
corresponds to a whole equivalence class of teleparallel geometries
$(\Sigma^{(WSW)}_{\tau},{}^3\theta
^{(a)}_{(P)})$, according to which coframe is chosen as the preferred OT one.

In Ref.\cite{p35c}b) it is pointed out that there exists a natural (of
elliptic type) gauge-fixing for the choice of a special OT coframe
${}^3\theta^{(a)}_{(P)}$: these are three conditions (one is the
cyclic condition), which determine a special orthonormal coframe on a
3-manifold \footnote{I.e. they determine the 3 Euler angles of the dual frame
with respect to a standard frame chosen as an identity cross section
in the orthonormal frame bundle $F(\Sigma^{(WSW)}_{\tau})$.} once
appropriate boundary conditions are fixed. For asymptotically flat
3-manifolds there is a certain boundary condition such that, when the
first de Rahm cohomology group $H_1(\Sigma^{(WSW)}_{\tau})
=0$ vanishes, a certain closed 1-form is globally exact in this
gauge and determines a function $F_{(P)}$ up to a constant, which,
suitably normalized at infinity, is the best definition of the
generalization of the Newton potential\cite{p35c}c). With
this gauge\cite{p35c}c), one gets a locally positive representation
for the Hamiltonian density allowing a new, {\it strictly tensorial} (in
contrast to Witten's spinor method\cite{p27}) proof of positive energy
for Einstein's theory of gravity. Given an orthonormal coframe
${}^3\theta^{(a)}$, these gauge conditions  become a nonlinear
second-order elliptic system for the rotation matrix defining an OT
coframe ${}^3\theta^{(a)}_{(P)}=R^{(a)}{}_{(b)}{}^3
\theta^{(b)}$. In Ref.\cite{p35c}b) it is shown that the associated
linearized problem has a unique solution if $d{}^3\theta^{(a)}$ is not
too large and the second deRahm cohomology group
$H_2(\Sigma^{(WSW)}_{\tau})=0$ vanishes (for asymptotically flat
spaces one should use the first paper in Ref.\cite{soluz}). In
Ref.\cite{dimakis} it is shown that for 3-manifolds the gauge
conditions are essentially equivalent to the {\it linear} Dirac equation,
for which unique solutions exist. Hence for 3-manifolds special OT
coframes exist except possibly at those (isolated) points where the
Dirac spinor vanishes.

Coming back to the rest-frame instant form of metric gravity defined in 
Section VIII, one has that
the asymptotic transition functions from arbitrary coordinates on
$M^4$ to WSW hypersurfaces $\Sigma^{(WSW)}_{\tau}$ are 

\begin{eqnarray}
 {\hat b}^{(\mu )}_{(\infty )l}&\approx& \epsilon l^{(\mu )}_{(\infty
 )}\approx -\epsilon b^{(\mu )}_{(\infty )\tau},
 \nonumber \\
 {\hat b}^{(\mu )}_{(\infty ) r}&=& b^{(\mu )}_{(\infty )r}=
 \epsilon^{(\mu )}_r(u(p_{(\infty )})),\nonumber \\
 {\hat b}^l_{(\infty )(\mu )}&=&  l_{(\infty )(\mu )}=
-\epsilon b^{\tau}_{(\infty )(\mu )},
 \nonumber \\
 {\hat b}^s_{(\infty )(\mu )}&=&b^s_{(\infty )(\mu )}-
{\tilde \lambda}_s(\tau )b^{\tau}
 _{(\infty )(\mu )} \approx b^s_{(\infty )(\mu )},\nonumber \\
  &&with\nonumber \\
  b^{(\mu )}_{(\infty )A}(\tau ) &\equiv& L^{(\mu )}{}_{(\nu
)=A}(p_{(\infty )},{\buildrel \circ \over p}_{(\infty )}) =
\epsilon^{(\mu )}_A(u(p_{(\infty )})),\nonumber \\
 &&{}\nonumber \\
 \epsilon l^{(\mu )}_{(\infty )}&=&
 \epsilon^{(\mu )}_o(u(p_{(\infty )}))=u^{(\mu )}(p_{(\infty )})
 \approx {{{\hat P}^{(\mu )}_{ADM}}\over {\epsilon_{(\infty )}}},\nonumber \\
 \epsilon_{(\infty )}\approx M_{ADM}&=&\sqrt{\epsilon {\hat P}^2_{ADM}},
\quad\quad
 S^{(\mu )(\nu )}_{(\infty )}\equiv {\hat S}^{(\mu )(\nu )}_{ADM}.
\label{XII4}
\end{eqnarray}

Given the previous boundary conditions on the triads [${}^3e^{(WSW)r}_{(a)}
\rightarrow {}^3e^{(WSW)}_{(\infty )}{}^r_{(a)}
=\delta_{(a)}^r$] and cotriads [${}^3e^{(WSW)}_{(a)r} \rightarrow 
{}^3e^{(WSW)}_{(\infty ) (a)r}=\delta_{(a)r}$], we have the following
associated asymptotic tetrads on $\Sigma^{(WSW)}_{\tau}$

\begin{eqnarray}
{}^4E^{(WSW)}_{(\infty )}{}^{(\mu )}_{(o)} \delta^{\mu}_{(\mu )}&=&
{}^4_{(\Sigma )}{\check E}^{(WSW)}_{(\infty )}{}^{(\mu )}_{(o)}
\delta^{\mu}_{(\mu )}=
\delta^{\mu}_{(\mu )} l^{(\mu )}_{(\infty )}=\delta^{\mu}_{(\mu )} 
{\hat b}^{(\mu )}_{(\infty )l}
 \equiv \delta^{\mu}_{(\mu )} b^{(\mu )}_{(\infty )\tau}= \delta^{\mu}_{(\mu )}
u^{(\mu )}(p_{(\infty )}),\nonumber \\
 {}^4E^{(WSW)}_{(\infty)}{}^{(\mu )}_{(a)} \delta^{\mu}_{(\mu )} &=&
 {}^4_{(\Sigma )}{\check E}^{(WSW)}_{(\infty )}{}^{(\mu )}_{(a)} 
\delta^{\mu}_{(\mu )}=
 \delta^{\mu}_{(\mu )} {\hat b}^{(\mu )}_{(\infty )s}\, 
{}^3e^{(WSW)}_{(\infty )}{}^s_{(a)}
 \equiv \delta^{\mu}_{(\mu )} b^{(\mu )}_{(\infty )s}\, 
{}^3e^{(WSW)}_{(\infty )}{}^s_{(a)}=
 \nonumber \\
 &=&\delta^{\mu}_{(\mu )} b^{(\mu )}_{(\infty )s} \delta^s_{(a)}=
 \delta^{\mu}_{(\mu )}
\epsilon^{(\mu )}_s(u(p_{(\infty )})) \delta^s_{(a)},\nonumber \\
&&{}\nonumber \\
 {}^4E^{(WSW)}_{(\infty )}{}^A_{(o)}&=& {}^4_{(\Sigma )}{\check 
{\tilde E}}^{(WSW)}
 _{(\infty )}{}^A_{(o)}= (-\epsilon ; 0),\nonumber \\
 {}^4E^{(WSW)}_{(\infty )}{}^a_{(a)} &=& {}^4_{(\Sigma )}{\check 
{\tilde E}}^{(WSW)}
 _{(\infty )}{}^A_{(a)}= (0; \delta^r_{(a)}),\nonumber \\
 &&{}\nonumber \\
 {}^4E^{(WSW)}_{(\infty )}{}^{(o)}_{(\mu )} \delta^{(\mu )}_{\mu} &=&
 {}^4_{(\Sigma )}{\check E}^{(WSW)}_{(\infty )}{}^{(o)}_{(\mu )} 
\delta^{(\mu )}_{\mu}=
 \delta^{(\mu )}_{\mu}  l_{(\infty )(\mu )} =\delta^{(\mu )}_{\mu}
 {\hat b}^l_{(\infty )(\mu )}= -\epsilon \delta^{(\mu )}_{\mu} 
b^{\tau}_{(\infty )(\mu )},
 \nonumber \\
 {}^4E^{(WSW)}_{(\infty )}{}^{(a)}_{(\mu )} \delta^{(\mu )}_{\mu} &=&
 {}^4_{(\Sigma )}{\check E}^{(WSW)}_{(\infty )}{}^{(a)}_{(\mu )} 
\delta^{(\mu )}_{\mu}=
 \delta^{(\mu )}_{\mu} {\hat b}^s_{(\infty )(\mu )}\,
 {}^3e^{(WSW)}_{(\infty )}{}^{(a)}_s \equiv \delta^{(\mu )}_{\mu}
 b^s_{(\infty )(\mu )} \delta^{(a)}_s,\nonumber \\
 &&{}\nonumber \\
 {}^4E^{(WSW)}_{(\infty )}{}^{(o)}_A &=& {}^4_{(\Sigma )}{\check 
{\tilde E}}^{(WSW)}
 _{(\infty )}{}^{(o)}_A \equiv (-\epsilon ; 0),\nonumber \\
 {}^4E^{(WSW)}_{(\infty )}{}^{(a)}_A &=& {}^4_{(\Sigma )}{\check 
{\tilde E}}^{(WSW)}
 _{(\infty )}{}^{(a)}_A \equiv (0; {}^3e^{(WSW)}_{(\infty )}{}_{(a)r}=
\delta_{(a)r}).
\label{XII5}
\end{eqnarray}

The embeddings $z^{\mu}(\tau ,\vec \sigma )$ of $R^3$ into $M^4$
associated with WSW spacelike hypersurfaces
$\Sigma^{(WSW)}_{\tau}$ in the rest-frame instant form of tetrad
gravity are restricted to assume the same form at spatial infinity of
those in Minkowski spacetime identifying the Wigner hyperplanes in the
rest-frame instant form [see Eq.(\ref{a6})]

\begin{eqnarray}
z^{\mu}(\tau ,\vec \sigma ) &{\rightarrow}_{r \rightarrow \infty}&\,
\delta^{\mu}_{(\mu )} z^{(\mu )}_{(\infty )}(\tau ,\vec \sigma ),\nonumber \\
 &&{}\nonumber \\
 z^{(\mu )}_{(\infty )}(\tau ,\vec \sigma ) &=& x^{(\mu )}_{(\infty )}(\tau )
 + \epsilon^{(\mu )}_r(u(p_{(\infty )})) \sigma^r =\nonumber \\
 &=& x^{(\mu )}_{(\infty )}(0) +u^{(\mu )}(p_{(\infty )}) \tau +
 \epsilon^{(\mu )}_r(u(p_{(\infty )})) \sigma^r.
\label{XII6}
\end{eqnarray}

By using the notation

\begin{eqnarray}
l^{\mu}&=& \epsilon {\hat b}^{\mu}_l ={{\epsilon}\over {-\epsilon +n}}
[ b^{\mu}_{\tau}- n^r b^{\mu}_r]={1\over {\sqrt{{}^3g}}}
\epsilon^{\mu}_{\alpha\beta\gamma}\, {}^4_{(\Sigma )}{\check E}^{(WSW)}{}^{\alpha}_{(1)}\,
{}^4_{(\Sigma )}{\check E}^{(WSW)}{}^{\beta}_{(2)}\, {}^4_{(\Sigma
)}{\check E}^{(WSW)}{}^{\gamma}_{(3)},\nonumber \\
\epsilon^{\mu}_r &=& b^{\mu}_s \, {}^3e^{(WSW)}{}^s_{(a)} \delta_{(a)r}\, \rightarrow \,
\delta^{\mu}_{(\mu )} b^{(\mu )}_{(\infty )s} \delta^s_{(a)}\delta_{(a)r}=
\delta^{\mu}_{(\mu )} b^{(\mu )}_{(\infty ) r},\nonumber \\
 &&{}\nonumber \\
 {\hat b}^{\mu}_r&=& b^{\mu}_r,\nonumber \\
 {\hat b}^l_{\mu}&=& l_{\mu} =(-\epsilon +n) b^{\tau}_{\mu}=(-\epsilon +n)
 \partial_{\mu} \tau (z),\nonumber \\
 {\hat b}^r_{\mu}&=& b^r_{\mu} +n^r b^{\tau}_{\mu},
\label{XII7}
\end{eqnarray}

\noindent we get the following expression for the embedding

\begin{eqnarray}
z^{\mu}_{(WSW)}(\tau ,\vec \sigma)&=& \delta^{\mu}_{(\mu )} x^{(\mu
)}_{(\infty )}(0)+ l^{\mu}(\tau ,\vec \sigma ) \tau
+\epsilon^{\mu}_r(\tau ,\vec \sigma ) \sigma^r =\nonumber \\
 &=&x^{\mu}_{(\infty )}(0)+ l^{\mu}(\tau ,\vec \sigma ) \tau + b^{\mu}_s(\tau ,\vec \sigma )
 \, {}^3e^{(WSW)}{}^s_{(a)}(\tau ,\vec \sigma ) \delta_{(a)r} \sigma^r=\nonumber \\
 &=& x^{\mu}_{(\infty )}(0) +b^{\mu}_A(\tau ,\vec \sigma ) F^A(\tau ,\vec \sigma ),\nonumber \\
  &&{}\nonumber \\
  &&F^{\tau}(\tau ,\vec \sigma )= {{\tau}\over {-\epsilon +n(\tau ,\vec \sigma )}},
  \nonumber \\
  && F^s(\tau ,\vec \sigma )={}^3e^{(WSW)}{}^s_{(a)}(\tau ,\vec \sigma ) \delta
  _{(a)r} \sigma^r -{{n^s(\tau ,\vec \sigma )}\over {-\epsilon +n(\tau ,\vec \sigma )}} \tau ,
\label{XII8}
\end{eqnarray}

\noindent
with $x^{(\mu )}_{(\infty )}(0)$ arbitrary \footnote{It reflects the
arbitrariness of the absolute location of the origin of asymptotic
coordinates (and, therefore, also of the {\it external} center of mass
${\tilde x}^{(\mu )}_{(\infty )}(0)$) near spatial infinity.}. See Ref.
\cite{bai} and its interpretation of the center of mass in general
relativity (this paper contains the main references on the problem
starting from Dixon's definition\cite{dixon}): $x^{(\mu )}_{(\infty
)}(\tau )$ may be interpreted as the arbitrary {\it reference} (or
{\it central}) timelike worldline of this paper.

From Eqs.(\ref{XII8}) we can find the equations for determining the
transition coefficients $b^{\mu}_A(\tau ,\vec \sigma )={{\partial
z^{\mu}_{(WSW)}(\tau ,\vec \sigma )}\over {\partial \sigma^A}}$ and
therefore the coordinate transformation $x^{\mu} \mapsto \sigma^A$
from general 4-coordinates to adapted 4-coordinates

\begin{eqnarray}
b^{\mu}_A &=& {{\partial z^{\mu}_{(WSW)}}\over {\partial \sigma^A}}=
b^{\mu}_B {{\partial F^B}\over {\partial \sigma^A}} + {{\partial
b^{\mu}_B}\over {\partial \sigma^A}} F^B,\nonumber \\
 &&{}\nonumber \\
 &&A_A{}^B = \delta_A^B - {{\partial F^B}\over {\partial \sigma^A}},\nonumber \\
 &&{}\nonumber \\
 F^B {{\partial b^{\mu}_B}\over {\partial \sigma^A}} &=& A_A{}^B b^{\mu}_B,\nonumber \\
 &&or\quad b^{\mu}_b = (A^{-1})_B{}^a F^C {{\partial b^{\mu}_C}\over {\partial \sigma^A}}.
\label{XII9}
\end{eqnarray}

The coordinates $\sigma^A$, for instance the special 3-orthogonal coordinates,
for the 3+1 splitting of $M^4$ with leaves $\Sigma^{(WSW)}_{\tau}$ replace
the {\it standard PN coordinates} ($x^{\mu}(o)$ is the arbitrary origin)
and should tend to them in the Post-Newtonian approximation.

Moreover, from the equation $\partial_{\mu}
\tau (z)=l_{\mu}(z)/[-\epsilon +n(z)]$ we could determine the function 
$\tau (z)$ associated
with this class of globally hyperbolic spacetimes. The WSW
hypersurface $\Sigma^{(WSW)}_{\tau}$ associated with the given
solution is the set of points $z^{\mu}(\tau ,\vec
\sigma )$ such that $\tau (z) = \tau$.

In conclusion it turns out that with WSW Minkowski-compatible
foliations with spacelike hypersurfaces $\Sigma^{(WSW)}_{\tau}$ ,
preferred adapted tetrads and cotetrads are associated. Therefore,
there are {\it preferred geometrical observers} associated with the
leaves $\Sigma^{(WSW)}_{\tau}$ of a WSW foliation, which are
determined by both the intrinsic and extrinsic (${}^3K$) geometry of
these $\Sigma_{\tau}^{(WSW)}$'s.

Therefore, there are {\it preferred ADM Eulerian observers}

\beq
{}^4_{(\Sigma )}{\check E}^{(WSW)}{}^{\mu}_{(\alpha )}=\Big( l^{\mu};
b^{\mu}_s\, {}^3e^{(WSW)}{}^s_{(a)} \Big)
\left[ {}^4_{(\Sigma )}{\check{\tilde E}}^A_{(\alpha )}
= \Big( {1\over {-\epsilon +n}} (1; -n^r) ;
(0; {}^3e^{(WSW)}{}^r_{(a)}) \Big) \right] .
\label{XII10}
\eeq

They should be used as
{\it conventional celestial reference system (CCRS)} $S_I$ based on an
extragalactic radio-source catalogue system \cite{frame}
: this is a conventional definition of inertiality with respect to
rotations \footnote{The tabulated {\it right ascensions} and 
{\it declinations} and, in the case of a star catalogue, the 
{\it proper motions (ephemerides)}
define the reference axes of CCRS. The axes are chosen in such a way
that at a basic epoch  they coincide in optimal
approximation with the {\it mean equatorial frame} defined by the {\it mean
celestial pole} and the {\it mean dynamical equinox}; these are
non-relativistic definitions which can be applied to the asymptotic
triads; in the relativistic case one considers the {\it proper reference
frame of a single observer}, represented as a tetrad propagated along
the worldline of the observer by Fermi-Walker transport: the time axis
of the tetrad is the timelike worldline of the observer, while the
three space axes are spacelike geodesics (Fermi normal coordinates).}.

\vfill\eject

\section{Conclusions.}

In this paper it has been shown that with suitable boundary conditions at 
spatial infinity compatible with Christodoulou-Klainermann spacetimes and 
following Dirac's ideas on asymptotically flat metric gravity, it is possible
to define the {\it rest-frame  instant form} of 
dynamics also for metric gravity and not only for parametrized Minkowski 
theories. In particular,
it turns out that in this approach there are {\it dynamical preferred timelike
accelerated observers} tending to inertial observers at spatial infinity
(the {\it fixed stars}).
A clarification on the interpretation of Hamiltonian gauge transformations
and of observables in metric gravity has been given.

This Hamiltonian  approach, oriented towards the canonical reduction to the
physical degrees of freedom of the gravitational field, violates the
geometrical structure of general relativity breaking general covariance 
\footnote{But in a  way associated with the privileged presymplectic Darboux 
bases naturally selected by the Shanmugadhasan canonical transformations.}.
It avoids the {\it spacetime problem} with the choice of the privileged
WSW foliations and  it allows
the deparametrization of general relativity and a soldering with parametrized
Minkowski theories (and parametrized Newton theories for $c\, \rightarrow
\infty$) and to make contact with the kinematical framework, which will be
used\cite{india,bari} to find the Tomonaga-Schwinger asymptotic states
needed for relativistic bound states (the Fock asymptotic states have
no control on the relative times of the asymptotic particles). The
problem whether general covariance may be recovered at the quantum
level has to be attacked only after having seen if this minimal
quantization program can work.

What is still lacking is the explicit construction of a quasi-Shanmugadhasan
canonical transformation and of a canonical basis of Dirac's observables.
This task will be faced starting from tetrad gravity and then deducing the
results for metric gravity. Tetrad gravity is to be preferred for the
following reasons: i) the  configuration variables describe accelerated 
timelike observers, namely the tetrads carry the same type of information of
the embeddings in parametrized Minkowski theories; ii) it couples
naturally to fermions; iii) the supermomentum constraints may be replaced
with SO(3) Yang-Mills Gauss laws, which are easier to be solved.

Since in the Dirac-Bergmann canonical reduction of metric gravity spin
networks do not show up (but they could be hidden in the non-tensorial
character of the Dirac observables $r_{\bar a}$, $\pi_{\bar a}$ still
to be explored), it is not clear which could be the overlap with
Ashtekar-Rovelli-Smolin program \cite{ashte} for spatially either compact
or non-compact
spacetimes, which is generally covariant but only after having fixed
the lapse and shift functions (so that it is not clear how one can
rebuild the spacetime from the 3-geometries) and replaces local
variables of the type $r_{\bar a}(\tau ,\vec \sigma )$ with global
holonomies of the 3-spin connection over closed 3-loops.

 Let us now make some comments on the quantization of tetrad gravity
in this scheme in which general covariance is completely broken having
completely fixed all the gauges. See Ref.\cite{gr13} for an updated
discussion of quantization problems in canonical gravity (and
Ref.\cite{ku1} for the quantization of parametrized theories).

The quantization of the rest-frame instant form of metric gravity in a
completely fixed gauge like, after a quasi-Shanmugadhasan canonical
transformation, the 3-orthogonal gauge with the natural gauge fixing 
$\pi_{\phi}(\tau ,\vec \sigma ) \approx 0$ by using the mathematical time 
parameter $T_{(\infty )}\equiv \tau$
(the rest-frame time of the {\it external} decoupled point particle clock) on 
the Wigner-Sen-Witten hypersurfaces should be done with the following steps:
\hfill\break
a) Assume to have found either the exact or an approximate solution of the
classical reduced Lichnerowicz equation $\phi = \phi (r_{\bar a},\pi_{\bar a})$
and to have evaluated the associated weak ADM 4-momentum ${\hat P}^A_{ADM,R}=
{\hat P}^A_{ADM,R}[r_{\bar a},\pi_{\bar a},\phi (r_{\bar a},\pi_{\bar a})]$.
\hfill\break
b) On each WSW hypersurface $\Sigma^{(WSW)}_{\tau} \approx R^3$ replace the
Hamiltonian gravitational field physical degrees of freedom $r_{\bar a}(\tau
,\vec \sigma )$, $\pi_{\bar a}(\tau ,\vec \sigma )$ with operators ${\hat r}
_{\bar a}(\tau ,\vec \sigma )=r_{\bar a}(\tau ,\vec \sigma )$, ${\hat \pi}
_{\bar a}(\tau ,\vec \sigma )=i{{\delta}\over {\delta r_{\bar a}(\tau ,\vec
\sigma )}}$ (Schroedinger representation) on some Hilbert space.\hfill\break
c) Write the functional Schroedinger wave equation

\begin{equation}
i {{\partial}\over {\partial \tau}} \Psi (\tau ,\vec \sigma |r_{\bar a}] =
{\hat P}^{(op)\, \tau}_{ADM,R}[r_{\bar a},{\hat \pi}_{\bar a},\phi (r_{\bar a},
{\hat \pi}_{\bar a})] \Psi (\tau ,\vec \sigma |r_{\bar a}],
\label{XIII1}
\end{equation}

\noindent plus the 3 conditions defining the rest frame

\begin{equation}
{\hat P}^{(op)\, r}_{ADM,R}[r_{\bar a},{\hat \pi}_{\bar a},\phi (r_{\bar a},
{\hat \pi}_{\bar a})] \Psi (\tau ,\vec \sigma |r_{\bar a}] =0,
\label{XIII2}
\end{equation}

\noindent after having chosen (if possible!) an ordering such that $[ {\hat P}
^{(op)\, A}_{ADM,R}, {\hat P}^{(op)\, B}_{ADM,R} ] =0$.
Let us remark that at this
stage it could be useful the suggestion of Ref.\cite{dc13} that the
unphysical space of these functionals does not need to be a Hilbert space and
that, in it, the observables need not to be self-adjoint operators (these
properties must hold only in the physical space with the physical scalar
product). This Schroedinger equation has not
an {\it internal Schroedinger interpretation} since neither {\it 
Misner internal intrinsic time} nor {\it York internal extrinsic time} nor 
any function like the {\it Komar-Bergmann individuating fields} are the time: 
it does not use the superhamiltonian constraint (like the Wheeler-De Witt 
equation) but the derived weak ADM energy.

The scalar product associated with this Schroedinger equation defines the
Hilbert space and the operators ${\hat P}^{(op)A}_{ADM,R}$
should be self-adjoint
with respect to it. Since there are the 3 conditions coming from the 3 first
class constraints defining the rest frame, the physical Hilbert space of the
wave functionals $\Psi_{phys}$ solution of Eq.(\ref{XIII2}) will have an 
induced
physical scalar product which depends on the functional form of the constraints
${\hat P}^r_{ADM,R}\approx 0$ as it can be shown explicitly in
finite-dimensional examples \cite{lll,longhi}, so that it is not given by a
system-independent rule.

Another possibility is to add and quantize also the gauge fixings ${\vec
\sigma}_{ADM}\approx 0$. In this case one could impose the second class
constraints in the form
$< \Psi |\sigma^{(op)r}_{ADM}| \Psi > =0$, $< \Psi | {\hat P}^{(op)r}_{ADM,R}
| \Psi > =0$ and look whether it is possible to define a Gupta-Bleuler
procedure.

The best would be to be able to find the
canonical transformation $r_{\bar a}(\tau ,\vec \sigma )$, $\pi_{\bar a}(\tau
,\vec \sigma )$ $\mapsto$ ${\vec \sigma}_{ADM}$, ${\hat {\vec P}}_{ADM,R}$,
$R_{\bar a}(\tau ,\vec \sigma )$, $\Pi_{\bar a}(\tau ,\vec \sigma )$
[$R_{\bar a}$, $\Pi_{\bar a}$ being relative variables], since in this case
we would quantize only the final relative variables:

\begin{eqnarray}
\Psi_{phys}&=& \tilde \Psi (\tau ,\vec \sigma |R_{\bar a}],\nonumber \\
i{{\partial}\over {\partial \tau}} \Psi_{phys}&=& {\hat E}^{(op)}_{ADM}[R_
{\bar a},{\hat \Pi}_{\bar a}=i{{\delta}\over {\delta R_{\bar a}}}] \Psi_{phys},
\nonumber \\
&&with\quad {\hat E}_{ADM}={\hat P}^{\tau}_{ADM,R}[r_{\bar a},\pi_{\bar a},\phi
(r_{\bar a},\pi_{\bar a})] {|}_{{\vec \sigma}_{ADM}={\hat {\vec P}}_{ADM,R}=0}.
\label{XIII3}
\end{eqnarray}

Let us remark that many aspects of the problem of time in quantum gravity
\cite{kuchar1} would be avoided: i) there would be no {\it multiple choice
problem} since there is only one mathematical
time variable $T_{(\infty )}=\tau$;
ii) the problem of {\it functional evolution} would be reduced to find an
ordering such that $[ {\hat P}^{(op)\, A}_{ADM,R},
{\hat P}^{(op)\, B}_{ADM,R} ] =0$
; iii) the {\it Hilbert space problem} is not there because we do not have the
Wheeler-De Witt equation but an ordinary Schroedinger equation; iv)
there is a physical ultraviolet cutoff (the M\o ller radius) like in
parametrized Minkowski theories which could help in regularization
problems.

Naturally, general covariance is completely broken and everything is
defined only on the Wigner-Sen-Witten foliation associated with the
natural gauge fixing $\pi_{\phi}(\tau ,\vec \sigma )\approx 0$. If we would
do the same quantization procedure in 3-normal coordinates on their
WSW hypersurfaces associated with the corresponding natural gauge
fixing $\pi_{\phi \, normal}(\tau ,\vec \sigma )
\approx 0$, we would get a different physical Hilbert space whose being
unitarily equivalent to the one in 3-orthogonal coordinates is a completely
open problem.

However, Refs.\cite{tova} point towards the possible existence of a generic 
obstruction to the quantization of field theory formulated on arbitrary
spacelike hypersurfaces like in the Tomonoga-Schwinger point of view:
if the initial and final Cauchy hypersurfaces are not isometric, the quantum
evolution cannot be implemented in a unitary way. Therefore, notwithstanding
the possibility of having consistent quantizations for each 3+1 splitting
of spacetime, the quantization associated with different 3+1 splittings may
be inequivalent. There could be a {\it generalized Unruh effect} connected
with the transition from a congruence of timelike accelerated observers to
another one both in flat and curved spacetimes.

If this quantization can be done,  the completely gauge-fixed 4-metric
${}^4g_{AB}$ on the mathematical manifold $M^4$ would become an
operator ${}^4{\hat g}_{AB}(\tau ,\vec \sigma |r_{\bar a},{\hat
\pi}_{\bar a}]$ with the implication of a quantization of the Dirac
observables associated with 3-volumes (the volume element Dirac
observable is the solution $\phi$ of the reduced Lichnerowicz equation
for $\pi_{\phi}=0$), 2-areas and lengths. Let us remark that these
quantities would not a priori commute among themselves: already at the
classical level there is no reason that they should have vanishing
Dirac brackets (however, two quantities with compact disjoint supports
relatively spacelike would have vanishing Dirac brackets).

If the quantization can be made meaningful, the quantum Komar-Bergmann
individuating fields would lead to a quantization of the {\it physical
coordinates} for the spacetime $M^4$. This will give a quantum spacetime
connected with {\it non commutative geometry} approaches.

Let us also remark that instead of using a solution of the classical reduced
Lichnerowicz equation with $\pi_{\phi}(\tau ,\vec \sigma )=0$, one could use
weak ADM 4-momentum ${\hat P}^{(op) A}_{ADM,R}[r_{\bar a},{\hat \pi}_{\bar a},
\phi^{(op)}]$ with $\phi^{(op)}$ an operatorial solution of a quantum
operatorial reduced Lichnerowicz equation (not a quantum constraint
on the states but the quantization of the classical Lichnerowicz equation
with $\pi_{\phi}=0$ after having gone to Dirac brackets).

Finally, let us observe that even if our approach is more complicated
than Ashtekar and string ones, it opens the possibility of a unified
description of the four interactions after having 
coupled  the standard SU(3)xSU(2)xU(1) model to tetrad gravity and how
to make the canonical reduction of the complete theory. 
The problem of which choice to make for the function space
of the fields associated with the four interactions will require to
understand whether the Gribov ambiguity is only a mathematical
obstruction to be avoided \footnote{In metric gravity this would eliminate the
3-isometries and in Yang-Mills theory the gauge symmetries and the
gauge copies.} or whether there is some physics in it \footnote{In this case 
one should learn how to make the canonical reduction in presence of gauge
symmetries, gauge copies and 3-isometries.}.

Even if it is too early to understand whether our approach can be
useful either from a computational point of view (like numerical
gravity) or for the search of exact solutions, we felt the necessity
to revisit the Hamiltonian formulation of metric gravity with its
intrinsic naturalness for the search of the physical degrees of
freedom of any gauge theory and for the formulation of quantization
rules so that one can have a clear idea of the meaning of the gauge
fixings and the possibility to have an insight on the role of the
gauge degrees of freedom in the realm of exact solutions where
traditionally one starts with suitable parametrizations of the line
element $ds^2$ and then uses symmetries to simplify the mathematics.
For instance, when a known solution of Einstein's equation can be
transformed to 3-orthogonal coordinates, it should give informations
on the solutions of the reduced Lichnerowicz equation and on the
associated lapse and shift functions.

ACKNOWLEDGMENTS

I thank Prof.H.Nicolai for his friendly hospitality at he Max Planck Einstein 
Institute, where this work was completed. I thank Prof. M.Pauri 
for clarifying discussions on the interpretational problems, 
Prof.C.Isham, Prof.K.Kuchar, Prof P.Hajicek, Prof.C.Rovelli and 
Prof. M.J.Gotay for constructive criticism at various
stages of this work. I also thank DrR.De Pietri for his help in checking the 
results of some calculations.

\vfill\eject

\appendix

\section{Parametrized Minkowski Theories and the Rest-Frame Instant
         Form of Dynamics.}

Let us review the main aspects of parametrized Minkowski theories,
following Refs.\cite{lus1,crater,india}, where there is a complete
treatment of the isolated system composed by N scalar charged positive
energy particles plus the electromagnetic field, in which the use of 
Grassmann-valued electric charges as a semiclassical approximation allows 
to obtain the regularization of the Coulomb self-energies.

The starting point
was Dirac's\cite{dirac} reformulation of classical field theory on
spacelike hypersurfaces foliating Minkowski spacetime $M^4$. The
foliation is defined by an embedding $R\times
\Sigma \rightarrow M^4$, $(\tau ,\vec
\sigma ) \mapsto z^{(\mu )}(\tau ,\vec \sigma )$ [$(\mu )$ are flat
Cartesian indices], with $\Sigma$ an abstract 3-surface diffeomorphic
to $R^3$ \footnote{It is the classical basis of Tomonaga-Schwinger quantum
field theory.}. In this way one gets a parametrized field theory with a
covariant 3+1 splitting of flat spacetime and already in a form suited
to the transition to general relativity in its ADM canonical
formulation \footnote{See also Ref.\cite{kuchar} , where a theoretical study of
this problem is done in curved spacetimes.}. The price is that one has
to add as new configuration variables  the embeddings $z^{(\mu )}(\tau
,\vec \sigma )$ identifying  the points of the spacelike hypersurface
$\Sigma_{\tau}$ \footnote{Only  the embeddings carry Lorentz indices; the 
scalar parameter $\tau$ labels the leaves of the foliation and $\vec \sigma$
are curvilinear coordinates on $\Sigma_{\tau}$.} and then to define the
fields on $\Sigma_{\tau}$ so that they know  the hypersurface
$\Sigma_{\tau}$ of $\tau$-simultaneity \footnote{For a Klein-Gordon field $\phi
(x)$, this new field is $\tilde \phi (\tau ,\vec \sigma )=\phi (z(\tau
,\vec \sigma ))$: it contains the nonlocal information about the
embedding, namely the associated notion of {\it equal time}.}. 

The notation $\sigma^{A}=(\tau ,\sigma^{\check r})$of Refs.\cite{lus1,crater}
is used. The $z^{(\mu )}_{A}(\sigma )= \partial z^{(\mu )(\sigma )}/\partial 
\sigma^{A}$ are flat cotetrad fields on Minkowski
spacetime [i.e. ${}^4\eta^{(\mu )(\nu )}=z^{(\mu )}_A\, {}^4g^{AB}\,
z^{(\nu )}_B$ with ${}^4g^{AB}$ the inverse of ${}^4g_{AB}$] with the
$z^{(\mu )}_r$'s tangent to $\Sigma_{\tau}$. In metric gravity the
$z^{\mu}_{A}\not= \partial z^{\mu}/\partial
\sigma^{A}$ are not cotetrad fields since the holonomic coordinates
$z^{\mu}(\sigma )$ do not exist.

Then one rewrites the Lagrangian of the given isolated
system in the form required by the coupling to an external
gravitational field, makes the previous 3+1 splitting of Minkowski
spacetime and interprets all the fields of the system as the new
fields on $\Sigma_{\tau}$ (they are Lorentz scalars, having only
surface indices). Instead of considering the 4-metric as describing a
gravitational field (and therefore as an independent field as it is
done in metric gravity, where one adds the Hilbert action to the
action for the matter fields), here one replaces the 4-metric with the
the induced metric $g_{ AB}[z]
=z^{(\mu )}_{A}\eta_{(\mu )(\nu )}z^{(\nu )}_{B}$ on
$\Sigma_{\tau}$ [a functional of $z^{(\mu )}$] and considers the
embedding coordinates $z^{(\mu )}(\tau ,\vec \sigma )$ as independent
fields.

These extra independent fields $z^{(\mu )}(\tau ,\vec \sigma )$ allow to 
associate with each 3+1 splitting two congruences of timelike observers: i)
a non-rotating (surface-forming) one in which the accelerated observers have 
the normal $l^{(\mu )}(\tau ,\vec \sigma )$ to the embedded hypersurface
as unit 4-velocity; ii) a rotating accelerated one in which the unit 
4-velocity is $u^{(\mu )}(\tau ,\vec \sigma ) = z^{(\mu )}_{\tau}(\tau
,\vec \sigma ) / \sqrt{\epsilon \, {}^4g_{\tau\tau}(\tau ,\vec \sigma )}$.
Therefore, the Lagrangian density ${\cal L}(\tau ,\vec \sigma ) z^{(\mu )}, 
matter ]$ describes not only the given matter on arbitrary spacelike
hypersurfaces but also the accelerated timelike observers associated to them.
Let us remark that to have a similar description in general relativity, 
metric gravity has to be replaced with tetrad gravity, with the tetrads 
describing the observers.

The evolution vector is given by $z^{(\mu
)}_{\tau}=N_{[z](flat)}l^{(\mu )}+ N^{\check r}_{[z](flat)}z^{(\mu
)}_{\check r}$, where $l^{(\mu )}(\tau ,\vec \sigma )= \Big( 
{{\epsilon^{(\mu )}{}_{(\alpha )(\beta )(\gamma )} z^{(\alpha )}_{\check 1}
z^{(\beta )}_{\check 2} z^{(\gamma )}_{\check 3}}\over 
{\sqrt{\gamma}}}\Big) (\tau ,\vec \sigma )$  [$\gamma = 
|det\, {}^3g_{\check r\check s}|$] is the normal
to $\Sigma_{\tau}$ in $z^{(\mu )}(\tau ,\vec \sigma )$ and

\bea
N_{[z](flat)}(\tau ,\vec \sigma
)&=&\sqrt{{}^4g_{\tau\tau}-{}^3\gamma^{\check r\check s}\, {}^4g_{\tau
\check r}\, {}^4g_{\tau \check s}}=\sqrt{{}^4g
/{}^3\gamma},\nonumber \\
 N_{[z](flat) \check r}(\tau ,\vec \sigma )&=&{}^3g
_{\check r\check s}(\tau ,\vec \sigma )N^{\check s}_{[z](flat)}(\tau ,\vec
\sigma )={}^4g_{\tau \check r},
\label{a1}
\eea

\noindent are the flat lapse and shift functions
defined through the metric like in metric gravity [here ${}^3g^{\check
r\check u}\, {}^4g_{\check u\check s}=\delta^{\check r}_{\check s}$ and
${}^4g=|det\, {}^4g_{AB}|$];
however, in Minkowski spacetime they are not independent variables but
functionals of $z^{(\mu )}(\tau ,\vec \sigma )$.

{}From this Lagrangian, besides a Lorentz-scalar form of the constraints
of the given system, one gets four extra primary first class constraints
which imply the independence of the description from the choice of the
foliation with spacelike hypersufaces:

\beq
{\cal H}_{(\mu )}(\tau ,\vec
\sigma )=\rho_{(\mu )}(\tau ,\vec \sigma )-l_{(\mu )}(\tau ,\vec \sigma )
T_{system}^{\tau\tau}(\tau ,\vec \sigma )-z_{\check r (\mu )}(\tau ,\vec
\sigma )T_{system}^{\tau \check r}(\tau ,\vec \sigma ) \approx 0,
\label{a2}
\eeq

\noindent
where $T_{system}^{\tau\tau}(\tau ,\vec \sigma )$, $T_{system}^{\tau
\check r}(\tau ,\vec \sigma )$, are the components of the
energy-momentum tensor in the holonomic coordinate system on
$\Sigma_{\tau}$ corresponding to the energy- and momentum-density of
the isolated system. One can check that hese four constraints satisfy an 
Abelian Poisson algebra, $\quad \lbrace {\cal H}_{(\mu )}(\tau ,\vec
\sigma ), {\cal H}_{(\nu )}(\tau ,{\vec \sigma}^{'}) \rbrace
=0$, being solved in 4-momenta $\rho_{(\mu )}(\tau ,\vec \sigma )$
conjugate to the embedding variables $z^{(\mu )}(\tau ,\vec \sigma )$.

The Dirac Hamiltonian is

\beq
H_D = H_{(c)} + \int d^3\sigma
\lambda^{(\mu )}(\tau ,\vec \sigma ){\cal H}_{(\mu )} (\tau ,\vec
\sigma )+ (\mbox{\it system-dependent\, primary\, constraints}),
\label{a3}
\eeq

\noindent
with $\lambda^{(\mu )}(\tau ,\vec \sigma )$ arbitrary Dirac multipliers
[$H_{(c)}$ is the canonical part]. By using ${}^4\eta^{(\mu )(\nu
)}=[l^{(\mu )}l^{(\nu )}- z^{(\mu )}_{\check r}\, {}^3g^{\check
r\check s} z^{(\nu )}_{\check s}](\tau ,
\vec \sigma )$ we can write

\bea
\lambda_{(\mu )}(\tau ,\vec
\sigma ){\cal H}^{(\mu )}(\tau ,\vec \sigma )&=&[(\lambda_{(\mu )}l^{(\mu )})
(l_{(\nu )}{\cal H}^{(\nu )})-(\lambda_{(\mu )}z^{(\mu )}_{\check r})({}^3g
^{\check r\check s} z_{\check s (\nu )}{\cal H}^{(\nu )})](\tau ,\vec 
\sigma )\,
{\buildrel  {def} \over =}\nonumber \\
 &{\buildrel {def} \over =}&\, N_{(flat)}(\tau ,\vec
\sigma ) (l_{(\mu )}{\cal H}^{(\mu )})(\tau ,\vec \sigma )
-N_{(flat) \check r}(\tau ,\vec \sigma ) ({}^3g^{\check r\check s} z_{\check
s (\nu )}{\cal H}^{(\nu )})(\tau ,\vec \sigma ),
\label{a4}
\eea

\noindent with the (nonholonomic form of
the) constraints $(l_{(\mu )}{\cal H}^{(\mu )})(\tau ,\vec \sigma )\approx 0$,
$({}^3g^{\check r\check s} z_{\check s (\mu )} {\cal H}^{(\mu )})(\tau ,\vec
\sigma )\approx 0$, satisfying the universal Dirac algebra
(\ref{II19}). In this way new  flat lapse and shift functions

\begin{eqnarray}
N_{(flat)}(\tau ,\vec \sigma )&=& \lambda_{(\mu )}(\tau ,\vec \sigma )
l^{(\mu )}(\tau ,\vec \sigma ),\nonumber \\
N_{(flat) \check r}(\tau ,\vec \sigma )&=& \lambda_{(\mu )}(\tau ,\vec \sigma )
z^{(\mu )}_{\check r}(\tau ,\vec \sigma ).
\label{a5}
\end{eqnarray}

\noindent have been defined. They have the same content of the arbitrary 
Dirac multipliers
$\lambda_{(\mu )}(\tau ,\vec \sigma )$, namely they multiply primary
first class constraints satisfying the Dirac algebra. In Minkowski
spacetime they are quite distinct from the previous lapse and shift
functions $N_{[z](flat)}$, $N_{[z](flat) \check r}$, defined starting
from the metric. Only with the use of the Hamilton equations $z^{(\mu
)}_{\tau}(\tau ,\vec \sigma )\, {\buildrel \circ \over =}\, \{ z^{(\mu
)}(\tau ,\vec \sigma ), H_D \} = \lambda^{(\mu )}(\tau ,\vec \sigma )$
we get $N_{[z](flat)}\, {\buildrel \circ \over =}\, N_{(flat)}$,
$N_{[z](flat)\check r}\, {\buildrel \circ \over =}\, N_{(flat)\check
r}$.

In ADM metric gravity, where the coordinates $z^{\mu}(\tau ,\vec
\sigma )$ do not exist, the lapse and shift functions defined starting
from the 4-metric are also the coefficient of secondary first class 
constraints satisfying the Dirac algebra without any use of the equation of 
motion in the canonical part (\ref{II18}) of the Hamiltonian.

Therefore, when  arbitrary 3+1 splittings of the spacetime
with arbitrary spacelike hypersurfaces are given, the descriptions of metric 
gravity plus matter and  the parametrized Minkowski
description of the same matter do not seem to follow the same pattern.
However, the situation changes if the allowed 3+1 splittings of
spacetime in ADM metric gravity are restricted to have the leaves
approaching Minkowski spacelike hyperplanes at spatial infinity and if
parametrized Minkowski theories are restricted either to spacelike
hyperplanes or to hypersurfaces tending to spacelike hyperplanes at spatial
infinity.

The restriction of parametrized Minkowski theories to flat hyperplanes
in Minkowski spacetime  is done by adding the gauge-fixings\cite{lus1}

\beq
z^{(\mu )}(\tau ,\vec \sigma )- x^{(\mu )}_s(\tau )-b^{(\mu )}_{\check r}(\tau
)\sigma^{\check r} \approx 0. 
\label{a6}
\eeq

\noindent Here $x^{(\mu )}_s(\tau )$
denotes a point on the hyperplane $\Sigma_{\tau}$ chosen as an
arbitrary origin; the $b^{(\mu )}_{\check r}(\tau )$'s form an
orthonormal triad at $x^{(\mu )}_s(\tau )$ and the $\tau$-independent
normal to the family of spacelike hyperplanes is $l^{(\mu )}=b^{(\mu
)}_{\tau}=\epsilon^{(\mu )}{}
_{(\alpha )(\beta )(\gamma )}b^{(\alpha )}_{\check 1}(\tau )b^{(\beta )}
_{\check 2}(\tau )b^{(\gamma )}_{\check 3}(\tau )$. Each hyperplane is
described by 10 configuration variables, $x^{(\mu )}_s(\tau )$, plus the 6
independent degrees of freedom contained in the triad $b^{(\mu )}_{\check r}
(\tau )$, and by the 10 conjugate momenta: $p^{(\mu )}_s$ and 6 variables 
hidden
in a spin tensor $S^{(\mu )(\nu )}_s$\cite{lus1}. With these 20 canonical
variables it is possible to build 10 Poincar\'e generators ${\bar p}^{(\mu )}_s
=p^{(\mu )}_s$, ${\bar J}^{(\mu )(\nu )}_s=x^{(\mu )}_sp^{(\nu )}_s-x^{(\nu )}
_sp^{(\mu )}_s+S^{(\mu )(\nu )}_s$.

After the restriction to spacelike hyperplanes the piece $\int
d^3\sigma \lambda^{(\mu )}(\tau ,\vec \sigma ) {\cal H}_{(\mu )}(\tau
,\vec \sigma )$ of the Dirac Hamiltonian (\ref{a3}) is reduced to

\beq
{\tilde \lambda}^{(\mu )}(\tau ){\tilde {\cal H}}_{(\mu )}(\tau )
-{1\over 2}{\tilde \lambda}^{(\mu )(\nu )}(\tau ){\tilde {\cal H}}_{(\mu
)(\nu )}(\tau ),
\label{a7}
\eeq

\noindent because the time constancy of the gauge-fixings (\ref{a3}) implies 
$\lambda_{(\mu )}(\tau ,\vec \sigma )={\tilde
\lambda}_{(\mu )}(\tau )+{\tilde \lambda}_{(\mu )(\nu )}(\tau )b^{(\nu )}
_{\check r}(\tau )\sigma^{\check r}$ with ${\tilde \lambda}^{(\mu )}(\tau )=
-{\dot x}
^{(\mu )}_s(\tau )$, ${\tilde \lambda}^{(\mu )(\nu )}(\tau )=-{\tilde \lambda}
^{(\nu )(\mu )}(\tau )={1\over 2}\sum_{\check r}[{\dot b}^{(\mu )}_{\check r}
b^{(\nu )}_{\check r}-b^{(\mu )}_{\check r}{\dot b}^{(\nu )}_{\check r}](\tau 
)$ [$\, \, \, \dot {}$ means $d/d\tau$].
Since at this stage we have $z^{(\mu )}_{\check r}(\tau
,\vec \sigma )\approx b^{(\mu )}_{\check r}(\tau )$, so that $z^{(\mu )}
_{\tau}(\tau ,\vec \sigma )\approx
N_{[z](flat)}(\tau ,\vec \sigma ) l^{(\mu )}(\tau ,\vec
\sigma )+N^{\check r}
_{[z](flat)}(\tau ,\vec \sigma )b^{(\mu )}_{\check r}(\tau ,\vec
\sigma )\approx {\dot x}^{(\mu )}_s(\tau )+{\dot b}^{(\mu )}_{\check r}(\tau )
\sigma^{\check r}=-{\tilde \lambda}^{(\mu )}(\tau )-{\tilde
\lambda}^{(\mu )(\nu )}(\tau )b_{\check r (\nu )}(\tau )\sigma^{\check r}$,
it is only now that we get the coincidence of the two definitions of
flat lapse and shift functions independently from the equations of
motion, i.e.

\beq
N_{[z](flat)}(\tau ,\vec \sigma )\approx N_{(flat)}(\tau ,\vec \sigma ),\quad
\quad N_{[z](flat) \check r}(\tau ,\vec \sigma )\approx N_{(flat)\check 
r}(\tau ,\vec \sigma ).
\label{a8}
\eeq

The description on arbitrary foliations with spacelike hyperplanes is
independent from the choice of the foliation, due to the remaining 10
first class constraints

\bea
{\tilde {\cal H}}^{(\mu )}(\tau )&=&\int d^3\sigma {\cal H}^{(\mu )}
(\tau ,\vec \sigma )=p^{(\mu )}_s - P^{(\mu )}_{sys}=p^{(\mu
)}_s-\nonumber \\
 &-&[total\, momentum\,
of\, the\, system\, inside\, the\, hyperplane]^{(\mu )}\approx
0,\nonumber \\
 &&{}\nonumber \\
 {\tilde {\cal H}}^{(\mu )(\nu )}(\tau )&=&
 b^{(\mu )}_{\check r}(\tau ) \int d^3\sigma \, \sigma
^{\check r}{\cal H}^{(\nu )}(\tau ,\vec \sigma )-b^{(\nu )}_{\check r}(\tau )
\int d^3\sigma \, \sigma^{\check r} {\cal H}^{(\mu )}(\tau ,\vec \sigma )=
S_s^{(\mu )(\nu )} - S_{sys}^{(\mu )(\nu )}=\nonumber \\
 &=&S^{(\mu )(\nu )}_s-[intrinsic\, angular\, momentum\, of\, the\, 
system\nonumber \\
 &&inside\, the\, hyperplane]^{(\mu )(\nu )}=S^{(\mu )(\nu )}_s -\nonumber \\
 &-&(b^{(\mu )}_{\check r}(\tau )l^{(\nu )}-b^{(\nu )}_{\check r}(\tau )
l^{(\mu )}
)[boost\, part\, of\, system's\, angular\, momentum]^{\tau \check
r}-\nonumber  \\
 &-&(b^{(\mu )}_{\check r}(\tau )b^{(\nu )}_{\check
s}(\tau )-b^{(\nu )}_{\check r}(\tau )b^{(\mu )}_{\check s}(\tau
))[spin\, part\, of\, system's\, angular\, momentum]^{\check r\check
s}\approx \nonumber \\
 &\approx& 0.
\label{a9}
\eea

Therefore, on spacelike hyperplanes in Minkowski spacetime we have

\begin{eqnarray}
N_{(flat)}(\tau ,\vec \sigma )&=&\lambda_{(\mu )}(\tau ,\vec \sigma )l^{(\mu )}
(\tau ,\vec \sigma ) \mapsto \nonumber \\
&\mapsto& N_{(flat)}(\tau ,\vec \sigma )=N_{[z](flat)}(\tau ,\vec \sigma )=
\nonumber \\
&&=-{\tilde \lambda}
_{(\mu )}(\tau )l^{(\mu )}-l^{(\mu )}{\tilde \lambda}_{(\mu )(\nu )}(\tau )b
^{(\nu )}_{\check s}(\tau ) \sigma^{\check s},\nonumber \\
N_{(flat)\, \check r}(\tau
,\vec \sigma )&=&\lambda_{(\mu )}(\tau ,\vec \sigma )z^{(\mu )}_{\check r}
(\tau ,\vec \sigma ) \mapsto \nonumber \\
&\mapsto& N_{(flat )}(\tau ,\vec \sigma )=N_{[z](flat)\check r}
(\tau ,\vec \sigma )=\nonumber \\
&&=-{\tilde \lambda}
_{(\mu )}(\tau )b^{(\mu )}_{\check r}(\tau )-b^{(\mu )}_{\check 
r}(\tau ){\tilde
\lambda}_{(\mu )(\nu )}(\tau ) b^{(\nu )}_{\check s}(\tau ) \sigma^{\check s}.
\label{a10}
\end{eqnarray}

This is the main difference from the treatment of parametrized Minkowski
theories given in Refs.\cite{isha}: there, in the phase action (no
configuration action is defined),  one uses $N_{[z](flat)}$, 
$N_{[z](flat)\check r}$ in place of $N_{(flat)}$, $N_{(flat)\check r}$ also 
on arbitrary spacelike hypersurfaces and not only on spacelike hyperplanes.

At this stage the embedding canonical variables $z^{(\mu )}(\tau ,\vec
\sigma )$, $\rho_{(\mu )}(\tau ,\vec \sigma )$ are reduced to:\hfill\break
\hfill\break
i) $x^{(\mu )}_s(\tau ), p^{(\mu )}_s$ [$\{ x^{(\mu )}_s,p^{(\nu )}_s\} =
-{}^4\eta^{(\mu )(\nu )}$], parametrizing the arbitrary origin of the 
coordinates on the
family of spacelike hyperplanes. The four constraints ${\cal H}^{(\mu )}(\tau )
\approx p_s^{(\mu )} -p_{sys}^{(\mu )}\approx0$  say that
$p_s^{(\mu )}$ is determined by the 4-momentum of the isolated system.
\hfill\break
ii) $b^{(\mu )}_A(\tau )$ \footnote{The $b^{(\mu )}_r(\tau )$'s are
three orthogonal spacelike unit vectors generating the fixed
$\tau$-independent timelike unit normal $b^{(\mu )}_{\tau}=l^{(\mu )}$
to the hyperplanes.} and $S^{(\mu )(\nu )}_s=-S^{(\nu )(\mu )}_s$ with
the orthonormality constraints $b^{(\mu )}_A\, {}^4\eta_{(\mu )(\nu )}
b^{(\nu )}_B={}^4\eta_{AB}$. The non-vanishing Dirac brackets
enforcing the orthonormality constraints \cite{hanson,lus1} for the
$b^{(\mu )}_A$'s are

\bea
\{ b^{(\rho )}_A, S^{(\mu )(\nu )}_s \}&=&{}^4\eta^{(\rho )(\mu )} b^{(\nu )}_A
-{}^4\eta^{(\rho )(\nu )} b^{(\mu )}_A,\nonumber \\
 \{ S^{(\mu )(\nu )}_s,S^{(\alpha )(\beta )}_s \} &=& C^{(\mu )(\nu )(\alpha
)(\beta )}_{(\gamma )(\delta )} S^{(\gamma )(\delta )}_s,
\label{a11}
\eea

\noindent with $C^{(\mu )(\nu
)(\alpha )(\beta )}_{(\gamma )(\delta )}$ the structure constants of the
Lorentz algebra. Then one has that $p^{(\mu )}_s$, $J^{(\mu )(\nu )}_s=x
^{(\mu )}_sp^{(\nu )}_s-x^{(\nu )}_sp^{(\mu )}_s+S^{(\mu )(\nu )}_s$, satisfy
the algebra of the Poincar\'e group, with $S^{(\mu )(\nu )}_s$ playing the
role of the spin tensor. The other six constraints $
{\cal H}^{(\mu )(\nu )}(\tau )\approx S^{(\mu )(\nu )}
_s-S^{(\mu )(\nu )}_{sys}\approx 0$ say that
$S_s^{(\mu )(\nu )}$ coincides the spin tensor of the isolated system.

Let us remark that, for each configuration of an isolated system with
timelike total 4-momentum there is a privileged family of hyperplanes
(the {\it Wigner hyperplanes} orthogonal to $p^{(\mu )}_s$, existing when
$\epsilon p^2_s > 0$) corresponding to the intrinsic rest-frame of the
isolated system. If we choose these hyperplanes with suitable gauge
fixings to the constraints ${\tilde {\cal H}}^{(\mu )(\nu )}(\tau
)\approx 0$ \cite{lus1}, we remain with only the four constraints
${\cal H}^{(\mu )}(\tau )\approx 0$, which can be rewritten as

\bea
\sqrt{\epsilon p^2_s} &\approx& [invariant\, mass\, of\, the\,
isolated\, system\, under\, investigation]= M_{sys}; \nonumber \\
 &&{}\nonumber \\
 {\vec p}_{sys}&=&[3-momentum\, of\, the\, isolated\, system\, inside\, the\,
Wigner\, hyperplane]\approx 0.
\label{a12}
\eea

There is no more a restriction on $p_s^{(\mu )}$, because $u^{(\mu
)}_s(p_s)=p^{(\mu )}_s/\sqrt{\epsilon p^2_s}$ gives the orientation of
the Wigner hyperplanes containing the isolated system with respect to
an arbitrary given external observer.

In this special gauge we have $b^{(\mu )}_A\equiv L^{(\mu )}{}_A(p_s,{\buildrel
\circ \over p}_s)$ (the standard Wigner boost for timelike Poincar\'e orbits),
$S_s^{(\mu )(\nu )}\equiv S_{sys}^{(\mu )(\nu )}$, ${\tilde
\lambda}_{(\mu )(\nu )}(\tau )\equiv 0$.

In general, there is the problem that in the gauges where ${\tilde
\lambda}_{(\mu )(\nu )}(\tau )$ are
different from zero the foliations with leaves $\Sigma_{\tau}$
associated to arbitrary 3+1 splittings of Minkowski spacetime are
geometrically ``ill-defined" at spatial infinity so that the
variational principle describing the isolated system could make sense
only for those 3+1 splittings having these part of the Dirac's
multipliers vanishing. The problem is that, since on hyperplanes
${\dot l}^{(\mu )}=0$ and $l^{(\mu )}\, b_{\check r(\mu )}(\tau )=0$
imply $l^{(\mu )} {\dot b}_{\check r(\mu )}(\tau )=0$, then the
analogue of Eqs.(\ref{III5}) implies ${\tilde \lambda}_{\tau \check
r}(\tau )=0$ (i.e. only three ${\tilde \lambda}_{(\mu )(\nu )}(\tau )$
independent) on spacelike hyperplane, because otherwise Lorentz boosts
can create crossing of the leaves of the foliation. This points toward
the necessity of making  the reduction from arbitrary spacelike
hypersurfaces either directly to the Wigner hyperplanes (instead 
the reduction described above is done in two steps) or  to spacelike 
hypersurfaces approaching asymptotically Wigner hyperplanes 
\footnote{Asymptotically we must
fix the gauge freedom generated by the spin part of Lorentz boosts,
see Eq.(\ref{a9}); how this can be done before the restriction to
spacelike hyperplanes has still to be studied.} to avoid
inconsistencies.

Therefore till now, the 3+1 splittings of Minkowski spacetime whose
leaves are Wigner hyperplanes are the only ones for which the
foliation is well defined at spatial infinity (both the induced proper
time interval and shift functions are finite there).

The only remaining canonical variables describing the Wigner
hyperplane are the non-covariant Newton-Wigner-like canonical
{\it external} center-of-mass coordinate ${\tilde x}^{(\mu )}_s(\tau )$
\footnote{It lives on the Wigner hyperplanes; see Eq.(\ref{VIII3})for its
expression.} and $p^{(\mu )}_s$. Now 3 degrees of freedom of the
isolated system, an {\it internal} center-of-mass 3-variable ${\vec
\sigma}_{sys}$ defined inside the Wigner hyperplane and conjugate to
${\vec p}_{sys}$, become gauge variables. The natural gauge fixing is
${\vec \sigma}_{sys}\approx 0$:  in this way the {\it internal} 3-center of
mass is put into the origin $x^{(\mu )}_s(\tau )=z^{(\mu )}(\tau ,\vec
\sigma =0)$ of the Wigner hyperplane, and only the {\it external} 
${\tilde x}^{(\mu )}(\tau )$ remains: it plays the role of a kinematical 
external 4-center of
mass for the isolated system and may be interpreted as a decoupled
observer with his parametrized clock (point particle clock). All the
fields living on the Wigner hyperplane are now either Lorentz scalar
or with their 3-indices transforming under Wigner rotations (induced
by Lorentz transformations in Minkowski spacetime) as any Wigner spin
1 index. Let us remark that the constant $x^{(\mu )}_s(0)$ [and,
therefore, also ${\tilde x}
^{(\mu )}_s(0)$] is arbitrary, reflecting the arbitrariness in the absolute
location of the origin of the {\it internal} coordinates on each hyperplane in
Minkowski spacetime.

One obtains in this way a new kind of instant form of the dynamics,
the  {\it Wigner-covariant 1-time rest-frame instant
form}\cite{lus1,india} with a universal breaking of Lorentz covariance
restricted to ${\tilde x}^{(\mu )}_s$ independently from the given
isolated system. It is the special relativistic generalization of the
non-relativistic separation of the center of mass from the relative
motion [$H={{ {\vec P}^2}\over {2M}}+H_{rel}$]. The role of the center
of mass is taken by the Wigner hyperplane, identified by the point
${\tilde x}^{(\mu )} (\tau )$ and by its normal $p^{(\mu )}_s$.

The invariant mass $M_{sys}$ of the system, which is also the
{\it internal} energy generator of the isolated system, replaces the
non-relativistic Hamiltonian $H_{rel}$ for the relative degrees of
freedom, after the addition of the gauge-fixing $T_s-\tau
\approx 0$ \footnote{It identifies the time parameter $\tau$, labelling the
leaves of the foliation,  with the Lorentz scalar time of the center
of mass in the rest frame, $T_s=p_s\cdot {\tilde x}_s/M_{sys}$;
$M_{sys}$  generates the evolution in this time.}: it happens like with
the frozen Hamilton-Jacobi theory, in which the time evolution can be
reintroduced by using the energy generator of the Poincar\'e group as
Hamiltonian.

After the gauge fixings $T_s-\tau \approx 0$, the embedding of the
Wigner hyperplane into Minkowski spacetime is

\beq
z^{(\mu )}(\tau ,\vec \sigma ) = x^{(\mu )}_s(\tau ) + \epsilon^{(\mu
)}_r(u(p_s)) \sigma^r = x^{(\mu )}_s(0) + u^{(\mu )}(p_s) \tau +
\epsilon^{(\mu )}_r(u(p_s)) \sigma^r,
\label{a13}
\eeq

\noindent where $x^{(\mu )}_s(0)$ is an arbitrary point and $\epsilon^{(\mu
)}_r(u(p_s)) =L^{(\mu )}{}_r(p_s,{\buildrel \circ \over p}_s)$.

Finally, when fields are present, to identify  the natural gauge-fixings
to eliminate the three 1st class constraints ${\vec p}_{sys}\approx 0$, 
one needs to find a rest-frame canonical basis containing the {\it internal}
3-center-of-mass [${\vec  \sigma}_{sys}$] and relative variables for
fields (in analogy to particles). A basis with a {\it center of phase}
has already been found for a real Klein-Gordon field both in the
covariant approach\cite{lon} and on spacelike hypersurfaces
\cite{mate}. In this case also the {\it internal} center of mass has been
found, but not yet a canonical basis containing it.

The determination of ${\vec \sigma}_{sys}$ may be done with the group
theoretical methods of Ref.\cite{pauri}: given a realization on the
phase space of a given system of the ten Poincar\'e generators one can
build three 3-position variables only in terms of them, which in our
case of a system on the Wigner hyperplane with ${\vec p}_{sys}\approx
0$ are: i) a canonical 3-center of mass (the {\it internal} center of
mass ${\vec \sigma}_{sys}$); ii) a non-canonical {\it internal} M\o ller 
3-center of energy ${\vec \sigma}^{(E)}_{sys}$; iii) a non-canonical 
{\it internal} Fokker-Price 3-center of inertia ${\vec \sigma}^{(FP)}_{sys}$. 
Due to ${\vec p}_{sys}\approx 0$, we have ${\vec \sigma}_{sys} \approx {\vec
\sigma}^{(FP)}_{sys} \approx {\vec \sigma}^{(E)}_{sys} =
\{ boost\,\, generator / energy \}$. By adding the
gauge fixings ${\vec \sigma}_{sys}\approx 0$ one can show that the
origin $x_s^{(\mu )}(\tau )$ becomes  simultaneously the Dixon center
of mass of an extended object and both the Pirani and Tulczyjew
centroids \footnote{See Ref. \cite{mate} for the application of these methods
to find the center of mass of a configuration of the Klein-Gordon
field after the preliminary work of Ref.\cite{lon}.}. With similar
methods,  see also Refs.\cite{crater,iten},one can construct three
{\it external} collective 4-positions (all located on the Wigner
hyperplane): i) the {\it external} canonical non-covariant center of mass
${\tilde x}_s^{(\mu )}$; ii) the {\it external} non-canonical and
non-covariant M\o ller center of energy $R^{(\mu )}_s$; iii) the
{\it external} covariant non-canonical Fokker-Price center of inertia
$Y^{(\mu )}_s$ (when there are the gauge fixings ${\vec
\sigma}_{sys}\approx 0$ it  coincides with the origin $x^{(\mu
)}_s$). It turns out that the Wigner hyperplane is the natural setting
for the study of the Dixon multipoles of extended relativistic
systems\cite{dixon} and for defining the canonical relative variables
with respect to the center of mass. The Wigner hyperplane with its
natural Euclidean metric structure offers a natural solution to the
problem of boost for lattice gauge theories and realizes explicitly
the Machian aspect of dynamics that only relative motions are
relevant.

In the rest-frame instant form there are two realizations of the Poincar\'e 
algebra: i) a degenerate {\it internal} one with generators $M_{sys}$, 
${\vec p}_{sys}\approx 0$, ${\vec S}_{sys}$, ${\vec K}_{sys}$ \footnote{One 
can take ${\vec \sigma}_{sys}=-{{{\vec K}_{sys}}\over {M_{sys}}}$ as gauge
{\it internal} 3-center of mass.}; ii) an {\it external} one with generators 
$p_s^{\mu}$, $J_s^{ij}={\tilde x}^i_sp^j_s-{\tilde x}^j_sp^i_s + S^{ij}_{sys}$,
$J^{oi}_s={\tilde x}^o p^i_s - {\tilde x}^i_s \sqrt{M^2_{sys}+{\vec p}_s^2} -
{{S^{ij} p^j_s}\over {M_{sys}+\sqrt{M^2_{sys}+{\vec p}^2_s}}}$ (it is 
independent from the boosts $S^{oi}_{sys}$).

\vfill\eject


\begin{references}


\bibitem{dirac}P.A.M.Dirac, Can.J.Math. {\bf 2}, 129 (1950); "Lectures on
Quantum Mechanics", Belfer Graduate School of Science, Monographs
Series (Yeshiva University, New York, N.Y., 1964).
\bibitem{ber}J.L.Anderson and P.G.Bergmann, Phys.Rev. {\bf 83}, 1018 (1951).
P.G.Bergmann and J.Goldberg, Phys.Rev. {\bf 98}, 531 (1955).
\bibitem{lu1}L.Lusanna, a) Phys.Rep. {\bf 185}, 1 (1990).
 b) Riv. Nuovo Cimento {\bf 14}, n.3, 1 (1991).
 c) J.Math.Phys. {\bf 31}, 428 and  2126 (1990).
 d) Int.J.Mod.Phys. {\bf A8}, 4193 (1993). e) Comtemp.Math. {\bf 132}, 531 
(1992). f) M.Chaichian, D.Louis Martinez and L.Lusanna, Ann.Phys.(N.Y.)
{\bf 232}, 40 (1994).
\bibitem{haux}M.Henneaux, Phys.Rep. {\bf 126}, 1 (1985). M.Henneaux and
C.Teitelboim, ``Quantization of Gauge Systems" (Princeton University
Press, Princeton, 1992).
\bibitem{india}L.Lusanna, ``Towards a Unified Description of the Four
Interactions in Terms of Dirac-Bergmann Observables", invited
contribution to the book ``Quantum Field Theory: a 20th Century Profile'', 
of the Indian National Science Academy, ed.A.N.Mitra, foreward F.J.Dyson
(Hindustan Book Agency, New Delhi, 2000) (HEP-TH/9907081). ``Tetrad
Gravity and Dirac's Observables", talk given at the Conf. ``Constraint
Dynamics and Quantum Gravity 99", Villasimius 1999
(GR-QC/9912091).``The Rest-Frame Instant Form of Dynamics and Dirac's
Observables", talk given at the Int.Workshop``Physical Variables in
Gauge Theories", Dubna 1999.
\bibitem{re}L.Lusanna, ``Solving Gauss' Laws and Searching Dirac Observables
for the Four Interactions", talk at the ``Second Conf. on Constrained
Dynamics and Quantum Gravity", S.Margherita Ligure 1996, eds. V.De
Alfaro, J.E.Nelson, G.Bandelloni, A.Blasi, M.Cavagli\`a and
A.T.Filippov, Nucl.Phys. (Proc.Suppl.) {\bf B57}, 13 (1997)
(HEP-TH/9702114). ``Unified Description and Canonical Reduction to
Dirac's Observables of the Four Interactions", talk at the
Int.Workshop ``New non Perturbative Methods and Quantization on the
Light Cone', Les Houches School 1997, eds. P.Grang\'e, H.C.Pauli,
A.Neveu, S.Pinsky and A.Werner (Springer, Berlin, 1998)
(HEP-TH/9705154). ``The Pseudoclassical Relativistic Quark Model in
the Rest-Frame Wigner-Covariant Gauge", talk at the Euroconference
QCD97, ed. S.Narison, Montpellier 1997, Nucl.Phys. (Proc. Suppl.) {\bf
B64}, 306 (1998).
\bibitem{mol}C.M\o ller, Ann.Inst.H.Poincar\'e {\bf 11}, 251 (1949); ``The
Theory of Relativity" (Oxford Univ.Press, Oxford, 1957).
\bibitem{dira2}P.A.M.Dirac, {\it Rev.Mod.Phys.} {\bf 21} (1949) 392.
\bibitem{lus1}L.Lusanna, Int.J.Mod.Phys. {\bf A12}, 645 (1997).
\bibitem{sha}S.Shanmugadhasan, J.Math.Phys. {\bf 14}, 677 (1973).
L.Lusanna, Int.J.Mod.Phys. {\bf A8}, 4193 (1993). 
\bibitem{lich}A.Lichnerowicz, C.R.Acad.Sci.Paris, Ser. A, {\bf 280}, 523 
(1975).
W.Tulczyiew, Symposia Math. {\bf 14}, 247 (1974). N.Woodhouse, ``Geometric
Quantization" (Clarendon, Oxford, 1980). J.\`Sniatycki, Ann.Inst.
H.Poincar\`e {\bf 20}, 365 (1984). G.Marmo, N.Mukunda and J.Samuel, Riv.Nuovo
Cimento {\bf 6}, 1 (1983). M.J.Bergvelt and E.A.De Kerf, Physica {\bf 139A},
101 and 125 (1986). B.A.Dubrovin, M.Giordano, G.Marmo and A.Simoni,
Int.J.Mod.Phys. {\bf 8}, 4055 (1993).
\bibitem{go}M.J.Gotay, J.M.Nester and G.Hinds, J.Math.Phys. {\bf 19}, 2388
(1978). M.J.Gotay and J.M.Nester, Ann.Inst.Henri Poincar$\grave e$
{\bf A30}, 129 (1979) and {\bf A32}, 1 (1980). M.J.Gotay and J.$\grave
S$niatycki, Commun. Math.Phys. {\bf 82}, 377 (1981). M.J.Gotay,
Proc.Am.Math.Soc. {\bf 84}, 111 (1982); J.Math.Phys. {\bf 27}, 2051
(1986).
\bibitem{kulk}J.A.Schouten and W.V.D.Kulk, ``Pfaff's Problem and Its 
Generalizations'' (Clarendon, Oxford, 1949).
\bibitem{kulk1}S.Lie, ``Theorie der Transformation Gruppe'', Vol.II 
(B.G.Teubner, Leipzig, 1890). A.R.Forsyth, ``Theory of Differential 
Equations'', Vol.V, Ch.IX (Dover, New York, 1959). L.P.Eisenhart,
``Continuous Groups of Transformations'' (Dover, New York, 1961). R.O.Fulp
and J.A.Marlin, Pacific J.Math. {\bf 67}, 373 (1976); Rep.Math.Phys.
{\bf 18}, 295 (1980).
\bibitem{adm}R.Arnowitt, S.Deser and C.W.Misner, Phys.Rev. {\bf 117}, 1595
(1960); in ``Gravitation: an Introduction to Current Research",
ed.L.Witten (Wiley, New York, 1962).
\bibitem{isha}C.J.Isham and K.Kuchar, Ann.Phys.(N.Y.) {\bf 164}, 288 and 316
(1984). K.Kuchar, Found.Phys. {\bf 16}, 193 (1986).
\bibitem{reg}T.Regge and C.Teitelboim, Ann.Phys.(N.Y.) {\bf 88}, 286 (1974).
\bibitem{reg1}R.Beig and \'O Murchadha, Ann.Phys.(N.Y.) {\bf 174}, 463 (1987).
\bibitem{p3}A.Ashtekar and R.O.Hansen, J.Math.Phys. {\bf 19}, 1542 (1978).
A.Ashtekar, ``Asymptotic Structure of the Gravitational Field at
Spatial Infinity", in ``General Relativity and Gravitation", Vol. 2, ed.A.Held
(Plenum, New York, 1980); in ``General Relativity and Gravitation" (GRG10),
eds.B.Bertotti, F.de Felice and A.Pascolini (Reidel, Dordrecht, 1984).
\bibitem{mcca}P.J.McCarthy, J.Math.Phys. {\bf 13}, 1837 (1972); Proc.Roy.Soc.
London {\bf A330}, 517 (1972) and {\bf A333}, 317 (1973); Phys.Rev.Lett. 
{\bf 29}, 817 (1972). P.J.McCarthy and M.Crampin, Proc.Roy.Soc.London 
{\bf A335}, 301 (1973).
\bibitem{p10}J.Winicour, ``Angular Momentum in General Relativity", in 
``General Relativity and Gravitation", vol.2, ed.A.Held (Plenum, New York, 
1980).
\bibitem{wald}R.M.Wald, ``General Relativity" (Chicago Univ.Press, Chicago,
1984).
\bibitem{ckl}D.Christodoulou and S.Klainerman, ``The Global Nonlinear
Stability of the Minkowski Space" (Princeton, Princeton, 1993).
\bibitem{p22}P.A.M.Dirac, Canad.J.Math. {\bf 3}, 1 (1951).
\bibitem{fermi}Y.Choquet-Bruhat, A.Fischer and J.E.Marsden, ``Maximal
Hypersurfaces and Positivity of Mass", LXVII E.Fermi Summer School of
Physics ``Isolated Gravitating Systems in General Relativity", ed.
J.Ehlers (North-Holland, Amsterdam, 1979).
\bibitem{p28}J.Frauendiener,Class.Quantum Grav. {\bf 8}, 1881 (1991).
\bibitem{sen}A.Sen, J.Math.Phys. {\bf 22}, 1781 (1981); Phys.Lett. {119B},
89 (1982).
\bibitem{p27}E.Witten, Commun.Math.Phys. {\bf 80}, 381 (1981).
\bibitem{be}P.G.Bergmann, Rev.Mod.Phys. {\bf 33}, 510 (1961).
\bibitem{rich}H.Friedrich, ``Einstein's equation and geometric asymptotics",
talk at GR15, Pune, GR-QC/9804009; J.Geom.Phys. {\bf 24}, 83 (1998).
\bibitem{p15}H.Friedrich, ``On the Conformal Structure of Gravitational Fields
in the Large", in ``Highlights in Gravitation and Cosmology",
eds.B.R.Iyer, A.Kembhavi, J.V.Narlikar and C.V.Vishveshwara (Cambridge
Univ.Press, Cambridge, 1988); ``Asymptotic Structure of Space-Time",
in ``Recent Advances in General Relativity", eds.A.I.Janis and
J.R.Porter (Birkhauser, Basel, 1992). ``Calculating Asymptotic
Quantities near Spacelike and Null Infinity from Cauchy Data"
(GR-QC/9911103).
\bibitem{frauen}J.Frauendiener, ''Conformal Infinity'',
 Online Journal Living Reviews in Relativity {\bf 3}, n. 4 (2000)
(http://www.livingreviews.org/Articles/Volume3/2000-4frauendiener).
\bibitem{bicak}J.Bic\' ak, ``Radiative Spacetimes: Exact Approaches", in
``Relativistic Gravitation and Gravitational Radiation", Les Houches 1995,
eds. J.A.Marck and J.P.Lasota (Cambridge Univ.Press, Cambridge, 1997).
\bibitem{bk}P.G.Bergmann and A.Komar, Int.J.Theor.Phys. {\bf 5}, 15 (1972).
\bibitem{conf}A.Lichnerowicz, J.Math.Pure Appl. {\bf 23}, 37 (1944).
Y.Faures-Bruhat,\hfill\break C.R.Acad.Sci.Paris {\bf 226}, 1071 (1948);
J.Rat.Mech.Anal. {\bf 5}, 951 (1956); ``The Cauchy Problem" in
``Gravitation: An Introduction to Current Research", ed.L.Witten
(Wiley, New York, 1962).
\bibitem{york} J.W.York jr, Phys.Rev.Lett. {\bf 26}, 1656 (1971); {\bf 28},
1082 (1972). J.Math.Phys. {\bf 13}, 125 (1972); {\bf 14}, 456 (1972).
Ann.Ins.H.Poincar\'e {\bf XXI}, 318 (1974). N.O'Murchadha and J.W.York
jr, J.Math.Phys. {\bf 14}, 1551 (1972). Phys.Rev. {\bf D10}, 428
(1974).
\bibitem{yoyo}J.W.York jr., ``Kinematics and Dynamics of General Relativity",
in ``Sources of Gravitational Radiation", Battelle-Seattle Workshop
1978, ed.L.L.Smarr (Cambridge Univ.Press, Cambridge, 1979). A.Qadir
and J.A.Wheeler, ``York's Cosmic Time Versus Proper Time", in ``{}From
SU(3) to Gravity", Y.Ne'eman's festschrift, eds. E.Gotsma and G.Tauber
(Cambridge Univ.Press, Cambridge, 1985).
\bibitem{ciuf}I.Ciufolini and J.A.Wheeler, ``Gravitation and Inertia" 
(Princeton Univ.Press, Princeton, 1995).
\bibitem{yorkmap}J.Isenberg and J.E.Marsden, J.Geom.Phys. {\bf 1}, 85 (1984).
\bibitem{russo1}L.Lusanna and S.Russo, Tetrad Gravity: I) A New Formulation,
Firenze Univ. preprint (GR-QC/9807072).
\bibitem{russo2}L.Lusanna and S.Russo, Tetrad Gravity: II) Dirac's
Observables, Firenze Univ. preprint (GR-QC/9807073).
\bibitem{russo3}R.De Pietri and L.Lusanna, ``Tetrad Gravity III: Asymptotic
Poincar\'e Charges, the Physical Hamiltonian and Void Spacetimes",
Firenze Univ. preprint 1999 (GR-QC/9909025).
\bibitem{einst}A.Einstein, letter of January 3rd 1916 in `Albert Einstein and
Michele Besso Correspondence 1903-1955', ed. P.Speziali (Hermann,
Paris, 1972); `Relativity and the Problem of Space' in `Relativity:
the Special and General Theory' (Crown, New York, 1961). M.Jammer,
`Concepts of Space' (Harvard Univ.Press, Cambridge, 1954).
\bibitem{stachel}J.Stachel, ``The Meaning of General Covariance", in
``Philosophical Problems of the Internal and External Worlds", Essays
in the Philosophy of A.Gr\"unbaum, eds. J.Earman, A.I.Janis,
G.J.Massey and N.Rescher (Pittsburgh Univ. Press, Pittsburgh, 1993).
``How Einstein Discovered General Relativity: a Historical Tale with
Some Contemporary Morals", in Proc. GR11, ed. M.A.H.MacCallum
(Cambridge Univ.Press, Cambridge, 1987).
\bibitem{rov}C.Rovelli, Class.Quanyum Grav. {\bf 8}, 297 and 317 (1991).
\bibitem{komar}A.Komar, Phys.Rev. {\bf 111}, 1182 (1958). P.G.Bergmann
and A.Komar, Phys.Rev.Lett. {\bf 4}, 432 (1960).
\bibitem{mtw}C.W.Misner, K.S.Thorne and J.A.Wheeler, Gravitation (Freeman,
New York, 1973).
\bibitem{wei}S.Weinberg, ``Gravitation and Cosmology" (J.Wiley, New York, 
1972).
\bibitem{in}J.Isenberg and J.Nester, ``Canonical Gravity", in ``General
Relativity and Gravitation", vol.1, ed.A.Held (Plenum, New York,
1980).
\bibitem{ish}C.J.Isham, ``Canonical Quantum Gravity and the Problem of Time",
in ``Integrable Systems, Quantum Groups and Quantum Field Theories",
eds.L.A.Ibort and M.A.Rodriguez, Salamanca 1993 (Kluwer, London, 1993);
``Conceptual and Geometrical Problems in Quantum Gravity", in ``Recent Aspects
of Quantum Fields", Schladming 1991, eds. H.Mitter and H.Gausterer
(Springer, Berlin, 1991); ``Prima Facie Questions in Quantum Gravity" and
``Canonical Quantum Gravity and the Question of Time", in ``Canonical Gravity:
{}From Classical to Quantum", eds. J.Ehlers and H.Friedrich (Springer, Berlin,
1994).
\bibitem{cho}Y.Choquet-Bruhat and J.W.York jr., ``The Cauchy Problem", in
``General Relativity and Gravitation", vol.1, ed. A.Held (Plenum, New
York, 1980).
\bibitem{rendal}A.D.Rendall, ``Local and Global Existence Theorems for the
Einstein Equations'', Online Journal Living Reviews in Relativity {\bf 1},
n.4 (1998) and {\bf 3}, n.1 (2000),
http://www.livingreviews.org/Articles/Volume3/2000-1rendall (GR-QC/0001008).
\bibitem{hfr}H.Friedrich and A.D.Rendall, ``The Cauchy Problem for Einstein 
Equations'', in ``Einstein's Field Equations and their Physical 
Interpretation'', ed. B.G.Schmidt (Springer, Berlin, 2000).
\bibitem{tei}C.Teitelboim, ``The Hamiltonian Structure of Space-Time", in
``General Relativity and Gravitation", ed.A.Held, Vol.I (Plenum, New
York, 1980).
\bibitem{kuchar}K.Kuchar, J.Math.Phys. {\bf 17}, 777, 792, 801 (1976); 
{\bf 18}, 1589 (1977).
\bibitem{wa}J.Lee and R.M.Wald, J.Math.Phys. {\bf 31}, 725 (1990).
\bibitem{ppons}J.M.Pons and L.Shepley, Class.Quantum Grav. {\bf 12}, 1771
(1995)(GR-QC/9508052); J.M.Pons, D.C.Salisbury and L.C.Shepley, Phys. Rev. 
{\bf D55}, 658 (1997)(GR-QC/9612037).
\bibitem{giap} R.Sugano, Y.Kagraoka and T.Kimura, Int.J.Mod.Phys. {A7}, 61 
(1992).
\bibitem{lusa}L.Lusanna, Int.J.Mod.Phys. {\bf A10}, 3531 and 3675 (1995).
\bibitem{lus3}D.Alba and L.Lusanna,
Int.J.Mod.Phys.{\bf A13}, 3275 (1998) (HEP-TH/9705156).
\bibitem{p14}R.Beig, ``Asymptotic Structure of Isolated Systems", in
``Highlights in Gravitation and Cosmology", eds.B.R.Iyer, A.Kembhavi,
J.V.Narlikar and C.V.Vishveshwara (Cambridge Univ.Press, Cambridge, 1988).
\bibitem{p2}R.Penrose, Phys.Rev.Lett. {\bf 10}, 66 (1963); Proc.Roy.Soc.London
{\bf A284}, 159 (1965).
\bibitem{p0}R.Geroch and G.T.Horowitz, Phys.Rev.Lett. {\bf 40}, 203 (1978).
\hfill\break
R.Geroch and B.C.Xanthopoulous, J.Math.Phys. {\bf 19}, 714 (1978).
\bibitem{p1}R.Geroch, J.Math.Phys. {\bf 13}, 956 (1972); in ``Asymptotic
Structure of Space-Time", eds. P.Esposito and L.Witten (Plenum, New York, 
1976).
\bibitem{p4}P.Sommers, J.Math.Phys. {\bf 19}, 549 (1978).
\bibitem{p5}A.Ashtekar and J.D.Romano, Class.Quantum Grav. {\bf 9}, 1069 
(1992).
\bibitem{p6}P.Cru\'sciel, J.Math.Phys. {\bf 30}, 2094 (1990).
\bibitem{p7}P.G.Bergmann, Phys.Rev. {\bf 124}, 274 (1961). 
A.Ashtekar, Found.Phys. {\bf 15}, 419 (1985).
\bibitem{p16}J.Winicour, ``Radiative Space-Times: Physical Properties and
Parameters", in ``Highlights in Gravitation and Cosmology", eds.B.R.Iyer, 
A.Kembhavi, J.V.Narlikar and C.V.Vishveshwara (Cambridge Univ.Press, 
Cambridge, 1988).
\bibitem{p8}R.Beig and B.G.Schmidt, Commun.Math.Phys. {\bf 87}, 65 (1982).
\hfill\break
R.Beig, Proc.Roy.Soc.London {\bf A391}, 295 (1984).
\bibitem{p9}H.Bondi, Nature {\bf 186}, 535 (1960). \hfill\break
H.Bondi, M.G.van der Burg
and A.W.K.Metzner, Proc.Roy.Soc.London {\bf A269}, 21 (1962).\hfill\break
R.K.Sachs,
Proc.Roy.Soc.London {\bf A264}, 309 (1962) and {\bf A270}, 103 (1962); 
Phys.Rev. {\bf 128}, 2851 (1962).
\bibitem{p11}A.Ashtekar and A.Magnon, J.Math.Phys. {\bf 25}, 2682 (1984).
\bibitem{reg2}L.Andersson, J.Geom.Phys. {\bf 4}, 289 (1987).
\bibitem{p12}N.O'Murchadha, J.Math.Phys. {\bf 27}, 2111 (1986).
\bibitem{chris}D.Christodoulou and N.\'O Murchadha, Commun.Math.Phys. {\bf 80},
271 (1981).
\bibitem{ym}J.Tafel and A.Trautman, J.Math.Phys. {\bf 24}, 1087 (1983).
S.Schlieder, Nuovo Cimento {\bf A63}, 137 (1981). B.D.Bramson,
Proc.Roy.Soc.London {\bf A341}, 463 (1975).
\bibitem{moncr}V.Moncrief, J.Math.Phys. {\bf 20}, 579 (1979). M.Cantor,
Bull.Am.Math.Soc. {\bf 5}, 235 (1981).
\bibitem{hh}S.W.Hawking and G.T.Horowitz, Class.Quantum Grav. {\bf 13}, 1487
(1996).
\bibitem{p13}P.T.Chru\'sciel, Commun.Math.Phys. {\bf 120}, 233 (1988).
\bibitem{reg3}T.Thiemann, Class.Quantum Grav. {\bf 12}, 181 (1995).
\bibitem{reg4}V.O.Solov'ev, Theor.Math.Phys, {\bf 65}, 1240 (1985);
 Sov.J.Part.Nucl.{\bf 19}, 482 (1988).
\bibitem{p18}D.Marolf, Class,Quantum Grav, {\bf 13}, 1871 (1996).
\bibitem{p19}A.Ashtekar, ``New Perspectives in Canonical Gravity" (Bibliopolis,
Napoli, 1988). H.A.Kastrup and T.Thiemann, Nucl.Phys. {\bf B399}, 211
(1993) and {\bf B425}, 665 (1994). K.Kuchar, Phys.Rev. {\bf D50}, 3961
(1994). A.P.Balachandran, L.Chandar and A.Momen, ``Edge states in
canonical gravity", Syracuse Univ. preprint SU-4240-610 1995
(GR-QC/9506006).
\bibitem{p20}J.Barbour, ``General Relativity as a Perfectly Machian Theory",
in ``Mach's Principle: {}From Newton's Bucket to Quantum Gravity", eds.
J.B.Barbour and H.Pfister, Einstein's Studies n.6 (BirkH\"auser,
Boston, 1995).
\bibitem{p21}D.Giulini, C.Kiefer and H.D.Zeh, ``Symmetries, Superselection
Rules and Decoherence", Freiburg Univ. preprint THEP-94/30 1994 (GR-QC
/9410029). J.Hartle, R.Laflamme and D.Marolf, Phys.Rev. {\bf D51}, 7007 (1995).
\bibitem{giul}D.Giulini, Helv.Phys.Acta {\bf 68}, 86 (1995).
\bibitem{witt}B.S.De Witt, Phys.Rev. {\bf 160}, 1113 (1967).
\bibitem{dew}B.S.De Witt, Phys.Rev. {\bf 162}, 1195 (1967); ``The Dynamical
Theory of Groups and Fields" (Gordon and Breach, New York, 1967) and
in ``Relativity, Groups and Topology", Les Houches 1963, eds. C.De
Witt and B.S.De Witt (Gordon and Breach, London, 1964); ``The
Spacetime Approach to Quantum Field Theory", in ``Relativity, Groups
and Topology II", Les Houches 1983, eds. B.S.DeWitt and R.Stora
(North-Holland, Amsterdam, 1984). B.S.De Witt and R.W.Brehme,
Ann.Phys.(N.Y.) {\bf 9}, 220 (1960).
\bibitem{p17}
R.Schoen and S.T.Yau, Phys.Rev.Lett. {\bf 43}, 1457 (1979);
Commun.Math. Phys. {\bf 65}, 45 (1979) and {\bf 79}, 231 (1980).
E.Witten, Commun.Math.Phys. {\bf 80}, 381 (1981). D.M.Brill and
P.S.Jang, ``The Positive Mass Conjecture", in ``General Relativity and
Gravitation", Vol. 1, ed.A.Held (Plenum, New York, 1980).
Y.Choquet-Bruhat, ``Positive Energy Theorems", in ``Relativity, Groups
and Topology II", Les Houches XL 1983, eds. B.S.DeWitt and R.Stora
(North-Holland, Amsterdam, 1984). G.T.Horowitz, ``The Positive Energy
Theorem and its Extensions", in ``Asymptotic Behaviour of Mass and
Spacetime Geometry", ed.F.J.Flaherty, Lecture Notes Phys.202
(Springer, Berlin, 1984). M.J.Perry, ``The Positive Mass Theories and
Black Holes", in ``Asymptotic Behaviour of Mass and Spacetime
Geometry", ed.F.J.Flaherty, Lecture Notes Phys.202 (Springer, Berlin,
1984).
\bibitem{crater}H.Crater and L.Lusanna, ``The Rest-Frame Darwin Potential
from the Lienard-Wiechert Solution in the Radiation Gauge", Firenze
univ. preprint 2000 (HEP-TH/0001046).
\bibitem{iten}D.Alba, L.Lusanna and M.Pauri, ``Dynamical Body Frames, 
Orientation-Shape Variables and Canonical Spin Bases for the Nonrelativistic
N-Body Problem'', Firenze Univ. preprint 2000 (HEP-TH/0011014); ``Center of 
Mass, Rotational Kinematics and Multipolar Expansions for the Relativistic and
Non-Relativistic N-Body Problems in the Rest-Frame Instant Form", in
preparation.
\bibitem{mate}L.Lusanna and M.Materassi, ``The Canonical Decomposition in
Collective and Relative Variables of a Klein-Gordon Field in the
Rest-Frame Wigner-Covariant Instant Form", 
Int.J.Mod.Phys. {\bf A15}, 2821 (2000) (HEP-TH/9904202).
\bibitem{lon}G.Longhi and M.Materassi, ``A Canonical Realization of the BMS
Algebra", J.Math.Phys. {\bf 40}, 480 (1999) (HEP-TH/9803128);
``Collective and Relative Variables for a Classical Klein-Gordon
Field", Int.J.Mod.Phys. {\bf A14}, 3397 (1999) (HEP-TH/9890024).
\bibitem{ll}L.Landau and E.Lifschitz, ``The Classical Theory of Fields"
(Addison-Wesley, Cambridge, 1951).
\bibitem{positive}D.M.Brill and P.S.Jang, ``The Positive Mass Conjecture", in
``General Relativity and Gravitation", Vol. 1, ed.A.Held (Plenum, New
York, 1980).
\bibitem{stephani}H.Stephani, ``General Relativity" (Cambridge Univ.Press,
Cambridge, 1996).
\bibitem{fadde}L.D.Faddeev, Sov.Phys.Usp. {\bf 25}, 130 (1982).
\bibitem{trautm}A.Trautman, in ``Gravitation, an Introduction to Current
Research", ed.L.Witten (Wiley, New York, 1962).
\bibitem{pauri}M.Pauri and M.Prosperi, J.Math.Phys. {\bf 16}, 1503 (1975).
\bibitem{soffel}M.H.Soffel, ``Relativity in Astrometry, Celestial Mechanics
and Geodesy" (Springer, Berlin, 1989).
\bibitem{bini}R.T.Jantzen, P.Carini and D.Bini, Ann.Phys.(N.Y.) {\bf 215}, 1
(1992).
\bibitem{gr13}K.Kuchar, ``Canonical Quantum Gravity" in
``General Relativity and Gravitation" Int.Conf. GR13, Cordoba
(Argentina) 1992, eds. R.J.Gleiser, C.N.Kozameh and O.M.Moreschi (IOP,
Bristol, 1993).
\bibitem{beig}R.Beig, ``The Classical Theory of Canonical General Relativity",
in ``Canonical Gravity: {}From Classical to Quantum", Bad Honnef 1993,
eds. J.Ehlers and H.Friedrich, Lecture Notes Phys. 434 (Springer,
Berlin, 1994).
\bibitem{kku}K.Kuchar, Phys.Rev. {\bf D4}, 955 (1971); J.Math.Phys. {\bf 11},
3322 (1970); {\bf 13}, 768 (1972).
\bibitem{misner}C.W.Misner, Phys.Rev.Lett. {\bf 22}, 1071 (1969); Phys.Rev.
{\bf 186}, 1319 and 1328 (1969).
\bibitem{kuchar1}K.Kuchar, ``Time and Interpretations of Quantum Gravity", in
Proc.4th Canadian Conf. on ``General Relativity and Relativistic
Astrophysics", eds. G.Kunstatter, D.Vincent and J.Williams (World
Scientific, Singapore, 1992).
\bibitem{ku1}K.Kuchar, ``Canonical Methods of Quantization", in ``Quantum
Gravity 2", eds.C.J.Isham, R.Penrose and D.W.Sciama (Clarendon Press,
Oxford, 1981).
\bibitem{dc11}R.F.Baierlein, D.H.Sharp and J.A.Wheeler, Phys.Rev.{\bf 126},
1864 (1962).
\bibitem{paren}R.Parentani, ``The Notions of Time and Evolution in Quantum
Cosmology", 1997, GR-QC/9710130.
\bibitem{kiefer}C.Kiefer, ``The Semiclassical Approximation to Quantum Gravity"
in ``Canonical Gravity - from Classical to Quantum", ed.J.Ehlers
(Springer, Berlin, 1994). ``Semiclassical Gravity and the Problem of
Time", in Proc. Cornelius Lanczos Int.Centenary Conf., eds. M.Chu,
R.Flemmons, D.Brown and D.Ellison (SIAM, 1994). Nucl.Phys. {\bf B475},
339 (1996).
\bibitem{york1}Y.Choquet-Bruhat, J.Isenberg and J.W.York Jr., ``Einstein
Constraints on Asymptotically Euclidean Manifolds", GR-QC/9906095.
A.Anderson, Y.Choquet-Bruhat and J.York Jr., ``Einstein's Equations
and Equivalent Dynamical Systems", GR-Qc/9907099 and ``Curvature-Based
Hyperbolic Systems for General Relativity", talk at the 8th
M.Grossmann Meeting (Jerusalem, Israel, 1997), GR-QC/9802027.
A.Anderson and J.W.York Jr., Phys.Rev.Letters {\bf 81}, 1154 and 4384
(1999). J.W.York Jr., Phys.Rev.Letters {\bf 82}, 1350 (1999).
\bibitem{bafo}R.Bartnik and G.Fodor, Phys.Rev. {\bf D48}, 3596 (1993).
\bibitem{giusa}D.Giulini, J.Math.Phys. {\bf 40}, 2470 (1999).
\bibitem{dc1}L.Lusanna, Nuovo Cimento, {\bf 65B}, 135 (1981).
\bibitem{longhi}G.Longhi and L.Lusanna, Phys.Rev. {\bf D34},
3707 (1986).
\bibitem{i1}J.Isenberg,Phys.Rev.Lett. {\bf 59},2389 (1987).
\bibitem{i2}J.Isenberg, Class.Quantum Grav. {\bf 12}, 2249 (1995).
\bibitem{im}J.Isenberg and V.Moncrief, Class.Quantum Grav. {\bf 13}, 1819
(1996).
\bibitem{brill1}R.Bartnik, Commun.Math.Phys. {\bf 117}, 615 (1988).
D.Brill, in Proc.Third Marcel Grossman Meeting, ed.H.Ning (North-Holland,
Amsterdam,1982).
\bibitem{dd}P.A.M.Dirac, Rev.Mod.Phys. {\bf 21}, 392 (1949).
\bibitem{gaida}R.P.Gaida, Yu.B.Kluchkovsky and V.I.Tretyak, Theor.Math.Phys.
{\bf 55}, 372 (1983); in ``Constraint's Theory and Relativistic
Dynamics", eds. G.Longhi and L.Lusanna (World Scientific, Singapore,
1987).
\bibitem{mo}V.Moncrief, J.Math.Phys. {\bf 16}, 1556 (1965). J.Arms,
A.Fischer and J.E.Marsden, C.R.Acad.Sci.Paris {A281}, 517 (1975).
J.Arms, Acta Phys.Pol. {\bf B17}, 499 (1986) and Contep.Math. {\bf
81}, 99 (1988). J.M.Arms, M.J.Gotay anf G.Jennings, Adv.Math. {\bf
79}, 43 (1990).
\bibitem{whe}J.A.Wheeler, ``Geometrodynamics and the Issue of the Final State",
in ``Relativity, Groups and Topology", Les Houches 1963, eds. B.S.De Witt and
C.De Witt (Gordon and Breach, London, 1964). ``Superspace and the Nature of
Quantum Geometrodynamics", in Battelle Rencontres 1967, eds. C.De Witt and
J.A.Wheeler (Gordon and Breach, New York, 1968).
\bibitem{fis}A.E.Fischer, ``The Theory of Superspace", in ``Relativity", eds.
M.Carmeli, L.Fickler and L.Witten (Plenum, New York, 1970);
Gen.Rel.Grav. {\bf 15}, 1191 (1983); J.Math.Phys. {\bf 27}, 718
(1986). M.Rainer, ``The Moduli Space of Local Homogeneous
3-Geometries", talk at the Pacific Conf. on Gravitation and Cosmology,
Seoul 1996.
\bibitem{ing}S.Timothy Swift, J.Math.Phys. {\bf 33}, 3723 (1992); {\bf 34},
3825 and 3841 (1993).
\bibitem{arms}J.M.Arms, J.E.Marsden and V.Moncrief, Commun.Math.Phys. {\bf 78},
 455 (1981).
\bibitem{mw}R.F.Marzke and J.A.Wheeler, in ``Gravitation and Relativity", eds.
H.Y.Chiu and W.F.Hoffman (Benjamin, New York, 1964).
\bibitem{pirani}J.Ehlers, F.A.E.Pirani and A.Schild, ``The Geometry of
Free-Fall and Light Propagation" in ``General Relativity, Papers in Honor of 
J.L.Synge", ed. L.O'Raifeartaigh (Oxford Univ.Press, London, 1972).
\bibitem{brown}J.D.Brown and K.Kuchar, Phys.Rev. {\bf D51}, 5600 (1995).
\bibitem{wittt}B.S.De Witt, in ``Gravitation", ed. L.Witten (Wiley, New York,
1962).
\bibitem{fmm}
A.E.Fischer and J.E.Marsden, ``The Initial Value Problem and the
Dynamical Formulation of General Relativity", in ``General Relativity. An
Einstein Centenary Survey", eds. S.W.Hawking and W.Israel (Cambridge 
Univ.Press, Cambridge, 1979).  A.E.Fischer, J.E.Marsden and V.Moncrief,
Ann.Inst.H.Poincar\'e {\bf A33}, 147 (1980). J.M.Arms, J.E.Marsden and
V.Moncrief, Ann.Phys.(N.Y.) {\bf 144}, 81 (1982).
\bibitem{monc}V.Moncrief, J.Math.Phys. {\bf 16}, 493 and 1556 (1975); {\bf 17},
1893 (1976). Phys.Rev. {\bf D18}, 983 (1978).
\bibitem{simul}P.Havas, Gen.Rel.Grav. {\bf 19}, 435 (1987).
R.Anderson, I.Vetharaniam and G.E.Stedman, Phys.Rep. {\bf 295}, 93
(1998).
\bibitem{dc8}P.H\'aj\'ichek, J.Math.Phys. {\bf 36}, 4612 (1995); Class.Quantum
Grav. {\bf 13}, 1353 (1996); Nucl.Phys. (Proc.Suppl.) {\bf B57}, 115
(1997). P.H\'aj\'ichek, A.Higuchi and J.Tolar, J.Math.Phys. {\bf 36},
4639 (1995). C.J.Isham and P.H\'aj\'ichek, J.Math.Phys. {\bf 37}, 3505
and 3522 (1996).
\bibitem{torre}C.G.Torre, Phys.Rev. {\bf D48}, R2373 (1993).
\bibitem{but}J.Butterfield and C.J.Isham, ``Spacetime and the Philosophical
Challenge of Quantum Gravity", Imperial College preprint
(GR-QC/9903072).
\bibitem{feng}S.S.Feng and C.G. Huang, Int.J.Theor.Phys. {\bf 36}, 1179 (1997).
\bibitem{gede}J.G\'eh\'eniau and R.Debever, in `Jubilee of Relativity Theory',
eds. A.Mercier and M.Kervaire, Bern 1955, Helvetica Physica Acta Supplementum 
IV (Birkh\"auser, Basel, 1956).
\bibitem{elma}G.F.R.Ellis and D.R.Matravers, Gen.Rel.Grav. {\bf 27}, 777
(1995).
\bibitem{zala}R.Zalaletdinov, R.Tavakol and G.F.R.Ellis, Gen.Rel.Grav. {\bf
28}, 1251 (1996).
\bibitem{ro1}C.Rovelli, Phys.Rev. {\bf D42}, 2638 (1990), {\bf D43}, 442 
(1991)and {\bf D44}, 1339 (1991).
\bibitem{ro2}I.D.Lawrie and R.J.Epp, Phys.Rev. {\bf D53}, 7336 (1996).
\bibitem{thermo}C.Rovelli, Class.Quantum Grav. {\bf 10}, 1549 and 1567 (1993).
A.Connes and C.Rovelli, Class. Quantum Grav. {\bf 11}, 2899 (1994).
\bibitem{brk}J.D.Brown and K.Kuchar, Phys.Rev. {\bf D51}, 5600 (1995).
\bibitem{dc12}J.B.Barbour, Class.Quantum Grav. {\bf 11}, 2853 and 2875 (1994).
\bibitem{dc120}G.M.Clemence, Rev.Mod.Phys. {\bf 29}, 2 (1957).
J.Kovalevski, I.I.Mueller and B.Kolaczek, ``Reference Frames in Astronomy and
Geophysics", pp.355 and 367 (Kluwer, Dordrecht, 1989).
\bibitem{paur}R.DePietri, L.Lusanna and M.Pauri, Class.Quantum Grav. {\bf 12},
219 (1995).
\bibitem{paur1}R.DePietri, L.Lusanna and M.Pauri, Class.Quantum Grav. {\bf 12},
255 (1995).
\bibitem{dc9}C.M.Will, ``Theory and Experiment in Gravitational Physics",
rev.ed. (Cambridge Univ.Press, Cambridge, 1993). S.G.Turyshev,
``Relativistic Navigation: A Theoretical Foundation", NASA/JPL No
96-013 (GR-QC/9606063).
\bibitem{dam}T.Damour, ``Selected Themes in Relativistic Gravity", in
``Relativistic Gravitation and Gravitational Radiation", Les Houches 1995,
eds. J.A.Marck and J.P.Lasota (Cambridge Univ.Press, Cambridge, 1997).
\bibitem{dam1}L.Blanchet and T.Damour, Ann.Inst.H.Poincar\'e {\bf 50}, 377
(1989).L.Blanchet, T.Damour and G.Sch\"afer, Mon.Not.R.Astr.Soc. {\bf
242}, 289 (1990). T.Damour, M.Soffel and C.Xu, Phys.Rev. {\bf D43},
3273 (1991); {\bf D45}, 1017 (1992); {\bf D47}, 3124 (1993); {\bf
D49}, 618 (1994).
\bibitem{dc2}G.Longhi, L.Lusanna and J.M.Pons, J.Math.Phys. {\bf 30}, 1893
(1989).
\bibitem{dc3}L.Lusanna, ``Multitemporal Relativistic Particle Mechanics: a
Gauge Theory Without Gauge-Fixings", in Proc. IV M.Grossmann Meeting,
ed.R.Ruffini, (Elsevier, Amsterdam, 1986).
\bibitem{dc4}W.Unruh and R.Wald, Phys.Rev. {\bf D40}, 2598 (1989).
\bibitem{dc7}J.B.Hartle, Class.Quantum Grav. {\bf 13}, 361 (1996).
\bibitem{dc6}P.H\'aj\'ichek, Phys.Rev. {\bf D44}, 1337 (1991).
W.Unruh, in ``Gravitation: a Banff Summer Institute", eds. R.Mann and
P.Wesson (World Scientific, Singapore, 1991).
\bibitem{nowak}L.Lusanna and D.Nowak-Szczepaniak, ``The Rest-Frame Instant
Form of Relativistic Perfect Fluids with Equation of State $\rho =\rho
(n,s)$ and of Non-Dissipative Elastic Materials", to appear in 
Int.J.Mod.Phys. A (HEP-TH/0003095).
\bibitem{p23}C.M\o ller, Ann.Phys.(N.Y.) {\bf 12}, 118 (1961); in Proc.Int.
School of Physics E.Fermi, Course XX (Academic Press, New York, 1962).
\bibitem{p24}F.A.E.Pirani, ``Gauss' Theorem and Gravitational Energy", in ``Les
Theories Relativistes de la Gravitation", Proc.Int.Conf. at Royaumont 1959,
eds. A.Lichnerowicz and M.A.Tonnelat (CNRS, Paris, 1962).
\bibitem{gold}J.N.Goldberg, Phys.Rev. {\bf D37},2116 (1988).
\bibitem{p25}N.Rosen, Phys.Rev. {\bf 57}, 147 (1940); Ann.Phys.(N.Y.) {\bf 22},
1 (1963); Found.Phys. {\bf 15}, 998 (1986); in ``{}From SU(3) to Gravity",
Y.Ne'eman's festschrift, eds.E.Gotsman and G.Tauber (Cambridge Univ.Press,
Cambridge, 1985); in ``Topological Properties and Global Structure of
Space-Time", eds.P.G.Bergmann and V.de Sabbata (Plenum, New York, 1986).
\bibitem{p26}A.Ashtekar and G.T.Horowitz, J.Math.Phys. {\bf 25}, 1473 (1984).
\bibitem{spinor}A.Sen,  Int.J.Theor.Phys. {\bf 21}, 1 (1982).
P.Sommers, J.Math.Phys. {\bf 21}, 2567 (1980).
\bibitem{spinor1}A.Ashtekar, ``New Perspectives in Canonical Gravity"
(Bibliopolis, Napoli, 1988).
\bibitem{rindler}R.Penrose and W.Rindler, ``Spinors and Space-Time" vol.1 and 2
(Cambridge Univ.Press, Cambridge,  1986).
\bibitem{p29}G.A.J.Sparling, ``Differential Ideals and the Einstein Vacuum
Equations", Pittsburgh Univ. preprint 1983;
``Twistors, Spinors and the Einstein Vacuum
Equations", Pittsburgh Univ. preprint 1984; ``Twistor Theory and the
Characterization of Fefferman Conformal Structures", Pittsburgh Univ.
preprint 1984; ``A Development of the Theory of Classical Supergravity",
Pittsburgh Univ.preprint 1988.
\bibitem{p30}
L.J.Mason and J.Frauendiener, ``The Sparling 3-form, Ashtekar Variables and
Quasi-Local Mass", in ``Twistors in Mathematical Physics", eds.T.Bailey and
R.Baston (Cambridge Univ.Press, Cambridge, 1990).
\bibitem{p35a} J.M.Nester, Phys.Lett. {\bf 83A}, 241 (1981).
W.Israel and J.M.Nester, Phys.Lett. {\bf 85A}, 259 (1981). J.M.Nester,
in ``Asymptotic Behaviour of Mass and Spacetime Geometry",
ed.F.J.Flaherty, Lecture Notes Phys.202 (Springer, Berlin, 1984).
J.M.Nester, Phys.Lett. {\bf 139A}, 112 (1989). J.M.Nester,
Class.Quantum Grav. {\bf 11}, 983 (1994).
\bibitem{he}S.W.Hawking and G.F.R. Ellis, ``The Large Scale Structure of
Spacetime" (Cambridge Univ.Press, Cambridge, 1973).
\bibitem{p35b}J.M.Nester,
Mod.Phys.Lett. {\bf A6}, 2655 (1991). Phys.Lett. {\bf 203A}, 5
(1995); Gen.Rel.Grav. {\bf 27}, 115 (1995).
\bibitem{p31}R.Penrose, Proc.Roy.Soc.London {\bf A381}, 53 (1982).
L.J.Mason, Class.Quantum Grav. {\bf 6}, L7 (1989).
\bibitem{p32}J.Frauendiener, Class.Quantum Grav. {\bf 6}, L237 (1989).
\bibitem{emp}A.Einstein, Sitzungsber.Preuss.Akad.Wiss.Phys.Math.Kl. {\bf 42},
1111 (1916).
\bibitem{empt}J.Goldberg, ``Invariant Transformations, Conservation Laws and
Energy-Momentum", in `General Relativity and Gravitation', ed.A.Held (Plenum,
New York, 1980).
\bibitem{p33}M.Dubois-Violette and J.Madore, Commun.Math.Phys. {\bf 108},
213 (1987).
\bibitem{p34}J.Frauendiener, Gen.Rel.Grav. {\bf 22}, 1423 (1990).
\bibitem{soluz}Y.Choquet-Bruhat and D.Christodoulou, Acta Math. {\bf 146},
129 (1981). O.Reula, J.Math.Phys. {\bf 23}, 810 (1982). O.Reula and
K.Todd, J.Math.Phys. {\bf 25}, 1004 (1984). T.Parker and C.H.Taub,
Commun.Math.Phys. {\bf 84}, 223 (1982). T.Parker, Commun.Math.Phys.
{\bf 100}, 471 (1985). P.Biz\'on and E.Malec, Class.Quantum Grav. {\bf
3}, L123 (1986).
\bibitem{horotod}G.T.Horowitz and P.Tod, Commun.Math.Phys. {\bf 85}, 429 
(1982).
\bibitem{bergq}G.Bergqvist, Class.Quantum Grav. {\bf 11}, 2545 (1994); Phys.
Rev. {\bf D48}, 628 (1993).
\bibitem{horope}G.T.Horowitz and M.J.Perry, Phys.Rev.Lett. {\bf 48}, 371 
(1982).
\bibitem{p35c}
a) J.M.Nester, Class.Quantum Grav. {\bf 5}, 1003 (1988). W.H.Cheng,
D.C.Chern and J.M.Nester, Phys.Rev. {\bf D38}, 2656 (1988). b)
J.M.Nester, J.Math.Phys. {\bf 30}, 624 (1980) and {\bf 33}, 910
(1992). c) J.M.Nester, Int.J.Mod.Phys. {\bf A4}, 1755 (1989).
J.M.Nester, Class.Quantum Grav. {\bf 8}, L19 (1991).
\bibitem{dimakis}A.Dimakis and F.M\"uller-Hoissen, Phys.Lett. {\bf 142A}, 73
(1989).
\bibitem{bai}I.Bailey and W.Israel, Ann.Phys. (N.Y.) {\bf 130}, 188 (1980).
\bibitem{frame} eds. J.Kovalevvsky, I.I.Mueller and B.Kolaczek, ``Reference
Frames in Astronomy and Geophysics" (Kluwer, Dordrecht, 1989).
\bibitem{bari}L.Lusanna, ``The N-body Problem in Tetrad Gravity: a First Step
towards the Unified Description of the Four Interactions", talk given
at the XI Int.Conf. ``Problems of Quantum Field Theory", Dubna 1998,
and at the III W.Fairbank Meeting and I ICRA Network Workshop, ``The
Lense-Thirring Effect", Roma-Pescara 1998 (GR-QC/9810036).
\bibitem{ashte}A.Ashtekar, Phys.Rev.Lett. {\bf 57}, 2244 (1986); ``New
Perspectives in Canonical Gravity" (Bibliopolis, Naples, 1988); ``Lectures on
Non-Perturbative Canonical Gravity" (World Scientific, Singapore, 1991);
``Quantum Mechanics of Riemannian Geometry", 
http://vishnu.nirvana.phys.psu.edu/riem${}_{-}$qm/riem${}_{-}$qm.html.
C.Rovelli and L.Smolin, Nucl.Phys. {\bf B331}, 80 (1990); {\bf B442}, 593
(1995). C.Rovelli, ''Loop Quantum Gravity", Living Reviews in Relativity
http://www.livingreviews.org/Articles/Volume1/1998-1rovelli.
\bibitem{dc13}K.Kuchar, Phys.Rev. {\bf D34}, 3031 and 3044 (1986).
\bibitem{lll}L.Lusanna, ``Classical Observables of Gauge Theories from the
Multitemporal Approach", in ``Mathematical Aspects of Classical Field Theory",
Seattle 191, Contemporary Mathematics {\bf 132}, 531 (1992).
\bibitem{tova}C.G.Torre and M.Varadarajan, Class.Quantum Grav. {\bf 16}, 2651
(1999); Phys.Rev. {\bf D58}, 064007 (1998).
\bibitem{hanson}A.J.Hanson and T.Regge, Ann.Phys. (N.Y.) {\bf 87}, 498 (1974).
A.J.Hanson, T.Regge and C.Teitelboim, ``Constrained Hamiltonian Systems", in
Contributi del Centro Linceo Interdisciplinare di Scienze Matematiche, Fisiche
e loro Applicazioni, n.22 (Accademia Nazionale dei Lincei, Roma, 1975).
\bibitem{dixon}W.G.Dixon, J.Math.Phys. {\bf 8}, 1591 (1967). ``Extended Objects
in General Relativity: their Description and Motion", in ``Isolated
Gravitating Systems in General Relativity", ed. J.Ehlers
(North-Holland, Amsterdam, 1979).




\end{references}
\end{document}